\documentclass[aps,pra,onecolumn,superscriptaddress,nolongbibliography]{revtex4-2}
\usepackage[colorlinks=true,breaklinks=true,linkcolor=blue,citecolor=red,urlcolor=magenta]{hyperref}
\usepackage{graphicx,xcolor}
\usepackage{bbm}
\usepackage[version=4]{mhchem}
\usepackage{mathtools}
\newcommand*\vett[1]{{\bf{#1}}}
\newcommand{\pose}[1]{\vett{r}_{#1}}
\newcommand{\posn}[1]{\vett{R}_{#1}}
\newcommand{\crt}[1]{\hat{c}_{#1}^\dagger}
\newcommand{\dst}[1]{\hat{c}_{#1}^{\phantom{\dagger}}}
\newcommand{\rot}[2]{R_{\lowercase{#1}}\left(#2\right)}
\newcommand{\rotUC}[2]{R_{#1}\left(#2\right)}
\newcommand{\rotnoangle}[2]{R_{\lowercase{#1}}}
\newcommand{\energy}{E}
\newcommand{\wfn}{\Psi}
\newcommand{\subspacewfn}{\tilde{\Psi}}
\newcommand{\subspaceenergy}{\tilde{E}}
\newcommand{\pertsubspaceenergy}{\tilde{\tilde{E}}}
\newcommand{\calcwfn}{\Phi}
\newcommand{\vac}{\mbox{\O}}
\newcommand{\ket}[1]{|#1\rangle}
\newcommand{\identity}{\mathbbm{1}}
\newcommand{\complex}{\mathbbm{C}}
\newcommand{\order}{\mathcal{O}}
\newcommand{\technical}[1]{#1}
\newcommand{\etal}{{\textit{et al}}}

\begin{document}

\title{Subspace methods for electronic structure simulations on quantum computers}

\author{Mario Motta}
\address{IBM Quantum, IBM Research - Almaden, San Jose, CA 95120, USA}

\author{William Kirby}
\address{IBM Quantum, IBM Research - Cambridge, Cambridge, MA 02142, USA}

\author{Ieva Liepuoniute}
\address{IBM Quantum, IBM Research - Almaden, San Jose, CA 95120, USA}

\author{Kevin J. Sung}
\address{IBM Quantum, T. J. Watson Research Center, Yorktown Heights, NY 10598, USA}

\author{Jeffrey Cohn}
\address{IBM Quantum, IBM Research - Almaden, San Jose, CA 95120, USA}

\author{Antonio Mezzacapo}
\address{IBM Quantum, T. J. Watson Research Center, Yorktown Heights, NY 10598, USA}

\author{Katherine Klymko}
\address{National Energy Research Scientific Computing Center (NERSC), Lawrence Berkeley National Laboratory, Berkeley, CA 94720, USA}

\author{Nam Nguyen}
\address{Integrated Vehicle Systems, Applied Mathematics, Boeing Research \& Technology, Huntington Beach, CA 92647, USA}

\author{Nobuyuki Yoshioka}
\address{Department of Applied Physics, The University of Tokyo, 7-3-1 Hongo, Bunkyo-ku, Tokyo, Japan}

\author{Julia E. Rice}
\address{IBM Quantum, IBM Research - Almaden, San Jose, CA 95120, USA}

\begin{abstract}
Quantum subspace methods (QSMs) are a class of quantum computing algorithms where the time-independent Schr\"{o}dinger equation for a quantum system is projected onto a subspace of the underlying Hilbert space. This projection transforms the Schr\"{o}dinger equation into an eigenvalue problem determined by measurements carried out on a quantum device. The eigenvalue problem is then solved on a classical computer, yielding approximations to ground- and excited-state energies and wavefunctions.

QSMs are examples of hybrid quantum-classical methods, where a quantum device supported by classical computational resources is employed to tackle a problem.
QSMs are rapidly gaining traction as a strategy to simulate electronic wavefunctions on quantum computers, and thus their design, development, and application is a key research field at the interface between quantum computation and electronic structure.

In this review, we provide a self-contained introduction to QSMs, with emphasis on their application to the electronic structure of molecules. We present the theoretical foundations and applications of QSMs, and we discuss their implementation on quantum hardware, illustrating the impact of noise on their performance.

\end{abstract}

\maketitle
%\tableofcontents

The simulation of ground- and excited-state properties of electronic systems is an important application of quantum computing algorithms~\cite{bauer2020quantum,georgescu2014quantum,cao2019quantum,cerezo2020variational}.
In recent years, new approaches based on the projection of the Schr\"{o}dinger equation onto a subspace of the many-electron Hilbert space have been proposed to extract Hamiltonian eigenpairs~\cite{mcclean2017hybrid,colless2018computation,huggins2020non,motta2020determining,ollitrault2020quantum,parrish2019quantum,stair2020multireference}. These methods, which we will collectively refer to as quantum subspace methods (QSMs), are rapidly emerging as a prominent and promising class of quantum computing algorithms for near-term and fault-tolerant quantum devices.

This review provides a survey of the state of the art of QSMs. It is aimed at practitioners of electronic structure (ES) interested in familiarising themselves with quantum computation and QSMs in particular, as well as at practitioners of quantum computation (QC) interested in the simulation of electronic structure.

\begin{figure}[t!]
\centering
\includegraphics[width=0.65\textwidth]{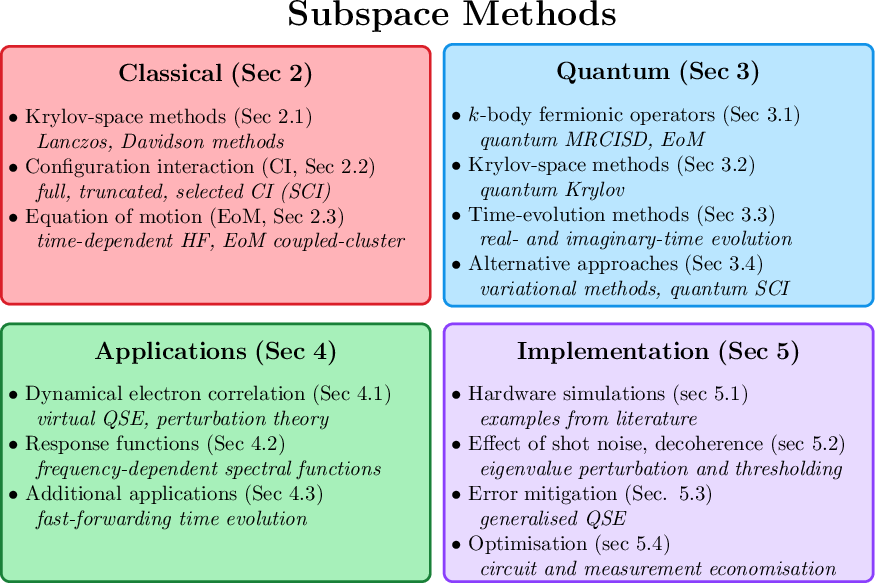}
\caption{Structure of this review. Abbreviations indicate configuration interaction (CI), Hartree-Fock (HF), equation of motion (EOM), multireference CI with singles and doubles (MRCISD), quantum subspace expansion (QSE), selected CI (SCI).}
\label{fig:structure}
\end{figure}

This work begins in Section \ref{sec:basic}, with a review of basic concepts of ES and QC, that readers can choose to read based on their background and expertise.
Afterward, Section \ref{sec:classical}  describes classical subspace methods, grouping them in algorithms based on the notions of Krylov space, configuration interaction (CI), and equation-of-motion (EOM), as shown in Fig.~\ref{fig:structure}. This Section is primarily destined for practitioners of QC interested in surveying subspace methods for ES simulations on classical computers. It also provides the notation and defines the concepts used in the remainder of the review.
In Section \ref{sec:quantum}, QSMs are reviewed. These are divided into algorithms based on $k$-body fermionic operators, the Hamiltonian Krylov space, and quantum subroutines implementing real- and imaginary-time evolution.
Sections \ref{sec:classical} and \ref{sec:quantum} are interconnected, and discuss the computational cost (especially the scaling of simulation on classical/quantum computers) and accuracy (especially the convergence properties) of classical and quantum subspace methods.
Section \ref{sec:applications} presents some applications of QSMs. It illustrates the concerted use of classical and quantum computers, for example to account for static and dynamical electronic correlation and simulate spectral functions.
Section \ref{sec:implementation} focuses on the implementation of QSMs on quantum computing devices. It emphasises the importance and the challenges posed by finite statistical samples (shot noise) and decoherence, the economisation of quantum circuits and measurements to contain the computational cost of QSMs, and opportunities for the mitigation of errors affecting quantum computers based on the formalism of QSMs.
Finally, conclusions and perspectives are given in Section \ref{sec:perspectives}.

\section{Basic concepts}
\label{sec:basic}

In this Section, we provide readers with basic concepts of ES (Subsection \ref{sec:es_basic}) and QC (Subsection \ref{sec:qc_basic}). Readers may
consult these Subsections at their discretion, based on their background and expertise.
Readers interested in ES methods for classical computers may consult textbooks Refs.~\cite{roos2005multiconfigurational,shavitt2009many,levine2009quantum,helgaker2013molecular,jensen2017introduction,martin2020electronic} and reviews Refs.~\cite{friesner2005ab,bartlett2007coupled,helgaker2012recent}. Readers interested in QC applied to ES may consult textbooks Refs.~\cite{nielsen2010quantum,benenti2004principles,manenti2023quantum} and reviews Refs.~\cite{bauer2020quantum,georgescu2014quantum,cao2019quantum,cerezo2020variational,motta2021emerging} respectively.

\subsection{The electronic structure problem}
\label{sec:es_basic}

Our starting point is the molecular Born-Oppenheimer Hamiltonian written in second quantisation
and in a basis of orthonormal one-electron orbitals 
$\{ \varphi_p  \}_{p=1}^M$,
\begin{equation}
\label{eq:es_hamiltonian}
\hat{H} = E_{\mathrm{nuc}} + \sum_{\substack{p r \\ \sigma}} h_{p r} \crt{p \sigma} \dst{r \sigma} + \sum_{\substack{p r q s \\ \sigma\tau}} \frac{(pr|qs)}{2} \crt{p \sigma} \crt{q \tau} \dst{s\tau}  \dst{r \sigma}
\;,
\end{equation}
where indices {$p,r,q,s$} label spatial orbitals in a finite orthonormal basis set, $\sigma,\tau \in \{ \uparrow,\downarrow \}$ are spin indices, and $\crt{p \sigma}/\dst{r \sigma}$ creates/destroys an electron in orbital $p/r$ with spin $\sigma/\tau$. The internuclear electrostatic interaction energy is
\begin{equation}
E_{\mathrm{nuc}} = \sum_{a<b}^{N_{nuc}} \frac{Z_a Z_b}{\| \posn{a} - \posn{b} \|} 
\;,
\end{equation}
where $\posn{a}$ and $Z_a$ are the position and atomic number of nucleus $a$, and the coefficients 
\begin{equation}
\begin{split}
h_{p r} &= \int d \pose{} \, \varphi^*_{p} (\pose{}) \, \left[ - \frac{1}{2} \, \frac{\partial^2}{\partial \vett{r}^2}  - \sum_{a=1}^{N_{nuc}} \frac{Z_a}{ \| \vett{r} - \posn{a} \| } \right] \, \varphi_{r}(\pose{}) \;,
\\
(pr|qs) &= \int d \pose{1} \int d \pose{2} \,  \frac{ \varphi^*_{p} (\pose{1}) \varphi_{r} (\pose{1}) \, \varphi^*_q (\pose{2}) \varphi_{s} (\pose{2}) }{ \| \pose{1} - \pose{2} \| } \;,
\end{split}
\end{equation}
describe the one-electron part of the Hamiltonian and the electron-electron electrostatic interaction respectively. 
Atomic units are used throughout and the numbers of spin-up and spin-down electrons, nuclei, and orbitals are denoted by $N_\uparrow$, $N_\downarrow$, $N_{nuc}$ and $M$ respectively.
The electronic structure problem consists in solving for the ground (i.e. the lowest-energy) and low-lying excited states of the Hamiltonian Eq.~\eqref{eq:es_hamiltonian},
\begin{equation}
\label{eq:es_problem}
\hat{H} | \wfn_\mu \rangle = \energy_\mu | \wfn_\mu \rangle
\;,\; \mu = 0 \dots D-1 \;,
\end{equation}
where $D$ is the dimension of the Hilbert space on which $\hat{H}$ acts (for a system of $N_\uparrow$ spin-up and $N_\downarrow$ spin-down electrons in $M$ orbitals, $D = \binom{M}{N_\uparrow} \binom{M}{N_\downarrow}$).
In this work, we focus on subspace methods, a family of linear variational methods to approximately solve Eq.~\eqref{eq:es_problem} for a quantum-chemical many-electron system.

\subsubsection{Hartree-Fock method}

The exact solution of the Schr\"{o}dinger equation ~\eqref{eq:es_problem} for a molecule with more than two electrons is a formidable problem. Within the Born-Oppenheiner approximation, the difficulty originates from the two-electron terms of ~\eqref{eq:es_hamiltonian}, which introduce correlations in the motion of electrons under the potential generated by nuclei. For this reason, the Schr\"{o}dinger equation is solved approximately. A simple approximation is the Hartree-Fock (HF) method, in which the wave function is a Slater determinant
\begin{equation}
| \calcwfn_C \rangle = \prod_{i=1}^{N_\downarrow} \crt{c_i \downarrow} \prod_{i=1}^{N_\uparrow} \crt{c_i \uparrow} | \vac \rangle
\;,
\end{equation}
where $| \vac \rangle$ is the vacuum state, and orbitals $|c_i \rangle = \sum_p C_{pi} | \varphi_p \rangle$ are determined by minimising the energy
\begin{equation}
E_C = 
\frac{\langle \calcwfn_C | \hat{H} | \calcwfn_C \rangle}
{\langle \calcwfn_C | \calcwfn_C \rangle}
\;,
\end{equation}
yielding the Hartree-Fock energy, $E_{\mathrm{HF}} = \min_C E_C$, and wavefunction $|\calcwfn_{\mathrm{HF}}\rangle$. This procedure defines a set of orthonormal molecular orbitals, $|c_p \rangle$, divided into occupied ($p\leq \max(N_\uparrow,N_\downarrow)$, denoted by indices $i,j,k,l$)
and virtual
($p>\max(N_\uparrow,N_\downarrow)$, denoted by indices $a,b,c,d$), and a set of Slater determinants of the form
\begin{equation}
\label{eq:configuration}
| {\bf{x}} \rangle 
= 
\prod_{p\sigma} 
\left( \crt{c_p \sigma} \right)^{x_{p\sigma}} 
| \vac \rangle
\;,\;
x_{p\sigma} \in \{0,1\}^{2 M}
\;,\;
\sum_p x_{p \sigma} = N_\sigma
\;.
\end{equation}
We will call these Slater determinants \technical{configurations},
a term that should not be confused with configuration state functions (i.e. eigenfunctions of the total spin operator, which are generally linear combinations of Slater determinants).
Configurations span the Hilbert space of $(N_\uparrow, N_\downarrow)$ electrons in $M$ spatial orbitals, and approximate electronic ground and excited states. 

\subsubsection{Electronic correlation}
\label{sec:correlation}

The eigenstates of Eq.~\eqref{eq:es_problem} cannot be expressed as single determinants.
Recognising this fact, L\"{o}wdin~\cite{lowdin1958correlation} introduced the concept of electron correlation energy, $E_{\mathrm{corr}}$, as the difference between the exact non-relativistic ground-state energy of the molecule and $E_{\mathrm{HF}}$.
Conventionally, correlation energy is divided into static and dynamical, as initially proposed by Sinanoglu~\cite{sinanoglu1964many}.
\technical{Dynamical correlation} arises when the ground-state wavefunction,
$| \wfn_0 \rangle = \sum_{{\bf{x}}} c_{{\bf{x}}} |{\bf{x}} \rangle$, can be qualitatively approximated by the HF configuration, $| c_{{\bf{x}}_\mathrm{HF}}| \gg | c_{{\bf{x}}} |$ for all configurations ${\bf{x}} \neq {\bf{x}}_\mathrm{HF}$, in presence of smaller corrections from other configurations.
Wavefunctions with this property are called single-reference, signalling that no individual configuration mixes significantly with the HF configuration when representing the ground-state wavefunction as a linear combination of the form $| \wfn_0 \rangle = \sum_{{\bf{x}}} c_{{\bf{x}}} |{\bf{x}} \rangle$.

\begin{figure}[t!]
\centering
\includegraphics[width=0.35\textwidth]{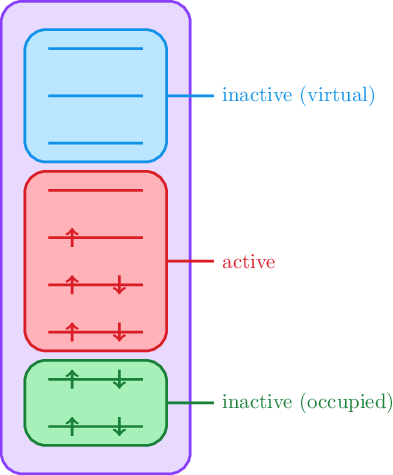}
\caption{Schematic representation of an active space of 5 electrons in 4 orbitals. Spin-up/down electrons are represented by up/down-pointing arrows. Active, inactive occupied, and inactive virtual orbitals are shown in green, red, and blue respectively.}
\label{fig:active}
\end{figure}

Static correlation, on the other hand, occurs when the ground state wavefunction cannot be qualitatively approximated by a single Slater determinant. Wavefunctions with this property are called multi-reference, signalling that the Hartree-Fock configuration interacts significantly with other low-energy configurations.
Accounting for this near-degeneracy effect requires diagonalising an appropriate secular matrix $\langle {\bf{x}} | \hat{H} |{\bf{y}} \rangle$, where the configurations ${\bf{x}}$ and ${\bf{y}}$ arise from all possible occupations of a set of active orbitals by a set of active electrons, i.e. an active space (see Fig.~\ref{fig:active}). For molecules containing first- and second-row atoms, it is desirable to include all valence orbitals (including bonding, nonbonding, and antibonding orbitals) in the active space, and to optimize active-space orbitals to 
self-consistency~\cite{ruedenberg1976mcscf,mok1996dynamical}.
Static correlation is essential for the accurate representation of e.g. a molecule's dissociation into its constituent atoms. Dynamical correlation is essential for the accurate determination of chemical properties at any molecular geometry. This effect is important at both long range, where it describes dispersion, and at short range. In the latter case, it is associated with the behaviour of the wavefunction as two electrons approach each other~\cite{kato1957eigenfunctions}. 
Properly accounting for dynamical correlation requires wavefunctions that explicitly incorporate electronic distances or the transfer of electronic correlation from wavefunctions to operators through canonical transformations~\cite{kong2012explicitly}.

It should be noted that today, dynamical correlation energy usually refers to the difference between the exact energy and a reference energy, often known as the energy of a zeroth order reference wavefunction (within a given one-particle basis set). For situations dominated by a single configuration, the reference energy is thus the Hartree-Fock energy. However, for situations with static correlation, the zeroth order reference energy corresponds to that of the multi-reference wavefunction required to describe the low-lying electronic states.

\subsection{Basic quantum computing concepts} 
\label{sec:qc_basic}

The building blocks of quantum computers are qubits: a qubit is a physical system whose states are described by unit vectors in a two-dimensional Hilbert space $\mathcal{H} \simeq \complex^2$. A system of $N_q$ qubits, also called a register, has states described by unit vectors in the Hilbert space $\mathcal{H}_{N_q} = \mathcal{H}^{\otimes N_q}$. An orthonormal basis of the Hilbert space $\mathcal{H}_{N_q}$ is given by the following vectors, called {computational basis states},
\begin{equation}
\label{eq:qcomputational_basis}
\ket{ \vett{z} } 
= 
\bigotimes_{\ell=0}^{N_q-1} \ket{ z_\ell }
=
\ket{ z_{n-1} \dots z_0 } 
= 
\ket{ z } 
\;,\;
\vett{z} \in \{0,1\}^{N_q}
\;,\;
z = \sum_{\ell=0}^{N_q-1} z_\ell \, 2^\ell \in \{0 \dots 2^{N_q-1} \}
\;.
\end{equation}
Starting from a register of $N_q$ qubits initialised in the state $\ket{\vett{0}} \in \mathcal{H}_{N_q}$, an $N_q$-qubit state $\ket{\calcwfn}$ can be prepared applying single- and multi-qubit unitary transformations, or gates.

\subsubsection{Single-qubit gates}

Single-qubit Pauli operators
\begin{equation}
\identity 
=
\left( 
\begin{array}{cc}
1 & 0 \\
0 & 1 \\
\end{array} \right)
\;,\;
X
=
\left( 
\begin{array}{cc}
0 & 1 \\
1 & 0 \\
\end{array} \right)
\;,\;
Y
=
\left( 
\begin{array}{rr}
0 & -i \\
i &  0 \\
\end{array} \right)
\;,\;
Z
=
\left( 
\begin{array}{rr}
1 &  0 \\
0 & -1 \\
\end{array} \right)
\;,
\end{equation}
are very important in quantum computation. A single qubit can be prepared in a generic state $\ket{\calcwfn}$ by initialisation in $\ket{\vett{0}}$ and application of single-qubit Pauli rotations, i.e. unitary transformations of the form 
\begin{equation}
\rot{\sigma}{\theta} = e^{-i \frac{\theta}{2} \sigma}
\;,\;
\sigma \in \{X,Y,Z\}
\;,\;
\theta \in [0,2\pi)
\;.
\end{equation}
In matrix form, with $\alpha = \cos(\theta/2)$, $\beta = \sin(\theta/2)$, and $\gamma = e^{i \theta/2}$,
\begin{equation}
\rot{X}{\theta} = 
\left( 
\begin{array}{rr}
\alpha &  -i \beta \\
-i\beta & \alpha \\
\end{array} \right)
\;,\;
\rot{Y}{\theta} = 
\left( 
\begin{array}{rr}
\alpha &  - \beta \\
\beta & \alpha \\
\end{array} \right)
\;,\;
\rot{Z}{\theta} = 
\left( 
\begin{array}{rr}
\gamma^* & 0 \\
0 & \gamma \\
\end{array} \right)
\;.
\end{equation}
It is also useful to recall that $\rot{\sigma}{\theta} = \cos(\theta/2) \identity - i \sin(\theta/2) \sigma$.
A generic single-qubit gate with unit determinant, $U$, can be written as
\begin{equation}
U = \rot{Z}{\theta_2} \rot{Y}{\theta_1} \rot{Z}{\theta_0}
=
\left( 
\begin{array}{rr}
e^{-i \frac{\theta_0+\theta_2}{2}}
\cos(\theta_1/2) &
-e^{i \frac{\theta_0-\theta_2}{2}}
\sin(\theta_1/2) \\
e^{-i \frac{\theta_0-\theta_2}{2}}
\sin(\theta_1/2) & e^{i \frac{\theta_0+\theta_2}{2}}
\cos(\theta_1/2) \\
\end{array} \right)
\end{equation}
for suitable angles $\theta_0,\theta_1,\theta_2$, in what is called a ZYZ decomposition~\cite{barenco1995elementary} (one can think of these angles as equivalent to Euler angles up to the double-cover of $SO(3)$ by $SU(2)$, see e.g.~\cite{woit2017quantum}).

For some superconducting qubit architectures, the available 1-qubit gates are the ``square-root of X'' and phase gates (the latter are applied in a virtual way by changing the phase of subsequent electromagnetic pulses~\cite{krantz2019quantum}, which reduces both the number of gates in the circuit and the errors, since virtual phase gates are implemented on classical software and not as physical operations), respectively
\begin{equation}
\sqrt{X} 
=
e^{i\pi/4} \rot{X}{\pi/2}
=
\left( 
\begin{array}{rr}
\frac{1+i}{2} & \frac{1-i}{2} \\
\frac{1-i}{2} & \frac{1+i}{2} \\
\end{array} \right)
\;,\;
\mathsf{P}(\lambda) 
=
e^{-i\lambda/2}
\rot{Z}{\lambda}
=
\left( 
\begin{array}{rr}
1 & 0 \\
0 & e^{i\lambda} \\
\end{array} \right)
\;.
\end{equation}
Combining $\sqrt{X}$ and phase gates one can construct $X$ and $Y$ rotations, e.g. $\rot{Y}{\theta} = \sqrt{X} \rot{Z}{\theta} \sqrt{X}^\dagger$.
Similarly, a generic single-qubit gate with unit determinant can be written in the following VZ decomposition, as
\begin{equation}
\begin{split}
\mathsf{U}_3(\theta,\phi,\lambda) &= 
\left( 
\begin{array}{rr}
\cos(\theta/2) &  -e^{i\lambda} \sin(\theta/2) \\
e^{i\phi}  \cos(\theta/2) & e^{i (\phi+\lambda)} \cos(\theta/2) \\
\end{array} \right) \\
&=
e^{-i \frac{\pi+\theta}{2}}
\mathsf{P}(\phi+\pi)
\sqrt{X}
\mathsf{P}(\theta+\pi)
\sqrt{X}
\mathsf{P}(\lambda)
\;.
\end{split}
\end{equation}
Important single-qubit gates are the Hadamard, S, and T gates, respectively
\begin{equation}
\mathrm{Had} = \frac{1}{\sqrt{2}}
\left( 
\begin{array}{rr}
1 & 1 \\
1 & -1 \\
\end{array} \right)
\;,\;
S = \mathsf{P}(\pi/2)
\;,\;
T = \mathsf{P}(\pi/4)
\;.
\end{equation}

\subsubsection{Two-qubit gates} 
\begin{figure}[t!]
\centering
\includegraphics[width=0.5\textwidth]{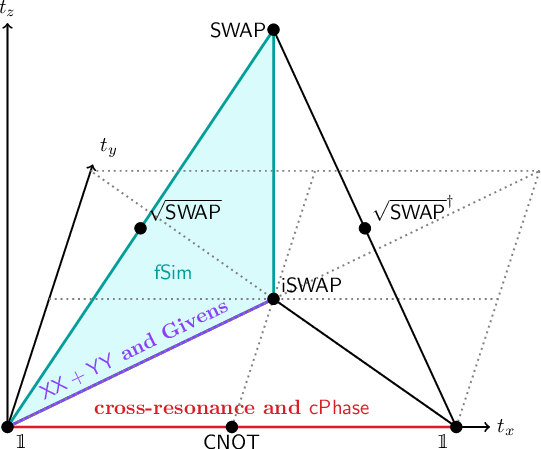}
\caption{Location of some important 2-qubit gates in the Weyl chamber. Gates are defined in the main text, black points indicate parameter-free gates, and colored lines/surfaces indicate gates with one/two parameters.}
\label{fig:weyl}
\end{figure}

Single-qubit gates are not sufficient for universal quantum computation, since they cannot entangle qubits.
This goal is accomplished by combining single- and two-qubit gates. An important example is the canonical gate
\begin{equation}
\mathsf{Can}(t_x,t_y,t_z) =e^{- i \frac{\pi}{2} (t_x X \otimes X + t_y Y \otimes Y + t_z Z \otimes Z)} = \rot{XX}{\pi t_x} \rot{YY}{\pi t_y} \rot{ZZ}{\pi t_z}
\;,
\end{equation}
where the parameters $t_x,t_y,t_z$ lie in the so-called~\cite{zhang2003geometric} Weyl chamber $1/2 \geq t_x \geq t_y \geq t_z \geq 0$ $\cup$ $1/2 \geq (1-t_x) \geq t_y \geq t_z \geq 0$, shown in Fig.~\ref{fig:weyl}. Any 2-qubit gate can be decomposed into a canonical gate and single-qubit gates by the so-called KAK decomposition~\cite{zhang2003geometric,zhang2004optimal,blaauboer2008analytical,watts2013metric,crooks2020gates}. The following well-known two-qubit gates, for example, are given (up to a global phase) by
\begin{equation}
\begin{split}
\mathsf{cNOT} &= 
\left( \rot{Z}{- \frac{\pi}{2}} \rot{Y}{- \frac{\pi}{2}}  \otimes \rot{X}{- \frac{\pi}{2}} \right) \mathsf{Can}(1/2,0,0) \left( \rot{Y}{\frac{\pi}{2}}  \otimes \identity \right) \;, \\
\mathsf{SWAP} &= \mathsf{Can}(1/2,1/2,1/2) \;, \\
\mathsf{iSWAP} &= (Z \otimes Z)\mathsf{Can}(1/2,1/2,0) \;, \\
\sqrt{\mathsf{SWAP}} &= \mathsf{Can}(1/4,1/4,1/4) \;, \\
\end{split}
\end{equation}
and correspond to precise points in the Weyl chamber, shown in Fig.~\ref{fig:weyl}.
The cross-resonance and $\mathsf{fSim}$ gates, native to superconducting devices with fixed and tunable frequencies respectively~\cite{paraoanu2006microwave,rigetti2010fully,yan2018tunable,foxen2020demonstrating}, are given by 
\begin{equation}
\begin{split}
\rot{ZX}{\theta} &= 
(\identity \otimes \mathrm{Had}) \mathsf{Can}(\theta/\pi,0,0) (\identity \otimes \mathrm{Had}) \;, \\
\mathsf{fSim}(\theta,\phi) &= 
\left( \rot{Z}{-\frac{\phi}{2}} \otimes \rot{Z}{-\frac{\phi}{2}} \right) 
\mathsf{Can}(\theta/\pi,\theta/\pi,\phi/(2\pi)) \;. \\
\end{split}
\end{equation}
The following $\mathsf{XX+YY}$, Givens, and controlled-phase unitaries, very important in the simulation of fermionic systems (see Subsection \ref{sec:fermions_second}), are given by
\begin{equation}
\begin{split}
\mathsf{V}_{\mathsf{XX+YY}}(\theta,\beta) &= ( \identity \otimes \rot{Z}{\beta} ) 
\mathsf{Can}(\theta/\pi,\theta/\pi,0) ( \identity \otimes \rot{Z}{-\beta} )  \;, \\
\mathsf{G}(\theta) &= (T \otimes T^\dagger) \mathsf{Can}(\theta/\pi,\theta/\pi,0) (T^\dagger \otimes T) \;, \\
\mathsf{cP}(\theta) &= \left( \rot{Z}{-\frac{\theta}{2}} \otimes \rot{Z}{-\frac{\theta}{2}} \mathrm{Had} \right) \mathsf{Can}(\theta/(2\pi),0,0) (\identity \otimes \mathrm{Had}) \;. \\
\end{split}
\end{equation}

\subsubsection{Multi-qubit Pauli operators} These operators are defined as
\begin{equation}
\hat{\sigma}_{{\bf{m}}} = \hat{\sigma}_{m_{N_q-1}} \otimes \dots \otimes \hat{\sigma}_{m_0}
\;,\;
\hat{\sigma}_m \in \{ \identity , X, Y, Z \}
\;.
\end{equation}
Pauli operators are conceptually useful and important since they form a basis for the space of linear operators on $\mathcal{H}_{N_q}$. This fact allows us to represent, either exactly or approximately, multi-qubit unitaries as products of exponentials of Pauli operators.
The exponential of a Pauli operator $P = \hat{\sigma}_{\bf{m}}$ such that $\hat{\sigma}_m \neq \identity$ for all $m$ can be applied to a register of qubits~\cite{seeley2012bravyi} by first mapping it onto the exponential of a Pauli-$Z$ operator,
\begin{equation}
\rotUC{ P }{\theta}
= 
e^{ -\frac{i \theta}{2} P } 
= 
\hat{V}^\dagger 
e^{ - \frac{i \theta}{2} Z \otimes \dots \otimes Z } \hat{V}
\;,\;
\hat{V} 
= 
\bigotimes_{\ell=0}^{N_q-1} \hat{A}_{m_\ell}
\;,\;
\hat{A}_{m}^\dagger 
\hat{\sigma}_m 
\hat{A}_{m}
= 
Z
\;,
\end{equation}
and then by introducing a ladder of $\mathsf{cNOT}$ gates that maps $Z \otimes \dots \otimes Z$ onto $Z_{N_q-1}$,
\begin{equation}
\rotUC{ P }{\theta}
= 
\hat{V}^\dagger \hat{W}^\dagger \left( \rot{Z}{\theta} \right)_{N_q-1} \hat{W} \hat{V}
\;,\;
\hat{W}
=
\prod_{\ell=0}^{N_q-2} \mathsf{cNOT}_{\ell,\ell+1}
\;.
\end{equation}

\subsubsection{Quantum measurements}
\label{sec:meas}

Quantum computers allow for measurement of one or more qubits in the computational basis, Eq.~\eqref{eq:qcomputational_basis}. Measuring a register of $N_q$ qubits prepared in a state $|\calcwfn\rangle$ a number $N_s$ of times yields a collection of samples, or ``shots'', $\{ \vett{z}_i \}_{i=0}^{N_s-1}$, which are binary strings of length $N_q$, $\vett{z}_i \in \{0,1\}^{N_q}$, statistically independent and distributed according to $p(\vett{z}) = | \langle \vett{z} | \calcwfn \rangle |^2$. The expectation value of a diagonal operator, i.e. any operator of the form $\hat{B} = \sum_\vett{z} f(\vett{z}) | \vett{z} \rangle \langle \vett{z}|$ can be estimated as $\langle \psi | \hat{B} | \psi \rangle = \mu \pm \sigma/\sqrt{N_s}$ with
\begin{equation}
\label{eq:sample}
\mu = \frac{1}{N_s} 
\sum_{i=0}^{N_s-1} f( \vett{z}_i )
\;,\;
\sigma^2
= 
\frac{1}{N_s-1}
\sum_{i=0}^{N_s-1} \left( f( \vett{z}_i ) - \mu \right)^2
\;.
\end{equation}
For example, $\hat{B} = Z \otimes \dots \otimes Z$ has $f(\vett{z}) = \prod_{\ell=0}^{N_q-1} (-1)^{z_\ell}$. 
Non-diagonal operators can be measured by prepending a unitary to a computational basis measurement. For example, a generic Pauli operator can be expressed as
\begin{equation}
\hat{\sigma}_{{\bf{m}}} 
=
\hat{V}^\dagger {(Z \otimes \dots \otimes Z)} \hat{V}
=
\hat{V}^\dagger \left[ \sum_{ \vett{z} } f(\vett{z}) \,
 | \vett{z} \rangle \langle \vett{z} | \right] \hat{V}
\;,\;
f(\vett{z}) = \prod_{\ell=0}^{N_q-1} (-1)^{z_\ell}
\;,
\end{equation}
and its expectation value can be computed by applying Eq.~\eqref{eq:sample} to measurement outcomes drawn from the probability distribution $p(\vett{z}) = | \langle \vett{z} | \hat{V} | \calcwfn \rangle|^2$.
Since a generic Hermitian operator can be written as a linear combination of Pauli operators, its expectation value can be computed by measuring Pauli operators only.

Statistical uncertainties in measurement is a crucial aspect of quantum computation. 
Quantum algorithms should be understood and formulated in terms of random variables, and their outcomes should be accompanied by carefully estimated statistical uncertainties. These aspects cannot be overlooked in the implementation and design of quantum algorithms, including QSMs.

\subsubsection{Noisy quantum devices} 
\label{sec:noisy}

Noisy quantum devices are subject to decoherence (i.e., unwanted interaction with the environment) and imperfect implementation of quantum operations (i.e., gates and measurements).
Qubits undergo relaxation and dephasing on timescales known as $T_1$ and $T_2$ respectively which, for superconducting qubits, are on the scale of $10^{2}$ to $10^{3} \, \mu s$.
Single-qubit gates are fast and precise operations, with duration $T_{1q} \simeq 20 \, ns$ and error $\varepsilon_{1q} \simeq 0.1 \%$, whereas two-qubit operations and measurements have duration $T_{2q} \simeq 200 \, ns$, $T_{m} \simeq 700 \, ns$ and error $\varepsilon_{2q} \simeq 1 \%$, $\varepsilon_{m} \simeq 1 \%$ respectively~\cite{bronn2017fast,jurcevic2021demonstration}.
The qubit coherence times define the timescales over which they lose quantum information to decoherence, and the gate times and errors define the duration and accuracy of a computation: in the absence of error correction, the duration of a computation should not exceed the qubit coherence time, and the number of gates should be such that the accumulation of error does not prevent algorithms from yielding accurate results. 

It should also be noted that many quantum computer architectures have limited qubit connectivity, i.e. two-qubit gates can only be applied across certain pairs of qubits. This fact conflicts with the quantum circuit model which allows general 2-qubit interactions and thus implicitly assumes a completely connected network of qubits. To remedy this, Beals \etal~\cite{beals2013efficient} demonstrated the use of $\mathsf{SWAP}$ gates, which can be inserted into the quantum circuit to enable it to be executed on a quantum computer with limited qubit connectivity (see Fig.~\ref{fig:lowrank2}b for an example). An important optimization problem is how to insert these gates to run a quantum circuit while minimising the adverse effect of these additional gates on the performance of the algorithm (runtime and error)~\cite{fowler2004implementation,maslov2007linear,kutin2006shor,brierley2015efficient,herbert2018depth}.

Two parameters, respectively called {width and depth}, are often used to characterise the cost of a quantum circuit. Width refers to the number of qubits that comprise the circuit, $N_q$. 
Depth is the maximal length $d$ of a path from the input (qubit initialization) to the output (measurement operation) of a circuit.
Width and depth are both limiting factors in the execution of quantum algorithms (large width corresponds to many qubits, and large depth to many operations carried out sequentially), and depth is an important computational bottleneck when the physical duration of a circuit, roughly approximated by $d T_{2q}$, is comparable with the coherence time of a qubit.

Techniques for the mitigation of readout~\cite{temme2017error,nation2021scalable} and gate~\cite{viola1998dynamical,biercuk2009optimized,kandala2018extending} errors exist, which alleviate these bottlenecks and, along with continued improvement in device manufacturing and operation and research in algorithm design and refinement, are fundamental activities towards practical quantum computation.

\subsubsection{Qubit mappings for fermions}
\label{sec:fermions_second}

The Fock space of electrons occupying $M$ spatial orbitals has the same dimension, $4^M$, as the Hilbert space of $N_q=2M$ qubits. Therefore, it is possible to construct a one-to-one mapping between the two spaces. There are combinatorially many ways to map a quantum system to a set of qubits~\cite{wu2002qubits,batista2004algebraic} and, since fermions exhibit non-locality of their state space, due to their antisymmetric exchange statistics, any representation 
of fermionic systems on collections of qubits must introduce non-local structures~\cite{bravyi2002fermionic}. Here, we consider the Jordan-Wigner (JW) transformation 
\cite{jordan1993paulische,abrams1997simulation,ortiz2001quantum,somma2002simulating}, 
that maps electronic configurations with generic particle number, Eq.~\eqref{eq:configuration}, onto computational basis states, Eq.~\eqref{eq:qcomputational_basis},
\begin{equation}
\label{eq:jwstate}
\big( \crt{M-1,\downarrow} \big)^{x_{M-1,\downarrow}}
\dots
\big( \crt{0,\uparrow} \big)^{x_{0,\uparrow}}
| \vac \rangle
\mapsto 
| \vett{x} \rangle 
\;,
\end{equation} 
and fermionic creation and annihilation operators ($\crt{p}$ and $\dst{p}$ respectively)
onto non-local qubit operators of the form
\begin{equation}
\label{eq:jwgivescar}
\begin{split}
\crt{p\sigma} \mapsto 
\frac{X_{p\sigma} - i Y_{p\sigma}}{2}
\otimes Z_{p\sigma-1} \otimes \dots \otimes Z_0
\equiv
S^{(+)}_{p\sigma} \, Z^{p\sigma-1}_0
\;, \\
\dst{p\sigma} \mapsto 
\frac{X_{p\sigma} + i Y_{p\sigma}}{2}
\otimes Z_{p\sigma-1} \otimes \dots \otimes Z_0
\equiv
S^{(-)}_{p\sigma} \, Z^{p\sigma-1}_0
\;.
\end{split}
\end{equation} 
In Eqs.~\eqref{eq:jwstate} and \eqref{eq:jwgivescar} we used $Z^p_r = \otimes_{l=r}^p Z_l$ to denote a tensor product of Pauli $Z$ operators acting on qubits $r$ to $p$, and we used the following notation for qubit indices,
\begin{equation}
p\sigma = 
\left\{
\begin{array}{ll}
p & \sigma \, = \, \uparrow \\
p+M & \sigma \, = \, \downarrow \\
\end{array}
\right.
\end{equation} 
The non-locality of these operators is required to preserve canonical anticommutation relations between creation and destruction operators. The main limitation of the JW transformation is that the qubit representation of $\crt{p\sigma}$ has $\order(M)$ non-locality~\cite{aspuru2005simulated,whitfield2011simulation}, which immediately translates to $k$-body fermionic operators. Another limitation is that, since the JW transformation operates in the Fock space (i.e. it allows the description of states with any particle number, spin, and point group symmetry), on noisy quantum devices the conservation of particle number, spin, and point group symmetry (an important requirement in typical ES calculations) is not guaranteed.

The qubit representation of Eq.~\eqref{eq:es_hamiltonian} can be derived by using Eq.~\eqref{eq:jwgivescar} to represent each product of creation and destruction operators as a linear combination of Pauli operators. For the one-body part $\hat{H}_1 = \sum_{pr,\sigma} h_{pr} \crt{p\sigma} \dst{r\sigma}$, assuming real-valued coefficients, one has
\begin{equation}
\hat{H}_1 
\to 
\sum_{p\sigma} \frac{h_{pp}}{2} (1-Z_{p\sigma})
+ 
\sum_{p<r,\sigma} \frac{h_{pr}}{2} Z^{p\sigma-1}_{r\sigma+1} 
(X_{p\sigma} X_{r\sigma} + Y_{p\sigma} Y_{r\sigma}) 
= 
\sum_\alpha c_\alpha \sigma_{\vett{m}_\alpha} \;,
\end{equation} 
where we used the notation $(p\uparrow) = p$, $(p\downarrow) = p+M$ for qubit indices. This expression, combined with Eq.~\eqref{eq:sample}, immediately shows that $\hat{H}_1$ can be measured Pauli-by-Pauli. A step of time evolution under $\hat{H}_1$ can similarly be approximated with e.g. a Trotter product formula,
\begin{equation}
e^{-i \Delta t \hat{H}_1 } \to 
\prod_\alpha e^{-i \Delta t P_\alpha } + \order(\Delta t^2) \;.
\end{equation}
For the two-body part, one has a similar but more complicated expression~\cite{whitfield2011simulation}. 

\begin{figure}[t!]
\centering
\includegraphics[width=0.65\textwidth]{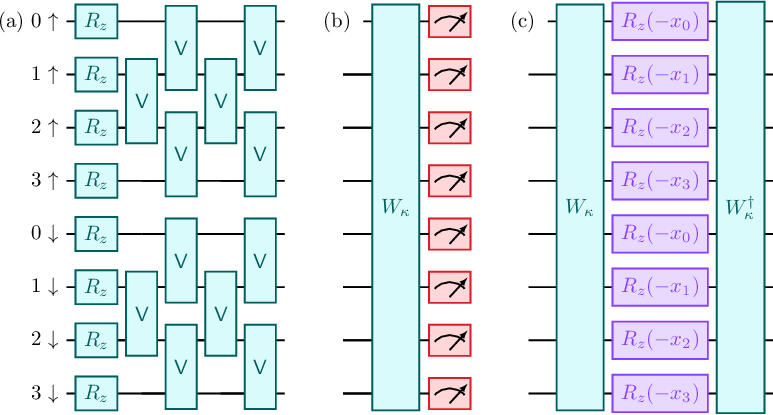}
\caption{Left (a): Implementation of a change-of-basis unitary $e^{\hat{K}}$ as a circuit  $W_\kappa$ comprising $2M$ $\rotnoangle{Z}{\theta}$ gates and $M(M-1)$ $\mathsf{V_{XX+YY}}$ gates arranged in $M$ layers (marked as $\rotnoangle{Z}{\theta}/\mathsf{V}$ and omitting parameters to avoid clutter), illustrated for a system of $M=4$ spatial orbitals. Center (b): measurement of $\hat{H}_1$ using the circuit $W_\kappa$ (teal block) and a computational basis measurement (red meter symbols). Right (c): time evolution under $\hat{H}_1$ using the circuit $W_\kappa$ and a single layer of single-qubit $Z$ rotations (purple blocks) with $\vett{x}$ defined in Eq.~\eqref{eq:1b_evo}.
}
\label{fig:lowrank}
\end{figure}

Alternatively, $\hat{H}_1$ can be diagonalised by the exponential of a one-body operator,
\begin{equation}
\hat{H}_1 = e^{-\hat{K}}
\left( \sum_{p\sigma} \eta_{p} \, \crt{p\sigma} \dst{p\sigma} \right) 
e^{\hat{K}}
\;,\;
\hat{K} = \sum_{p<r \sigma} \kappa_{pr} (\crt{p\sigma} \dst{r\sigma} - \crt{r\sigma} \dst{p\sigma} )
\;,
\end{equation}
where $\eta_p$ are the eigenvalues of the matrix with elements $h_{pr}$. The qubit representation of $\hat{H}_1$ is
\begin{equation}
\hat{H}_1 \to
W_\kappa^\dagger 
\left( \sum_{p\sigma} \frac{\eta_p}{2} (1-Z_{p\sigma}) \right) 
W_\kappa 
\;,
\end{equation}
where $W_\kappa$ is a unitary that can be implemented with a quantum circuit (see Fig.~\ref{fig:lowrank}a) containing $\order(M^2)$ $\mathsf{XX+YY}$ gates, $\order(M)$ depth, and requiring linear qubit connectivity only~\cite{reck1994experimental,clements2016optimal,jiang2018quantum,motta2021low,motta2023bridging}. As a consequence, $\hat{H}_1$ can be measured as shown in Fig.~\ref{fig:lowrank}b and time evolution can then be exactly implemented up to a global phase (see Fig.~\ref{fig:lowrank}c) as
\begin{equation}
\label{eq:1b_evo}
e^{-i t \hat{H}_1 } \to 
W_\kappa^\dagger 
\,
U_1(t \boldsymbol{\eta})
\,
W_\kappa 
\;,\;
U_1(\vett{x}) 
= 
\bigotimes_{p\sigma} \big( \rot{Z}{-x_p} \big)_{p\sigma}
\;.
\end{equation}

\begin{figure}[t!]
\centering
\includegraphics[width=0.8\textwidth]{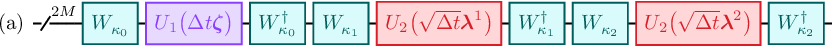}
\includegraphics[width=0.8\textwidth]{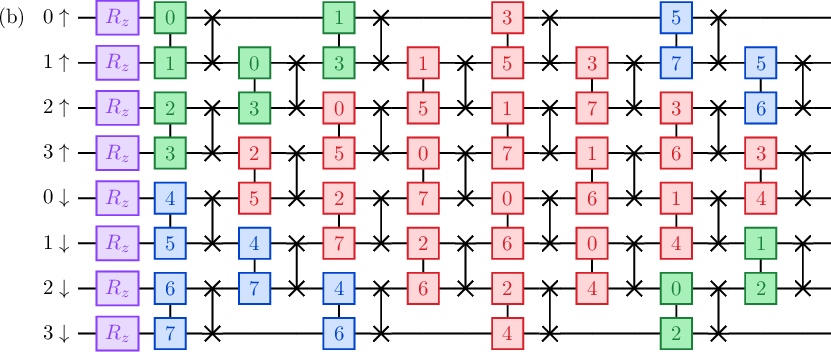}
\caption{Top (a): quantum circuit implementing a step of time evolution under the ES Hamiltonian, 
with a Trotter product formula based on a low-rank decomposition of the two-body part, Eq.~\eqref{eq:lowrank_trotter_2}. The circuits $W_{\kappa}$ and $U_1\big(\vett{x}\big)$ are defined as in Fig.~\ref{fig:lowrank}, and the circuit $U_2\big(\vett{x}\big)$ is shown in the bottom panel (b) for $M=4$ spatial orbitals, with purple blocks labeling single-qubit $Z$ rotations, and 
green/blue/red blocks connected by vertical black lines labeling controlled-phase rotations implementing $\uparrow,\uparrow$/$\downarrow,\downarrow$/$\uparrow,\downarrow$ terms in Eq.~\eqref{eq:lowrank_trotter_2}. Note the use of a $\mathsf{SWAP}$ network to implement two-qubit gates acting on distant qubits assuming linear device connectivity only, and the fact that the $\mathsf{SWAP}$ network inverts the qubit order.
}
\label{fig:lowrank2}
\end{figure}

Similarly, the two-body part of the Hamiltonian can be mapped on a qubit operator using a low-rank approximation~\cite{motta2021low}. One can write the electron repulsion integral (ERI) with a density fitting~\cite{
Whitten:1973:4496,
Dunlap:1977:81,
Dunlap:1979:3396,
Feyereisen:1993:359,
Komornicki:1993:1398,
Vahtras:1993:514,
Rendell:1994:400,
Kendall:1997:158,
Weigend:2002:4285} or Cholesky decomposition~\cite{
Beebe:1977:683,
Roeggen:1986:154,
Koch:2003:9481,
Aquilante:2007:194106,
Aquilante:2009:154107,
motta2019efficient,
peng2017highly} as
\begin{equation}
\frac{(pr|qs)}{2} \simeq \sum_{\gamma=1}^{N_\gamma} L^\gamma_{pr} L^\gamma_{qs} 
\;,\;
N_\gamma = \order(M)
\;,
\end{equation}
and obtain
\begin{equation}
\label{eq:elecstruct_qubit_lr_1}
\hat{H} = E_\mathrm{nuc} + \hat{J}_1 + \sum_{\gamma=1}^{N_\gamma} \hat{L}_\gamma^2 
\;,\;
\hat{L}_\gamma = \sum_{pr,\sigma} L^\gamma_{pr} \crt{p\sigma} \dst{r\sigma}
\;,
\end{equation}
where $\hat{J}_1 = \sum_{pr,\sigma} \big( h_{pr} - \sum_{q\gamma} L^\gamma_{pq} L^\gamma_{qr} \big) \crt{p\sigma} \dst{r\sigma}$ and
each $\hat{L}_\gamma$ is real-valued and symmetric due to the 8-fold symmetry of the ERI. The qubit representation of Eq.~\eqref{eq:elecstruct_qubit_lr_1} is 
\begin{equation}
\label{eq:elecstruct_qubit_lr_2}
\hat{H} 
\to 
E_\mathrm{nuc} 
+ 
W^\dagger_{\kappa_0} 
\Big( \sum_{p\sigma} \zeta_p \, n_{p\sigma} \Big) 
W_{\kappa_0} 
+ 
\sum_{\gamma=1}^{N_\gamma} W_{\kappa_\gamma}^\dagger 
\Big( \sum_{\substack{pq \\ \sigma\tau}} 
\lambda_p^\gamma \lambda^\gamma_q \, n_{p\sigma} n_{q\tau} \Big) 
W_{\kappa_\gamma}
\;,\;
\end{equation}
where $n_{p\sigma} = (1-Z_{p\sigma})/2$, $\zeta_p$ ($\lambda_p^\gamma$) are the eigenvalues of $\hat{J}_1$ ($\hat{L}_\gamma$) and $W_{\kappa_0}$ ($W_{\kappa_\gamma}$) is a circuit representation of the unitary diagonalising $\hat{J}_1$ ($\hat{L}_\gamma$). The operator Eq.~\eqref{eq:elecstruct_qubit_lr_2} can be measured term-by-term, and a step of time evolution under $\hat{H}$ can be approximated with e.g. a Trotter product formula, 
\begin{equation}
\label{eq:lowrank_trotter}
e^{-i \Delta t \hat{H}} 
\to 
\prod_{\gamma=1}^{N_\gamma} 
\left(
W^\dagger_{\kappa_{\gamma}}
U_2(\sqrt{\Delta t} \, \boldsymbol{\lambda}^\gamma)
W_{\kappa_{\gamma}} \right)
W^\dagger_{\kappa_0}
U_1(\Delta t \boldsymbol{\zeta})
W_{\kappa_0}
+ \order(\Delta t^2)
\;.
\end{equation}
The circuit in Eq.~\eqref{eq:lowrank_trotter} comprises $N_\gamma+2$ change-of-basis circuits, a layer of single-qubit $Z$ rotations, and $N_\gamma$ circuits of the form
\begin{equation}
\label{eq:lowrank_trotter_2}
U_2(\boldsymbol{x})
=
\prod_{pq} \left( \mathsf{cP}_{2 x_p x_q} \right)_{p\uparrow,q\downarrow}
\prod_{ \substack{p<q \\ \sigma}} \left( \mathsf{cP}_{2 x_p x_q} \right)_{p\sigma,q\sigma}
U_1(\boldsymbol{x}^2)
\;.
\end{equation}
The circuit $U_2(\boldsymbol{x})$ has $M^2+M(M-1)$ $\mathsf{cP}$ gates and $2M$ single-qubit $Z$ rotations, depth $2M+1$, and requires all-to-all qubit connectivity, see Fig.~\ref{fig:lowrank2}.

\section{Classical subspace methods}
\label{sec:classical}

Subspace methods construct finite-dimensional approximations to the eigenvalue equation Eq.~\eqref{eq:es_problem}, an approach called Galerkin's method~\cite{evans2010partial}. Specifically, given a set of many-electron wavefunctions $\vett{v}_0 \dots \vett{v}_{n-1}$, we seek a function of the form
$| \subspacewfn^{(n)}_\mu \rangle = \sum_{\beta=0}^{n-1} C_{\beta\mu} |\vett{v}_\beta \rangle$ that solves the projection of the Schr\"odinger equation Eq.~\eqref{eq:es_problem} onto the subspace $V_n = \mathrm{Span}(\vett{v}_0 \dots \vett{v}_{n-1})$, i.e., for each $\alpha,\mu\in[n]\coloneqq\{0,1,...,n-1\}$,
\begin{equation}
\label{eq:es_problem_subspace}
\sum_{\beta=0}^{n-1} \langle \vett{v}_\alpha | \hat{H} | \vett{v}_\beta \rangle \; C_{\beta\mu} 
= 
\subspaceenergy^{(n)}_\mu 
\sum_{\beta=0}^{n-1} \langle \vett{v}_\alpha | \vett{v}_\beta \rangle \; C_{\beta\mu}
\;.
\end{equation}
The generalised eigenvalue equation (GEEV)~\eqref{eq:es_problem_subspace} will be compactly written as $H C = S C \subspaceenergy^{(n)}$, where $H$ and $S$ denote the Hamiltonian and overlap matrices, respectively, whose entries are
\begin{equation}
\label{eq:subspace_matrices}
    H_{\alpha\beta} \coloneqq \langle \vett{v}_\alpha | \hat{H} | \vett{v}_\beta \rangle, \quad S_{\alpha\beta} \coloneqq \langle \vett{v}_\alpha | \vett{v}_\beta \rangle.
\end{equation}
The approximate eigenvalues $\subspaceenergy^{(n)}_\mu$ and eigenvectors $| \subspacewfn^{(n)}_\mu \rangle$ differ from the exact ones since the approximate eigenvectors are restricted to the subspace $V_n$.
The quality of a subspace method depends on several factors: 
(i) the nature of the basis vectors $| \vett{v}_\beta \rangle$, 
(ii) the dimension $n$ of the employed subspace,
(iii) the cost (and associated error) 
of computing the matrices $H$, $S$ in Eq.~\eqref{eq:subspace_matrices}, and
(iv) the numeric stability of Eq.~\eqref{eq:es_problem_subspace}, connected with the condition number of the overlap matrix $S$.

\subsection{Krylov space methods}
\label{sec:krylov_space}

A Krylov space is associated to a matrix-vector pair $(A,\vett{v}_0)$, and is the subspace spanned by powers of $A$ applied to $\vett{v}_0$, $\mathrm{K}_n(A,\vett{v}_0) = \mathrm{Span}(\vett{v}_0 \dots A^{n-1} \vett{v}_0)$.
In the context of classical numerical eigensolvers, $A$ is often chosen to be the matrix whose spectrum one wishes to study~\cite{saad2011numerical}.
To study the low-energy eigenpairs of a Hamiltonian, one applies powers of the Hamiltonian to a trial vector~\cite{liesen2013krylov},
$| \vett{v}_\alpha \rangle = \hat{H}^\alpha |  \vett{v}_0 \rangle$, $\alpha=0 \dots n-1$. We will often call this ubspace the ``Hamiltonian Krylov space''.
The overlap and Hamiltonian matrices are ${S_{\alpha\beta} = \langle \vett{v}_0 | \hat{H}^{\alpha+\beta} |  \vett{v}_0 \rangle = f_{\alpha+\beta}}$ and ${H_{\alpha\beta} = \langle  \vett{v}_0 | \hat{H}^{\alpha+\beta+1} |  \vett{v}_0 \rangle = f_{\alpha+\beta+1}}$,
where $f_\ell = \langle  \vett{v}_0 | \hat{H}^\ell |  \vett{v}_0 \rangle$ and $\ell=0 \dots 2n-1$.

An intuitive justification of the Krylov space comes from the notion of imaginary-time evolution (ITE): for a Hamiltonian with a non-degenerate ground state $| \wfn_0 \rangle$, any trial state $| \vett{v}_0 \rangle$ non-orthogonal to $| \wfn_0 \rangle$ is mapped onto the ground state by ITE for a sufficiently long time $\tau>0$,
\begin{equation}
\label{eq:ite}
| \calcwfn_\tau \rangle 
= \frac{ e^{-\tau \hat{H}} |  \vett{v}_0 \rangle }
{ \| e^{-\tau \hat{H}} \vett{v}_0 \| } 
= 
\frac{ \sum_{\mu} e^{-\tau (\energy_\mu - \energy_0)} 
| \wfn_\mu \rangle \langle \wfn_\mu |  \vett{v}_0 \rangle}
{ \sqrt{ \sum_{\mu} e^{- 2 \tau (\energy_\mu - \energy_0)} 
|\langle \wfn_\mu | \vett{v}_0 \rangle|^2} } 
\;,\;
\lim_{\tau \to \infty} 
| \calcwfn_\tau \rangle
=
| \wfn_0 \rangle
\;.
\end{equation}
As $e^{-\tau \hat{H}}$ can be approximated by a truncated Taylor series,
${e^{-\tau \hat{H}} \simeq \sum_{\alpha=0}^{n-1} \frac{(-\tau)^\alpha}{\alpha!} \hat{H}^\alpha}$,
the state $| \calcwfn_\tau \rangle$ can be approximated by a linear combination of vectors $\hat{H}^\alpha |  \vett{v}_0 \rangle$, i.e. by a vector in the Hamiltonian Krylov space.

Eq.~\eqref{eq:ite} suggests that the quality of the $n$-dimensional Hamiltonian Krylov space is affected by several factors, such as the overlap between the initial state $\vett{v}_0$ and the ground state $\wfn_0$, $\langle \wfn_0 |  \vett{v}_0 \rangle$, and the spectral gap, $\energy_1-\energy_0$, whose inverse is the decay rate of excited-state contributions in Eq.~\eqref{eq:ite}.
A more formal result is the Kaniel-Paige inequality~\cite{kaniel1966estimates,paige1971computation},
\begin{equation}
\label{eq:kaniel_paige}
0 
\leq \subspaceenergy_0^{(n)} - \energy_0 
\leq (\energy_{D-1} - \energy_0) 
\left[ \frac{ \tan \theta(  \wfn_0 ,  \vett{v}_0) }{T_{n-1}(\gamma_0)} \right]^2 
\;,
\end{equation}
where $\energy_0 < \dots < \energy_{D-1}$ and $\wfn_0 \dots \wfn_{D-1}$ are the eigenpairs of $\hat{H}$, $\theta(\wfn_0 , \vett{v}_0)$ is the angle between $\wfn_0$ and 
$\vett{v}_0$, $T_{n-1}$ is the $(n-1)$-th Chebyshev polynomial, and $\gamma_0 = 1 + 2 \frac{\energy_1 -\energy_0}{\energy_{D-1} - \energy_1}$ is related to the spectral gap of $\hat{H}$.
The left-hand side of Eq.~\eqref{eq:kaniel_paige} shows that the Krylov method produces an upper bound for the ground-state eigenvalue. On the right-hand side, $\tan$ shows that the bound is tighter when $\vett{v}_0$ is closer to $\wfn_0$. Furthermore, as $\lim_{n \to \infty} T_{n-1}(1+x) = \infty$ for all $x>0$, increasing the dimension of the Krylov space results in a progressively more accurate estimation of  $\energy_0$. This property holds provided that the spectral gap is positive.
Expanding this analysis, the Saad inequality~\cite{saad1980rates} quantifies the accuracy of the eigenvalue approximations yielded by a Krylov space for excited states.
For a generic eigenvector $\wfn_\mu$ non-orthogonal to $\vett{v}_0$,
\begin{equation}
\label{eq:kaniel_paige_saad}
0 \leq 
\subspaceenergy_\mu^{(n)} - \energy_\mu 
\leq 
(\energy_{D-1} - \energy_\mu) 
\left[ \frac{ L_\mu^{(n)} \tan \theta( \wfn_\mu,\vett{v}_0) }{T_{n-1-\mu}(\gamma_\mu)} \right]^2 
\;,
\end{equation}
where $\gamma_\mu = 1 + \frac{\energy_{\mu+1} - \energy_\mu}{\energy_{D-1} -\energy_{\mu+1}}$, 
and $L_\mu^{(n)} = \prod_{\nu<\mu} \frac{\energy_{D-1} - \subspaceenergy^{(n)}_\nu}{\energy_\mu - \subspaceenergy^{(n)}_\nu}$.
Not only do these results show that the best approximate eigenvalues represented in the Krylov space converge, they converge extremely quickly: since ${T_n(\gamma)\ge\frac{1}{2}(\gamma+\sqrt{\gamma^2-1})^n}$ for $\gamma\ge1$, the upper bounds in Eqs.~\eqref{eq:kaniel_paige} and \eqref{eq:kaniel_paige_saad} converge exponentially with the Krylov space dimension provided the corresponding gap condition holds.

From a numerical standpoint, the Krylov method has an important limitation: the condition number of the overlap matrix grows exponentially with subspace dimension $n$.
This fact is expressed by the Beckermann-Townsend inequality~\cite{beckermann2019bounds}
stating that, for any two integers $j,k$ such that $0 \leq j+2k \leq n-1$,
\begin{equation}
\frac{ \sigma_j(S) }{ \sigma_{j+2k}(S) } \geq \frac{1}{4} \exp\left[ \frac{\pi^2}{4 \log\left( \frac{4 (n- n\%2) }{\pi} \right) } \right]^{k-n \%2} \;,
\end{equation}
where $\sigma_\ell(S)$ is the $\ell$-th singular value of the overlap matrix $S$ and $\%$ denotes modulo 2. Choosing $n$ odd,  $j=0$, and $2k=n-1$, shows that the condition number is lower-bounded by a quantity,
\begin{equation}
\label{eq:townsend}
\mathrm{cond}(S) = \frac{\sigma_0(S)}{\sigma_{n-1}(S)} \geq \frac{1}{4} \exp\left[ \frac{\pi^2}{4 \log\left( \frac{4 (n-1) }{ \pi} \right) } \right]^{\frac{n-1}{2}-1} 
\;,
\end{equation}
that diverges very rapidly with $n$. This may be intuitively understood as a trade-off for achieving the exponential convergence of the lowest energy in the subspace toward the true ground-state energy. However, as we will see in Section \ref{sec:lanczos}, this ill-conditioning can be largely mitigated by constructing an orthonormal basis for the Krylov space.

\subsubsection{Lanczos method}
\label{sec:lanczos}
Lanczos is a specific variant of Krylov subspace methods. As $n$ increases, Krylov vectors tend to become almost linearly dependent, see the divergence of $\mathrm{cond}(S)$ in Eq.~\eqref{eq:townsend}, leading to ill-conditioning. In the Lanczos method~\cite{lanczos1952solution}, the objective is to create an orthonormal basis $| \vett{q}_\alpha \rangle$ for the Hamiltonian Krylov space, such that $H_{\alpha\beta} = \langle  \vett{q}_\alpha | \hat{H} |  \vett{q}_\beta \rangle$ is tridiagonal. More specifically, one constructs Lanczos vectors $| \vett{q}_\alpha \rangle$ according to
\begin{equation}
\label{eq:lanczos}
\begin{split}
| \vett{q}_0 \rangle 
&= 
| \vett{v}_0 \rangle
\;, \\
| \vett{q}_1 \rangle 
&\propto 
\Big( \identity - | \vett{q}_0 \rangle \langle \vett{q}_0 | \Big)
\hat{H} | \vett{q}_0 \rangle
\;, \\
| \vett{q}_{\alpha+1} \rangle 
&\propto 
\Big( \identity - | \vett{q}_\alpha \rangle \langle \vett{q}_\alpha | - | \vett{q}_{\alpha-1} \rangle \langle \vett{q}_{\alpha-1} | \Big) 
\hat{H} | \vett{q}_\alpha \rangle
\;, \\
\end{split}
\end{equation}
where the proportionality symbols indicate that the $| \vett{q}_\alpha \rangle $ should be normalized.
We can see that the resulting $H$ is tridiagonal, and the basis is orthogonal, as follows.
By construction, $\hat{H}| \vett{q}_{\alpha} \rangle$ is in the span of $\{| \vett{q}_{\alpha+1} \rangle,| \vett{q}_{\alpha} \rangle,| \vett{q}_{\alpha-1} \rangle,...,| \vett{q}_{0} \rangle\}$.
Assuming as an inductive hypothesis that $\{| \vett{q}_{\alpha} \rangle,| \vett{q}_{\alpha-1} \rangle,...,| \vett{q}_{0} \rangle\}$ are orthogonal, an equivalent statement is that
\begin{equation}
    H_{\beta\gamma}=\langle \vett{q}_{\beta} | \hat{H} | \vett{q}_{\gamma} \rangle
\end{equation}
is upper-Hessenberg up to dimension $\alpha+1$, i.e. $H_{\beta\gamma}=0$ whenever $\beta>\gamma+1$.
However, $H$ is also Hermitian (since $\hat{H}$ is), which implies that it is tridiagonal in the basis of $| \vett{q}_{\alpha} \rangle$.
Hence, $\hat{H}| \vett{q}_{\alpha} \rangle$ is actually guaranteed to lie in the span of $\{| \vett{q}_{\alpha+1} \rangle,| \vett{q}_{\alpha} \rangle,| \vett{q}_{\alpha-1} \rangle\}$, so the partial orthogonalization in the last line of Eq.~\eqref{eq:lanczos} is in fact sufficient to orthogonalize with respect to all previous basis vectors.

Finally, it can be proved by induction over $n$ that the Lanczos and Krylov bases span identical subspaces.
While the vectors $| \vett{q}_\alpha \rangle$ are orthonormal assuming arithmetic operations are carried out exactly, computational simulations use floating-point arithmetic, which results in loss of orthonormality and spurious eigenpairs~\cite{cullum2002lanczos,saad2011numerical}. Practical implementations of the Lanczos algorithm mitigate its numerical instability by preventing orthogonality loss through repeated re-orthogonalisation of each newly generated vector 
with all the previously generated ones~\cite{ojalvo1970vibration}.

\subsubsection{Davidson method}

The standard diagonalisation algorithm in classical electronic structure is the Davidson method~\cite{davidsorq1975theiterative,morgan1986generalizations,vogiatzis2017pushing}. Unlike the Krylov subspace method, the Davidson method iteratively extends an $n$-dimensional subspace $V_n = \mathrm{Span}[ \vett{v}_0 \dots \vett{v}_{n-1} ]$ by adding a vector $| \vett{w} \rangle$ to its basis. This vector is chosen such that $\langle \vett{w} | \vett{v}_\alpha \rangle = 0$ and the approximation for the ground state of $\hat{H}$, restricted to the subspace $V_{n+1} = \mathrm{Span}[  \vett{v}_0 \dots \vett{v}_{n-1} \vett{w} ]$, is as accurate as possible.
The extension of $V_n$ is guided by a linear transformation $\hat{P}$ called a \technical{preconditioner}. A simple and widespread example is the Jacobi or diagonal preconditioner \cite{sleijpen2000jacobi}, which is efficient for diagonally-dominant matrices. Given the best approximation $| \subspacewfn^{(n)}_0 \rangle$ to the ground state in $V_n$ and the corresponding eigenvalue $\subspaceenergy^{(n)}_0$, let us assume that one seeks an eigenstate of the form $| \vett{w} \rangle = \alpha | \subspacewfn^{(n)}_0 \rangle + \beta | \vett{e}_\ell \rangle$, where $| \vett{e}_\ell \rangle$ is the $\ell$-th element of the canonical basis. The coefficients $\alpha,\beta$, which are the solution of a simple $2\times2$ eigenvalue equation, can be expanded to first order in $\langle \vett{e}_\ell | \subspacewfn^{(n)}_0 \rangle$ to give
\begin{equation}
| \vett{w} \rangle 
= 
| \subspacewfn^{(n)}_0 \rangle
+
| \vett{e}_\ell \rangle \langle \vett{e}_\ell | \hat{P} | \vett{r} \rangle
\;,\;
| \vett{r} \rangle = \left[ \hat{H} - \subspaceenergy^{(n)}_0 \right] | \subspacewfn^{(n)}_0 \rangle
\;,\;
\hat{P} = \sum_\ell \frac{ | \vett{e}_\ell \rangle \langle \vett{e}_\ell | }{\langle \vett{e}_\ell | \hat{H} | \vett{e}_\ell \rangle - \subspaceenergy^{(n)}_0}
\;.
\end{equation}
The vector $| \vett{r} \rangle$ and the operator $\hat{P}$ are called the residue and the diagonal preconditioner respectively.
In the Davidson method with the Jacobi preconditioner, one applies the preconditioner to the residual vector, yielding $| \vett{w} \rangle = | \subspacewfn^{(n)}_0 \rangle + \hat{P} | \vett{r} \rangle$, and then expands the subspace by orthonormalising $| \vett{w} \rangle$ against ${\bf{v}}_0 \dots {\bf{v}}_{n-1}$. There exist alternatives to the diagonal preconditioner illustrated here, which are suited to non-diagonally-dominant matrices, such as the ``pspace'' preconditioner of Olsen \etal~\cite{olsen1990passing}.

\subsection{Configuration interaction (CI) methods}

CI methods are a type of subspace method for ES calculations. CI methods operate within a subspace of the Hilbert space spanned by a collection of electronic configurations, i.e. Slater determinants of the form Eq.~\eqref{eq:configuration}. Interaction means constructing linear combinations, 
$| \subspacewfn_\mu \rangle = \sum_{\alpha \in A} C_{\alpha \mu} | \vett{x}_\alpha \rangle$, of electronic 
configurations $\vett{x}_\alpha$ drawn from a set $A$, which may be predefined (as in truncated CI approaches) or adaptively constructed (as in selected CI approaches).

\subsubsection{Full and truncated CI}

If the set $A$ includes all possible configurations of the appropriate symmetry, the resulting method is called full CI (FCI) and exactly solves the electronic Schr\"{o}dinger equation, Eq.~\eqref{eq:es_problem}. Exact diagonalizaton in an active space is called complete active space (CAS) CI, and CASCI with variational optimization of the active-space orbitals is called CASSCF.
The FCI method has many attractive characteristics, especially that it exactly solves the Schr\"{o}dinger equation. 
However, the computational cost of FCI grows combinatorially with the numbers of electrons and orbitals, $|A| = D = \binom{M}{N_\uparrow} \binom{M}{N_\downarrow}$ and thus, notwithstanding considerable progress, an exact FCI approach is feasible only for relatively small basis sets~\cite{vogiatzis2017pushing,dachsel1998out,evangelisti1996one,ben1998benchmark,de2010utilizing}.
Motivated by this observation, attempts were made to search for approximate FCI approaches by a judicious choice of the configurations~\cite{sinanouglu1962many,hinze1967multi,shavitt1977method}.

The first term in the expansion of the exact ground state onto configurations, $| \wfn_\mu \rangle = \sum_{\vett{x}} C_{\vett{x}\mu} | \vett{x} \rangle$ where $| \vett{x} \rangle$ denotes a configuration as in Eq.~\eqref{eq:configuration}, is often the HF state. The other terms can be characterised by the number of electronic excitations from occupied to virtual orbitals in the HF state (singles, doubles, triples, quadruples, etc).
The CI space may be truncated by retaining configurations with a finite number of excitations, to save computational resources. Well-established examples of truncated CI methods are CIS (single excitations), 
and CISD (singles and doubles excitations)~\cite{bender1969studies}.
An important limitation of truncated CI methods is their lack of size-consistency, i.e. the energy $\energy(A+B)$ of two infinitely-separated systems $A$ and $B$ is not equal to $\energy(A)+\energy(B)$~\cite{pople1987quadratic}. Size consistency is generally regarded as being more important than the provision of an energy upper-bound. Therefore, CISD results are often modified using various corrections, such as the one introduced by Langhoff and Davidson~\cite{langhoff1974configuration}, to make the energies approximately size-consistent, although this adjustment sacrifices variationality.

\subsubsection{Selected CI}

Selected CI methods rely on the same principle as the usual CI approaches. However, in selected CI methods, determinants are not chosen solely based on the number of electronic excitations. Instead, they are adaptively selected from  the entire set of determinants based on their estimated contribution to the FCI wavefunction. This is because, even inside a predefined set of determinants, only a fraction significantly contributes to the wave function~\cite{bytautas2009priori,anderson2018breaking}.

The first multireference selected-CI algorithm going beyond singles and doubles was the CIPSI (perturbatively selected configuration interaction scheme) method of Huron \etal~\cite{huron1973iterative,buenker1974individualized,buenker1975energy}, who proposed to iteratively select external determinants (i.e. determinants which are not present in the variational set) using a perturbative criterion.
Recent years have witnessed the development of various selected-CI approaches. Though based on the original CIPSI method, these approaches feature significant variations in the core idea as well as in the algorithm, in part motivated by the availability of more powerful computational hardware. These include: (i) stochastic and semi-stochastic approaches such as the FCI quantum Monte Carlo~\cite{booth2009fermion,cleland2010communications,guther2020neci} and the heat-bath CI approach~\cite{holmes2016heat,sharma2017semistochastic,li2018fast}; (ii) purely variational approaches such as the iterative and static-dynamic-static CI~\cite{liu2016ici,liu2016sds}; (iii) size-consistent approximate FCI approaches such as full coupled-cluster reduction (FCCR)~\cite{xu2018full} and the many-body expansion FCI (MBE-FCI)~\cite{eriksen2018many,eriksen2019many,eriksen2019generalized}.

\subsection{The equation-of-motion approach}

The equation of motion (EOM) approach~\cite{rowe1968equations} is an alternative subspace-construction technique, that remedies some of the limitations of truncated CI methods (e.g. the lack of size-extensitivity and size-consistency of ground-state energies and lack of size-intensivity of excitation energies).
Starting from the Schr\"{o}dinger equation, Eq.~\eqref{eq:es_problem}, EOM introduces a set of excitation operators $\hat{Q}^\dagger_\mu = | \wfn_\mu \rangle \langle \wfn_0 |$ with the properties:
(i) $\hat{Q}^\dagger_\mu | \wfn_0 \rangle = | \wfn_\mu \rangle$,
(ii) $\hat{Q}_\mu | \wfn_0 \rangle = 0$,
and (iii) $\langle \wfn_0 | \hat{Q}^\dagger_\mu = 0$.
Eq.~\eqref{eq:es_problem} is then rewritten as
\begin{equation}
[ \hat{H} , \hat{Q}^\dagger_\mu ] | \wfn_0 \rangle 
= 
\Delta \energy_\mu \hat{Q}^\dagger_\mu | \wfn_0 \rangle 
\;,
\end{equation}
where $\Delta \energy_\mu = \energy_\mu-\energy_0$ are excitation energies. 
Taking the overlap with a generic state $\langle \Psi_0 | \delta \hat{Q}$ leads to the equation
\begin{equation}
\label{eq:EOM}
\langle \wfn_0 | [ \delta \hat{Q} , [ \hat{H} , \hat{Q}^\dagger_\mu ] ] | \wfn_0 \rangle 
= 
\Delta \energy_\mu \langle \wfn_0 | [ \delta \hat{Q} , \hat{Q}^\dagger_\mu ] | \wfn_0 \rangle
\;,
\end{equation}
using the fact that $\langle \wfn_0 | [ \hat{H} , \hat{Q}^\dagger_\mu ]$ and $\langle \wfn_0 | \hat{Q}^\dagger_\mu$ are both zero.
Eq.~\eqref{eq:EOM} is called an EOM for the excitation operators $\hat{Q}^\dagger_\mu$ and energies $\Delta \energy_\mu$. It can be turned into a GEEV, and then numerically solved, by expanding the excitation operators as $\hat{Q}^\dagger_\mu = \sum_J X^\mu_J \hat{F}_J - Y^\mu_J \hat{F}^\dagger_J$ where the $\hat{F}_J$ are some set of many-electron operators and $X^\mu_J,Y^\mu_J$ are coefficients.
Then combining the equations associated to each assignment of $\delta \hat{Q} = \hat{F}_I$ or $\delta \hat{Q} = \hat{F}_I^\dagger$ for each $I$, and replacing the exact unknown ground state $\wfn_0$ with an approximation $\calcwfn$, yields the matrix equation
\begin{equation}
\label{eq:EOM_Y}
\left( 
\begin{array}{cc}
M & Q \\
Q^* & M^* \\
\end{array}
\right)
\left( 
\begin{array}{cc}
X^\mu \\
Y^\mu \\
\end{array}
\right)
=
\Delta \subspaceenergy_\mu
\left( 
\begin{array}{cc}
V & W \\
-W^* & -V^* \\
\end{array}
\right)
\left( 
\begin{array}{cc}
X^\mu \\
Y^\mu \\
\end{array}
\right)
\end{equation}
with
\begin{equation}
\label{eq:EOM_matrices}
\begin{array}{ll}
V_{IJ} = \phantom{-} \langle \calcwfn | [ \hat{F}^\dagger_I , \hat{F}_J^{\phantom{{\dagger}}} ] | \calcwfn \rangle \; , &
M_{IJ} = \phantom{-} \langle \calcwfn | [ \hat{F}^\dagger_I , [ \hat{H} , \hat{F}_J^{\phantom{{\dagger}}} ] ] | \calcwfn \rangle \; , \\
\\
W_{IJ} = - \langle \calcwfn | [ \hat{F}^\dagger_I , \hat{F}^\dagger_J ] | \calcwfn \rangle \; , &
Q_{IJ} = - \langle \calcwfn | [ \hat{F}^\dagger_I , [ \hat{H} , \hat{F}^\dagger_J ] ] | \calcwfn \rangle \; . \\
\end{array}
\end{equation}
The Tamm-Dancoff (TDA) approximation to Eq.~\eqref{eq:EOM_Y} is obtained by forcing $Y^\mu = 0$,  and leads to the simplified eigenvalue equation $M X^\mu = \Delta \subspaceenergy_\mu V X^\mu$.
The cost and accuracy of an EOM calculation depends on two factors: the state $\Phi$, which may not be an accurate approximation for the ground state in statically correlated systems, and the truncation of electronic excitations connecting ground and excited states (e.g. singles and doubles), which may not always be sufficient for multireference excited states.

\subsubsection{Time-dependent Hartree-Fock and equation-of-motion coupled-cluster}

Notable examples, summarised in Table \ref{table:eom}, are the random phase approximation (RPA) or time-dependent HF (TD-HF) method~\cite{bohm1951collective,sawada1957correlation,pines1952collective,bohm1953collective,mclachlan1964time}, the CIS method (which is the TDA of TD-HF)~\cite{bonche1976one,dreuw2005single},
and the EOM-CCSD method for electronic excitations~\cite{stanton1993equation}.
EOM-CCSD can also be formulated for electron-removing excitations (IP)~\cite{prasad1985some,datta1993consistent,stanton1994analytic,krylov2008equation}, 
and electron-attaching excitations (EA)~\cite{nooijen1995equation}.
EOM-CCSD is a widespread method to compute response and excited-state properties, from energy differences like singlet-triplet gaps, ionisation potentials, and electron affinities, 
to quasiparticle properties, Green's functions, the density of states, and spectral functions~\cite{wang2020excitons,gao2020electronic,lange2021improving}.
Some of the advantages associated with the EOM-CCSD formalism are its
theoretical rigour, the accuracy and correct scaling behavior of
energy differences, and the ability to systematically
improve results. However, standard quantum chemistry
methods such as EOM-CCSD sometimes face challenges in
a quantitative determination of excited states and their
properties, notably for same-symmetry conical intersections~\cite{thomas2021complex,kohn2007can,yarkony2012nonadiabatic,bernardi1997role}
and when the ground state has a prominent multi-reference
character~\cite{schmidt1998construction,kohn2013state,mahapatra1998state,roos1996multiconfigurational}.
Since quantum algorithms are projected to provide accurate ground-state wavefunctions, even in the case of statically-correlated systems,
they can improve these features of EOM-CCSD with practical computational expenses.

\begin{table}[t!]
\centering
\begin{tabular}{cccc}
\hline\hline
name & $\calcwfn$ & $\hat{F}_I$ & TDA \\ 
\hline
RPA (TD-HF) & $\calcwfn_{\mathrm{HF}}$ & $\{ \crt{a\sigma} \dst{i\sigma} \}_{ai\sigma}$ & no \\
CIS & $\calcwfn_{\mathrm{HF}}$ & $\{ \crt{a\sigma} \dst{i\sigma} \}_{ai\sigma}$ & yes \\
IP-EOM-CCSD & $\calcwfn_{\mathrm{CCSD}}$ & $\{ \dst{i\sigma} \}_{a\sigma} \cup \{ \dst{i\sigma} \crt{a\tau} \dst{j\tau} \}_{aij\sigma\tau}$ & no \\
EA-EOM-CCSD & $\calcwfn_{\mathrm{CCSD}}$ & $\{ \crt{a\sigma} \}_{a\sigma} \cup \{ \crt{a\sigma} \crt{b\tau} \dst{i\tau} \}_{abi\sigma\tau}$ & no \\
EE-EOM-CCSD & $\calcwfn_{\mathrm{CCSD}}$ & $\{ \crt{a\sigma} \dst{i\sigma} \}_{ai\sigma} \cup \{ \crt{a\sigma} \crt{b\tau} \dst{j\tau} \dst{i\sigma} \}_{abij\sigma\tau}$ & no \\
\hline\hline
\end{tabular}
\caption{Examples of EOM-based classical subspace methods. $\calcwfn_{\mathrm{HF}}/\calcwfn_{\mathrm{CCSD}}$ denotes the Hartree-Fock/coupled-cluster with singles and doubles ground-state, $ij$/$ab$ label occupied/unoccupied spatial orbitals in the Hartree-Fock state, and $\sigma\tau$ label spin polarisations.}
\label{table:eom}
\end{table}

% orbital optimization

\subsection{Nonlinear subspace methods}

Although they are somewhat less closely tied to existing quantum subspace methods, it is worth mentioning the existence of subspace methods for nonlinear eigenvalue problems~\cite{zhou2006parallel,zhou2006selfconsistent}. These methods find natural applications in HF or other SCF (self-consistent field) calculations.
The high-level idea of these algorithms is to track a subspace that is iteratively evolved along with the SCF updates.
By doing so, the outer SCF iteration and the inner subspace iteration are combined into a single loop, which can in principle lead to substantial computational savings, although these methods are quite modern so practical evidence is somewhat limited.
For a recent summary of these methods and their theoretical analysis, see~\cite{saad2016analysis}.

\section{Quantum subspace methods}
\label{sec:quantum}

\begin{figure}[t!]
\centering
\includegraphics[width=0.7\textwidth]{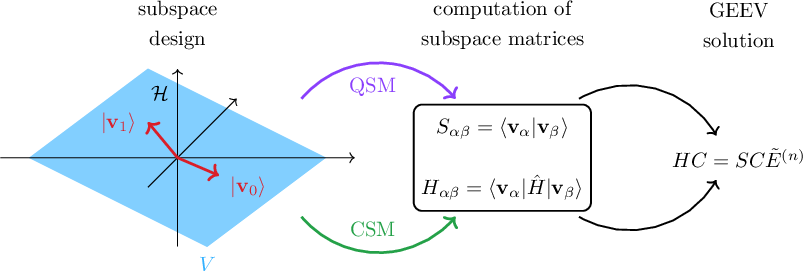}
\caption{Schematics of quantum and classical subspace methods (QSM, CSM). The search for Hamiltonian eigenpairs is restricted from a full Hilbert space $\mathcal{H}$ to a subspace $V$ spanned by vectors $\{ | \vett{v}_\alpha \rangle \}_{\alpha=0}^{n-1}$ (left). Overlap and Hamiltonian matrices are computed using a classical or quantum computer (middle, enclosed in a rounded rectangle, with QSM/CSM abbreviating quantum/classical subspace method). A generalised eigenvalue equation (GEEV) is solved on a classical computer, yielding approximate Hamiltonian eigenpairs (right).
}
\label{fig:qsms}
\end{figure}

The quantum subspace expansion (QSE) method~\cite{mcclean2017hybrid,colless2018computation,huggins2020non,motta2020determining,parrish2019quantum,stair2020multireference,takeshita2020increasing,cohn2021quantum,seki2021powermethod,yoshioka2022generalized,cortes2022quantum,klymko2022real,baek2022say,tkachenko2022davidson,kirby2023exact} constructs a variational subspace spanned by a set of states $\{ | \vett{v}_\alpha \rangle \}_{\alpha=0}^{n-1}$, as sketched in Fig.~\ref{fig:qsms}. Hamiltonian eigenstates are approximated by a linear combination of basis states, $| \subspacewfn^{(n)}_\mu \rangle = \sum_\alpha C_{\alpha\mu} | \vett{v}_\alpha \rangle$. To obtain the expansion coefficients $C_{\alpha\mu}$, one computes matrix
elements $S_{\alpha\beta} = \langle \vett{v}_\alpha | \vett{v}_\beta \rangle$ and $H_{\alpha\beta} = \langle \vett{v}_\alpha | \hat{H} | \vett{v}_\beta \rangle$ in the chosen basis and solves the GEEV $H C = S C \subspaceenergy^{(n)}$. Within QSE, a quantum device is used to compute the matrix elements $H_{\alpha\beta}$ and $S_{\alpha\beta}$, and a classical computer to solve the GEEV and obtain approximate eigenpairs $\{ \subspaceenergy^{(n)}_\mu, | \subspacewfn^{(n)}_\mu \rangle \}_\mu$. The quantum circuits required to compute $H_{\alpha\beta}$ and $S_{\alpha\beta}$ depend on the choice of the basis $\{ | \vett{v}_\alpha \rangle \}_{\alpha=0}^{n-1}$. Important families of QSMs will be illustrated in Subsections \ref{sec:qse_fermionic}, \ref{sec:qse_time}, and \ref{sec:qse_power}, and alternative approaches will be discussed in Subsection \ref{sec:alternative}.

The main advantage of QSE (and other subspace methods) over the well-established variational quantum eigensolver (VQE) is that QSE does not require a non-linear parameter optimisation, which is an NP-hard problem~\cite{bittel2021training}, as part of the eigenvalue approximation.
While both QSE and VQE employ quantum and classical computers in synergy, the nature of their interaction is profoundly different: in VQE, each update of the quantum circuit parameters requires a new call of the quantum computer, while in QSE all the circuits required to measure $H_{\alpha\beta}$ and $S_{\alpha\beta}$ can be sent to the quantum computer in just one call. Furthermore, QSE circuits can be naturally parallelised over multiple quantum computers (or groups of qubits within a quantum computer).
A limitation of QSE is that the eigenstates $| \subspacewfn^{(n)}_\mu \rangle$ are never actually stored on a quantum device, so computing properties after a QSE calculation generally requires additional measurements.

As mentioned above, the accuracy and computational cost of a QSE calculation depends on the choice of the basis states $\{ | \vett{v}_\alpha \rangle \}_{\alpha=0}^{n-1}$. However, no specific prescription is provided for this selection, similar to the original Galerkin's method. In this sense, QSE can be regarded as a family of QSMs, differentiated by the choice of the basis states. We will now discuss representative QSMs, starting from those based on $k$-body fermionic operators (e.g., single and double electronic excitations), then moving to those based on real- and imaginary-time evolution under the Hamiltonian, and polynomials of the Hamiltonian.

\subsection{QSMs based on k-body fermionic operators}
\label{sec:qse_fermionic}

The simplest example of a QSM is QSE based on $k$-body fermionic operators, e.g. single- and double-excitation operators applied to an initial state $|\calcwfn\rangle$,
\begin{equation}
\label{eq:quantum_mrcisd}
\begin{split}
| \subspacewfn^{(n)}_\mu \rangle &= \Big[ 
C_\mu + 
\sum_{ai,\sigma} C^{ai,\sigma}_\mu \, \crt{a\sigma} \dst{i\sigma} + 
\sum_{aibj,\sigma\tau} C_\mu^{aibj,\sigma\tau} \, \crt{a\sigma} \crt{b\tau} \dst{j\tau} \dst{i\sigma} \Big] | \calcwfn \rangle 
= \sum_\alpha C_{\alpha\mu} \, \hat{O}_\alpha | \calcwfn \rangle \;,
\end{split}
\end{equation}
with $\hat{O}_\alpha \in \{ \identity \} \cup \{ \crt{a\sigma} \dst{i\sigma} \}_{ai\sigma} \cup \{ \crt{a\sigma} \crt{b\tau} \dst{j\tau} \dst{i\sigma} \}_{abij\sigma\tau}$. This flavor of QSE can be regarded to as a multi-reference CISD method (MRCISD) where the wavefunction $| \calcwfn \rangle$, prepared on a quantum device, is not a single Slater determinant but a correlated electronic state.

\begin{figure}[t!]
\includegraphics[width=0.9\textwidth]{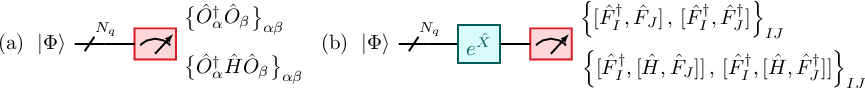}
\caption{Quantum circuits used in the multi-reference CISD (a) and self-consistent quantum EOM (b) methods. Within MRCISD a register of qubits is prepared in a state $| \calcwfn \rangle$, and operators defined in Eq.~\eqref{eq:quantum_mrcisd} are measured (red meter blocks). Within quantum EOM, a different set of operators is measured, with $\hat{F}_I$ defined as in Table \ref{table:eom}. In the case of self-consistent quantum EOM, a unitary operator $e^{\hat{X}}$ is applied to the register before the measurement (teal block). The expectation values of the operators in the top/bottom row define the overlap/Hamiltonian matrices of the method, respectively.}
\label{fig:qse_mrcisd}
\end{figure}

QSE based on single- and double-excitation operators has had significant success in the last few years, see e.g. Section \ref{sec:implementation} for a list of implementations on quantum hardware. As sketched in Fig.~\ref{fig:qse_mrcisd}, it requires measuring a set of operators on a register of qubits prepared in the state $| \calcwfn \rangle$. Importantly, it does not increase the depth of the quantum circuit required to prepare $| \calcwfn \rangle$ and measure the target operators. This feature is especially beneficial on near-term hardware limited by qubit coherence times and errors of two-qubit gates. Furthermore, it typically leads to well-conditioned eigenvalue equations.

However, it is important to note that the overhead of measurements is substantial in quantum MRCISD. This method requires estimating $k$-body reduced density matrices (RDMs) of order higher than $k=2$, which poses a significant computational bottleneck. We will discuss this issue further in Section \ref{sec:economisation}.

Furthermore, a significant drawback of the quantum MRCISD approach, which it shares with the classical CISD and MRCISD methods, is the lack of size-intensivity in the computed excitation energies. This limitation can result in quantitative errors and non-physical predictions.
This may become a severe limitation when QSE is applied to larger systems and/or the underlying ground-state wavefunction is imprecise.
We emphasise the existence of alternatives to $k$-body fermionic operators in the construction of a subspace, for example, Pauli operators~\cite{colless2018computation} and elements of the Hamiltonian~\cite{bharti2021iterative,lim2021fast}.

\subsubsection{Quantum equation-of-motion}

In search of a size-intensive alternative to MRCISD, the quantum EOM method (qEOM) was proposed by Ollitrault \etal~\cite{ollitrault2020quantum} for
electronic excitation energies (EEs). 
qEOM is essentially a transposition of the EOM-CC approach to quantum computers. A quantum algorithm is used to produce an approximation $| \calcwfn \rangle$ for the ground-state of the Hamiltonian, and the matrix elements Eq.~\eqref{eq:EOM_matrices} are measured on the quantum computer over the state $| \calcwfn \rangle$.
While qEOM provides good agreement for EEs with the exact results obtained by exact diagonalisation~\cite{ollitrault2020quantum}, it does not necessarily satisfy the vacuum annihilation conditions $\hat{Q}^\dagger_\mu | \calcwfn \rangle = 0$ ensuring that the ground-state wavefunction cannot be de-excited. 
This may result in the appearance of large errors when the formalism is extended to calculate properties such as IPs and EAs. Moreover, the qEOM
method, like QSE based on single- and double-excitation operators, requires high-body RDMs which significantly increases the measurement cost. To remedy this limitation, Asthana \etal~\cite{asthana2023quantum} proposed a self-consistent generalisation of qEOM, where excitation operators have the form $\hat{G}_I = e^{-\hat{X}} \hat{F}_I e^{\hat{X}}$ where $\hat{X}$ is a linear combination of single- and double-excitations,
\begin{equation}
\hat{X} = \sum_{ai, \sigma} x^a_i \, \crt{a\sigma} \dst{i\sigma} + \sum_{abij,\sigma\tau} 
 x^{ab}_{ij} \, \crt{a\sigma} \crt{b\tau} \dst{j\tau} \dst{i\sigma} - \mathrm{h.c.}
 \;,
\end{equation} 
and $\hat{F}_I$ as in Table \ref{table:eom}. The self-consistent qEOM method satisfies the vacuum annihilation conditions, produces size-intensive and real-valued energy differences between ground and excited/charges states, and requires the measurement of 1- and 2-body RDMs.
Notwithstanding these desirable features, qEOM shares with quantum MRCISD a significant measurement overhead and, as sketched in Fig.~\ref{fig:qse_mrcisd}, deeper quantum circuits, due to the need of applying $e^{\hat{X}}$ to $| \calcwfn \rangle$.

\subsection{Hamiltonian Krylov-space methods}
\label{sec:qse_power}

In the previous Subsection, we explored QSMs based on the application of $k$-body fermionic operators to an initial state. These methods are compelling for near-term devices as they do not require deep quantum circuits, but they suffer from high measurement costs and accuracy limitations akin to those of classical truncated CI calculations. Alternatively, there are QSMs based on Krylov spaces, and a specific example is a Chebyshev quantum Krylov. However, constructing a Krylov space generated by powers of the Hamiltonian on a quantum computer is non-trivial. This is because the Hamiltonian and its powers are non-unitary operations, and thus cannot be trivially mapped onto a quantum circuit. The use of block-encoding unitaries was proposed to circumvent this limitation.

\subsubsection{Chebyshev quantum Krylov}
\label{sec:chebyshev_lanczos}

The Chebyshev quantum Krylov method is a quantum implementation of the classical Krylov space~\cite{kirby2023exact}.
This method is exact in the sense that it has no algorithmic error in the construction of the Krylov space, i.e., approximations such as Trotterization or other approximate time evolution algorithms are not required.
The method exploits the fact that the classical Krylov space generated by powers of the Hamiltonian is mathematically equivalent to the subspace generated by any basis for polynomials of the Hamiltonian (powers may be viewed as the monomial basis).

\begin{figure}[t!]
\centering
\includegraphics[width=0.75\textwidth]{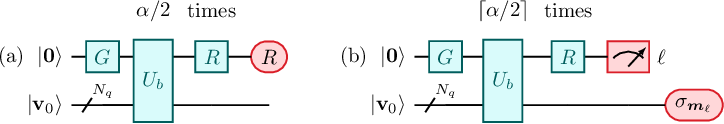}
\caption{Quantum circuits used in the quantum Krylov method.
For $\alpha=0 \dots n-1$ even/odd, we apply the left/right circuit.
As defined in the text, $U_b$ block-encodes the Hamiltonian $\hat{H}$, 
$G$ prepares the state that identifies the block containing $\hat{H}$, 
and $R$ reflects around that state. In panel (b), when the ancillae collapse in the state $|\ell\rangle$ upon measurement, the Pauli operator  $\sigma_{\boldsymbol{m}_\ell}$ is measured on the main qubit register.
}
\label{fig:krylov}
\end{figure}

The method is based on the notion of block-encoding~\cite{low2019hamiltonian} of a Hamiltonian $\hat{H}$, i.e. a unitary operator $\hat{U}_b$ acting on an extended Hilbert space and such that
\begin{equation}
\langle G , \cdot | \hat{U}_b | G , \cdot \rangle 
=
\langle \cdot | \hat{H} | \cdot \rangle \;,
\end{equation}
for some state $|G \rangle$ (in the remainder of this paragraph we will omit operator hats to avoid clutter). An example of block-encoding is the following: for a Hamiltonian operator $H = \sum_\ell h_\ell \, \sigma_{\boldsymbol{m}_\ell}$ where $\sum_\ell | h_\ell | = \sum_\ell h_\ell = 1$ and $\sigma_{\boldsymbol{m}_\ell}$ are Pauli operators, the unitary
$U_b = \sum_\ell | \ell \rangle \langle \ell | \otimes \sigma_{\boldsymbol{m}_\ell}$, in conjunction with the state $| G \rangle = \sum_\ell \sqrt{h_\ell} | \ell \rangle = G | \boldsymbol{0} \rangle$, defines a block encoding for $\hat{H}$.
Under the assumption that the block encoding is self-inverse, $U_b^2 = \identity$, one has that
\begin{equation}
\langle G , \cdot | (R U_b)^\alpha | G , \cdot \rangle 
=
\langle \cdot | T_\alpha(H) | \cdot \rangle \;,
\end{equation}
where $R  = \big[ 2 | G \rangle \langle G| - \identity \big] \otimes \identity$ is a reflection operator and $T_\alpha$ the 
$\alpha$-th Chebyshev polynomial of the first kind. In other words, 
$(R U_b)^\alpha$ is a block encoding of $T_\alpha(H)$. 
Since the Chebyshev polynomials $T_\alpha$ with $\alpha = 0 \dots n-1$ are a basis for polynomials of degree less than $n$, the subspace spanned by the states $| \vett{v}_\alpha \rangle = T_\alpha(H) | \vett{v}_0 \rangle$,
$k=0 \dots n-1$, coincides with the $n$-dimensional Krylov space generated by powers of the Hamiltonian. The overlap and Hamiltonian matrix elements are given by
\begin{equation}
S_{\alpha\beta} = \frac{f_{\alpha+\beta}+f_{|\alpha-\beta|}}{2}
\;,\;
H_{\alpha\beta} = \frac{f_{\alpha+\beta+1}+f_{|\alpha+\beta-1|}+f_{|\alpha-\beta|+1}+f_{|\alpha-\beta-1|}}{4}
\;,
\end{equation}
where the function $f_\alpha = \langle \vett{v}_0 | T_\alpha(H) | \vett{v}_0 \rangle$ can be expressed as
\begin{equation}
f_\alpha =
\begin{cases}
\langle \vett{v}_0 | (U_bR)^{\alpha/2} R (R U_b)^{\alpha/2} | \vett{v}_0 \rangle & \alpha \; \mathrm{even} \\
\langle \vett{v}_0 | (U_bR)^{\lfloor \alpha/2 \rfloor} U_b (R U_b)^{\lfloor \alpha/2 \rfloor} | \vett{v}_0 \rangle & \alpha \; \mathrm{odd} \\
\end{cases}
\end{equation}
and measured with the circuits in Fig.~\ref{fig:krylov}.

This quantum Krylov method is a technique to construct a Krylov space on a quantum computer. It addresses the challenge posed by the non-unitarity of Hamiltonian powers using a block-encoding unitary transformation. Due to its relationship with the classical Lanczos method, it is provably convergent to the ground state (see Section \ref{sec:krylov_space}) and can benefit from heuristic quantum computing algorithms providing initial states having high overlap with the ground state. Furthermore, it offers an important advantage in terms of the memory required to store the Krylov space information, i.e. it bypasses the need to store combinatorially large CI strings in memory.
On the other hand, like the classical Lanczos method, it may be numerically ill-conditioned (see Section \ref{sec:krylov_space}) and sensitive to shot noise (see
Section \ref{sec:shot_noise} for a detailed discussion). Furthermore, the use of a block-encoding unitary requires ancillae and deep quantum circuits, making it more suited for future quantum devices~\cite{kirby2023exact}.

\subsubsection{Gaussian-power quantum Krylov}
\label{sec:GaussianPower_lanczos}
This method constructs a subspace using the conventional power function acting on the Hamiltonian, as described in Section \ref{sec:krylov_space}. However, the reference state $|{\bf{v}}_0\rangle$ is now replaced with the state $e^{-\frac{1}{2} (H-E_0)^2 \tau^2} |{\bf{v}}_0\rangle$. In other words, each of the basis states $|{\bf{v}}_\alpha \rangle$ is generated as follows:

\begin{equation}
\label{eq:GaussianPower_krylov}
| {\bf{v}}_\alpha \rangle = (\hat{H} - E_0)^{\alpha} e^{-\frac{1}{2} (H-E_0)^2 \tau^2 } | {\bf{v}}_0 \rangle, \;  \alpha=0 \dots n-1
\end{equation}
where $E_0$ is an arbitrary constant of choice. This is known as the Gaussian-power basis \cite{ZhangGaussianPowerKrylov}. The spectral norm for each basis vector $| {\bf{v}}_\alpha \rangle$ can be bounded as $ || |{\bf{v}}_\alpha \rangle ||_2 \leq \big( \dfrac{n-1}{e \tau^2} \big)^{\frac{n-1}{2}}$. Therefore, when $e \tau^2 \geq n-1$, the spectral norm of $|{\bf{v}}_\alpha \rangle$ decreases exponentially with respect to $\alpha$, and the rate is parameterized by the parameter $\tau$. This is useful since the exact projected matrices $H$ and $S$ are unknown due to the statistical errors that arise from finite sampling. The exponential suppression of the spectral norm of $|{\bf{v}}_\alpha \rangle$ is the main factor leading to a significant decrease in statistical errors and minimizes the number of measurements required in this approach.

One way to realise Eq. \eqref{eq:GaussianPower_krylov} on a quantum computer is through the method of Linear Combination of Unitaries (LCU), as proposed by \cite{ZhangGaussianPowerKrylov}. In this approach, the author showed that the generating function $(\hat{H} - E_0)^{\alpha} e^{-\frac{1}{2} (\hat{H}-E_0)^2 \tau^2}$ in Eq. \eqref{eq:GaussianPower_krylov} can be rewritten as
\begin{equation}
(\hat{H} - E_0)^{\alpha} e^{-\frac{1}{2} (\hat{H}-E_0)^2 \tau^2}
=
\left( \frac{i}{\sqrt{2 \tau^2}} \right)^\alpha
\int_{-\infty}^{\infty} dt \, H_{\alpha}\Big( \dfrac{t}{\sqrt{2 \tau^2}} \Big) \, g_{\tau} (t) \, e^{-it(H-E_0)} \;,
\end{equation} 
where $H_{\alpha}(u)$ are Hermite polynomials, and $g_{\tau} (t) = e^{-t^2/2\tau^2}/\sqrt{2 \pi \tau^2}$. In this representation, the only function that depends on the Hamiltonian is the real-time evolution operator. Several techniques, e.g. Trotterisation and particularly those discussed in Section \ref{sec:real_time_evolution}, can be employed for this purpose. Another alternative realization of Eq. \eqref{eq:GaussianPower_krylov} on a quantum computer is through the technique of block encoding.

\subsection{QSMs based on time evolution}
\label{sec:qse_time}

Here we present a third family of QSMs, that make use of real- or imaginary-time evolution operations to construct subspaces. These methods are the Quantum Filter Diagonalisation (QFD) and Quantum Lanczos (QLanczos). 

\subsubsection{Quantum Filter Diagonalisation}
\label{sec:real_time_evolution}
This method constructs a subspace by applying the time evolution operator to an initial state~\cite{parrish2019quantum},
\begin{equation}
| {\bf{v}}_\alpha \rangle = e^{-i t_\alpha \hat{H}} | {\bf{v}}_0 \rangle
\;,\;  \alpha=0 \dots n-1
\end{equation}
where the times $t_\alpha$ are a set of times often, though not necessarily, given by $t_\alpha = \alpha \Delta t$ for some time step $\Delta t>0$.
Note that technically the subspace is a Krylov space only if $t_\alpha = \alpha \Delta t$, since then the operators generating the subspace are powers of the unitary $e^{-i \Delta t \hat{H}}$. In the remainder of the Section, we will assume this construction is used.
The same basic idea is known elsewhere in the literature as \technical{quantum subspace diagonalization}~\cite{epperly2022theory} and \technical{variational quantum phase estimation}~\cite{klymko2022real}; herein we will use QFD since that was the name conferred in the original paper proposing this method~\cite{parrish2019quantum}.
A natural and compelling generalisation of QFD is the multireference selected quantum Krylov (MRSQK) algorithm~\cite{stair2020multireference}, where time evolution is applied to a set of initial states,
\begin{equation}
| {\bf{v}}_{\alpha k} \rangle = e^{-i t_\alpha \hat{H}} | {\bf{v}}_{0 k} \rangle
\;,\; \alpha=0 \dots n-1
\;,\; k = 0 \dots n_{k}-1
\;.
\end{equation}
QFD is a particularly compelling method because it possesses a formal error analysis~\cite{epperly2022theory}, even accounting for noise on the quantum device.

Before discussing that analysis, we mention a common pitfall in thinking about the error of QFD.
For small $\Delta t$, the QFD Krylov space is approximately equal to the classical Krylov space (which we use to refer to the Krylov space generated by powers of the Hamiltonian) up to high-order terms,
\begin{equation}
\label{eq:intuitive_krylov}
\sum_\beta c_\beta | {\bf{v}}_{\beta} \rangle 
= 
\sum_{\alpha=0}^{n-1} 
\left[ \sum_{\beta=0}^{n-1} M_{\alpha\beta} c_\beta \right] 
\hat{H}^\alpha | {\bf{v}}_0 \rangle + \order(\Delta t^n)
\;,\; M_{\alpha\beta} = \frac{(-i \beta \Delta t)^\alpha}{\alpha !}
\;,
\end{equation}
as the matrix $M$ is invertible (it is the product of a Vandermonde and a diagonal matrix).
Since the residual terms in $\order(\Delta t^n)$ vanish as $\Delta t$ is taken to zero, one might be tempted to run the algorithm for some very small $\Delta t$ since the errors resulting from classical Krylov methods provably converge~\cite{kaniel1966estimates,paige1971computation,saad1980rates}, see Section \ref{sec:krylov_space}.
However, for very small $\Delta t$, the QFD subspace approaches linear dependence (time-evolution operators approximate the identity operator closely). Such a scenario induces ill-conditioning and thus complicated tradeoffs in the choice of $\Delta t$.

However, those tradeoffs turn out to be illusory, essentially because the residual term in \eqref{eq:intuitive_krylov} does not actually reduce the quality of the lowest-energy state in the QFD subspace, so $\Delta t$ does not have to be pushed as close to zero as possible.
This fact was demonstrated in Epperly \etal~\cite{epperly2022theory} (in particular Theorem 3.1). Choosing $\Delta t = \frac{\pi}{\Delta \energy_L}$ and $t_\alpha = \frac{\pi \alpha}{\Delta \energy_L}$, $\alpha = - n/2 \dots n/2$, where $\Delta \energy_L$ is the energy difference between the ground state and some excited state with index $0 < L < D-1$, one has 
\begin{equation}
0 
\leq 
\subspaceenergy^{(n)}_0 - \energy_0 
\leq
\left( 1 + \frac{\pi \Delta \energy_1}{\Delta \energy_L} \right)^{-n} 
  8 \sum_{\mu = 1}^L     \Delta \energy_\mu \frac{w_\mu}{w_0} 
+ 2 \sum_{\mu=L+1}^{D-1} \Delta \energy_\mu \frac{w_\mu}{w_0} \; ,
\end{equation}
$w_\mu = | \langle \wfn_\mu | {\bf{v}}_0 \rangle |^2$.
The error $\subspaceenergy^{(n)}_0 - \energy_0$ contains
two terms, arising from the components of $|{\bf{v}}_0 \rangle$ on eigenstates with energies respectively below and above $\energy_L$.
The first term decreases exponentially with subspace dimension $n$, albeit at a rate determined by the ratio $\frac{\Delta \energy_1}{\Delta \energy_L}$, meaning that larger QFD subspaces remove unwanted low-energy components of $|{\bf{v}}_0 \rangle$ as long as the system has a non-zero gap. The second is independent of $n$, meaning that high-energy components of $|{\bf{v}}_0 \rangle$ cannot be eliminated by increasing the dimension of the QFD subspace, but only by choosing a finer mesh of times $t_\alpha$ (equivalent to increasing $L$) or a trial state with support on the low-energy subspace ($w_\mu \simeq 0$ for $\mu > L$).

By choosing $L=D-1$ we get an upper bound on the ground state energy approximation that is free from conditions on the support of the trial state, obtaining
\begin{equation}
0 
\leq 
\subspaceenergy^{(n)}_0 - \energy_0 
\leq
\left( 1 + \frac{\pi \Delta \energy_1}{\Delta \energy_{D-1}} \right)^{-n} 
  8 \sum_{\mu = 1}^{D-1} \Delta \energy_\mu \frac{w_\mu}{w_0} .
\end{equation}
This shows that the smallest $\Delta t$ should ever be chosen is ${\Delta t = \frac{\pi}{\Delta \energy_{D-1}} \ge \frac{\pi}{\|H\|}}$, where $\|\cdot\|$ is the spectral norm (but could in practice be replaced by more easily computed upper bounds).
The convergence of the method is thus not improved by making $\Delta t$ smaller past this point, even if ill-conditioning were not a concern. We note that, in the presence of noise and imperfect time-evolution simulation, the above choice of $\Delta t$ may no longer be optimal, and it should instead be chosen heuristically.
Roughly, the relation of the analysis of Epperly~\etal~\cite{epperly2022theory} to the classical analysis of Kaniel, Paige, and Saad~\cite{kaniel1966estimates,paige1971computation,saad1980rates} is that while the latter is based on finding approximate projectors among polynomials of the Hamiltonian (which are elements of the classical Krylov space), the former is based on finding approximate projectors among trigonometric polynomials of the Hamiltonian (which are elements of the QFD space), without any need to approximate to the classical Krylov space along the way.
This analysis is discussed further in Subsection \ref{sec:shot_noise}.

\begin{figure}[t!]
\centering
\includegraphics[width=0.4\textwidth]{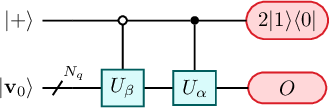}
\caption{Quantum circuits used in the QFD method to compute $O_{\alpha\beta} = \langle \vett{v}_\alpha | \hat{O} | \vett{v}_\beta \rangle$. An ancilla is prepared in the state  $|+\rangle = (|0\rangle+|1\rangle)/\sqrt{2}$, and the unitary transformation $|0\rangle\langle0| \otimes \hat{U}_{\beta} + |1\rangle\langle1| \otimes \hat{U}_{\alpha}$ is applied, where $\hat{U}_{\alpha} = e^{-i t_\alpha \hat{H}}$ and $\hat{U}_\beta = e^{-i t_\beta \hat{H}}$ are time-evolution unitaries (or approximations thereof) controlled by the state of the ancilla. The operator $2|1\rangle\langle0| \otimes \hat{O}$ is then measured, yielding an estimate of $O_{\alpha\beta}$. For $\hat{O} = \identity/\hat{H}$ one obtains overlap/Hamiltonian matrix elements respectively.
}
\label{fig:qfd}
\end{figure}

From a quantum computing perspective, QFD is appealing because it is based on real-time evolution, an operation that can be simulated on a quantum computer with polynomial cost and controllable accuracy (more formally, it lies in the
complexity class BQP). The method lies between the variational quantum eigensolver (VQE) and the phase estimation algorithm (PEA) in terms of required quantum circuit resources and conceptual simplicity.
Compared to the Chebyshev quantum Krylov method discussed in Sec.~\ref{sec:chebyshev_lanczos}, QFD is appealing because time-evolutions admit relatively near-term-friendly approximate implementations.

On the other hand, QFD leads to an eigenvalue equation that is in general ill-conditioned in practice, much like in classical Krylov methods (see Subsection~\ref{sec:krylov_space}), although the Beckermann-Townsend inequality~\cite{beckermann2019bounds} does not apply since in QFD the Krylov space is not generated by powers of a Hermitian matrix.
This ill-conditioning is aggravated by the presence of decoherence and shot noise, although it can be ameliorated by a thresholding procedure~\cite{epperly2022theory}, as discussed in Subsection \ref{sec:shot_noise}. Furthermore, time evolution is an expensive operation for the electronic structure Hamiltonian, as noted in Subsection \ref{sec:fermions_second}, and the implementation of QFD in general requires controlled application of a time-evolution circuit, as sketched in Fig.~\ref{fig:qfd}.

In addition to the theoretical analysis in~\cite{epperly2022theory}, the accuracy of QFD and the dependence of QFD energies on time step, subspace dimension, and shot noise, were extensively analyzed numerically by Klymko \etal and Shen \etal~\cite{klymko2022real,shen2023real}. 
These papers also introduced and analyzed the so-called unitary version of QFD, where the GEEV can be written in a unitary form, requiring only measurement of overlap matrix elements, and restoring the Toeplitz structure of the GEEV in the case of Trotterized time evolution with a uniform grid. 
Additionally, QFD's asymptotic computational cost was reduced by means of low-rank approximations by Cohn \etal~\cite{cohn2021quantum} and stochastic compilation of the time evolution subroutine was proposed by Stair \etal~\cite{stair2023stochastic}. In Subsection \ref{sec:economisation} we examine the issue of controlled time evolution, and describe procedures to bypass this expensive operation under suitable assumptions.

\subsubsection{Quantum Lanczos based on imaginary-time evolution}

The quantum Lanczos (QLanczos) method constructs a subspace of vectors by applying an imaginary-time evolution (ITE) operator to an initial state,
\begin{equation}
| {\bf{v}}_\alpha \rangle = \frac{ e^{- \alpha \Delta \tau \hat{H}} | {\bf{v}}_0 \rangle }{ \| e^{- \alpha \Delta \tau \hat{H}} | {\bf{v}}_0 \rangle \| }
\;,\;  \alpha=0 \dots n-1 \;,
\end{equation}
where $\Delta \tau > 0$ is an imaginary-time step. Like QFD, QLanczos is formally connected with the classical Krylov space in the sense of Eq.~\eqref{eq:intuitive_krylov}.
However, an important difference between QFD and QLanczos is that the latter is based on an operation, the ITE, that is neither unitary nor linear. On a fault-tolerant quantum device, non-linear operations can be implemented using techniques like quantum signal processing~\cite{low2017optimal,dong2021efficient,wang2022energy,lin2022lecture}.
On a near-term device, ITE can be approximated by variational ansatz-based simulations~\cite{mcardle2019variational}, or by the quantum ITE (QITE) algorithm~\cite{motta2020determining}. The latter approximates a step of ITE on a wavefunction $\calcwfn$ by the action of a quantum circuit $\hat{U}(\theta)$,
\begin{equation}
\frac{ e^{- \Delta \tau \hat{H}} |  \calcwfn \rangle }{ \| e^{-\Delta \tau \hat{H}} \calcwfn \| } \simeq \hat{U}(\boldsymbol{\theta}) | \calcwfn \rangle
\;.
\end{equation}
Both members of the equation above are expanded to the first order around the identity,
\begin{equation}
\Big[ \identity - \Delta \tau \big( \hat{H} - E_\calcwfn \big) \Big] | \calcwfn \rangle
\simeq
\Big[ \identity + i \sum_m \theta_m \hat{G}_m \Big] | \calcwfn \rangle
\;,
\end{equation}
where $E_\calcwfn = \langle \calcwfn | \hat{H} | \calcwfn \rangle$. The parameters $\theta_m$ are determined solving the equation $A \boldsymbol{\theta} = \boldsymbol{b}$, where $A_{mn} = \langle \calcwfn | \{ \hat{G}_m , \hat{G}_n \} | \calcwfn \rangle$ and $b_m = - i \Delta \tau \langle \calcwfn | [ \hat{H}, \hat{G}_m ] | \calcwfn \rangle$.
For a $k$-local spin Hamiltonian~\cite{motta2020determining}, a natural choice is $\hat{U}(\boldsymbol{\theta}) = \prod_m \exp( i x_m \sigma_m)$, where $\sigma_m$ are spatially local Pauli operators. For electronic systems~\cite{gomes2020}, a compelling alternative is the use of one- and two-body fermionic operators,
\begin{equation}
\hat{U}(\boldsymbol{\theta}) = \exp\left( \sum_{ai} \theta^a_i \crt{a\sigma} \dst{i\sigma} + \sum_{aibj} \theta^{ab}_{ij} \crt{a\sigma} \crt{b\tau} \dst{j\tau} \dst{i\sigma} - \mathrm{h.c.} \right)
\;.
\end{equation}
A limitation of QITE is the growth of circuit depth with number of imaginary-time steps, which can be ameliorated e.g. using the step-merged approach of Gomes \etal~\cite{gomes2020}, where the approximation $\hat{U}(\boldsymbol{\theta}_1) \hat{U}(\boldsymbol{\theta}_2) \simeq \hat{U}(\boldsymbol{\theta}_1+\boldsymbol{\theta}_2)$ is made, in order to implement QITE with a fixed-depth circuit, albeit with a possible loss of accuracy.

On the other hand, QLanczos does not require ancillae and controlled operations, because matrix elements can be written in terms of norms and expectation values,
\begin{equation}
\langle \vett{v}_\alpha | \vett{v}_\beta \rangle = \frac{n^2_{\frac{\alpha+\beta}{2}}}{n_\alpha n_\beta}
\;,\;
\langle \vett{v}_\alpha | \hat{H} | \vett{v}_\beta \rangle = \frac{n^2_{\frac{\alpha+\beta}{2}}}{n_\alpha n_\beta}
h_{\frac{\alpha+\beta}{2}}
\;,\;
\end{equation}
with $n_\alpha = \| e^{- \alpha \Delta \tau \hat{H}} | {\bf{v}}_0 \rangle \|$
and $h_\alpha = \langle \vett{v}_\alpha | \hat{H} | \vett{v}_\alpha \rangle$, and these quantities can be evaluated without ancillae and controlled operations as the ITE unfolds~\cite{motta2020determining}.

\subsection{Alternative approaches}
\label{sec:alternative}

In the previous Subsections, we discussed representative examples of QSMs. Research in the design, implementation, and refinement of QSMs extends beyond these classes of methods. 
In this Subsection, we describe alternative QSMs.

\subsubsection{Subspace-based variational quantum simulations} 

Electronic ground- and excited-states can be approximated by variational quantum simulations, exemplified by the variational quantum eigensolver (VQE)~\cite{peruzzo2014variational}, wherein the ground-state wavefunction, $|\wfn_0 \rangle$, and energy, $\energy_{0}$, are approximated by variationally optimising a parameterised wavefunction ansatz $|\calcwfn(\boldsymbol{\theta})\rangle$,
\begin{equation}
E_{\mathrm{VQE}} 
=
\min_{\boldsymbol{\theta}} \, \langle \calcwfn(\boldsymbol{\theta})|H| \calcwfn(\boldsymbol{\theta})\rangle
=
\min_{\boldsymbol{\theta}} \, E(\boldsymbol{\theta})
\;.
\end{equation}
The energy $E(\boldsymbol{\theta})$ is evaluated on a quantum computer, and parameters $\boldsymbol{\theta}$ are optimised on a classical computer.
However, the quality of a VQE calculation depends on the ansatz and the convergence of the optimisation procedure. Literature~\cite{ollitrault2020quantum, Rice2021, Barkoutsos2018, Grimsley2019, Sokolov2020ooUCCSD, Gao2021} indicates that VQE applied to small active spaces can yield energies close to those of CASCI, but also that it can require a computational cost prohibitive for near-term devices~\cite{Motta2020transcorrelated} and feature symmetry-breaking, non-differentiable potential-energy curves, and exponentially expensive parameter optimisation when hardware-efficient ansatzes are used~\cite{d2023challenges}.
QSMs can be constructed starting from a wavefunction generated by a VQE simulation, making them a natural and compelling approach to enhance the quality of variational simulations.

\begin{figure}[t!]
\centering
\includegraphics[width=0.65\textwidth]{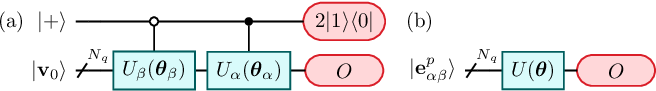}
\caption{Quantum circuits used in the (a) NO-VQE and (b) MC-VQE methods. NO-VQE requires an ancilla and controlled unitaries, whereas MC-VQE requires the preparation of superposition states of the form $|\vett{e}^p_{\alpha\beta}\rangle = (|\vett{e}_\alpha \rangle + i^p |\vett{e}_\beta \rangle)/\sqrt{2}$ with $p=0,1,2,3$. Measuring the operator $\hat{O} = \identity/\hat{H}$ in NO-VQE yields overlap/Hamiltonian matrix elements respectively. Within MC-VQE, the overlap matrix is equal to the identity due to the orthonormality of the states $|\vett{e}_\alpha\rangle$, so that only Hamiltonian measurements are required.
}
\label{fig:mcvqe}
\end{figure}

An example of this improvement is the multistate-contracted variational quantum eigensolver (MC-VQE)~\cite{parrish2021analytical}, where a parametrised Ansatz $\hat{U}(\boldsymbol{\theta})$ is applied to a linear combination of states $| \vett{e}_\alpha \rangle$ that are e.g. qubit representations of electronic configurations,
\begin{equation}
| \calcwfn(\boldsymbol{\theta},\vett{c}) \rangle = \hat{U}(\boldsymbol{\theta}) \sum_\alpha c_\alpha | \vett{e}_\alpha \rangle 
\;.
\end{equation}
The parameters $\boldsymbol{\theta},\vett{c}$ are jointly optimised. In particular, optimisation of coefficients $\vett{c}$ for a given $\boldsymbol{\theta}$ is a ground-state search in the subspace spanned by the states $\hat{U}(\boldsymbol{\theta})| \vett{e}_\alpha \rangle$.
In a similar vein, the non-orthogonal VQE (NO-VQE) method~\cite{huggins2020non} proposes to construct the variational ansatz
\begin{equation}
| \calcwfn(\boldsymbol{\theta},\vett{c}) \rangle = \sum_\alpha c_\alpha | \calcwfn_\alpha(\boldsymbol{\theta}_\alpha) \rangle
\;,\;
| \calcwfn_\alpha(\boldsymbol{\theta}_\alpha) \rangle
=
\hat{U}_\alpha(\boldsymbol{\theta}_\alpha) | \vett{v}_0 \rangle\;,
\end{equation}
where the parameters $\boldsymbol{\theta}_\alpha$ are optimised in an outer loop and the coefficients $c_\alpha$ are determined by a ground-state search in the subspace spanned by the states $| \calcwfn_\alpha(\boldsymbol{\theta}_\alpha) \rangle$. NO-VQE can be considered a generalisation of MC-VQE; on the other hand, it requires a Hadamard test (circuit in Fig.~\ref{fig:mcvqe}a),
whereas the quantum circuits required by MC-VQE are of the form in Fig.~\ref{fig:mcvqe}b (no Hadamard test). The non-orthogonal quantum eigensolver (NOQE) is a modification~\cite{baek2022say} of NO-VQE that relies on a quantum computer to synthesise wavefunctions of the form $|\calcwfn_\alpha(\boldsymbol{\theta}_\alpha) \rangle$, where parameters are suggested by classical perturbative calculations on top of spin-unrestricted Slater determinants, and to compute overlap and Hamiltonian matrices to solve for Hamiltonian eigenstates in the subspace spanned by the states $|\calcwfn_\alpha(\boldsymbol{\theta}_\alpha) \rangle$, without further parameter optimisation. NOQE can be considered a quantum transposition of the classical non-orthogonal configuration interaction (NOCI) method~\cite{sundstrom2014non,yost2016size}.

\subsubsection{Quantum selected CI} 
Quantum computing algorithms can be used to inform classical subspace calculations.
An interesting example is the quantum selected CI method by Kanno \etal~\cite{kanno2023quantum}, where an approximation $\tilde{\calcwfn}$ for the ground state of a quantum system is prepared on a quantum computer, for example using a variational method. Measuring a register of qubits prepared in $| \calcwfn \rangle = \sum_{{\bf{x}}} c_{{\bf{x}}} |{{\bf{x}}} \rangle$ yields a bitstring ${{\bf{x}}}$ with probability ${p_{{\bf{x}}} = |c_{{\bf{x}}}|^2}$. Executing the measurement $N$ times yields bitstrings ${{{\bf{x}}}_m, m=1 \dots N}$. Having sampled those bitstrings, one can construct a matrix $H_{mn} = \langle {{\bf{x}}}_m | H | {{\bf{x}}}_n \rangle$ on a classical computer using the Slater-Condon rules, and diagonalise it on a classical computer, to find another approximation $| \tilde{\calcwfn} \rangle = \sum_{m=1}^N \tilde{c}_{{\bf{x}}_m} |{{\bf{x}}}_m \rangle$ for the ground state. This algorithm is variational, and intrinsically resistant to decoherence. 
While the probability distribution $p_{{\bf{x}}}$ of an actual simulation may differ from $|c_{{\bf{x}}}|^2$ due to decoherence, $| \tilde{\calcwfn} \rangle$ is prepared on a classical computer. This algorithm presents several challenges and opportunities for refinement. 
First, ensuring the generation of continuous potential energy curves: when independent finite samples are drawn from the probability distribution $p_{{\bf{x}}}$ along a potential energy curve, discrepancies in the corresponding bitstrings can result in irregular potential energy curves. Second, addressing sampling inefficiency: in cases involving dynamically correlated wavefunctions, $p_{{\bf{x}}}$ tends to be concentrated around specific bitstrings, leading to repetitive sampling and reduced efficiency. Third, developing cost-effective and systematic approaches to modify the probability distribution $p_{{\bf{x}}}$ in such a way as to prepare low-energy states $| \tilde{\Phi} \rangle$.

\subsubsection{Eigenvalue continuation}

Eigenvector Continuation (EC) as a QSM method that considers a family of Hamiltonians, $\hat{H}_{\boldsymbol{\lambda}}$, that depend on one or more parameters ${\boldsymbol{\lambda}}$ \cite{francis2022subspace,mejuto2023quantum}. This situation is encountered in ES when ${\boldsymbol{\lambda}}$ is e.g. a set of nuclear coordinates or an external field applied to a molecule. A set of approximations $\calcwfn_{{\boldsymbol{\lambda}}_\alpha}$ for low-energy states of the Hamiltonian at different points ${{\boldsymbol{\lambda}}_\alpha}$ of the parameter space are chosen as the subspace basis. In other words, the ground state of $\hat{H}_{\boldsymbol{\lambda}}$ is approximated as a linear combinaton of the form $\sum_\alpha c_\alpha | \calcwfn_{{\boldsymbol{\lambda}}_\alpha}  \rangle$, and the coefficients
$c_\alpha$ are determined using the standard QSE method. EC allows to construct a subspace using physics-informed basis states, thereby allowing accurate calculations of low-energy eigenvalues.
In general, EC requires a Hadamard test to measure overlap and Hamiltonan matrices (see Section~\ref{sec:economisation} for further discussion). For accurate results, the Hamiltonian and overlap matrices need to be measured accurately, and the overlap matrix needs to be well-conditioned thanks to a suitable choice of the ${\boldsymbol{\lambda}}$ parameters, or regularized via e.g. a threshold procedure (see Section~\ref{sec:shot_noise} for a discussion).

\section{Applications of quantum subspace methods}
\label{sec:applications}

Some of the most natural and widespread applications of QSMs are the calculations of ground and low-lying excited electronic states. QSMs can also be used to study various phenomena, assuming that these can be accurately approximated using electronic eigenfunctions from the chosen subspace. In this Section, we illustrate the description of dynamical electron correlation, the fast-forwarding of time evolution, and the computation of frequency-dependent response functions.

\subsection{Dynamical electronic correlation}

Near-term quantum devices are predominantly used to carry out active-space simulations in the context of electronic structure (see Section~\ref{sec:correlation} and Fig.~\ref{fig:active}). While active-space simulations can describe forms of static electronic correlation, they cannot recover the dynamical correlation (described in Section \ref{sec:correlation}). A similar situation is encountered in classical ES methods when an active-space calculation is conducted, e.g. with CASCI or CASSCF. Retrieval of dynamical correlation is then carried out by combining the active-space method with a low-level calculation on the full basis set, e.g. second-order perturbation theory.
While a straightforward possibility is to simulate a realistic basis set with larger quantum simulations, this strategy is not always compatible with near-term devices~\cite{Motta2020transcorrelated}.

The possibility of extending QSMs with post-processing on classical computers to account for dynamical electronic correlation has been recognised and is an active research area. In this section, we will present illustrative studies. First, starting from the Hamiltonian Eq.~\eqref{eq:es_hamiltonian}, we divide orbitals into core (indices $i,j,k \in C$), active (indices $t,u,v,w \in A$), and virtual (indices $a,b,c \in V$).
For simplicity, we consider core orbitals to be frozen, i.e., they are doubly occupied and electrons in them are never excited into other orbitals. Since the Hamiltonian with frozen-core orbitals can be transformed into one containing only active and virtual orbitals, we will now exclude the core space. Active orbitals are considered crucial because electrons in these orbitals exhibit static correlation, making them the focus of treatment on quantum computers. Virtual orbitals contribute to additional dynamical correlation.

A hybrid quantum-classical technique for dynamical correlation is virtual QSE~\cite{takeshita2020increasing}. This method starts from an approximation for the ground state of the system as $|\calcwfn,\vac\rangle$, where $|\calcwfn\rangle$ is an active-space wavefunction and $|\vac \rangle$ the vacuum state for virtual orbitals. Then, it defines a subspace spanned by states of the form $\hat{O}_\alpha |\calcwfn,\vac\rangle$, where $\hat{O}_\alpha$ belongs to the set
\begin{equation}
D = \Big\{ \crt{a\sigma} \dst{t\sigma} , \crt{a\sigma} \crt{b\tau} \dst{u\tau} \dst{t\sigma} \;,\; uv \in A , ab \in V, \tau\sigma \in \{ \uparrow,\downarrow \} \Big\}
\;.
\end{equation}
Overlap and Hamiltonian matrices are formed as in standard QSE. Evaluating these matrices requires tracing out virtual degrees of freedom using Wick's theorem~\cite{wick1950evaluation}, e.g.
\begin{equation}
S_{bu\tau,at\sigma} = \langle \calcwfn , \vac | \crt{u\tau} \dst{b\tau} \crt{a\sigma} \dst{t\sigma} | \calcwfn , \vac \rangle = \delta_{ab} \delta_{\sigma\tau} 
\langle \calcwfn | \crt{u\sigma} \dst{t\sigma} | \calcwfn \rangle
\;,
\end{equation}
where $\langle \calcwfn | \crt{u\sigma} \dst{t\sigma} | \calcwfn \rangle$ is an element of the active-space one-body density matrix. Detailed formulas and representative applications are given in Refs.~\cite{urbanek2020chemistry,takeshita2020increasing}. Virtual QSE allows accounting for dynamical correlation by allowing single and double excitations in the virtual space.
However, the quality of the resulting energies is comparable to that of an MRCISD calculation, and it requires computing high-order RDMs.

An alternative to virtual QSE is second-order perturbation theory~\cite{tammaro2023n}. This formalism is a quantum transposition of classical $N$-electron valence perturbation theory (NEVPT2)~\cite{angeli2001introduction,angeli2001n,sokolov2016time,sokolov2017time}.
The Hamiltonian is partitioned as $\hat{H} = \hat{H}_D + \hat{V}$, where $\hat{H}_D$ is the Dyall Hamiltonian~\cite{dyall1995choice}, i.e. the sum between the active space Born-Oppenheimer Hamiltonian and the restriction of the Fock operator to the external space, and $\hat{V} = \hat{H} - \hat{H}_D$ is treated as a perturbation. The second-order energy contribution can be written as
\begin{equation}
\label{eq:nevpt2}
- \Delta E_{\mathrm{PT2}} = \sum_{\nu \neq 0} \frac{| \langle \wfn_\nu | \hat{V} | \wfn_0 \rangle |^2}{\energy_\nu - \energy_0} \; ,
\end{equation}
where $(\wfn_\nu,\energy_\nu)$ are the eigenpairs of the Dyall Hamiltonian, and $\nu=0$ labels the ground state. Eq.~\eqref{eq:nevpt2} yields the exact (or uncontracted) NEVPT2. Implementing uncontracted NEVPT2 has a combinatorial cost with active-space size, due to the summation over excited states. This limitation can be remedied using strongly-contracted NEVPT2~\cite{krompiec2022strongly}, which requires high-order ground-state RDMs, or partially-contracted NEVPT2~\cite{tammaro2023n}, which approximates the sum over excited states. 
The latter is very naturally interfaced with QSMs. One can observe that the action of the perturbation over the ground state,
\begin{equation}
\label{eq:nevpt3}
\hat{V} | \calcwfn \rangle 
= 
\Bigg[ \sum_{\substack{a \\ \sigma}} \crt{a\sigma} \hat{O}^{(1)}_{a,\sigma} + \sum_{\substack{a<b \\ \sigma}} \crt{a\sigma}\crt{b\sigma} \hat{O}^{(2)}_{ab,\sigma} 
+ 
\sum_{ab} \crt{a\uparrow} \crt{b\downarrow} \hat{O}^{(3)}_{ab} \Bigg] | \Psi_0 \rangle
\;,
\end{equation}
removes a particle with spin $\sigma$, or two particles with identical spins $\sigma$, or two particles with opposite spin from the active space, exciting the remaining electrons through the action of suitable operators $\hat{O}^{(1)}_{a,\sigma}$, $\hat{O}^{(2)}_{ab,\sigma}$, $\hat{O}^{(3)}_{ab}$.
Therefore, one can restrict the summation in Eq.~\eqref{eq:nevpt2} to excited states with $(N_\uparrow-\Delta N_\uparrow, N_\downarrow-\Delta N_\downarrow)$ particles, where $(N_\uparrow, N_\downarrow)$ is the number of electrons in the active-space ground-state wavefunction and $\Delta N_\uparrow,\Delta N_\downarrow=0,1,2$ and $\Delta N_\uparrow+\Delta N_\downarrow=1,2$. The latter can be evaluated, for example, with active-space QSE calculation~\cite{tammaro2023n}, allowing for recovery of dynamical correlation with accuracy between strongly-contracted and uncontracted NEVPT2.

\subsection{Response functions}

An important application of QSMs is the computation of frequency-dependent response functions,
\begin{equation}
\label{eq:cabw}
C_{AB}(\omega) = \int_{-\infty}^{\infty} \frac{dt}{2\pi} \, e^{i\omega t} \, 
\langle \Psi_0 | \hat{A} e^{-it (\hat{H}-E_0)} \hat{B} | \Psi_0 \rangle
\;,
\end{equation}
where $\hat{A},\hat{B}$ are two operators. 

When $\hat{A} = \hat{B} = \vett{n} \cdot \hat{\boldsymbol{\mu}}$ is the component of the dipole moment along direction $\vett{n}$, Eq.~\eqref{eq:cabw} is called the dipole spectral function, and characterises the absorption of ultraviolet and visible light by a molecule~\cite{gordon1965molecular,boulet1982short,clerk2010introduction,vitale2015anharmonic,nascimento2016linear,goings2018real,li2020real}. In solid-state systems, when $\hat{A} = \dst{ \vett{p} \sigma}$ and $\hat{B} = \crt{ \vett{p} \sigma}$ for a plane-wave with momentum $\vett{p}$ and spin $\sigma$, Eq.~\eqref{eq:cabw} is called the quasiparticle spectral function and it is used to compute the cross-section of angular-resolved photoemission spectroscopy (ARPES) experiments~\cite{damascelli2004probing,nazarov2009quantum,patterson2010coherent,weathersby2015mega,fischer2016invited,buzzi2018probing}. Eq.~\eqref{eq:cabw} is therefore physically relevant, and challenging to compute as it involves a time-evolution operator. The Lehmann representation of Eq.~\eqref{eq:cabw},
\begin{equation}
\label{eq:cabw_lehmann}
C_{AB}(\omega) = \sum_\mu \delta(\omega-\Delta \energy_\mu) \langle \wfn_0 | \hat{A} | \wfn_\mu \rangle \langle \wfn_\mu | \hat{B} | \wfn_0 \rangle
\;,
\end{equation}
indicates that QSMs are natural strategies to approximate frequency-dependent response functions, by simply replacing the exact Hamiltonian eigenfunctions in Eq.~\eqref{eq:cabw_lehmann} with the approximations yielded by a quantum subspace calculation~\cite{colless2018computation,rizzo2022one,jamet2022quantum,motta2023quantum}.

\subsection{Additional applications}

\subsubsection{Fast-forwarding time evolution}

An important goal of quantum simulations is to understand how the properties of a physical system evolve over time. In the context of ES, the non-equilibrium character of time-dependent Hamiltonians (e.g., in the presence of an external, oscillating electromagnetic field) requires solving the time-dependent Schr\"{o}dinger equation
\begin{equation}
\label{eq:tdse}
i\hbar \frac{d}{dt} |\calcwfn (t)\rangle = \hat{H}(t) |\calcwfn (t)\rangle
\end{equation}
to access frequency-dependent polarisabilities and other optical properties~\cite{nascimento2016linear,goings2018real,li2020real,mccullough1969quantum,kosloff1988time,huber2011explicitly,kristiansen2020numerical}.

The simulation of Eq.~\eqref{eq:tdse} is a particularly compelling application for a quantum computer, as this problem lies in the complexity class BQP, meaning that quantum computers can approximate the state $|\calcwfn(t)\rangle$ with accuracy $\varepsilon$ at a polynomial cost in $t, \varepsilon^{-1}$, and system size. While the complexity of simulating $|\calcwfn(t)\rangle$ on a quantum computer is linear in time for a generic quantum system \cite{berry2007efficient,childs2009limitations}, there exist exceptions to this lower bound: certain Hamiltonians that can be efficiently diagonalized, e.g. frustration-free Hamiltonians \cite{gu2021fast} and free fermions (see Section \ref{sec:fermions_second}). Time evolution under those Hamiltonians can be simulated exactly at a cost independent of $t$.

For a generic quantum system, one can approximately solve for $|\calcwfn(t)\rangle$ by \technical{fast-forwarding} time evolution. Within the fast-forwarding approach, one approximately diagonalizes a short-time propagator $\hat{U}_{\Delta t} = e^{-i \Delta t \hat{H}}$ and uses knowledge of its eigenpairs to approximate a generic-time propagator $\hat{U}_{t}$ \cite{atia2017fast,cirstoiu2020variational}. While the error of the fast-forwarding procedure scales linearly with time \cite{cirstoiu2020variational,commeau2020variational}, accurate approximations may be obtained in specific situations: for example, if $|\calcwfn (0)\rangle$ has support over $n$ Hamiltonian eigenstates, $|\calcwfn (t)\rangle$ can be simulated exactly in the $n$-dimensional Hamiltonian Krylov space.

QSMs offer a natural avenue to study fast-forwarded time evolution \cite{cortes2022fast,klymko2022real}. Examples of the use of QSMs to fast-forward time evolution are the Subspace Variational Quantum Simulator (SVQS)~\cite{heya2019subspace}, fixed-state Variational Fast Forwarding (fs-VFF)~\cite{gibbs2021long}, and Classical-Quantum Fast Forwarding (CQFF)~\cite{lim2021fast}.

\section{Implementation}
\label{sec:implementation}

In this Section, we list some implementations on quantum hardware, outline important challenges posed by the hardware implementation of QSMs, and discuss recent research aimed at addressing these challenges.

\subsection{Hardware implementations}

Recent hardware simulations of QSMs for electronic structure applications are listed in Table \ref{table:demos}.
For each simulation, we provide details on the studied system and properties the QSM  used, the number of qubits, the depth and number of $\mathsf{cNOT}$ gates in the simulated circuits, along with the hardware used. The simulations listed in the table predominantly employed JW mapping as described in Subsection \ref{sec:fermions_second}, except for Ref.~\cite{colless2018computation,tammaro2023n,gao2021applications}, which used parity mapping and qubit-tapering techniques~\cite{bravyi2017tapering,setia2020reducing}, and Ref.~\cite{huang2023quantum}, which used a first-quantisation mapping. Refs.~\cite{motta2023quantum,castellanos2023quantum} used a qubit-reduction technique called entanglement forging~\cite{eddins2021doubling}.

Simulations display improving trends in qubit number, circuit depth, and number of $\mathsf{cNOT}$ gates. To make substantial progress in this domain requires (in addition to the continuous refinement of hardware manufacturing and control) a deeper understanding and effective resolution of the challenges posed by the implementation of QSMs on quantum hardware. The remainder of this section addresses some of these challenges and the corresponding research efforts.

\begin{table}
\begin{tabular}{ccccccc}
\hline\hline
Reference & System & Properties & Algorithm & $N_q$ & ($\mathsf{depth},\mathsf{cNOT}$) & Hardware \\
\hline
\cite{colless2018computation} & \ce{H2}         & gs/es    & MRCISD & 2 & (3,1)     & sc \\
\cite{gao2021applications}    & OLED            & gs/es    & qEOM   & 2 & (3,2)     & sc \\
\cite{dhawan2023quantum}      & \ce{H2}         & gs/es    & MRCISD & 4 & (9,6)     & sc \\
\cite{cohn2021quantum}        & stilbene        & gs/es    & QFD    & 5 & (102,78)  & sc \\
\cite{tammaro2023n}           & \ce{OH/OH-}     & gs       & MRCISD & 6 & (20,15)   & sc \\
\cite{Huang2022simulating}    & NV              & gs/es    & MRCISD & 4 & (7,2)     & sc \\
\cite{huang2023quantum}       & NV              & gs/es    & MRCISD & 4 & (20,14)   & sc \\
\cite{motta2023quantum}       & \ce{H3S+}       & gs/es/rf & MRCISD & 6 & (21,19)   & sc \\
\cite{Khan2022sim}            & \ce{CH4}        & gs/es/rf & MRCIS  & 6  & (12,7)    & ti \\
\cite{castellanos2023quantum} & \ce{C4H4N2}     & gs/es    & MRCISD & 8 & (32,31)   & sc \\
\hline\hline
\end{tabular}
\caption{List of hardware simulations of QSMs for electronic structure applications. The abbreviations ``gs/es/rf'' stand for ground state/excited states/response functions, the abbreviations ``sc/ti'' stand for superconducting/trapped-ion, and the abbreviations ``OLED/NV'' stand for organic light-emitting diode and nitrogen-vacancy. Algorithm abbreviations are defined in the main text.}
\label{table:demos}
\end{table}

\subsection{Effect of shot noise and decoherence}
\label{sec:shot_noise}

QSMs involve extracting matrix elements of the Hamiltonian between pairs of subspace basis states, as well as inner products of subspace basis states (see Fig.~\ref{fig:qsms}).
As discussed in Subsection \ref{sec:measurements}, due to the probabilistic nature of quantum operations, repeating quantum measurements on multiple copies of the quantum circuit is necessary for attaining accurate outcomes and probabilities.
Therefore, unlike classical SMs, where numerical errors only arise due to machine precision, QSMs are inherently accompanied by errors originating from a finite number of samples.
Moreover, as mentioned above, QSMs often do not employ orthogonal subspace bases, leading to ill-conditioning in the classical post-processing.
Although this also appears in some variants of classical SMs (see Subsection~\ref{sec:krylov_space}), in the noisier context of QSMs it can have the effect of amplifying errors in the generalised eigenvalue problem unless handled carefully.
The relationship between sampling noise (or noise from other sources in a quantum algorithm) and its effects on eigenvalue estimates from QSMs is a subject of intense research~\cite{epperly2022theory,lee2023sampling}.

Let us first consider the simplified situation of a QSM targeting an orthogonal subspace, i.e. $S = \identity$. In that situation, assuming the computed Hamiltonian matrix $\tilde{H}$ differs from the exact one $H$ by a perturbation $\Delta_H$, the Bauer-Fike theorem~\cite{bauer1960norms} provides an upper bound for the difference $\subspaceenergy_\mu-\energy_\mu$ between the computed and exact Hamiltonian eigenvalues,
\begin{equation}
\min_\nu |\subspaceenergy_\mu-\energy_\nu| \leq \mathrm{cond}_p(V) \| \Delta_H \|_p
\;,
\end{equation}
where $V$ is the eigenvector matrix that diagonalises $H$, $\mathrm{cond}_p(V) = \| V \|_p \| V^{-1} \|_p$, and $\| V \|_p$ is the $p$-norm of $V$. This result quantifies the intuitive fact that eigenvalue perturbations are related to the condition number of the Hamiltonian and the magnitude of the perturbation.

Compared with the eigenvalue problem, the generalised eigenvalue of a matrix pair $(H, S)$ tends to be more sensitive to noise. In a generic QSM, the matrices $(H, S)$ may be estimated incorrectly due to finite sampling, quantum hardware decoherence, and algorithm-dependent errors.
These errors result in perturbations $\Delta_H$, $\Delta_S$ of $H,S$.
As demonstrated in literature~\cite{lee2023sampling}, these perturbations can affect the solution of the GEEV,
\begin{equation}
\tilde{H} \tilde{C}_\mu = \tilde{S} \tilde{C}_\mu \pertsubspaceenergy_\mu^{(n)}
\end{equation}
where $\tilde{H} = H+\Delta_H$ and $\tilde{S} = S +\Delta_S$ are perturbed Hamiltonian and overlap matrices, respectively, $\pertsubspaceenergy_\mu^{(n)}$ is the $\mu$-th perturbed eigenvalue and $\tilde{C}_\mu$ the corresponding eigenvector.
Because solving the GEEV involves the calculation of $\tilde{S}^{-1/2}$, small singular values of $\tilde{S}$ amplify the noise in the matrix pair significantly. Such cases in which $\tilde{S}$ has small singular values are called ill-conditioned problems. 

A previous study by Mathias and Li~\cite{mathias2004definite} reported an improved perturbation theory for GEEVs, using a geometrical approach on the complex plane describing the quadratic form of the problem.
Subsequently, Epperly \etal~\cite{epperly2022theory} used perturbation theory to describe QFD perturbation with a real-time evolution ansatz.
They also proposed regularizing the GEEV, i.e., reducing the condition number of $\tilde{S}$, by a thresholding procedure in which the least significant eigenvectors of $\tilde{S}$ are projected out of both $\tilde{H}$ and $\tilde{S}$, and the corresponding dimensions removed.
This yields the matrices $\tilde{A} = V^\dagger_\varepsilon \tilde{H} V_\varepsilon$ and $\tilde{B} = V^\dagger_\varepsilon \tilde{S} V_\varepsilon$ where $V_\varepsilon$ is the matrix whose columns are the $n_\varepsilon \leq n$ eigenvectors of $\tilde{S}$ with eigenvalues above a threshold $\varepsilon>0$. Mathias and Li characterised the relationship between $\Delta_H$, $\Delta_S$ and $\pertsubspaceenergy_\mu^{(n)}$. Defining $(\tilde{A},\tilde{B}) = (A+\Delta_A,B+\Delta_B)$ and $\chi^2 = \| \Delta_A \|^2 + \| \Delta_B  \|^2$, and assuming that 
\begin{equation}
\begin{split}
2 n_\varepsilon^2 \chi^2 &\leq \lambda_\varepsilon^2 \;, \\
| \tan^{-1} \subspaceenergy_1^{(n_\varepsilon)} - \tan^{-1} \subspaceenergy_0^{(n_\varepsilon)} | &\geq
\sin^{-1} \frac{ n_\varepsilon \chi}{\lambda_\varepsilon} \; 
\end{split}
\end{equation}
where $\subspaceenergy_\mu^{(n_\varepsilon)}$ are the eigenvalues of the exact pair $(A,B)$ and $\lambda_\varepsilon$ the smallest singular value of $B$, it follows that the lowest eigenvalue of the perturbed pair $(\tilde{A},\tilde{B})$ satisfies
\begin{equation}
| \tan^{-1} \pertsubspaceenergy_0^{(n_\varepsilon)} - \tan^{-1} \subspaceenergy_0^{(n_\varepsilon)} |  \leq
\sin^{-1} \frac{ \sqrt{2} n_\varepsilon \chi}{d_0} \;,
\end{equation}
where $\subspaceenergy_0^{(n_\varepsilon)}$ is the lowest energy of the unperturbed subspace subject to the same thresholding procedure.
Here $d_0^{-1}$ is the condition number of $\tan^{-1} \subspaceenergy_0^{(n_\varepsilon)}$, given by
\begin{equation}
d_0 = | {\bf{x_0}}\cdot (A+iB) {\bf{x_0}} |,
\end{equation}
where ${\bf{x_0}}$ is the unit-norm eigenvector of $(A, B)$ with the lowest eigenvalue.
Epperly \etal~\cite{epperly2022theory} obtained a different bound, based on the perturbations $\Delta_H$, $\Delta_S$,
\begin{equation}
\label{eq:bound_epperly}
| \tan^{-1} \pertsubspaceenergy_0^{(n_\varepsilon)} - \tan^{-1} \subspaceenergy_0^{(n_\varepsilon)} |  \leq
\order\left( \eta^{1/(1+\alpha)} d_0^{-1} \right) \; ,
\end{equation}
where $\alpha$ is a constant ranging from 0 to $1/2$, and $\eta^2 = \| \Delta_H \|^2 + \| \Delta_S  \|^2$
is related to $\varepsilon$ and $\chi$ by $\chi \leq \order\left( \eta^{1/(1+\alpha)} /n \right)$ and
$\varepsilon = \order\left( \eta^{1/(1+\alpha)} \right)$.
In summary, the perturbation bound in Eq.~\eqref{eq:bound_epperly} indicates that the perturbation error is sublinear to the error matrix norms and condition number after the truncation of the basis. Thus, with additional information about the error matrix norms $\eta$, one can establish a sampling error analysis for QSMs.

Alternative approaches for post-processing the quantum data have been suggested which circumvent solving the GEEV entirely and thus don't require dealing with the ill-conditioning of the overlap matrix.
Inspiring by signal processing techniques, the quantum exponential least squares routine (QCELS)~\cite{ding2023even,ding2023simultaneous} solves a nonlinear least-squares problem to approximate the best amplitude and phase parameters fitting the measured overlap matrix data.
QCELS has been shown to converge quickly for ground state estimation when the initial state has large overlap with the ground state.
Another approach, observable dynamic mode decomposition (ODMD)~\cite{shen2023estimating},  builds off of standard dynamic mode decomposition (DMD)~\cite{mezic2004comparison, mezic2005spectral}, which was originally developing for approximating classical dynamical systems. 
ODMD requires measuring only the real or imaginary part of the overlap matrix and embedding this data into a pair of time shifted Hankel matrices.
These Hankel matrices can be used to construct the DMD matrix through a least squares formulation, whose eigenvalues approximate the ground state energy, avoiding the GEEV and thus improving conditioning and stability.
ODMD has shown to have better convergence than many competing methods in the case of low initial overlap with the ground state.

\subsection{Error mitigation and generalised QSE} 

In the previous Subsection, we discussed how decoherence alters the expectation values and variances of random variables sampled on a quantum computer, which in turn impacts the accuracy and precision of a simulation. To remedy this limitation, techniques for the mitigation of readout~\cite{temme2017error,nation2021scalable} and gate~\cite{viola1998dynamical,biercuk2009optimized,kandala2018extending} errors were introduced.

Remarkably, QSMs are known to possess error-mitigating properties~\cite{mcclean2020decoding}.
In particular, QSE was predicted and experimentally confirmed to approximate excited states and reduce errors by performing additional measurements and solving an eigenvalue problem~\cite{colless2018computation}. The observation that QSE measurement can be used to mitigate errors lies at the core of the generalised QSE approach~\cite{yoshioka2022generalized}. Consider a quantum circuit that would prepare a pure state $|\calcwfn \rangle$ in the absence of quantum noise, but instead prepares a density operator $\hat{\rho}$. Then, by measuring a set of operators $\{ \hat{\sigma}_\alpha \}_\alpha$, one can construct a state of the form
\begin{equation}\label{eq:gqse_ansatz}
\rho_{\mathrm{EM}} = \frac{\hat{A}^\dagger \hat{P} \hat{A}}{\mbox{Tr}[\hat{A}^\dagger \hat{P} \hat{A}]}
\;,\;
\hat{A} = \sum_\alpha C_\alpha \hat{\sigma}_\alpha
\;,
\end{equation}
where $\hat{P}$ is a positive operator that is practically taken as $\identity$ or $\hat{\rho}$, and the coefficients $\{ C_\alpha \}_\alpha$ are determined, if the goal is to approximate the ground state of a Hamiltonian, by solving the eigenvalue equation
\begin{equation}
\label{eq:gqse}
H C = S C \subspaceenergy
\;,\;
H_{\alpha\beta} = \mbox{Tr}[\hat{\sigma}_\alpha^\dagger \hat{P} \hat{\sigma}_\beta \hat{H}]
\;,\;
S_{\alpha\beta} = \mbox{Tr}[\hat{\sigma}_\alpha^\dagger \hat{P} \hat{\sigma}_\beta ]
\;.
\end{equation}
An alternative to the measurement of Pauli operators is the use of powers~\cite{yoshioka2022generalized,xiong2021quantum} of the density operators, 
$\rho_{\mathrm{EM}} = \sum_{\alpha,\beta=0}^{n-1} C_\alpha^* C_\beta \, \hat{P}^{\alpha+\beta}$, where $\hat{P} \in \{\identity,\hat{\rho}\}$ is chosen according to $(\alpha+\beta) \% 2$.
The reason behind this choice is the fact that raising $\hat{\rho}$ to powers
suppresses non-dominant eigenvalues, thereby mitigating errors. 
Another alternative is the fault subspace, where a target density operator
$\hat{\rho}$ can be prepared approximately, i.e. one can produce a set of
density operators $\hat{\rho}(\eta_\ell)$ characterised by variable error rates 
$\{ \eta_\ell \}_\ell$ corresponding to the amplification of achievable error rates
(e.g. by gate repetition, probabilistic error amplification, decoherence
amplification, or cross-talk boost). One can then produce a state of the form
$\rho_{\mathrm{EM}} \propto \hat{\sigma}^\dagger \hat{\sigma}$ where
$\hat{\sigma} = \sum_\ell C_\ell \, \hat{\rho}(\eta_\ell)$ for suitable
coefficients $C_\ell$~\cite{yoshioka2022generalized,ohkura2023leveraging}, and $\hat{P}=\identity$ is assumed in Eq.~\eqref{eq:gqse_ansatz}.
It should be noted that, when $\rho_{\mathrm{EM}}$ involves the $n$-th power of $\hat{\rho}$, an ancilla is required to prepare $\hat{\rho}^n$ via controlled $\mathsf{SWAP}$ operations over $n$ copies of $\hat{\rho}$ \cite{koczor2021exponential}. While it is challenging to require multiple copies of a state $\hat{\rho}$, this issue can be addressed by running deeper quantum circuits, i.e., of depth increased by a factor of $n$ \cite{yang2023dual}.

However, numerical evidence suggests generalised QSE is capable of mitigating
various stochastic, coherent, and algorithmic errors~\cite{yoshioka2022generalized},
making it a compelling error mitigation technique for quantum devices.
Of particular relevance is the possibility to measure, in addition to the Hamiltonian, symmetry operators, e.g. the total spin, $\hat{S}^2$, and its component along the $z$-axis, $\hat{S}_z$. This is because the measurement of symmetry operators allows modifying the coefficients $\{ C_\alpha \}_\alpha$, thus mitigating errors arising from symmetry breaking.

\subsection{Optimisation of quantum circuits}
\label{sec:economisation}

As sketched in Figs.~\ref{fig:krylov}, \ref{fig:qfd}, and \ref{fig:mcvqe}, implementing a QSM may require ancillae and deep circuits containing controlled unitary operations. Furthermore, a considerable overhead of quantum measurements may be needed, see e.g. Fig.~\ref{fig:qse_mrcisd}.
The economisation of these operations stands to impact the successful application of QSMs to ES problems. In this Subsection, we briefly describe some recent research aimed at economising quantum circuits and measurements.

\begin{figure}[t!]
\centering
\includegraphics[width=0.8\textwidth]{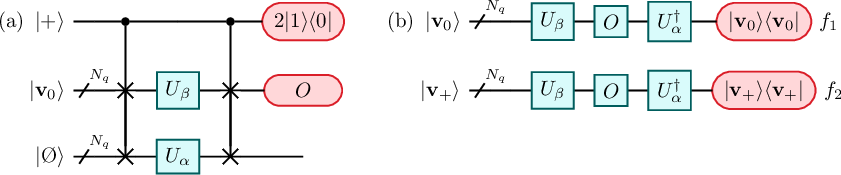}
\caption{Quantum circuit implementing the generalised $\mathsf{SWAP}$ network (a), and quantum circuits to measure matrix elements without Hadamard test (b).
}
\label{fig:economy}
\end{figure}

\subsubsection{Circumventing the Hadamard test}

QSMs may require the measurement of quantities of the form $O_{\alpha\beta} = \langle \vett{v}_0 | \hat{U}^\dagger_\alpha \hat{O} \hat{U}_\beta | \vett{v}_0 \rangle$, where $\hat{U}_\alpha$, $\hat{U}_\beta$
are unitary operators, and $\hat{O}$ is a Hermitian operator (e.g. the Hamiltonian or
the identity). Here, without loss of generality, we will assume that $\hat{O} = \hat{V}^\dagger \hat{\Delta} \hat{V}$ for a unitary $\hat{V}$ and a diagonal Pauli operator $\hat{\Delta}$.

In general, the measurement of $O_{\alpha\beta}$ requires an ancilla qubit and the application of the controlled versions of $\hat{U}_\alpha$ and $\hat{U}_\beta$ (see e.g. Figs.~\ref{fig:qfd} and \ref{fig:mcvqe}), a protocol known as Hadamard test in QC literature~\cite{somma2002simulating,aharonov2009polynomial}. The controlled $\hat{U}_\alpha$ and $\hat{U}_\beta$ operations are particularly expensive: every single-qubit and $\mathsf{cNOT}$ gate in a circuit representation of $\hat{U}_\alpha$ and $\hat{U}_\beta$ is replaced by a controlled-single-qubit operation (requiring 2 $\mathsf{cNOT}$ gates) and a Toffoli
gate (requiring 3 $\mathsf{cNOT}$ gates)~\cite{barenco1995elementary} respectively. Furthermore, unless the device has all-to-all connectivity, implementing the controlled unitaries involves a considerable overhead of $\mathsf{SWAP}$ gates, as discussed in Subsection \ref{sec:noisy}.

As a way to bypass the Hadamard test, Huggins
\etal proposed~\cite{huggins2020non} a generalised $\mathsf{SWAP}$ network, sketched in Fig.~\ref{fig:economy}a. This method requires the existence of a state $|\vac\rangle$ that is an eigenstate of $\hat{U}_{\ell}$ with $\ell=\alpha,\beta$ (a typical example in ES is the vacuum state). Under such assumption, the application of controlled $\hat{U}_\alpha$, $\hat{U}_\beta$ can be replaced with two controlled multi-qubit $\mathsf{SWAP}$ gates. While this method does not remove the need for an ancilla and controlled operations, it makes the computational cost of the computation of $O_{\alpha\beta}$
independent of the structure of $\hat{U}_\alpha$, $\hat{U}_\beta$.
A further improvement, sketched in Fig.~\ref{fig:economy}b, is due to Cortes and Gray~\cite{cortes2022quantum}. Under the same assumption, and the requirement that a superposition of the form $| \vett{v}_+ \rangle = \left( | \vett{v}_0 \rangle + |\vac\rangle \right)/\sqrt{2}$ can be prepared efficiently, one can evaluate $O_{\alpha\beta}$ from the quantities
\begin{equation}
\begin{split}
f_0 &= | \langle \vac | \hat{U}^\dagger_\alpha \hat{O} \hat{U}_\beta | \vac \rangle |^2 \;, \\
f_1 &= | \langle \vett{v}_0 | \hat{U}^\dagger_\alpha \hat{O} \hat{U}_\beta | \vett{v}_0 \rangle |^2 \;, \\
f_2 &= | \langle \vett{v}_+ | \hat{U}^\dagger_\alpha \hat{O} \hat{U}_\beta | \vett{v}_+ \rangle |^2 \;. \\
\end{split}
\end{equation}
Indeed, writing $f_2$ in terms of $|\vett{v}_0 \rangle$ and $| \vac \rangle$, and writing in polar form the complex numbers
$\langle\vac| \hat{U}^\dagger_\alpha \hat{O} \hat{U}_\beta |\vac\rangle = \sqrt{f_0} e^{i \theta_0}$
and
$O_{\alpha\beta} = \sqrt{f_1} e^{i\theta_1}$
(where the absolute values $\sqrt{f_0}$ and $\sqrt{f_1}$ are measurable and the phases $\theta_0$ and $\theta_1$  are respectively known and unknown), one obtains the relation
$4 f_2 = f_1 + f_0 + 2 \sqrt{f_0 f_1} \cos(\theta_1-\theta_0)$.

\subsubsection{Quantum measurements}
\label{sec:measurements}

An important challenge to overcome for the practicality of QSMs is improving the computational cost of quantum measurements. This is particularly important because the measurement requirements in QSMs are typically much higher than in other quantum computing algorithms (e.g. VQE) since one must measure the expectation values of multiple operators in addition to that of the electronic Hamiltonian.
Numerous quantum measurement techniques have been proposed recently, and their adoption and refinement stand to benefit the practicality of QSMs.

In the case of MRCISD, Takeshita \etal~\cite{takeshita2020increasing} noted
that the expectation values that one needs to measure in order to construct the 
subspace spanned by singles and doubles on top of a trial wavefunction are equal
to contractions between $k$-body density matrices with $k=1 \dots 6$ and suitable
coefficients defining the electronic Hamiltonian.
This high computational cost prompted them to propose the use of a cumulant approximation~\cite{mazziotti1998approximate} to express $k$-body density matrices as linear combinations of antisymmetrised products of cumulant operators, e.g.
\begin{equation}
\begin{split}
\rho^{(1)} &= \Delta^{(1)} \\
\rho^{(2)} &= \Delta^{(2)} + \Delta^{(1)} \wedge \Delta^{(1)} \\
\rho^{(3)} &= \Delta^{(3)} + \Delta^{(1)} \wedge \Delta^{(1)} \wedge \Delta^{(1)} + 3 
\Delta^{(2)} \wedge \Delta^{(1)} \\
\dots 
\end{split}
\end{equation}
where $\rho^{(k)}$ and $\Delta^{(k)}$ are the $k$-body density matrix and cumulant
respectively, and ``$\wedge$'' denotes an antisymmetrised tensor product. Measuring density matrices of order up to $l$ allows computation of cumulants of order up to $l$. By then requiring that cumulants of order $k > l$ vanish, one can approximate density matrices of order $k > l$ using cumulants of order up to $k$ only. 
The cumulant approximation reduces the computational cost of MRCISD calculations
but introduces approximations, that affect physical properties computed with MRCISD.
Furthermore, it is not generalizable to other QSMs.

A more general study was carried out by Choi \etal~\cite{choi2023measurement}.
Their starting point is the observation that the operators to be measured in a QSE calculation, $\hat{A}_d \in \{ \hat{O}^\dagger_\alpha \hat{H} \hat{O}_\beta, \hat{O}^\dagger_\alpha \hat{O}_\beta \}_{\alpha\beta}$, can be written as linear combinations of Pauli operators, $\hat{A}_d = \sum_{\vett{m} \in A_d} c_{\vett{m} d} \, \sigma_\vett{m}$, where $A_d$ is the set of indices labeling Pauli operators that appear in $\hat{A}_d$. The standard approach to economising such measurement is ``Pauli grouping'', wherein Pauli operators with indices  $A = \cup_d A_d$ are divided into sets $D_f$ of mutually commuting (and therefore jointly measurable) Pauli operators, $A = \cup_f D_f$ with $[\sigma_\vett{m},\sigma_\vett{n}]=0$ for all $\vett{m},\vett{n} \in D_f$. Expectation values can then be determined by writing operators as
$\hat{A}_d = \sum_f \sum_{\vett{m} \in A_d \cap D_f} c^{(f)}_{\vett{m}d} \, \sigma_\vett{m} = \sum_f \overline{A}^{(f)}_d$ (note that coefficients $c^{(f)}_{\vett{m}d}$ and $c_{\vett{m}d}$ may differ when the sets $D_f$ have non-zero intersection.
Each set $D_f$ has an associated unitary transformation $\hat{W}_f$ that turns Pauli operators in the set into $Z$ (diagonal) Pauli operators, which can be measured jointly, obtaining sample averages $\overline{\sigma}^{(f)}_\vett{m}$. By classical post-processing, one can obtain unbiased estimates of expectation values,
\begin{equation}
\overline{A}_d = \sum_f \sum_{\vett{m} \in A_d \cap D_f} c^{(f)}_{\vett{m}d} \, \overline{\sigma}^{(f)}_\vett{m} = \sum_f \overline{A}^{(f)}_d
\;.
\end{equation}
Since commuting groups are measured independently, $\mathrm{Var}[\overline{A}_d] = \sum_f \mathrm{Var}[\overline{A}^{(f)}_d]$ and this quantity can be maintained below a desired threshold $\varepsilon^2$ by gathering $M$ shots for each set, with $M = \varepsilon^{-2} \max_d \sum_f \mathrm{Var}[\overline{A}^{(f)}_d]$. It should be noted that, while maintaining variances below $\varepsilon^2$ does not guarantee that errors in the final eigenvalues are below $\varepsilon$, analyses based on first-order perturbation theory~\cite{yoshioka2022generalized} and more rigorous analysis involving the thresholding technique~\cite{epperly2022theory} show that the eigenvalue error is bounded by the condition number associated to the QSE eigenvalue problem and the dimension of the QSE matrices, see Subsection~\ref{sec:shot_noise}.

There are multiple ways of partitioning Pauli operators in commuting families, each leading to a specific measurement overhead. Common choices are qubit-algebra-based techniques, exploiting that a set of mutually commuting Pauli operators can be turned into Pauli $Z$ operators by a unitary transformation $\hat{W}_f$, as discussed above. The fully commuting (FC)~\cite{wootters1989optimal} and qubit-wise commuting (QWC) approaches~\cite{yen2020measuring} consider generic Clifford transformations and tensor products of single-qubit Clifford gates as diagonalising unitaries $\hat{W}_f$ respectively. While increasing the freedom of finding diagonalising unitaries leads to lower variance in FC versus QWC, it also increases quantum gate errors due to the presence of two-qubit gates.
The qubit-algebra-based classical shadows method (CS) randomly samples a set of Clifford transformations (in FC-CS) or tensor products of single-qubit Clifford transformations (in QWC-CS) to express the Hamiltonian as a linear combination of Pauli $Z$ operators~\cite{huang2020predicting,zhao2021fermionic}.
While such an approach is appropriate when the goal is to measure arbitrary Pauli operators, when one desires to measure a collection of target Pauli operators, the measurement overhead is reduced by prioritising unitaries that diagonalise the target Pauli operators, a reduction that is achieved in the derandomised version of QWC-CS (Derand)~\cite{huang2021efficient}.
Notable greedy algorithms for efficient measurement of Pauli operators are the sorted insertion~\cite{crawford2021efficient}, iterative coefficient splitting~\cite{yen2023deterministic}, and iterative measurement allocation~\cite{yen2023deterministic} methods.

Important alternatives to qubit-algebra-based techniques are fermionic-algebra-based techniques, which partition fermionic operators into linear combinations of terms diagonalizable by the exponential of a one-body operator. This operator is generally a non-Clifford transformation, but it is one that can be efficiently mapped~\cite{reck1994experimental,clements2016optimal,jiang2018quantum} onto a quantum circuit or a Majorana operator~\cite{bonet2020nearly,zhao2021fermionic}.
The former can be achieved by low- or full-rank decomposition techniques~\cite{motta2021low}, or by greedy approaches like the fluid fragment~\cite{choi2023fluid} method.

\section{Conclusions}
\label{sec:perspectives}

One way to approximate Hamiltonian eigenstates on classical and quantum computers is to select a set of basis states to form a subspace of the many-electron Hilbert space and project the time-independent Schr\"{o}dinger equation on the target subspace. After the subspace projection, a (generalised) eigenvalue problem is solved, yielding approximations to Hamiltonian eigenpairs.
While subspace methods have long been used in classical electronic structure methods, they can also be implemented efficiently as quantum-classical hybrid algorithms, which we refer to as quantum subspace methods (QSMs) here.
Within a QSM, a quantum computer is used to compute the subspace Hamiltonian and overlap matrix, and a classical one is used to subsequently solve a (generalised) eigenvalue problem.

In this review, we presented several recently proposed QSMs.
We illustrated QSMs involving the application of $k$-body fermionic operators and time-evolution operators to a trial state, as well as QSMs constructing a Krylov space. They differ from each other in terms of the subspace basis selection and implementation scheme, with implications on accuracy and computational cost.

We observed how QSMs can be used to approximate Hamiltonian eigenpairs, but also to inform variational quantum algorithms that simultaneously optimise the basis vectors and the expansion coefficients of approximate Hamiltonian eigenpairs. Furthermore, once a low-energy subspace has been determined, QSMs can be used to compute quantities like spectral functions, simulate Hamiltonian evolution, and account for dynamical electronic correlation in the context of hybrid quantum-classical algorithms.
QSMs are able to achieve these goals without increasing circuit depth but instead performing additional measurements.

The ability of QSMs to trade off coherent quantum resources for additional measurements is useful in making use of noisy quantum hardware, but also to implement innovative forms of error mitigation. These algorithms are not limited to near-term quantum hardware but are also promising candidates to study challenging instances of the ES problem on fault-tolerant quantum devices.
The emerging field of QSMs offers many research challenges and opportunities. These range from the design of new QSMs, to the characterisation of the cost and accuracy of existing QSMs, to the implementation of new and existing QSMs on quantum hardware.

We hope that the present review will be a useful resource for practitioners of electronic structure interested in familiarising themselves with quantum computation and quantum subspace methods, in particular, as well as for practitioners of quantum computation interested in the simulation of electronic structure.

\section*{Acknowledgment}

We acknowledge Hiromichi Nishimura and Gavin O. Jones for useful feedback about the manuscript.
N.Y. is supported by JST COI-NEXT Grant No. JPMJPF2221, JST CREST Grant No. JPMJCR23I4, JST ERATO Grant No. JPMJER2302, JST PRESTO Grant No. JPMJPR2119, and IBM Quantum. This research acknowledges resources of the National Energy Research Scientific Computing Center (NERSC), a U.S. Department of Energy Office of Science User Facility located at Lawrence Berkeley National Laboratory, operated under Contract No. DE-AC02-05CH11231. 

%\section*{References}

%\bibliographystyle{iopart-num}
%\bibliography{main}

\begin{thebibliography}{266}%
\makeatletter
\providecommand \@ifxundefined [1]{%
 \@ifx{#1\undefined}
}%
\providecommand \@ifnum [1]{%
 \ifnum #1\expandafter \@firstoftwo
 \else \expandafter \@secondoftwo
 \fi
}%
\providecommand \@ifx [1]{%
 \ifx #1\expandafter \@firstoftwo
 \else \expandafter \@secondoftwo
 \fi
}%
\providecommand \natexlab [1]{#1}%
\providecommand \enquote  [1]{``#1''}%
\providecommand \bibnamefont  [1]{#1}%
\providecommand \bibfnamefont [1]{#1}%
\providecommand \citenamefont [1]{#1}%
\providecommand \href@noop [0]{\@secondoftwo}%
\providecommand \href [0]{\begingroup \@sanitize@url \@href}%
\providecommand \@href[1]{\@@startlink{#1}\@@href}%
\providecommand \@@href[1]{\endgroup#1\@@endlink}%
\providecommand \@sanitize@url [0]{\catcode `\\12\catcode `\$12\catcode
  `\&12\catcode `\#12\catcode `\^12\catcode `\_12\catcode `\%12\relax}%
\providecommand \@@startlink[1]{}%
\providecommand \@@endlink[0]{}%
\providecommand \url  [0]{\begingroup\@sanitize@url \@url }%
\providecommand \@url [1]{\endgroup\@href {#1}{\urlprefix }}%
\providecommand \urlprefix  [0]{URL }%
\providecommand \Eprint [0]{\href }%
\providecommand \doibase [0]{https://doi.org/}%
\providecommand \selectlanguage [0]{\@gobble}%
\providecommand \bibinfo  [0]{\@secondoftwo}%
\providecommand \bibfield  [0]{\@secondoftwo}%
\providecommand \translation [1]{[#1]}%
\providecommand \BibitemOpen [0]{}%
\providecommand \bibitemStop [0]{}%
\providecommand \bibitemNoStop [0]{.\EOS\space}%
\providecommand \EOS [0]{\spacefactor3000\relax}%
\providecommand \BibitemShut  [1]{\csname bibitem#1\endcsname}%
\let\auto@bib@innerbib\@empty
%</preamble>
\bibitem [{\citenamefont {Bauer}\ \emph {et~al.}(2020)\citenamefont {Bauer},
  \citenamefont {Bravyi}, \citenamefont {Motta},\ and\ \citenamefont
  {Kin-Lic~Chan}}]{bauer2020quantum}%
  \BibitemOpen
  \bibfield  {author} {\bibinfo {author} {\bibfnamefont {B.}~\bibnamefont
  {Bauer}}, \bibinfo {author} {\bibfnamefont {S.}~\bibnamefont {Bravyi}},
  \bibinfo {author} {\bibfnamefont {M.}~\bibnamefont {Motta}},\ and\ \bibinfo
  {author} {\bibfnamefont {G.}~\bibnamefont {Kin-Lic~Chan}},\ }\href
  {https://pubs.acs.org/doi/10.1021/acs.chemrev.9b00829} {\bibfield  {journal}
  {\bibinfo  {journal} {Chem. Rev}\ }\textbf {\bibinfo {volume} {120}},\
  \bibinfo {pages} {12685} (\bibinfo {year} {2020})}\BibitemShut {NoStop}%
\bibitem [{\citenamefont {Georgescu}\ \emph {et~al.}(2014)\citenamefont
  {Georgescu}, \citenamefont {Ashhab},\ and\ \citenamefont
  {Nori}}]{georgescu2014quantum}%
  \BibitemOpen
  \bibfield  {author} {\bibinfo {author} {\bibfnamefont {I.~M.}\ \bibnamefont
  {Georgescu}}, \bibinfo {author} {\bibfnamefont {S.}~\bibnamefont {Ashhab}},\
  and\ \bibinfo {author} {\bibfnamefont {F.}~\bibnamefont {Nori}},\ }\href
  {https://journals.aps.org/rmp/abstract/10.1103/RevModPhys.86.153} {\bibfield
  {journal} {\bibinfo  {journal} {Rev. Mod. Phys}\ }\textbf {\bibinfo {volume}
  {86}},\ \bibinfo {pages} {153} (\bibinfo {year} {2014})}\BibitemShut
  {NoStop}%
\bibitem [{\citenamefont {Cao}\ \emph {et~al.}(2019)\citenamefont {Cao},
  \citenamefont {Romero}, \citenamefont {Olson}, \citenamefont {Degroote},
  \citenamefont {Johnson}, \citenamefont {Kieferov{\'a}}, \citenamefont
  {Kivlichan}, \citenamefont {Menke}, \citenamefont {Peropadre}, \citenamefont
  {Sawaya} \emph {et~al.}}]{cao2019quantum}%
  \BibitemOpen
  \bibfield  {author} {\bibinfo {author} {\bibfnamefont {Y.}~\bibnamefont
  {Cao}}, \bibinfo {author} {\bibfnamefont {J.}~\bibnamefont {Romero}},
  \bibinfo {author} {\bibfnamefont {J.~P.}\ \bibnamefont {Olson}}, \bibinfo
  {author} {\bibfnamefont {M.}~\bibnamefont {Degroote}}, \bibinfo {author}
  {\bibfnamefont {P.~D.}\ \bibnamefont {Johnson}}, \bibinfo {author}
  {\bibfnamefont {M.}~\bibnamefont {Kieferov{\'a}}}, \bibinfo {author}
  {\bibfnamefont {I.~D.}\ \bibnamefont {Kivlichan}}, \bibinfo {author}
  {\bibfnamefont {T.}~\bibnamefont {Menke}}, \bibinfo {author} {\bibfnamefont
  {B.}~\bibnamefont {Peropadre}}, \bibinfo {author} {\bibfnamefont {N.~P.}\
  \bibnamefont {Sawaya}}, \emph {et~al.},\ }\href
  {https://pubs.acs.org/doi/10.1021/acs.chemrev.8b00803} {\bibfield  {journal}
  {\bibinfo  {journal} {Chem. Rev}\ }\textbf {\bibinfo {volume} {119}},\
  \bibinfo {pages} {10856} (\bibinfo {year} {2019})}\BibitemShut {NoStop}%
\bibitem [{\citenamefont {Cerezo}\ \emph {et~al.}(2021)\citenamefont {Cerezo},
  \citenamefont {Arrasmith}, \citenamefont {Babbush}, \citenamefont {Benjamin},
  \citenamefont {Endo}, \citenamefont {Fujii}, \citenamefont {McClean},
  \citenamefont {Mitarai}, \citenamefont {Yuan}, \citenamefont {Cincio} \emph
  {et~al.}}]{cerezo2020variational}%
  \BibitemOpen
  \bibfield  {author} {\bibinfo {author} {\bibfnamefont {M.}~\bibnamefont
  {Cerezo}}, \bibinfo {author} {\bibfnamefont {A.}~\bibnamefont {Arrasmith}},
  \bibinfo {author} {\bibfnamefont {R.}~\bibnamefont {Babbush}}, \bibinfo
  {author} {\bibfnamefont {S.~C.}\ \bibnamefont {Benjamin}}, \bibinfo {author}
  {\bibfnamefont {S.}~\bibnamefont {Endo}}, \bibinfo {author} {\bibfnamefont
  {K.}~\bibnamefont {Fujii}}, \bibinfo {author} {\bibfnamefont {J.~R.}\
  \bibnamefont {McClean}}, \bibinfo {author} {\bibfnamefont {K.}~\bibnamefont
  {Mitarai}}, \bibinfo {author} {\bibfnamefont {X.}~\bibnamefont {Yuan}},
  \bibinfo {author} {\bibfnamefont {L.}~\bibnamefont {Cincio}}, \emph
  {et~al.},\ }\href {https://www.nature.com/articles/s42254-021-00348-9}
  {\bibfield  {journal} {\bibinfo  {journal} {Nat. Rev. Phys}\ ,\ \bibinfo
  {pages} {1}} (\bibinfo {year} {2021})}\BibitemShut {NoStop}%
\bibitem [{\citenamefont {McClean}\ \emph {et~al.}(2017)\citenamefont
  {McClean}, \citenamefont {Kimchi-Schwartz}, \citenamefont {Carter},\ and\
  \citenamefont {De~Jong}}]{mcclean2017hybrid}%
  \BibitemOpen
  \bibfield  {author} {\bibinfo {author} {\bibfnamefont {J.~R.}\ \bibnamefont
  {McClean}}, \bibinfo {author} {\bibfnamefont {M.~E.}\ \bibnamefont
  {Kimchi-Schwartz}}, \bibinfo {author} {\bibfnamefont {J.}~\bibnamefont
  {Carter}},\ and\ \bibinfo {author} {\bibfnamefont {W.~A.}\ \bibnamefont
  {De~Jong}},\ }\href
  {https://journals.aps.org/pra/abstract/10.1103/PhysRevA.95.042308} {\bibfield
   {journal} {\bibinfo  {journal} {Phys. Rev. A}\ }\textbf {\bibinfo {volume}
  {95}},\ \bibinfo {pages} {042308} (\bibinfo {year} {2017})}\BibitemShut
  {NoStop}%
\bibitem [{\citenamefont {Colless}\ \emph {et~al.}(2018)\citenamefont
  {Colless}, \citenamefont {Ramasesh}, \citenamefont {Dahlen}, \citenamefont
  {Blok}, \citenamefont {Kimchi-Schwartz}, \citenamefont {McClean},
  \citenamefont {Carter}, \citenamefont {De~Jong},\ and\ \citenamefont
  {Siddiqi}}]{colless2018computation}%
  \BibitemOpen
  \bibfield  {author} {\bibinfo {author} {\bibfnamefont {J.~I.}\ \bibnamefont
  {Colless}}, \bibinfo {author} {\bibfnamefont {V.~V.}\ \bibnamefont
  {Ramasesh}}, \bibinfo {author} {\bibfnamefont {D.}~\bibnamefont {Dahlen}},
  \bibinfo {author} {\bibfnamefont {M.~S.}\ \bibnamefont {Blok}}, \bibinfo
  {author} {\bibfnamefont {M.}~\bibnamefont {Kimchi-Schwartz}}, \bibinfo
  {author} {\bibfnamefont {J.}~\bibnamefont {McClean}}, \bibinfo {author}
  {\bibfnamefont {J.}~\bibnamefont {Carter}}, \bibinfo {author} {\bibfnamefont
  {W.}~\bibnamefont {De~Jong}},\ and\ \bibinfo {author} {\bibfnamefont
  {I.}~\bibnamefont {Siddiqi}},\ }\href
  {https://journals.aps.org/prx/abstract/10.1103/PhysRevX.8.011021} {\bibfield
  {journal} {\bibinfo  {journal} {Phys. Rev. X}\ }\textbf {\bibinfo {volume}
  {8}},\ \bibinfo {pages} {011021} (\bibinfo {year} {2018})}\BibitemShut
  {NoStop}%
\bibitem [{\citenamefont {Huggins}\ \emph {et~al.}(2020)\citenamefont
  {Huggins}, \citenamefont {Lee}, \citenamefont {Baek}, \citenamefont
  {O’Gorman},\ and\ \citenamefont {Whaley}}]{huggins2020non}%
  \BibitemOpen
  \bibfield  {author} {\bibinfo {author} {\bibfnamefont {W.~J.}\ \bibnamefont
  {Huggins}}, \bibinfo {author} {\bibfnamefont {J.}~\bibnamefont {Lee}},
  \bibinfo {author} {\bibfnamefont {U.}~\bibnamefont {Baek}}, \bibinfo {author}
  {\bibfnamefont {B.}~\bibnamefont {O’Gorman}},\ and\ \bibinfo {author}
  {\bibfnamefont {K.~B.}\ \bibnamefont {Whaley}},\ }\href
  {https://iopscience.iop.org/article/10.1088/1367-2630/ab867b} {\bibfield
  {journal} {\bibinfo  {journal} {New J. Phys}\ }\textbf {\bibinfo {volume}
  {22}},\ \bibinfo {pages} {073009} (\bibinfo {year} {2020})}\BibitemShut
  {NoStop}%
\bibitem [{\citenamefont {Motta}\ \emph
  {et~al.}(2020{\natexlab{a}})\citenamefont {Motta}, \citenamefont {Sun},
  \citenamefont {Tan}, \citenamefont {O’Rourke}, \citenamefont {Ye},
  \citenamefont {Minnich}, \citenamefont {Brand{\~a}o},\ and\ \citenamefont
  {Chan}}]{motta2020determining}%
  \BibitemOpen
  \bibfield  {author} {\bibinfo {author} {\bibfnamefont {M.}~\bibnamefont
  {Motta}}, \bibinfo {author} {\bibfnamefont {C.}~\bibnamefont {Sun}}, \bibinfo
  {author} {\bibfnamefont {A.~T.}\ \bibnamefont {Tan}}, \bibinfo {author}
  {\bibfnamefont {M.~J.}\ \bibnamefont {O’Rourke}}, \bibinfo {author}
  {\bibfnamefont {E.}~\bibnamefont {Ye}}, \bibinfo {author} {\bibfnamefont
  {A.~J.}\ \bibnamefont {Minnich}}, \bibinfo {author} {\bibfnamefont {F.~G.}\
  \bibnamefont {Brand{\~a}o}},\ and\ \bibinfo {author} {\bibfnamefont
  {G.~K.-L.}\ \bibnamefont {Chan}},\ }\href
  {https://www.nature.com/articles/s41567-019-0704-4} {\bibfield  {journal}
  {\bibinfo  {journal} {Nat. Phys}\ }\textbf {\bibinfo {volume} {16}},\
  \bibinfo {pages} {205} (\bibinfo {year} {2020}{\natexlab{a}})}\BibitemShut
  {NoStop}%
\bibitem [{\citenamefont {Ollitrault}\ \emph {et~al.}(2020)\citenamefont
  {Ollitrault}, \citenamefont {Kandala}, \citenamefont {Chen}, \citenamefont
  {Barkoutsos}, \citenamefont {Mezzacapo}, \citenamefont {Pistoia},
  \citenamefont {Sheldon}, \citenamefont {Woerner}, \citenamefont {Gambetta},\
  and\ \citenamefont {Tavernelli}}]{ollitrault2020quantum}%
  \BibitemOpen
  \bibfield  {author} {\bibinfo {author} {\bibfnamefont {P.~J.}\ \bibnamefont
  {Ollitrault}}, \bibinfo {author} {\bibfnamefont {A.}~\bibnamefont {Kandala}},
  \bibinfo {author} {\bibfnamefont {C.-F.}\ \bibnamefont {Chen}}, \bibinfo
  {author} {\bibfnamefont {P.~K.}\ \bibnamefont {Barkoutsos}}, \bibinfo
  {author} {\bibfnamefont {A.}~\bibnamefont {Mezzacapo}}, \bibinfo {author}
  {\bibfnamefont {M.}~\bibnamefont {Pistoia}}, \bibinfo {author} {\bibfnamefont
  {S.}~\bibnamefont {Sheldon}}, \bibinfo {author} {\bibfnamefont
  {S.}~\bibnamefont {Woerner}}, \bibinfo {author} {\bibfnamefont {J.~M.}\
  \bibnamefont {Gambetta}},\ and\ \bibinfo {author} {\bibfnamefont
  {I.}~\bibnamefont {Tavernelli}},\ }\href
  {https://journals.aps.org/prresearch/abstract/10.1103/PhysRevResearch.2.043140}
  {\bibfield  {journal} {\bibinfo  {journal} {Phys. Rev. Research}\ }\textbf
  {\bibinfo {volume} {2}},\ \bibinfo {pages} {043140} (\bibinfo {year}
  {2020})}\BibitemShut {NoStop}%
\bibitem [{\citenamefont {Parrish}\ and\ \citenamefont
  {McMahon}(2019)}]{parrish2019quantum}%
  \BibitemOpen
  \bibfield  {author} {\bibinfo {author} {\bibfnamefont {R.~M.}\ \bibnamefont
  {Parrish}}\ and\ \bibinfo {author} {\bibfnamefont {P.~L.}\ \bibnamefont
  {McMahon}},\ }\href {https://arxiv.org/abs/1909.08925} {\bibfield  {journal}
  {\bibinfo  {journal} {arXiv:1909.08925}\ } (\bibinfo {year}
  {2019})}\BibitemShut {NoStop}%
\bibitem [{\citenamefont {Stair}\ \emph {et~al.}(2020)\citenamefont {Stair},
  \citenamefont {Huang},\ and\ \citenamefont
  {Evangelista}}]{stair2020multireference}%
  \BibitemOpen
  \bibfield  {author} {\bibinfo {author} {\bibfnamefont {N.~H.}\ \bibnamefont
  {Stair}}, \bibinfo {author} {\bibfnamefont {R.}~\bibnamefont {Huang}},\ and\
  \bibinfo {author} {\bibfnamefont {F.~A.}\ \bibnamefont {Evangelista}},\
  }\href {https://pubs.acs.org/doi/10.1021/acs.jctc.9b01125} {\bibfield
  {journal} {\bibinfo  {journal} {J. Chem. Theory Comput}\ }\textbf {\bibinfo
  {volume} {16}},\ \bibinfo {pages} {2236} (\bibinfo {year}
  {2020})}\BibitemShut {NoStop}%
\bibitem [{\citenamefont {Roos}(2005)}]{roos2005multiconfigurational}%
  \BibitemOpen
  \bibfield  {author} {\bibinfo {author} {\bibfnamefont {B.~O.}\ \bibnamefont
  {Roos}},\ }in\ \href@noop {} {\emph {\bibinfo {booktitle} {Theory and
  Applications of Computational Chemistry}}}\ (\bibinfo  {publisher}
  {Elsevier},\ \bibinfo {year} {2005})\ pp.\ \bibinfo {pages}
  {725--764}\BibitemShut {NoStop}%
\bibitem [{\citenamefont {Shavitt}\ and\ \citenamefont
  {Bartlett}(2009)}]{shavitt2009many}%
  \BibitemOpen
  \bibfield  {author} {\bibinfo {author} {\bibfnamefont {I.}~\bibnamefont
  {Shavitt}}\ and\ \bibinfo {author} {\bibfnamefont {R.~J.}\ \bibnamefont
  {Bartlett}},\ }\href@noop {} {\emph {\bibinfo {title} {Many-body methods in
  chemistry and physics: MBPT and coupled-cluster theory}}}\ (\bibinfo
  {publisher} {Cambridge university press},\ \bibinfo {year}
  {2009})\BibitemShut {NoStop}%
\bibitem [{\citenamefont {Levine}\ \emph {et~al.}(2009)\citenamefont {Levine},
  \citenamefont {Busch},\ and\ \citenamefont {Shull}}]{levine2009quantum}%
  \BibitemOpen
  \bibfield  {author} {\bibinfo {author} {\bibfnamefont {I.~N.}\ \bibnamefont
  {Levine}}, \bibinfo {author} {\bibfnamefont {D.~H.}\ \bibnamefont {Busch}},\
  and\ \bibinfo {author} {\bibfnamefont {H.}~\bibnamefont {Shull}},\
  }\href@noop {} {\emph {\bibinfo {title} {Quantum chemistry}}},\ Vol.~\bibinfo
  {volume} {6}\ (\bibinfo  {publisher} {Pearson Prentice Hall, NJ},\ \bibinfo
  {year} {2009})\BibitemShut {NoStop}%
\bibitem [{\citenamefont {Helgaker}\ \emph {et~al.}(2013)\citenamefont
  {Helgaker}, \citenamefont {Jorgensen},\ and\ \citenamefont
  {Olsen}}]{helgaker2013molecular}%
  \BibitemOpen
  \bibfield  {author} {\bibinfo {author} {\bibfnamefont {T.}~\bibnamefont
  {Helgaker}}, \bibinfo {author} {\bibfnamefont {P.}~\bibnamefont
  {Jorgensen}},\ and\ \bibinfo {author} {\bibfnamefont {J.}~\bibnamefont
  {Olsen}},\ }\href@noop {} {\emph {\bibinfo {title} {Molecular
  electronic-structure theory}}}\ (\bibinfo  {publisher} {John Wiley \& Sons},\
  \bibinfo {year} {2013})\BibitemShut {NoStop}%
\bibitem [{\citenamefont {Jensen}(2017)}]{jensen2017introduction}%
  \BibitemOpen
  \bibfield  {author} {\bibinfo {author} {\bibfnamefont {F.}~\bibnamefont
  {Jensen}},\ }\href@noop {} {\emph {\bibinfo {title} {Introduction to
  computational chemistry}}}\ (\bibinfo  {publisher} {John wiley \& sons},\
  \bibinfo {year} {2017})\BibitemShut {NoStop}%
\bibitem [{\citenamefont {Martin}(2020)}]{martin2020electronic}%
  \BibitemOpen
  \bibfield  {author} {\bibinfo {author} {\bibfnamefont {R.~M.}\ \bibnamefont
  {Martin}},\ }\href@noop {} {\emph {\bibinfo {title} {Electronic structure:
  basic theory and practical methods}}}\ (\bibinfo  {publisher} {Cambridge
  university press},\ \bibinfo {year} {2020})\BibitemShut {NoStop}%
\bibitem [{\citenamefont {Friesner}(2005)}]{friesner2005ab}%
  \BibitemOpen
  \bibfield  {author} {\bibinfo {author} {\bibfnamefont {R.~A.}\ \bibnamefont
  {Friesner}},\ }\href {https://www.pnas.org/doi/abs/10.1073/pnas.0408036102}
  {\bibfield  {journal} {\bibinfo  {journal} {Proc. Natl. Acad. Sci. USA}\
  }\textbf {\bibinfo {volume} {102}},\ \bibinfo {pages} {6648} (\bibinfo {year}
  {2005})}\BibitemShut {NoStop}%
\bibitem [{\citenamefont {Bartlett}\ and\ \citenamefont
  {Musia{\l}}(2007)}]{bartlett2007coupled}%
  \BibitemOpen
  \bibfield  {author} {\bibinfo {author} {\bibfnamefont {R.~J.}\ \bibnamefont
  {Bartlett}}\ and\ \bibinfo {author} {\bibfnamefont {M.}~\bibnamefont
  {Musia{\l}}},\ }\href
  {https://journals.aps.org/rmp/abstract/10.1103/RevModPhys.79.291} {\bibfield
  {journal} {\bibinfo  {journal} {Rev. Mod. Phys}\ }\textbf {\bibinfo {volume}
  {79}},\ \bibinfo {pages} {291} (\bibinfo {year} {2007})}\BibitemShut
  {NoStop}%
\bibitem [{\citenamefont {Helgaker}\ \emph {et~al.}(2012)\citenamefont
  {Helgaker}, \citenamefont {Coriani}, \citenamefont {J{\o}rgensen},
  \citenamefont {Kristensen}, \citenamefont {Olsen},\ and\ \citenamefont
  {Ruud}}]{helgaker2012recent}%
  \BibitemOpen
  \bibfield  {author} {\bibinfo {author} {\bibfnamefont {T.}~\bibnamefont
  {Helgaker}}, \bibinfo {author} {\bibfnamefont {S.}~\bibnamefont {Coriani}},
  \bibinfo {author} {\bibfnamefont {P.}~\bibnamefont {J{\o}rgensen}}, \bibinfo
  {author} {\bibfnamefont {K.}~\bibnamefont {Kristensen}}, \bibinfo {author}
  {\bibfnamefont {J.}~\bibnamefont {Olsen}},\ and\ \bibinfo {author}
  {\bibfnamefont {K.}~\bibnamefont {Ruud}},\ }\href
  {https://doi.org/10.1021/cr2002239} {\bibfield  {journal} {\bibinfo
  {journal} {Chem. Rev}\ }\textbf {\bibinfo {volume} {112}},\ \bibinfo {pages}
  {543} (\bibinfo {year} {2012})}\BibitemShut {NoStop}%
\bibitem [{\citenamefont {Nielsen}\ and\ \citenamefont
  {Chuang}(2010)}]{nielsen2010quantum}%
  \BibitemOpen
  \bibfield  {author} {\bibinfo {author} {\bibfnamefont {M.~A.}\ \bibnamefont
  {Nielsen}}\ and\ \bibinfo {author} {\bibfnamefont {I.~L.}\ \bibnamefont
  {Chuang}},\ }\href@noop {} {\emph {\bibinfo {title} {Quantum computation and
  quantum information}}}\ (\bibinfo  {publisher} {Cambridge university press},\
  \bibinfo {year} {2010})\BibitemShut {NoStop}%
\bibitem [{\citenamefont {Benenti}\ \emph {et~al.}(2019)\citenamefont
  {Benenti}, \citenamefont {Casati}, \citenamefont {Rossini},\ and\
  \citenamefont {Strini}}]{benenti2004principles}%
  \BibitemOpen
  \bibfield  {author} {\bibinfo {author} {\bibfnamefont {G.}~\bibnamefont
  {Benenti}}, \bibinfo {author} {\bibfnamefont {G.}~\bibnamefont {Casati}},
  \bibinfo {author} {\bibfnamefont {D.}~\bibnamefont {Rossini}},\ and\ \bibinfo
  {author} {\bibfnamefont {G.}~\bibnamefont {Strini}},\ }\href@noop {} {\emph
  {\bibinfo {title} {Principles of quantum computation and information: A
  Comprehensive Textbook}}}\ (\bibinfo  {publisher} {World Scientific},\
  \bibinfo {year} {2019})\BibitemShut {NoStop}%
\bibitem [{\citenamefont {Manenti}\ and\ \citenamefont
  {Motta}(2023)}]{manenti2023quantum}%
  \BibitemOpen
  \bibfield  {author} {\bibinfo {author} {\bibfnamefont {R.}~\bibnamefont
  {Manenti}}\ and\ \bibinfo {author} {\bibfnamefont {M.}~\bibnamefont
  {Motta}},\ }\href@noop {} {\emph {\bibinfo {title} {Quantum Information
  Science}}}\ (\bibinfo  {publisher} {Oxford University Press},\ \bibinfo
  {year} {2023})\BibitemShut {NoStop}%
\bibitem [{\citenamefont {Motta}\ and\ \citenamefont
  {Rice}(2021)}]{motta2021emerging}%
  \BibitemOpen
  \bibfield  {author} {\bibinfo {author} {\bibfnamefont {M.}~\bibnamefont
  {Motta}}\ and\ \bibinfo {author} {\bibfnamefont {J.~E.}\ \bibnamefont
  {Rice}},\ }\href
  {https://wires.onlinelibrary.wiley.com/doi/abs/10.1002/wcms.1580} {\bibfield
  {journal} {\bibinfo  {journal} {WIREs Comput. Mol. Sci}\ ,\ \bibinfo {pages}
  {e1580}} (\bibinfo {year} {2021})}\BibitemShut {NoStop}%
\bibitem [{\citenamefont {L{\"o}wdin}(1958)}]{lowdin1958correlation}%
  \BibitemOpen
  \bibfield  {author} {\bibinfo {author} {\bibfnamefont {P.-O.}\ \bibnamefont
  {L{\"o}wdin}},\ }\href
  {https://onlinelibrary.wiley.com/doi/abs/10.1002/9780470143483.ch7}
  {\bibfield  {journal} {\bibinfo  {journal} {Adv. Chem. Phys}\ ,\ \bibinfo
  {pages} {207}} (\bibinfo {year} {1958})}\BibitemShut {NoStop}%
\bibitem [{\citenamefont {Sinanoglu}(1964)}]{sinanoglu1964many}%
  \BibitemOpen
  \bibfield  {author} {\bibinfo {author} {\bibfnamefont {O.}~\bibnamefont
  {Sinanoglu}},\ }\href
  {https://onlinelibrary.wiley.com/doi/10.1002/9780470143520.ch7} {\bibfield
  {journal} {\bibinfo  {journal} {Adv. Chem. Phys}\ ,\ \bibinfo {pages} {315}}
  (\bibinfo {year} {1964})}\BibitemShut {NoStop}%
\bibitem [{\citenamefont {Ruedenberg}\ and\ \citenamefont
  {Sundberg}(1976)}]{ruedenberg1976mcscf}%
  \BibitemOpen
  \bibfield  {author} {\bibinfo {author} {\bibfnamefont {K.}~\bibnamefont
  {Ruedenberg}}\ and\ \bibinfo {author} {\bibfnamefont {K.~R.}\ \bibnamefont
  {Sundberg}},\ }in\ \href
  {https://link.springer.com/chapter/10.1007/978-1-4757-1659-7_37} {\emph
  {\bibinfo {booktitle} {Quantum Science: Methods and Structure. A Tribute to
  Per-Olov L{\"o}wdin}}}\ (\bibinfo  {publisher} {Springer},\ \bibinfo {year}
  {1976})\ pp.\ \bibinfo {pages} {505--515}\BibitemShut {NoStop}%
\bibitem [{\citenamefont {Mok}\ \emph {et~al.}(1996)\citenamefont {Mok},
  \citenamefont {Neumann},\ and\ \citenamefont {Handy}}]{mok1996dynamical}%
  \BibitemOpen
  \bibfield  {author} {\bibinfo {author} {\bibfnamefont {D.~K.}\ \bibnamefont
  {Mok}}, \bibinfo {author} {\bibfnamefont {R.}~\bibnamefont {Neumann}},\ and\
  \bibinfo {author} {\bibfnamefont {N.~C.}\ \bibnamefont {Handy}},\ }\href
  {https://pubs.acs.org/doi/abs/10.1021/jp9528020} {\bibfield  {journal}
  {\bibinfo  {journal} {J. Phys. Chem}\ }\textbf {\bibinfo {volume} {100}},\
  \bibinfo {pages} {6225} (\bibinfo {year} {1996})}\BibitemShut {NoStop}%
\bibitem [{\citenamefont {Kato}(1957)}]{kato1957eigenfunctions}%
  \BibitemOpen
  \bibfield  {author} {\bibinfo {author} {\bibfnamefont {T.}~\bibnamefont
  {Kato}},\ }\href {https://doi.org/10.1002/cpa.3160100201} {\bibfield
  {journal} {\bibinfo  {journal} {Commun. Pure Appl. Math}\ }\textbf {\bibinfo
  {volume} {10}},\ \bibinfo {pages} {151} (\bibinfo {year} {1957})}\BibitemShut
  {NoStop}%
\bibitem [{\citenamefont {Kong}\ \emph {et~al.}(2012)\citenamefont {Kong},
  \citenamefont {Bischoff},\ and\ \citenamefont {Valeev}}]{kong2012explicitly}%
  \BibitemOpen
  \bibfield  {author} {\bibinfo {author} {\bibfnamefont {L.}~\bibnamefont
  {Kong}}, \bibinfo {author} {\bibfnamefont {F.~A.}\ \bibnamefont {Bischoff}},\
  and\ \bibinfo {author} {\bibfnamefont {E.~F.}\ \bibnamefont {Valeev}},\
  }\href {https://doi.org/10.1021/cr200204r} {\bibfield  {journal} {\bibinfo
  {journal} {Chem. Rev}\ }\textbf {\bibinfo {volume} {112}},\ \bibinfo {pages}
  {75} (\bibinfo {year} {2012})}\BibitemShut {NoStop}%
\bibitem [{\citenamefont {Barenco}\ \emph {et~al.}(1995)\citenamefont
  {Barenco}, \citenamefont {Bennett}, \citenamefont {Cleve}, \citenamefont
  {DiVincenzo}, \citenamefont {Margolus}, \citenamefont {Shor}, \citenamefont
  {Sleator}, \citenamefont {Smolin},\ and\ \citenamefont
  {Weinfurter}}]{barenco1995elementary}%
  \BibitemOpen
  \bibfield  {author} {\bibinfo {author} {\bibfnamefont {A.}~\bibnamefont
  {Barenco}}, \bibinfo {author} {\bibfnamefont {C.~H.}\ \bibnamefont
  {Bennett}}, \bibinfo {author} {\bibfnamefont {R.}~\bibnamefont {Cleve}},
  \bibinfo {author} {\bibfnamefont {D.~P.}\ \bibnamefont {DiVincenzo}},
  \bibinfo {author} {\bibfnamefont {N.}~\bibnamefont {Margolus}}, \bibinfo
  {author} {\bibfnamefont {P.}~\bibnamefont {Shor}}, \bibinfo {author}
  {\bibfnamefont {T.}~\bibnamefont {Sleator}}, \bibinfo {author} {\bibfnamefont
  {J.~A.}\ \bibnamefont {Smolin}},\ and\ \bibinfo {author} {\bibfnamefont
  {H.}~\bibnamefont {Weinfurter}},\ }\href
  {https://journals.aps.org/pra/abstract/10.1103/PhysRevA.52.3457} {\bibfield
  {journal} {\bibinfo  {journal} {Phys. Rev. A}\ }\textbf {\bibinfo {volume}
  {52}},\ \bibinfo {pages} {3457} (\bibinfo {year} {1995})}\BibitemShut
  {NoStop}%
\bibitem [{\citenamefont {Woit}(2017)}]{woit2017quantum}%
  \BibitemOpen
  \bibfield  {author} {\bibinfo {author} {\bibfnamefont {P.}~\bibnamefont
  {Woit}},\ }\href {https://www.math.columbia.edu/~woit/QM/qmbook.pdf} {\emph
  {\bibinfo {title} {Quantum Theory, Groups and Representations: An
  Introduction}}}\ (\bibinfo  {publisher} {Springer International Publishing},\
  \bibinfo {year} {2017})\BibitemShut {NoStop}%
\bibitem [{\citenamefont {Krantz}\ \emph {et~al.}(2019)\citenamefont {Krantz},
  \citenamefont {Kjaergaard}, \citenamefont {Yan}, \citenamefont {Orlando},
  \citenamefont {Gustavsson},\ and\ \citenamefont
  {Oliver}}]{krantz2019quantum}%
  \BibitemOpen
  \bibfield  {author} {\bibinfo {author} {\bibfnamefont {P.}~\bibnamefont
  {Krantz}}, \bibinfo {author} {\bibfnamefont {M.}~\bibnamefont {Kjaergaard}},
  \bibinfo {author} {\bibfnamefont {F.}~\bibnamefont {Yan}}, \bibinfo {author}
  {\bibfnamefont {T.~P.}\ \bibnamefont {Orlando}}, \bibinfo {author}
  {\bibfnamefont {S.}~\bibnamefont {Gustavsson}},\ and\ \bibinfo {author}
  {\bibfnamefont {W.~D.}\ \bibnamefont {Oliver}},\ }\href
  {https://pubs.aip.org/aip/apr/article/6/2/021318/570326} {\bibfield
  {journal} {\bibinfo  {journal} {Appl. Phys. Rev}\ }\textbf {\bibinfo {volume}
  {6}} (\bibinfo {year} {2019})}\BibitemShut {NoStop}%
\bibitem [{\citenamefont {Zhang}\ \emph {et~al.}(2003)\citenamefont {Zhang},
  \citenamefont {Vala}, \citenamefont {Sastry},\ and\ \citenamefont
  {Whaley}}]{zhang2003geometric}%
  \BibitemOpen
  \bibfield  {author} {\bibinfo {author} {\bibfnamefont {J.}~\bibnamefont
  {Zhang}}, \bibinfo {author} {\bibfnamefont {J.}~\bibnamefont {Vala}},
  \bibinfo {author} {\bibfnamefont {S.}~\bibnamefont {Sastry}},\ and\ \bibinfo
  {author} {\bibfnamefont {K.~B.}\ \bibnamefont {Whaley}},\ }\href
  {https://journals.aps.org/pra/abstract/10.1103/PhysRevA.67.042313} {\bibfield
   {journal} {\bibinfo  {journal} {Phys. Rev. A}\ }\textbf {\bibinfo {volume}
  {67}},\ \bibinfo {pages} {042313} (\bibinfo {year} {2003})}\BibitemShut
  {NoStop}%
\bibitem [{\citenamefont {Zhang}\ \emph {et~al.}(2004)\citenamefont {Zhang},
  \citenamefont {Vala}, \citenamefont {Sastry},\ and\ \citenamefont
  {Whaley}}]{zhang2004optimal}%
  \BibitemOpen
  \bibfield  {author} {\bibinfo {author} {\bibfnamefont {J.}~\bibnamefont
  {Zhang}}, \bibinfo {author} {\bibfnamefont {J.}~\bibnamefont {Vala}},
  \bibinfo {author} {\bibfnamefont {S.}~\bibnamefont {Sastry}},\ and\ \bibinfo
  {author} {\bibfnamefont {K.~B.}\ \bibnamefont {Whaley}},\ }\href
  {https://journals.aps.org/pra/abstract/10.1103/PhysRevA.69.042309} {\bibfield
   {journal} {\bibinfo  {journal} {Phys. Rev. A}\ }\textbf {\bibinfo {volume}
  {69}},\ \bibinfo {pages} {042309} (\bibinfo {year} {2004})}\BibitemShut
  {NoStop}%
\bibitem [{\citenamefont {Blaauboer}\ and\ \citenamefont
  {De~Visser}(2008)}]{blaauboer2008analytical}%
  \BibitemOpen
  \bibfield  {author} {\bibinfo {author} {\bibfnamefont {M.}~\bibnamefont
  {Blaauboer}}\ and\ \bibinfo {author} {\bibfnamefont {R.}~\bibnamefont
  {De~Visser}},\ }\href
  {https://iopscience.iop.org/article/10.1088/1751-8113/41/39/395307}
  {\bibfield  {journal} {\bibinfo  {journal} {J. Phys. A}\ }\textbf {\bibinfo
  {volume} {41}},\ \bibinfo {pages} {395307} (\bibinfo {year}
  {2008})}\BibitemShut {NoStop}%
\bibitem [{\citenamefont {Watts}\ \emph {et~al.}(2013)\citenamefont {Watts},
  \citenamefont {O’Connor},\ and\ \citenamefont {Vala}}]{watts2013metric}%
  \BibitemOpen
  \bibfield  {author} {\bibinfo {author} {\bibfnamefont {P.}~\bibnamefont
  {Watts}}, \bibinfo {author} {\bibfnamefont {M.}~\bibnamefont {O’Connor}},\
  and\ \bibinfo {author} {\bibfnamefont {J.}~\bibnamefont {Vala}},\ }\href
  {https://www.mdpi.com/1099-4300/15/6/1963} {\bibfield  {journal} {\bibinfo
  {journal} {Entropy}\ }\textbf {\bibinfo {volume} {15}},\ \bibinfo {pages}
  {1963} (\bibinfo {year} {2013})}\BibitemShut {NoStop}%
\bibitem [{\citenamefont {Crooks}(2020)}]{crooks2020gates}%
  \BibitemOpen
  \bibfield  {author} {\bibinfo {author} {\bibfnamefont {G.~E.}\ \bibnamefont
  {Crooks}},\ }\href {https://threeplusone.com/pubs/on_gates.pdf} {\bibinfo
  {title} {Gates, states, and circuits}} (\bibinfo {year} {2020})\BibitemShut
  {NoStop}%
\bibitem [{\citenamefont {Paraoanu}(2006)}]{paraoanu2006microwave}%
  \BibitemOpen
  \bibfield  {author} {\bibinfo {author} {\bibfnamefont {G.}~\bibnamefont
  {Paraoanu}},\ }\href
  {https://journals.aps.org/prb/abstract/10.1103/PhysRevB.74.140504} {\bibfield
   {journal} {\bibinfo  {journal} {Phys. Rev. B}\ }\textbf {\bibinfo {volume}
  {74}},\ \bibinfo {pages} {140504} (\bibinfo {year} {2006})}\BibitemShut
  {NoStop}%
\bibitem [{\citenamefont {Rigetti}\ and\ \citenamefont
  {Devoret}(2010)}]{rigetti2010fully}%
  \BibitemOpen
  \bibfield  {author} {\bibinfo {author} {\bibfnamefont {C.}~\bibnamefont
  {Rigetti}}\ and\ \bibinfo {author} {\bibfnamefont {M.}~\bibnamefont
  {Devoret}},\ }\href
  {https://journals.aps.org/prb/abstract/10.1103/PhysRevB.81.134507} {\bibfield
   {journal} {\bibinfo  {journal} {Phys. Rev. B}\ }\textbf {\bibinfo {volume}
  {81}},\ \bibinfo {pages} {134507} (\bibinfo {year} {2010})}\BibitemShut
  {NoStop}%
\bibitem [{\citenamefont {Yan}\ \emph {et~al.}(2018)\citenamefont {Yan},
  \citenamefont {Krantz}, \citenamefont {Sung}, \citenamefont {Kjaergaard},
  \citenamefont {Campbell}, \citenamefont {Orlando}, \citenamefont
  {Gustavsson},\ and\ \citenamefont {Oliver}}]{yan2018tunable}%
  \BibitemOpen
  \bibfield  {author} {\bibinfo {author} {\bibfnamefont {F.}~\bibnamefont
  {Yan}}, \bibinfo {author} {\bibfnamefont {P.}~\bibnamefont {Krantz}},
  \bibinfo {author} {\bibfnamefont {Y.}~\bibnamefont {Sung}}, \bibinfo {author}
  {\bibfnamefont {M.}~\bibnamefont {Kjaergaard}}, \bibinfo {author}
  {\bibfnamefont {D.~L.}\ \bibnamefont {Campbell}}, \bibinfo {author}
  {\bibfnamefont {T.~P.}\ \bibnamefont {Orlando}}, \bibinfo {author}
  {\bibfnamefont {S.}~\bibnamefont {Gustavsson}},\ and\ \bibinfo {author}
  {\bibfnamefont {W.~D.}\ \bibnamefont {Oliver}},\ }\href
  {https://journals.aps.org/prapplied/abstract/10.1103/PhysRevApplied.10.054062}
  {\bibfield  {journal} {\bibinfo  {journal} {Phys. Rev. Appl}\ }\textbf
  {\bibinfo {volume} {10}},\ \bibinfo {pages} {054062} (\bibinfo {year}
  {2018})}\BibitemShut {NoStop}%
\bibitem [{\citenamefont {Foxen}\ \emph {et~al.}(2020)\citenamefont {Foxen},
  \citenamefont {Neill}, \citenamefont {Dunsworth}, \citenamefont {Roushan},
  \citenamefont {Chiaro}, \citenamefont {Megrant}, \citenamefont {Kelly},
  \citenamefont {Chen}, \citenamefont {Satzinger}, \citenamefont {Barends}
  \emph {et~al.}}]{foxen2020demonstrating}%
  \BibitemOpen
  \bibfield  {author} {\bibinfo {author} {\bibfnamefont {B.}~\bibnamefont
  {Foxen}}, \bibinfo {author} {\bibfnamefont {C.}~\bibnamefont {Neill}},
  \bibinfo {author} {\bibfnamefont {A.}~\bibnamefont {Dunsworth}}, \bibinfo
  {author} {\bibfnamefont {P.}~\bibnamefont {Roushan}}, \bibinfo {author}
  {\bibfnamefont {B.}~\bibnamefont {Chiaro}}, \bibinfo {author} {\bibfnamefont
  {A.}~\bibnamefont {Megrant}}, \bibinfo {author} {\bibfnamefont
  {J.}~\bibnamefont {Kelly}}, \bibinfo {author} {\bibfnamefont
  {Z.}~\bibnamefont {Chen}}, \bibinfo {author} {\bibfnamefont {K.}~\bibnamefont
  {Satzinger}}, \bibinfo {author} {\bibfnamefont {R.}~\bibnamefont {Barends}},
  \emph {et~al.},\ }\href
  {https://journals.aps.org/prl/abstract/10.1103/PhysRevLett.125.120504}
  {\bibfield  {journal} {\bibinfo  {journal} {Phys. Rev. Lett}\ }\textbf
  {\bibinfo {volume} {125}},\ \bibinfo {pages} {120504} (\bibinfo {year}
  {2020})}\BibitemShut {NoStop}%
\bibitem [{\citenamefont {Seeley}\ \emph {et~al.}(2012)\citenamefont {Seeley},
  \citenamefont {Richard},\ and\ \citenamefont {Love}}]{seeley2012bravyi}%
  \BibitemOpen
  \bibfield  {author} {\bibinfo {author} {\bibfnamefont {J.~T.}\ \bibnamefont
  {Seeley}}, \bibinfo {author} {\bibfnamefont {M.~J.}\ \bibnamefont
  {Richard}},\ and\ \bibinfo {author} {\bibfnamefont {P.~J.}\ \bibnamefont
  {Love}},\ }\href {https://aip.scitation.org/doi/10.1063/1.4768229} {\bibfield
   {journal} {\bibinfo  {journal} {J. Chem. Phys}\ }\textbf {\bibinfo {volume}
  {137}},\ \bibinfo {pages} {224109} (\bibinfo {year} {2012})}\BibitemShut
  {NoStop}%
\bibitem [{\citenamefont {Bronn}\ \emph {et~al.}(2017)\citenamefont {Bronn},
  \citenamefont {Abdo}, \citenamefont {Inoue}, \citenamefont {Lekuch},
  \citenamefont {C{\'o}rcoles}, \citenamefont {Hertzberg}, \citenamefont
  {Takita}, \citenamefont {Bishop}, \citenamefont {Gambetta},\ and\
  \citenamefont {Chow}}]{bronn2017fast}%
  \BibitemOpen
  \bibfield  {author} {\bibinfo {author} {\bibfnamefont {N.~T.}\ \bibnamefont
  {Bronn}}, \bibinfo {author} {\bibfnamefont {B.}~\bibnamefont {Abdo}},
  \bibinfo {author} {\bibfnamefont {K.}~\bibnamefont {Inoue}}, \bibinfo
  {author} {\bibfnamefont {S.}~\bibnamefont {Lekuch}}, \bibinfo {author}
  {\bibfnamefont {A.~D.}\ \bibnamefont {C{\'o}rcoles}}, \bibinfo {author}
  {\bibfnamefont {J.~B.}\ \bibnamefont {Hertzberg}}, \bibinfo {author}
  {\bibfnamefont {M.}~\bibnamefont {Takita}}, \bibinfo {author} {\bibfnamefont
  {L.~S.}\ \bibnamefont {Bishop}}, \bibinfo {author} {\bibfnamefont {J.~M.}\
  \bibnamefont {Gambetta}},\ and\ \bibinfo {author} {\bibfnamefont {J.~M.}\
  \bibnamefont {Chow}},\ }in\ \href
  {https://iopscience.iop.org/article/10.1088/1742-6596/834/1/012003/meta}
  {\emph {\bibinfo {booktitle} {Journal of Physics: Conference Series}}},\
  Vol.\ \bibinfo {volume} {834}\ (\bibinfo {organization} {IOP Publishing},\
  \bibinfo {year} {2017})\ p.\ \bibinfo {pages} {012003}\BibitemShut {NoStop}%
\bibitem [{\citenamefont {Jurcevic}\ \emph {et~al.}(2021)\citenamefont
  {Jurcevic}, \citenamefont {Javadi-Abhari}, \citenamefont {Bishop},
  \citenamefont {Lauer}, \citenamefont {Bogorin}, \citenamefont {Brink},
  \citenamefont {Capelluto}, \citenamefont {G{\"u}nl{\"u}k}, \citenamefont
  {Itoko}, \citenamefont {Kanazawa} \emph
  {et~al.}}]{jurcevic2021demonstration}%
  \BibitemOpen
  \bibfield  {author} {\bibinfo {author} {\bibfnamefont {P.}~\bibnamefont
  {Jurcevic}}, \bibinfo {author} {\bibfnamefont {A.}~\bibnamefont
  {Javadi-Abhari}}, \bibinfo {author} {\bibfnamefont {L.~S.}\ \bibnamefont
  {Bishop}}, \bibinfo {author} {\bibfnamefont {I.}~\bibnamefont {Lauer}},
  \bibinfo {author} {\bibfnamefont {D.~F.}\ \bibnamefont {Bogorin}}, \bibinfo
  {author} {\bibfnamefont {M.}~\bibnamefont {Brink}}, \bibinfo {author}
  {\bibfnamefont {L.}~\bibnamefont {Capelluto}}, \bibinfo {author}
  {\bibfnamefont {O.}~\bibnamefont {G{\"u}nl{\"u}k}}, \bibinfo {author}
  {\bibfnamefont {T.}~\bibnamefont {Itoko}}, \bibinfo {author} {\bibfnamefont
  {N.}~\bibnamefont {Kanazawa}}, \emph {et~al.},\ }\href
  {https://iopscience.iop.org/article/10.1088/2058-9565/abe519/meta} {\bibfield
   {journal} {\bibinfo  {journal} {Quant. Sci. Tech}\ }\textbf {\bibinfo
  {volume} {6}},\ \bibinfo {pages} {025020} (\bibinfo {year}
  {2021})}\BibitemShut {NoStop}%
\bibitem [{\citenamefont {Beals}\ \emph {et~al.}(2013)\citenamefont {Beals},
  \citenamefont {Brierley}, \citenamefont {Gray}, \citenamefont {Harrow},
  \citenamefont {Kutin}, \citenamefont {Linden}, \citenamefont {Shepherd},\
  and\ \citenamefont {Stather}}]{beals2013efficient}%
  \BibitemOpen
  \bibfield  {author} {\bibinfo {author} {\bibfnamefont {R.}~\bibnamefont
  {Beals}}, \bibinfo {author} {\bibfnamefont {S.}~\bibnamefont {Brierley}},
  \bibinfo {author} {\bibfnamefont {O.}~\bibnamefont {Gray}}, \bibinfo {author}
  {\bibfnamefont {A.~W.}\ \bibnamefont {Harrow}}, \bibinfo {author}
  {\bibfnamefont {S.}~\bibnamefont {Kutin}}, \bibinfo {author} {\bibfnamefont
  {N.}~\bibnamefont {Linden}}, \bibinfo {author} {\bibfnamefont
  {D.}~\bibnamefont {Shepherd}},\ and\ \bibinfo {author} {\bibfnamefont
  {M.}~\bibnamefont {Stather}},\ }\href
  {https://royalsocietypublishing.org/doi/abs/10.1098/rspa.2012.0686}
  {\bibfield  {journal} {\bibinfo  {journal} {Proc. Roy. Soc. London A, Math.
  Phys. Sci}\ }\textbf {\bibinfo {volume} {469}},\ \bibinfo {pages} {20120686}
  (\bibinfo {year} {2013})}\BibitemShut {NoStop}%
\bibitem [{\citenamefont {Fowler}\ \emph {et~al.}(2004)\citenamefont {Fowler},
  \citenamefont {Devitt},\ and\ \citenamefont
  {Hollenberg}}]{fowler2004implementation}%
  \BibitemOpen
  \bibfield  {author} {\bibinfo {author} {\bibfnamefont {A.~G.}\ \bibnamefont
  {Fowler}}, \bibinfo {author} {\bibfnamefont {S.~J.}\ \bibnamefont {Devitt}},\
  and\ \bibinfo {author} {\bibfnamefont {L.~C.~L.}\ \bibnamefont
  {Hollenberg}},\ }\href {https://dl.acm.org/doi/10.5555/2011827.2011828}
  {\bibfield  {journal} {\bibinfo  {journal} {Quant. Info. Comput}\ }\textbf
  {\bibinfo {volume} {4}},\ \bibinfo {pages} {237–251} (\bibinfo {year}
  {2004})}\BibitemShut {NoStop}%
\bibitem [{\citenamefont {Maslov}(2007)}]{maslov2007linear}%
  \BibitemOpen
  \bibfield  {author} {\bibinfo {author} {\bibfnamefont {D.}~\bibnamefont
  {Maslov}},\ }\href
  {https://journals.aps.org/pra/abstract/10.1103/PhysRevA.76.052310} {\bibfield
   {journal} {\bibinfo  {journal} {Phys. Rev. A}\ }\textbf {\bibinfo {volume}
  {76}},\ \bibinfo {pages} {052310} (\bibinfo {year} {2007})}\BibitemShut
  {NoStop}%
\bibitem [{\citenamefont {Kutin}(2006)}]{kutin2006shor}%
  \BibitemOpen
  \bibfield  {author} {\bibinfo {author} {\bibfnamefont {S.~A.}\ \bibnamefont
  {Kutin}},\ }\href {https://arxiv.org/abs/quant-ph/0609001} {\bibfield
  {journal} {\bibinfo  {journal} {quant-ph/0609001}\ } (\bibinfo {year}
  {2006})}\BibitemShut {NoStop}%
\bibitem [{\citenamefont {Brierley}(2017)}]{brierley2015efficient}%
  \BibitemOpen
  \bibfield  {author} {\bibinfo {author} {\bibfnamefont {S.}~\bibnamefont
  {Brierley}},\ }\href {https://dl.acm.org/doi/10.5555/3179575.3179577}
  {\bibfield  {journal} {\bibinfo  {journal} {Quant. Info. Comput}\ }\textbf
  {\bibinfo {volume} {17}},\ \bibinfo {pages} {1096–1104} (\bibinfo {year}
  {2017})}\BibitemShut {NoStop}%
\bibitem [{\citenamefont {Herbert}(2018)}]{herbert2018depth}%
  \BibitemOpen
  \bibfield  {author} {\bibinfo {author} {\bibfnamefont {S.}~\bibnamefont
  {Herbert}},\ }\href {https://arxiv.org/abs/1805.12570} {\bibfield  {journal}
  {\bibinfo  {journal} {arXiv:1805.12570}\ } (\bibinfo {year}
  {2018})}\BibitemShut {NoStop}%
\bibitem [{\citenamefont {Temme}\ \emph {et~al.}(2017)\citenamefont {Temme},
  \citenamefont {Bravyi},\ and\ \citenamefont {Gambetta}}]{temme2017error}%
  \BibitemOpen
  \bibfield  {author} {\bibinfo {author} {\bibfnamefont {K.}~\bibnamefont
  {Temme}}, \bibinfo {author} {\bibfnamefont {S.}~\bibnamefont {Bravyi}},\ and\
  \bibinfo {author} {\bibfnamefont {J.~M.}\ \bibnamefont {Gambetta}},\ }\href
  {https://journals.aps.org/prl/abstract/10.1103/PhysRevLett.119.180509}
  {\bibfield  {journal} {\bibinfo  {journal} {Phys. Rev. Lett}\ }\textbf
  {\bibinfo {volume} {119}},\ \bibinfo {pages} {180509} (\bibinfo {year}
  {2017})}\BibitemShut {NoStop}%
\bibitem [{\citenamefont {Nation}\ \emph {et~al.}(2021)\citenamefont {Nation},
  \citenamefont {Kang}, \citenamefont {Sundaresan},\ and\ \citenamefont
  {Gambetta}}]{nation2021scalable}%
  \BibitemOpen
  \bibfield  {author} {\bibinfo {author} {\bibfnamefont {P.~D.}\ \bibnamefont
  {Nation}}, \bibinfo {author} {\bibfnamefont {H.}~\bibnamefont {Kang}},
  \bibinfo {author} {\bibfnamefont {N.}~\bibnamefont {Sundaresan}},\ and\
  \bibinfo {author} {\bibfnamefont {J.~M.}\ \bibnamefont {Gambetta}},\ }\href
  {https://journals.aps.org/prxquantum/abstract/10.1103/PRXQuantum.2.040326}
  {\bibfield  {journal} {\bibinfo  {journal} {PRX Quantum}\ }\textbf {\bibinfo
  {volume} {2}},\ \bibinfo {pages} {040326} (\bibinfo {year}
  {2021})}\BibitemShut {NoStop}%
\bibitem [{\citenamefont {Viola}\ and\ \citenamefont
  {Lloyd}(1998)}]{viola1998dynamical}%
  \BibitemOpen
  \bibfield  {author} {\bibinfo {author} {\bibfnamefont {L.}~\bibnamefont
  {Viola}}\ and\ \bibinfo {author} {\bibfnamefont {S.}~\bibnamefont {Lloyd}},\
  }\href {https://journals.aps.org/pra/abstract/10.1103/PhysRevA.58.2733}
  {\bibfield  {journal} {\bibinfo  {journal} {Phys. Rev. A}\ }\textbf {\bibinfo
  {volume} {58}},\ \bibinfo {pages} {2733} (\bibinfo {year}
  {1998})}\BibitemShut {NoStop}%
\bibitem [{\citenamefont {Biercuk}\ \emph {et~al.}(2009)\citenamefont
  {Biercuk}, \citenamefont {Uys}, \citenamefont {VanDevender}, \citenamefont
  {Shiga}, \citenamefont {Itano},\ and\ \citenamefont
  {Bollinger}}]{biercuk2009optimized}%
  \BibitemOpen
  \bibfield  {author} {\bibinfo {author} {\bibfnamefont {M.~J.}\ \bibnamefont
  {Biercuk}}, \bibinfo {author} {\bibfnamefont {H.}~\bibnamefont {Uys}},
  \bibinfo {author} {\bibfnamefont {A.~P.}\ \bibnamefont {VanDevender}},
  \bibinfo {author} {\bibfnamefont {N.}~\bibnamefont {Shiga}}, \bibinfo
  {author} {\bibfnamefont {W.~M.}\ \bibnamefont {Itano}},\ and\ \bibinfo
  {author} {\bibfnamefont {J.~J.}\ \bibnamefont {Bollinger}},\ }\href
  {https://www.nature.com/articles/nature07951} {\bibfield  {journal} {\bibinfo
   {journal} {Nature}\ }\textbf {\bibinfo {volume} {458}},\ \bibinfo {pages}
  {996} (\bibinfo {year} {2009})}\BibitemShut {NoStop}%
\bibitem [{\citenamefont {Kandala}\ \emph {et~al.}(2018)\citenamefont
  {Kandala}, \citenamefont {Temme}, \citenamefont {Corcoles}, \citenamefont
  {Mezzacapo}, \citenamefont {Chow},\ and\ \citenamefont
  {Gambetta}}]{kandala2018extending}%
  \BibitemOpen
  \bibfield  {author} {\bibinfo {author} {\bibfnamefont {A.}~\bibnamefont
  {Kandala}}, \bibinfo {author} {\bibfnamefont {K.}~\bibnamefont {Temme}},
  \bibinfo {author} {\bibfnamefont {A.~D.}\ \bibnamefont {Corcoles}}, \bibinfo
  {author} {\bibfnamefont {A.}~\bibnamefont {Mezzacapo}}, \bibinfo {author}
  {\bibfnamefont {J.~M.}\ \bibnamefont {Chow}},\ and\ \bibinfo {author}
  {\bibfnamefont {J.~M.}\ \bibnamefont {Gambetta}},\ }\href
  {https://www.nature.com/articles/s41586-019-1040-7} {\bibfield  {journal}
  {\bibinfo  {journal} {Nature}\ }\textbf {\bibinfo {volume} {567}},\ \bibinfo
  {pages} {491} (\bibinfo {year} {2018})}\BibitemShut {NoStop}%
\bibitem [{\citenamefont {Wu}\ and\ \citenamefont
  {Lidar}(2002)}]{wu2002qubits}%
  \BibitemOpen
  \bibfield  {author} {\bibinfo {author} {\bibfnamefont {L.-A.}\ \bibnamefont
  {Wu}}\ and\ \bibinfo {author} {\bibfnamefont {D.}~\bibnamefont {Lidar}},\
  }\href {https://aip.scitation.org/doi/10.1063/1.1499208} {\bibfield
  {journal} {\bibinfo  {journal} {J. Math. Phys}\ }\textbf {\bibinfo {volume}
  {43}},\ \bibinfo {pages} {4506} (\bibinfo {year} {2002})}\BibitemShut
  {NoStop}%
\bibitem [{\citenamefont {Batista}\ and\ \citenamefont
  {Ortiz}(2004)}]{batista2004algebraic}%
  \BibitemOpen
  \bibfield  {author} {\bibinfo {author} {\bibfnamefont {C.~D.}\ \bibnamefont
  {Batista}}\ and\ \bibinfo {author} {\bibfnamefont {G.}~\bibnamefont
  {Ortiz}},\ }\href
  {http://citeseerx.ist.psu.edu/viewdoc/download;jsessionid=9813919D5A55FCF21E625A112327FC7E?doi=10.1.1.130.6443&rep=rep1&type=pdf}
  {\bibfield  {journal} {\bibinfo  {journal} {Adv. Phys}\ }\textbf {\bibinfo
  {volume} {53}},\ \bibinfo {pages} {1} (\bibinfo {year} {2004})}\BibitemShut
  {NoStop}%
\bibitem [{\citenamefont {Bravyi}\ and\ \citenamefont
  {Kitaev}(2002)}]{bravyi2002fermionic}%
  \BibitemOpen
  \bibfield  {author} {\bibinfo {author} {\bibfnamefont {S.~B.}\ \bibnamefont
  {Bravyi}}\ and\ \bibinfo {author} {\bibfnamefont {A.~Y.}\ \bibnamefont
  {Kitaev}},\ }\href
  {https://www.sciencedirect.com/science/article/abs/pii/S0003491602962548}
  {\bibfield  {journal} {\bibinfo  {journal} {Ann. Phys}\ }\textbf {\bibinfo
  {volume} {298}},\ \bibinfo {pages} {210} (\bibinfo {year}
  {2002})}\BibitemShut {NoStop}%
\bibitem [{\citenamefont {Jordan}\ and\ \citenamefont
  {Wigner}(1993)}]{jordan1993paulische}%
  \BibitemOpen
  \bibfield  {author} {\bibinfo {author} {\bibfnamefont {P.}~\bibnamefont
  {Jordan}}\ and\ \bibinfo {author} {\bibfnamefont {E.~P.}\ \bibnamefont
  {Wigner}},\ }in\ \href {https://link.springer.com/article/10.1007/BF01331938}
  {\emph {\bibinfo {booktitle} {The Collected Works of Eugene Paul Wigner}}}\
  (\bibinfo  {publisher} {Springer},\ \bibinfo {year} {1993})\ pp.\ \bibinfo
  {pages} {109--129}\BibitemShut {NoStop}%
\bibitem [{\citenamefont {Abrams}\ and\ \citenamefont
  {Lloyd}(1997)}]{abrams1997simulation}%
  \BibitemOpen
  \bibfield  {author} {\bibinfo {author} {\bibfnamefont {D.~S.}\ \bibnamefont
  {Abrams}}\ and\ \bibinfo {author} {\bibfnamefont {S.}~\bibnamefont {Lloyd}},\
  }\href {https://doi.org/10.1103/PhysRevLett.79.2586} {\bibfield  {journal}
  {\bibinfo  {journal} {Phys. Rev. Lett}\ }\textbf {\bibinfo {volume} {79}},\
  \bibinfo {pages} {2586} (\bibinfo {year} {1997})}\BibitemShut {NoStop}%
\bibitem [{\citenamefont {Ortiz}\ \emph {et~al.}(2001)\citenamefont {Ortiz},
  \citenamefont {Gubernatis}, \citenamefont {Knill},\ and\ \citenamefont
  {Laflamme}}]{ortiz2001quantum}%
  \BibitemOpen
  \bibfield  {author} {\bibinfo {author} {\bibfnamefont {G.}~\bibnamefont
  {Ortiz}}, \bibinfo {author} {\bibfnamefont {J.~E.}\ \bibnamefont
  {Gubernatis}}, \bibinfo {author} {\bibfnamefont {E.}~\bibnamefont {Knill}},\
  and\ \bibinfo {author} {\bibfnamefont {R.}~\bibnamefont {Laflamme}},\ }\href
  {https://journals.aps.org/pra/abstract/10.1103/PhysRevA.64.022319} {\bibfield
   {journal} {\bibinfo  {journal} {Phys. Rev. A}\ }\textbf {\bibinfo {volume}
  {64}},\ \bibinfo {pages} {022319} (\bibinfo {year} {2001})}\BibitemShut
  {NoStop}%
\bibitem [{\citenamefont {Somma}\ \emph {et~al.}(2002)\citenamefont {Somma},
  \citenamefont {Ortiz}, \citenamefont {Gubernatis}, \citenamefont {Knill},\
  and\ \citenamefont {Laflamme}}]{somma2002simulating}%
  \BibitemOpen
  \bibfield  {author} {\bibinfo {author} {\bibfnamefont {R.}~\bibnamefont
  {Somma}}, \bibinfo {author} {\bibfnamefont {G.}~\bibnamefont {Ortiz}},
  \bibinfo {author} {\bibfnamefont {J.~E.}\ \bibnamefont {Gubernatis}},
  \bibinfo {author} {\bibfnamefont {E.}~\bibnamefont {Knill}},\ and\ \bibinfo
  {author} {\bibfnamefont {R.}~\bibnamefont {Laflamme}},\ }\href
  {https://journals.aps.org/pra/abstract/10.1103/PhysRevA.65.042323} {\bibfield
   {journal} {\bibinfo  {journal} {Phys. Rev. A}\ }\textbf {\bibinfo {volume}
  {65}},\ \bibinfo {pages} {042323} (\bibinfo {year} {2002})}\BibitemShut
  {NoStop}%
\bibitem [{\citenamefont {Aspuru-Guzik}\ \emph {et~al.}(2005)\citenamefont
  {Aspuru-Guzik}, \citenamefont {Dutoi}, \citenamefont {Love},\ and\
  \citenamefont {Head-Gordon}}]{aspuru2005simulated}%
  \BibitemOpen
  \bibfield  {author} {\bibinfo {author} {\bibfnamefont {A.}~\bibnamefont
  {Aspuru-Guzik}}, \bibinfo {author} {\bibfnamefont {A.~D.}\ \bibnamefont
  {Dutoi}}, \bibinfo {author} {\bibfnamefont {P.~J.}\ \bibnamefont {Love}},\
  and\ \bibinfo {author} {\bibfnamefont {M.}~\bibnamefont {Head-Gordon}},\
  }\href {https://science.sciencemag.org/content/309/5741/1704} {\bibfield
  {journal} {\bibinfo  {journal} {Science}\ }\textbf {\bibinfo {volume}
  {309}},\ \bibinfo {pages} {1704} (\bibinfo {year} {2005})}\BibitemShut
  {NoStop}%
\bibitem [{\citenamefont {Whitfield}\ \emph {et~al.}(2011)\citenamefont
  {Whitfield}, \citenamefont {Biamonte},\ and\ \citenamefont
  {Aspuru-Guzik}}]{whitfield2011simulation}%
  \BibitemOpen
  \bibfield  {author} {\bibinfo {author} {\bibfnamefont {J.~D.}\ \bibnamefont
  {Whitfield}}, \bibinfo {author} {\bibfnamefont {J.}~\bibnamefont
  {Biamonte}},\ and\ \bibinfo {author} {\bibfnamefont {A.}~\bibnamefont
  {Aspuru-Guzik}},\ }\href
  {https://www.tandfonline.com/doi/abs/10.1080/00268976.2011.552441} {\bibfield
   {journal} {\bibinfo  {journal} {Mol. Phys}\ }\textbf {\bibinfo {volume}
  {109}},\ \bibinfo {pages} {735} (\bibinfo {year} {2011})}\BibitemShut
  {NoStop}%
\bibitem [{\citenamefont {Reck}\ \emph {et~al.}(1994)\citenamefont {Reck},
  \citenamefont {Zeilinger}, \citenamefont {Bernstein},\ and\ \citenamefont
  {Bertani}}]{reck1994experimental}%
  \BibitemOpen
  \bibfield  {author} {\bibinfo {author} {\bibfnamefont {M.}~\bibnamefont
  {Reck}}, \bibinfo {author} {\bibfnamefont {A.}~\bibnamefont {Zeilinger}},
  \bibinfo {author} {\bibfnamefont {H.~J.}\ \bibnamefont {Bernstein}},\ and\
  \bibinfo {author} {\bibfnamefont {P.}~\bibnamefont {Bertani}},\ }\href
  {https://journals.aps.org/prl/abstract/10.1103/PhysRevLett.73.58} {\bibfield
  {journal} {\bibinfo  {journal} {Phys. Rev. Lett}\ }\textbf {\bibinfo {volume}
  {73}},\ \bibinfo {pages} {58} (\bibinfo {year} {1994})}\BibitemShut {NoStop}%
\bibitem [{\citenamefont {Clements}\ \emph {et~al.}(2016)\citenamefont
  {Clements}, \citenamefont {Humphreys}, \citenamefont {Metcalf}, \citenamefont
  {Kolthammer},\ and\ \citenamefont {Walmsley}}]{clements2016optimal}%
  \BibitemOpen
  \bibfield  {author} {\bibinfo {author} {\bibfnamefont {W.~R.}\ \bibnamefont
  {Clements}}, \bibinfo {author} {\bibfnamefont {P.~C.}\ \bibnamefont
  {Humphreys}}, \bibinfo {author} {\bibfnamefont {B.~J.}\ \bibnamefont
  {Metcalf}}, \bibinfo {author} {\bibfnamefont {W.~S.}\ \bibnamefont
  {Kolthammer}},\ and\ \bibinfo {author} {\bibfnamefont {I.~A.}\ \bibnamefont
  {Walmsley}},\ }\href {https://doi.org/10.1364/OPTICA.3.001460} {\bibfield
  {journal} {\bibinfo  {journal} {Optica}\ }\textbf {\bibinfo {volume} {3}},\
  \bibinfo {pages} {1460} (\bibinfo {year} {2016})}\BibitemShut {NoStop}%
\bibitem [{\citenamefont {Jiang}\ \emph {et~al.}(2018)\citenamefont {Jiang},
  \citenamefont {Sung}, \citenamefont {Kechedzhi}, \citenamefont
  {Smelyanskiy},\ and\ \citenamefont {Boixo}}]{jiang2018quantum}%
  \BibitemOpen
  \bibfield  {author} {\bibinfo {author} {\bibfnamefont {Z.}~\bibnamefont
  {Jiang}}, \bibinfo {author} {\bibfnamefont {K.~J.}\ \bibnamefont {Sung}},
  \bibinfo {author} {\bibfnamefont {K.}~\bibnamefont {Kechedzhi}}, \bibinfo
  {author} {\bibfnamefont {V.~N.}\ \bibnamefont {Smelyanskiy}},\ and\ \bibinfo
  {author} {\bibfnamefont {S.}~\bibnamefont {Boixo}},\ }\href
  {https://journals.aps.org/prapplied/abstract/10.1103/PhysRevApplied.9.044036}
  {\bibfield  {journal} {\bibinfo  {journal} {Phys. Rev. Appl}\ }\textbf
  {\bibinfo {volume} {9}},\ \bibinfo {pages} {044036} (\bibinfo {year}
  {2018})}\BibitemShut {NoStop}%
\bibitem [{\citenamefont {Motta}\ \emph {et~al.}(2021)\citenamefont {Motta},
  \citenamefont {Ye}, \citenamefont {McClean}, \citenamefont {Li},
  \citenamefont {Minnich}, \citenamefont {Babbush},\ and\ \citenamefont
  {Chan}}]{motta2021low}%
  \BibitemOpen
  \bibfield  {author} {\bibinfo {author} {\bibfnamefont {M.}~\bibnamefont
  {Motta}}, \bibinfo {author} {\bibfnamefont {E.}~\bibnamefont {Ye}}, \bibinfo
  {author} {\bibfnamefont {J.~R.}\ \bibnamefont {McClean}}, \bibinfo {author}
  {\bibfnamefont {Z.}~\bibnamefont {Li}}, \bibinfo {author} {\bibfnamefont
  {A.~J.}\ \bibnamefont {Minnich}}, \bibinfo {author} {\bibfnamefont
  {R.}~\bibnamefont {Babbush}},\ and\ \bibinfo {author} {\bibfnamefont {G.~K.}\
  \bibnamefont {Chan}},\ }\href
  {https://www.nature.com/articles/s41534-021-00416-z} {\bibfield  {journal}
  {\bibinfo  {journal} {npj Quantum Inf}\ }\textbf {\bibinfo {volume} {7}},\
  \bibinfo {pages} {1} (\bibinfo {year} {2021})}\BibitemShut {NoStop}%
\bibitem [{\citenamefont {Motta}\ \emph
  {et~al.}(2023{\natexlab{a}})\citenamefont {Motta}, \citenamefont {Sung},
  \citenamefont {Whaley}, \citenamefont {Head-Gordon},\ and\ \citenamefont
  {Shee}}]{motta2023bridging}%
  \BibitemOpen
  \bibfield  {author} {\bibinfo {author} {\bibfnamefont {M.}~\bibnamefont
  {Motta}}, \bibinfo {author} {\bibfnamefont {K.~J.}\ \bibnamefont {Sung}},
  \bibinfo {author} {\bibfnamefont {K.~B.}\ \bibnamefont {Whaley}}, \bibinfo
  {author} {\bibfnamefont {M.}~\bibnamefont {Head-Gordon}},\ and\ \bibinfo
  {author} {\bibfnamefont {J.}~\bibnamefont {Shee}},\ }\href
  {https://pubs.rsc.org/en/Content/ArticleLanding/2023/SC/D3SC02516K}
  {\bibfield  {journal} {\bibinfo  {journal} {Chem. Sci}\ }\textbf {\bibinfo
  {volume} {14}},\ \bibinfo {pages} {11213} (\bibinfo {year}
  {2023}{\natexlab{a}})}\BibitemShut {NoStop}%
\bibitem [{\citenamefont {Whitten}(1973)}]{Whitten:1973:4496}%
  \BibitemOpen
  \bibfield  {author} {\bibinfo {author} {\bibfnamefont {J.~L.}\ \bibnamefont
  {Whitten}},\ }\href {https://doi.org/NONE} {\bibfield  {journal} {\bibinfo
  {journal} {J. Chem. Phys}\ }\textbf {\bibinfo {volume} {58}},\ \bibinfo
  {pages} {4496} (\bibinfo {year} {1973})}\BibitemShut {NoStop}%
\bibitem [{\citenamefont {Dunlap}\ \emph {et~al.}(1977)\citenamefont {Dunlap},
  \citenamefont {Connolly},\ and\ \citenamefont {Sabin}}]{Dunlap:1977:81}%
  \BibitemOpen
  \bibfield  {author} {\bibinfo {author} {\bibfnamefont {B.~I.}\ \bibnamefont
  {Dunlap}}, \bibinfo {author} {\bibfnamefont {J.~W.~D.}\ \bibnamefont
  {Connolly}},\ and\ \bibinfo {author} {\bibfnamefont {J.~R.}\ \bibnamefont
  {Sabin}},\ }\href {https://doi.org/Transition-metal Atoms - Nickel Atom and
  Nickel H} {\bibfield  {journal} {\bibinfo  {journal} {Int. J. Quantum Chem.
  Symp}\ }\textbf {\bibinfo {volume} {12}},\ \bibinfo {pages} {81} (\bibinfo
  {year} {1977})}\BibitemShut {NoStop}%
\bibitem [{\citenamefont {Dunlap}\ \emph {et~al.}(1979)\citenamefont {Dunlap},
  \citenamefont {Connolly},\ and\ \citenamefont {Sabin}}]{Dunlap:1979:3396}%
  \BibitemOpen
  \bibfield  {author} {\bibinfo {author} {\bibfnamefont {B.~I.}\ \bibnamefont
  {Dunlap}}, \bibinfo {author} {\bibfnamefont {J.~W.~D.}\ \bibnamefont
  {Connolly}},\ and\ \bibinfo {author} {\bibfnamefont {J.~R.}\ \bibnamefont
  {Sabin}},\ }\href {https://doi.org/10.1063/1.438728} {\bibfield  {journal}
  {\bibinfo  {journal} {J. Chem. Phys}\ }\textbf {\bibinfo {volume} {71}},\
  \bibinfo {pages} {3396} (\bibinfo {year} {1979})}\BibitemShut {NoStop}%
\bibitem [{\citenamefont {Feyereisen}\ \emph {et~al.}(1993)\citenamefont
  {Feyereisen}, \citenamefont {Fitzgerald},\ and\ \citenamefont
  {Komornicki}}]{Feyereisen:1993:359}%
  \BibitemOpen
  \bibfield  {author} {\bibinfo {author} {\bibfnamefont {M.}~\bibnamefont
  {Feyereisen}}, \bibinfo {author} {\bibfnamefont {G.}~\bibnamefont
  {Fitzgerald}},\ and\ \bibinfo {author} {\bibfnamefont {A.}~\bibnamefont
  {Komornicki}},\ }\href
  {https://www.sciencedirect.com/science/article/abs/pii/000926149387156W?via%3Dihub}
  {\bibfield  {journal} {\bibinfo  {journal} {Chem. Phys. Lett}\ }\textbf
  {\bibinfo {volume} {208}},\ \bibinfo {pages} {359} (\bibinfo {year}
  {1993})}\BibitemShut {NoStop}%
\bibitem [{\citenamefont {Komornicki}\ and\ \citenamefont
  {Fitzgerald}(1993)}]{Komornicki:1993:1398}%
  \BibitemOpen
  \bibfield  {author} {\bibinfo {author} {\bibfnamefont {A.}~\bibnamefont
  {Komornicki}}\ and\ \bibinfo {author} {\bibfnamefont {G.}~\bibnamefont
  {Fitzgerald}},\ }\href {https://doi.org/http://dx.doi.org/10.1063/1.465054}
  {\bibfield  {journal} {\bibinfo  {journal} {J. Chem. Phys}\ }\textbf
  {\bibinfo {volume} {98}},\ \bibinfo {pages} {1398} (\bibinfo {year}
  {1993})}\BibitemShut {NoStop}%
\bibitem [{\citenamefont {Vahtras}\ \emph {et~al.}(1993)\citenamefont
  {Vahtras}, \citenamefont {Alml{\"o}f},\ and\ \citenamefont
  {Feyereisen}}]{Vahtras:1993:514}%
  \BibitemOpen
  \bibfield  {author} {\bibinfo {author} {\bibfnamefont {O.}~\bibnamefont
  {Vahtras}}, \bibinfo {author} {\bibfnamefont {J.}~\bibnamefont
  {Alml{\"o}f}},\ and\ \bibinfo {author} {\bibfnamefont {M.~W.}\ \bibnamefont
  {Feyereisen}},\ }\href {https://aip.scitation.org/doi/10.1063/1.2956507}
  {\bibfield  {journal} {\bibinfo  {journal} {Chem. Phys. Lett}\ }\textbf
  {\bibinfo {volume} {213}},\ \bibinfo {pages} {514} (\bibinfo {year}
  {1993})}\BibitemShut {NoStop}%
\bibitem [{\citenamefont {Rendell}\ and\ \citenamefont
  {Lee}(1994)}]{Rendell:1994:400}%
  \BibitemOpen
  \bibfield  {author} {\bibinfo {author} {\bibfnamefont {A.~P.}\ \bibnamefont
  {Rendell}}\ and\ \bibinfo {author} {\bibfnamefont {T.~J.}\ \bibnamefont
  {Lee}},\ }\href {https://doi.org/INTN} {\bibfield  {journal} {\bibinfo
  {journal} {J. Chem. Phys}\ }\textbf {\bibinfo {volume} {101}},\ \bibinfo
  {pages} {400} (\bibinfo {year} {1994})}\BibitemShut {NoStop}%
\bibitem [{\citenamefont {Kendall}\ and\ \citenamefont
  {Fruchtl}(1997)}]{Kendall:1997:158}%
  \BibitemOpen
  \bibfield  {author} {\bibinfo {author} {\bibfnamefont {R.~A.}\ \bibnamefont
  {Kendall}}\ and\ \bibinfo {author} {\bibfnamefont {H.~A.}\ \bibnamefont
  {Fruchtl}},\ }\href {https://doi.org/NONE} {\bibfield  {journal} {\bibinfo
  {journal} {Theor. Chem. Acc}\ }\textbf {\bibinfo {volume} {97}},\ \bibinfo
  {pages} {158} (\bibinfo {year} {1997})}\BibitemShut {NoStop}%
\bibitem [{\citenamefont {Weigend}(2002)}]{Weigend:2002:4285}%
  \BibitemOpen
  \bibfield  {author} {\bibinfo {author} {\bibfnamefont {F.}~\bibnamefont
  {Weigend}},\ }\href {https://doi.org/10.1039/B204199P} {\bibfield  {journal}
  {\bibinfo  {journal} {Phys. Chem. Chem. Phys}\ }\textbf {\bibinfo {volume}
  {4}},\ \bibinfo {pages} {4285} (\bibinfo {year} {2002})}\BibitemShut
  {NoStop}%
\bibitem [{\citenamefont {Beebe}\ and\ \citenamefont
  {Linderberg}(1977)}]{Beebe:1977:683}%
  \BibitemOpen
  \bibfield  {author} {\bibinfo {author} {\bibfnamefont {N.~H.~F.}\
  \bibnamefont {Beebe}}\ and\ \bibinfo {author} {\bibfnamefont
  {J.}~\bibnamefont {Linderberg}},\ }\href {https://doi.org/Two-electron
  Integrals in Molecular Calculations} {\bibfield  {journal} {\bibinfo
  {journal} {Int. J. Quantum Chem}\ }\textbf {\bibinfo {volume} {12}},\
  \bibinfo {pages} {683} (\bibinfo {year} {1977})}\BibitemShut {NoStop}%
\bibitem [{\citenamefont {Roeggen}\ and\ \citenamefont
  {Wisloff-Nilssen}(1986)}]{Roeggen:1986:154}%
  \BibitemOpen
  \bibfield  {author} {\bibinfo {author} {\bibfnamefont {I.}~\bibnamefont
  {Roeggen}}\ and\ \bibinfo {author} {\bibfnamefont {E.}~\bibnamefont
  {Wisloff-Nilssen}},\ }\href {https://doi.org/GEXT} {\bibfield  {journal}
  {\bibinfo  {journal} {Chem. Phys. Lett}\ }\textbf {\bibinfo {volume} {132}},\
  \bibinfo {pages} {154} (\bibinfo {year} {1986})}\BibitemShut {NoStop}%
\bibitem [{\citenamefont {Koch}\ \emph {et~al.}(2003)\citenamefont {Koch},
  \citenamefont {de~Meras},\ and\ \citenamefont {Pedersen}}]{Koch:2003:9481}%
  \BibitemOpen
  \bibfield  {author} {\bibinfo {author} {\bibfnamefont {H.}~\bibnamefont
  {Koch}}, \bibinfo {author} {\bibfnamefont {A.~S.}\ \bibnamefont {de~Meras}},\
  and\ \bibinfo {author} {\bibfnamefont {T.~B.}\ \bibnamefont {Pedersen}},\
  }\href {https://doi.org/10.1063/1.1578621} {\bibfield  {journal} {\bibinfo
  {journal} {J. Chem. Phys}\ }\textbf {\bibinfo {volume} {118}},\ \bibinfo
  {pages} {9481} (\bibinfo {year} {2003})}\BibitemShut {NoStop}%
\bibitem [{\citenamefont {Aquilante}\ \emph {et~al.}(2007)\citenamefont
  {Aquilante}, \citenamefont {Pedersen},\ and\ \citenamefont
  {Lindh}}]{Aquilante:2007:194106}%
  \BibitemOpen
  \bibfield  {author} {\bibinfo {author} {\bibfnamefont {F.}~\bibnamefont
  {Aquilante}}, \bibinfo {author} {\bibfnamefont {T.~B.}\ \bibnamefont
  {Pedersen}},\ and\ \bibinfo {author} {\bibfnamefont {R.}~\bibnamefont
  {Lindh}},\ }\href {https://doi.org/10.1063/1.2736701} {\bibfield  {journal}
  {\bibinfo  {journal} {J. Chem. Phys}\ }\textbf {\bibinfo {volume} {126}},\
  \bibinfo {pages} {194106} (\bibinfo {year} {2007})}\BibitemShut {NoStop}%
\bibitem [{\citenamefont {Aquilante}\ \emph {et~al.}(2009)\citenamefont
  {Aquilante}, \citenamefont {Gagliardi}, \citenamefont {Pedersen},\ and\
  \citenamefont {Lindh}}]{Aquilante:2009:154107}%
  \BibitemOpen
  \bibfield  {author} {\bibinfo {author} {\bibfnamefont {F.}~\bibnamefont
  {Aquilante}}, \bibinfo {author} {\bibfnamefont {L.}~\bibnamefont
  {Gagliardi}}, \bibinfo {author} {\bibfnamefont {T.~B.}\ \bibnamefont
  {Pedersen}},\ and\ \bibinfo {author} {\bibfnamefont {R.}~\bibnamefont
  {Lindh}},\ }\href {https://doi.org/10.1063/1.3116784} {\bibfield  {journal}
  {\bibinfo  {journal} {J. Chem. Phys}\ }\textbf {\bibinfo {volume} {130}},\
  \bibinfo {pages} {154107} (\bibinfo {year} {2009})}\BibitemShut {NoStop}%
\bibitem [{\citenamefont {Motta}\ \emph {et~al.}(2019)\citenamefont {Motta},
  \citenamefont {Shee}, \citenamefont {Zhang},\ and\ \citenamefont
  {Chan}}]{motta2019efficient}%
  \BibitemOpen
  \bibfield  {author} {\bibinfo {author} {\bibfnamefont {M.}~\bibnamefont
  {Motta}}, \bibinfo {author} {\bibfnamefont {J.}~\bibnamefont {Shee}},
  \bibinfo {author} {\bibfnamefont {S.}~\bibnamefont {Zhang}},\ and\ \bibinfo
  {author} {\bibfnamefont {G.~K.-L.}\ \bibnamefont {Chan}},\ }\href
  {https://pubs.acs.org/doi/abs/10.1021/acs.jctc.8b00996} {\bibfield  {journal}
  {\bibinfo  {journal} {J. Chem. Theory Comput}\ }\textbf {\bibinfo {volume}
  {15}},\ \bibinfo {pages} {3510} (\bibinfo {year} {2019})}\BibitemShut
  {NoStop}%
\bibitem [{\citenamefont {Peng}\ and\ \citenamefont
  {Kowalski}(2017)}]{peng2017highly}%
  \BibitemOpen
  \bibfield  {author} {\bibinfo {author} {\bibfnamefont {B.}~\bibnamefont
  {Peng}}\ and\ \bibinfo {author} {\bibfnamefont {K.}~\bibnamefont
  {Kowalski}},\ }\href {https://pubs.acs.org/doi/10.1021/acs.jctc.7b00605}
  {\bibfield  {journal} {\bibinfo  {journal} {J. Chem. Theory Comput}\ }\textbf
  {\bibinfo {volume} {13}},\ \bibinfo {pages} {4179} (\bibinfo {year}
  {2017})}\BibitemShut {NoStop}%
\bibitem [{\citenamefont {Evans}(2010)}]{evans2010partial}%
  \BibitemOpen
  \bibfield  {author} {\bibinfo {author} {\bibfnamefont {L.~C.}\ \bibnamefont
  {Evans}},\ }\href {https://books.google.com/books?id=Xnu0o\_EJrCQC} {\emph
  {\bibinfo {title} {Partial differential equations}}}\ (\bibinfo  {publisher}
  {American Mathematical Society},\ \bibinfo {year} {2010})\BibitemShut
  {NoStop}%
\bibitem [{\citenamefont {Saad}(2011)}]{saad2011numerical}%
  \BibitemOpen
  \bibfield  {author} {\bibinfo {author} {\bibfnamefont {Y.}~\bibnamefont
  {Saad}},\ }\href {https://doi.org/10.1137/1.9781611970739} {\emph {\bibinfo
  {title} {Numerical methods for large eigenvalue problems: revised edition}}}\
  (\bibinfo  {publisher} {SIAM},\ \bibinfo {year} {2011})\BibitemShut {NoStop}%
\bibitem [{\citenamefont {Liesen}\ and\ \citenamefont
  {Strakos}(2013)}]{liesen2013krylov}%
  \BibitemOpen
  \bibfield  {author} {\bibinfo {author} {\bibfnamefont {J.}~\bibnamefont
  {Liesen}}\ and\ \bibinfo {author} {\bibfnamefont {Z.}~\bibnamefont
  {Strakos}},\ }\href {https://academic.oup.com/book/36426} {\emph {\bibinfo
  {title} {Krylov subspace methods: principles and analysis}}}\ (\bibinfo
  {publisher} {Oxford University Press},\ \bibinfo {year} {2013})\BibitemShut
  {NoStop}%
\bibitem [{\citenamefont {Kaniel}(1966)}]{kaniel1966estimates}%
  \BibitemOpen
  \bibfield  {author} {\bibinfo {author} {\bibfnamefont {S.}~\bibnamefont
  {Kaniel}},\ }\href {https://doi.org/10.1090/S0025-5718-1966-0234618-4}
  {\bibfield  {journal} {\bibinfo  {journal} {Math. Comput}\ }\textbf {\bibinfo
  {volume} {20}},\ \bibinfo {pages} {369} (\bibinfo {year} {1966})}\BibitemShut
  {NoStop}%
\bibitem [{\citenamefont {Paige}(1971)}]{paige1971computation}%
  \BibitemOpen
  \bibfield  {author} {\bibinfo {author} {\bibfnamefont {C.~C.}\ \bibnamefont
  {Paige}},\ }\emph {\bibinfo {title} {The computation of eigenvalues and
  eigenvectors of very large sparse matrices}},\ \href
  {https://ethos.bl.uk/OrderDetails.do?uin=uk.bl.ethos.307848} {Ph.D. thesis},\
  \bibinfo  {school} {University of London} (\bibinfo {year}
  {1971})\BibitemShut {NoStop}%
\bibitem [{\citenamefont {Saad}(1980)}]{saad1980rates}%
  \BibitemOpen
  \bibfield  {author} {\bibinfo {author} {\bibfnamefont {Y.}~\bibnamefont
  {Saad}},\ }\href {https://doi.org/10.1137/0717059} {\bibfield  {journal}
  {\bibinfo  {journal} {SIAM J. Numer. Anal}\ }\textbf {\bibinfo {volume}
  {17}},\ \bibinfo {pages} {687} (\bibinfo {year} {1980})}\BibitemShut
  {NoStop}%
\bibitem [{\citenamefont {Beckermann}\ and\ \citenamefont
  {Townsend}(2019)}]{beckermann2019bounds}%
  \BibitemOpen
  \bibfield  {author} {\bibinfo {author} {\bibfnamefont {B.}~\bibnamefont
  {Beckermann}}\ and\ \bibinfo {author} {\bibfnamefont {A.}~\bibnamefont
  {Townsend}},\ }\href {https://doi.org/10.1137/19M1244433} {\bibfield
  {journal} {\bibinfo  {journal} {SIAM Rev}\ }\textbf {\bibinfo {volume}
  {61}},\ \bibinfo {pages} {319} (\bibinfo {year} {2019})}\BibitemShut
  {NoStop}%
\bibitem [{\citenamefont {Lanczos}(1952)}]{lanczos1952solution}%
  \BibitemOpen
  \bibfield  {author} {\bibinfo {author} {\bibfnamefont {C.}~\bibnamefont
  {Lanczos}},\ }\href
  {https://nvlpubs.nist.gov/nistpubs/jres/049/jresv49n1p33_A1b.pdf} {\bibfield
  {journal} {\bibinfo  {journal} {J. Res. Nat. Bur. Standards}\ }\textbf
  {\bibinfo {volume} {49}},\ \bibinfo {pages} {33} (\bibinfo {year}
  {1952})}\BibitemShut {NoStop}%
\bibitem [{\citenamefont {Cullum}\ and\ \citenamefont
  {Willoughby}(2002)}]{cullum2002lanczos}%
  \BibitemOpen
  \bibfield  {author} {\bibinfo {author} {\bibfnamefont {J.~K.}\ \bibnamefont
  {Cullum}}\ and\ \bibinfo {author} {\bibfnamefont {R.~A.}\ \bibnamefont
  {Willoughby}},\ }\href {https://doi.org/10.1137/1.9780898719192} {\emph
  {\bibinfo {title} {{L}anczos algorithms for large symmetric eigenvalue
  computations}}},\ Vol.~\bibinfo {volume} {1}\ (\bibinfo  {publisher} {SIAM},\
  \bibinfo {year} {2002})\BibitemShut {NoStop}%
\bibitem [{\citenamefont {Ojalvo}\ and\ \citenamefont
  {Newman}(1970)}]{ojalvo1970vibration}%
  \BibitemOpen
  \bibfield  {author} {\bibinfo {author} {\bibfnamefont {I.}~\bibnamefont
  {Ojalvo}}\ and\ \bibinfo {author} {\bibfnamefont {M.}~\bibnamefont
  {Newman}},\ }\href {https://arc.aiaa.org/doi/10.2514/3.5878} {\bibfield
  {journal} {\bibinfo  {journal} {AIAA Journal}\ }\textbf {\bibinfo {volume}
  {8}},\ \bibinfo {pages} {1234} (\bibinfo {year} {1970})}\BibitemShut
  {NoStop}%
\bibitem [{\citenamefont {Davidson}(1975)}]{davidsorq1975theiterative}%
  \BibitemOpen
  \bibfield  {author} {\bibinfo {author} {\bibfnamefont {E.~R.}\ \bibnamefont
  {Davidson}},\ }\href {https://doi.org/10.1016/0021-9991(75)90065-0}
  {\bibfield  {journal} {\bibinfo  {journal} {J. Comp. Phys}\ }\textbf
  {\bibinfo {volume} {17}},\ \bibinfo {pages} {87} (\bibinfo {year}
  {1975})}\BibitemShut {NoStop}%
\bibitem [{\citenamefont {Morgan}\ and\ \citenamefont
  {Scott}(1986)}]{morgan1986generalizations}%
  \BibitemOpen
  \bibfield  {author} {\bibinfo {author} {\bibfnamefont {R.~B.}\ \bibnamefont
  {Morgan}}\ and\ \bibinfo {author} {\bibfnamefont {D.~S.}\ \bibnamefont
  {Scott}},\ }\href {https://doi.org/10.1137/0907054} {\bibfield  {journal}
  {\bibinfo  {journal} {SIAM J. Sci. Stat. Comput}\ }\textbf {\bibinfo {volume}
  {7}},\ \bibinfo {pages} {817} (\bibinfo {year} {1986})}\BibitemShut {NoStop}%
\bibitem [{\citenamefont {Vogiatzis}\ \emph {et~al.}(2017)\citenamefont
  {Vogiatzis}, \citenamefont {Ma}, \citenamefont {Olsen}, \citenamefont
  {Gagliardi},\ and\ \citenamefont {De~Jong}}]{vogiatzis2017pushing}%
  \BibitemOpen
  \bibfield  {author} {\bibinfo {author} {\bibfnamefont {K.~D.}\ \bibnamefont
  {Vogiatzis}}, \bibinfo {author} {\bibfnamefont {D.}~\bibnamefont {Ma}},
  \bibinfo {author} {\bibfnamefont {J.}~\bibnamefont {Olsen}}, \bibinfo
  {author} {\bibfnamefont {L.}~\bibnamefont {Gagliardi}},\ and\ \bibinfo
  {author} {\bibfnamefont {W.~A.}\ \bibnamefont {De~Jong}},\ }\href
  {https://doi.org/10.1063/1.4989858} {\bibfield  {journal} {\bibinfo
  {journal} {J. Chem. Phys}\ }\textbf {\bibinfo {volume} {147}},\ \bibinfo
  {pages} {184111} (\bibinfo {year} {2017})}\BibitemShut {NoStop}%
\bibitem [{\citenamefont {Sleijpen}\ and\ \citenamefont {Van~der
  Vorst}(2000)}]{sleijpen2000jacobi}%
  \BibitemOpen
  \bibfield  {author} {\bibinfo {author} {\bibfnamefont {G.~L.}\ \bibnamefont
  {Sleijpen}}\ and\ \bibinfo {author} {\bibfnamefont {H.~A.}\ \bibnamefont
  {Van~der Vorst}},\ }\href
  {https://epubs.siam.org/doi/abs/10.1137/S0036144599363084} {\bibfield
  {journal} {\bibinfo  {journal} {SIAM review}\ }\textbf {\bibinfo {volume}
  {42}},\ \bibinfo {pages} {267} (\bibinfo {year} {2000})}\BibitemShut
  {NoStop}%
\bibitem [{\citenamefont {Olsen}\ \emph {et~al.}(1990)\citenamefont {Olsen},
  \citenamefont {J{\o}rgensen},\ and\ \citenamefont
  {Simons}}]{olsen1990passing}%
  \BibitemOpen
  \bibfield  {author} {\bibinfo {author} {\bibfnamefont {J.}~\bibnamefont
  {Olsen}}, \bibinfo {author} {\bibfnamefont {P.}~\bibnamefont
  {J{\o}rgensen}},\ and\ \bibinfo {author} {\bibfnamefont {J.}~\bibnamefont
  {Simons}},\ }\href {https://doi.org/10.1016/0009-2614(90)85633-N} {\bibfield
  {journal} {\bibinfo  {journal} {Chem. Phys. Lett}\ }\textbf {\bibinfo
  {volume} {169}},\ \bibinfo {pages} {463} (\bibinfo {year}
  {1990})}\BibitemShut {NoStop}%
\bibitem [{\citenamefont {Dachsel}\ \emph {et~al.}(1998)\citenamefont
  {Dachsel}, \citenamefont {Nieplocha},\ and\ \citenamefont
  {Harrison}}]{dachsel1998out}%
  \BibitemOpen
  \bibfield  {author} {\bibinfo {author} {\bibfnamefont {H.}~\bibnamefont
  {Dachsel}}, \bibinfo {author} {\bibfnamefont {J.}~\bibnamefont {Nieplocha}},\
  and\ \bibinfo {author} {\bibfnamefont {R.}~\bibnamefont {Harrison}},\ }in\
  \href {https://ieeexplore.ieee.org/document/1437328} {\emph {\bibinfo
  {booktitle} {SC'98: Proceedings of the 1998 ACM/IEEE Conference on
  Supercomputing}}}\ (\bibinfo {organization} {IEEE},\ \bibinfo {year} {1998})\
  p.~\bibinfo {pages} {41}\BibitemShut {NoStop}%
\bibitem [{\citenamefont {Evangelisti}\ \emph {et~al.}(1996)\citenamefont
  {Evangelisti}, \citenamefont {Bendazzoli}, \citenamefont {Ansaloni},
  \citenamefont {Dur{\`\i}},\ and\ \citenamefont {Rossi}}]{evangelisti1996one}%
  \BibitemOpen
  \bibfield  {author} {\bibinfo {author} {\bibfnamefont {S.}~\bibnamefont
  {Evangelisti}}, \bibinfo {author} {\bibfnamefont {G.~L.}\ \bibnamefont
  {Bendazzoli}}, \bibinfo {author} {\bibfnamefont {R.}~\bibnamefont
  {Ansaloni}}, \bibinfo {author} {\bibfnamefont {F.}~\bibnamefont
  {Dur{\`\i}}},\ and\ \bibinfo {author} {\bibfnamefont {E.}~\bibnamefont
  {Rossi}},\ }\href {https://doi.org/10.1016/0009-2614(96)00177-7} {\bibfield
  {journal} {\bibinfo  {journal} {Chem. Phys. Lett}\ }\textbf {\bibinfo
  {volume} {252}},\ \bibinfo {pages} {437} (\bibinfo {year}
  {1996})}\BibitemShut {NoStop}%
\bibitem [{\citenamefont {Ben-Amor}\ \emph {et~al.}(1998)\citenamefont
  {Ben-Amor}, \citenamefont {Evangelisti}, \citenamefont {Maynau},\ and\
  \citenamefont {Rossi}}]{ben1998benchmark}%
  \BibitemOpen
  \bibfield  {author} {\bibinfo {author} {\bibfnamefont {N.}~\bibnamefont
  {Ben-Amor}}, \bibinfo {author} {\bibfnamefont {S.}~\bibnamefont
  {Evangelisti}}, \bibinfo {author} {\bibfnamefont {D.}~\bibnamefont
  {Maynau}},\ and\ \bibinfo {author} {\bibfnamefont {E.}~\bibnamefont
  {Rossi}},\ }\href {https://doi.org/10.1016/S0009-2614(98)00289-9} {\bibfield
  {journal} {\bibinfo  {journal} {Chem. Phys. Lett}\ }\textbf {\bibinfo
  {volume} {288}},\ \bibinfo {pages} {348} (\bibinfo {year}
  {1998})}\BibitemShut {NoStop}%
\bibitem [{\citenamefont {De~Jong}\ \emph {et~al.}(2010)\citenamefont
  {De~Jong}, \citenamefont {Bylaska}, \citenamefont {Govind}, \citenamefont
  {Janssen}, \citenamefont {Kowalski}, \citenamefont {M{\"u}ller},
  \citenamefont {Nielsen}, \citenamefont {van Dam}, \citenamefont {Veryazov},\
  and\ \citenamefont {Lindh}}]{de2010utilizing}%
  \BibitemOpen
  \bibfield  {author} {\bibinfo {author} {\bibfnamefont {W.~A.}\ \bibnamefont
  {De~Jong}}, \bibinfo {author} {\bibfnamefont {E.}~\bibnamefont {Bylaska}},
  \bibinfo {author} {\bibfnamefont {N.}~\bibnamefont {Govind}}, \bibinfo
  {author} {\bibfnamefont {C.~L.}\ \bibnamefont {Janssen}}, \bibinfo {author}
  {\bibfnamefont {K.}~\bibnamefont {Kowalski}}, \bibinfo {author}
  {\bibfnamefont {T.}~\bibnamefont {M{\"u}ller}}, \bibinfo {author}
  {\bibfnamefont {I.~M.}\ \bibnamefont {Nielsen}}, \bibinfo {author}
  {\bibfnamefont {H.~J.}\ \bibnamefont {van Dam}}, \bibinfo {author}
  {\bibfnamefont {V.}~\bibnamefont {Veryazov}},\ and\ \bibinfo {author}
  {\bibfnamefont {R.}~\bibnamefont {Lindh}},\ }\href
  {https://pubs.rsc.org/en/content/articlelanding/2010/cp/c002859b} {\bibfield
  {journal} {\bibinfo  {journal} {Phys. Chem. Chem. Phys}\ }\textbf {\bibinfo
  {volume} {12}},\ \bibinfo {pages} {6896} (\bibinfo {year}
  {2010})}\BibitemShut {NoStop}%
\bibitem [{\citenamefont {Sinanoglu}(1962)}]{sinanouglu1962many}%
  \BibitemOpen
  \bibfield  {author} {\bibinfo {author} {\bibfnamefont {O.}~\bibnamefont
  {Sinanoglu}},\ }\href {https://doi.org/10.1063/1.1732596} {\bibfield
  {journal} {\bibinfo  {journal} {J. Chem. Phys}\ }\textbf {\bibinfo {volume}
  {36}},\ \bibinfo {pages} {706} (\bibinfo {year} {1962})}\BibitemShut
  {NoStop}%
\bibitem [{\citenamefont {Hinze}\ and\ \citenamefont
  {Roothaan}(1967)}]{hinze1967multi}%
  \BibitemOpen
  \bibfield  {author} {\bibinfo {author} {\bibfnamefont {J.}~\bibnamefont
  {Hinze}}\ and\ \bibinfo {author} {\bibfnamefont {C.~C.}\ \bibnamefont
  {Roothaan}},\ }\href {https://doi.org/10.1143/PTPS.40.37} {\bibfield
  {journal} {\bibinfo  {journal} {Prog. Theor. Phys}\ }\textbf {\bibinfo
  {volume} {40}},\ \bibinfo {pages} {37} (\bibinfo {year} {1967})}\BibitemShut
  {NoStop}%
\bibitem [{\citenamefont {Shavitt}(1977)}]{shavitt1977method}%
  \BibitemOpen
  \bibfield  {author} {\bibinfo {author} {\bibfnamefont {I.}~\bibnamefont
  {Shavitt}},\ }in\ \href
  {https://link.springer.com/chapter/10.1007/978-1-4757-0887-5_6} {\emph
  {\bibinfo {booktitle} {Methods of electronic structure theory}}},\ \bibinfo
  {editor} {edited by\ \bibinfo {editor} {\bibfnamefont {H.~F.}\ \bibnamefont
  {Schaefer}}}\ (\bibinfo  {publisher} {Springer},\ \bibinfo {year} {1977})\
  Chap.~\bibinfo {chapter} {6}, pp.\ \bibinfo {pages} {189--275}\BibitemShut
  {NoStop}%
\bibitem [{\citenamefont {Bender}\ and\ \citenamefont
  {Davidson}(1969)}]{bender1969studies}%
  \BibitemOpen
  \bibfield  {author} {\bibinfo {author} {\bibfnamefont {C.~F.}\ \bibnamefont
  {Bender}}\ and\ \bibinfo {author} {\bibfnamefont {E.~R.}\ \bibnamefont
  {Davidson}},\ }\href
  {https://journals.aps.org/pr/abstract/10.1103/PhysRev.183.23} {\bibfield
  {journal} {\bibinfo  {journal} {Phys. Rev}\ }\textbf {\bibinfo {volume}
  {183}},\ \bibinfo {pages} {23} (\bibinfo {year} {1969})}\BibitemShut
  {NoStop}%
\bibitem [{\citenamefont {Pople}\ \emph {et~al.}(1987)\citenamefont {Pople},
  \citenamefont {Head-Gordon},\ and\ \citenamefont
  {Raghavachari}}]{pople1987quadratic}%
  \BibitemOpen
  \bibfield  {author} {\bibinfo {author} {\bibfnamefont {J.~A.}\ \bibnamefont
  {Pople}}, \bibinfo {author} {\bibfnamefont {M.}~\bibnamefont {Head-Gordon}},\
  and\ \bibinfo {author} {\bibfnamefont {K.}~\bibnamefont {Raghavachari}},\
  }\href {https://doi.org/10.1063/1.453520} {\bibfield  {journal} {\bibinfo
  {journal} {J. Chem. Phys}\ }\textbf {\bibinfo {volume} {87}},\ \bibinfo
  {pages} {5968} (\bibinfo {year} {1987})}\BibitemShut {NoStop}%
\bibitem [{\citenamefont {Langhoff}\ and\ \citenamefont
  {Davidson}(1974)}]{langhoff1974configuration}%
  \BibitemOpen
  \bibfield  {author} {\bibinfo {author} {\bibfnamefont {S.~R.}\ \bibnamefont
  {Langhoff}}\ and\ \bibinfo {author} {\bibfnamefont {E.~R.}\ \bibnamefont
  {Davidson}},\ }\href {https://doi.org/10.1002/qua.560080106} {\bibfield
  {journal} {\bibinfo  {journal} {Int. J. Quantum Chem}\ }\textbf {\bibinfo
  {volume} {8}},\ \bibinfo {pages} {61} (\bibinfo {year} {1974})}\BibitemShut
  {NoStop}%
\bibitem [{\citenamefont {Bytautas}\ and\ \citenamefont
  {Ruedenberg}(2009)}]{bytautas2009priori}%
  \BibitemOpen
  \bibfield  {author} {\bibinfo {author} {\bibfnamefont {L.}~\bibnamefont
  {Bytautas}}\ and\ \bibinfo {author} {\bibfnamefont {K.}~\bibnamefont
  {Ruedenberg}},\ }\href {https://doi.org/10.1016/j.chemphys.2008.11.021}
  {\bibfield  {journal} {\bibinfo  {journal} {Chem. Phys}\ }\textbf {\bibinfo
  {volume} {356}},\ \bibinfo {pages} {64} (\bibinfo {year} {2009})}\BibitemShut
  {NoStop}%
\bibitem [{\citenamefont {Anderson}\ \emph {et~al.}(2018)\citenamefont
  {Anderson}, \citenamefont {Heidar-Zadeh},\ and\ \citenamefont
  {Ayers}}]{anderson2018breaking}%
  \BibitemOpen
  \bibfield  {author} {\bibinfo {author} {\bibfnamefont {J.~S.}\ \bibnamefont
  {Anderson}}, \bibinfo {author} {\bibfnamefont {F.}~\bibnamefont
  {Heidar-Zadeh}},\ and\ \bibinfo {author} {\bibfnamefont {P.~W.}\ \bibnamefont
  {Ayers}},\ }\href {https://doi.org/10.1016/j.comptc.2018.08.017} {\bibfield
  {journal} {\bibinfo  {journal} {Theor. Comp. Chem}\ }\textbf {\bibinfo
  {volume} {1142}},\ \bibinfo {pages} {66} (\bibinfo {year}
  {2018})}\BibitemShut {NoStop}%
\bibitem [{\citenamefont {Huron}\ \emph {et~al.}(1973)\citenamefont {Huron},
  \citenamefont {Malrieu},\ and\ \citenamefont
  {Rancurel}}]{huron1973iterative}%
  \BibitemOpen
  \bibfield  {author} {\bibinfo {author} {\bibfnamefont {B.}~\bibnamefont
  {Huron}}, \bibinfo {author} {\bibfnamefont {J.}~\bibnamefont {Malrieu}},\
  and\ \bibinfo {author} {\bibfnamefont {P.}~\bibnamefont {Rancurel}},\ }\href
  {https://doi.org/10.1063/1.1679199} {\bibfield  {journal} {\bibinfo
  {journal} {J. Chem. Phys}\ }\textbf {\bibinfo {volume} {58}},\ \bibinfo
  {pages} {5745} (\bibinfo {year} {1973})}\BibitemShut {NoStop}%
\bibitem [{\citenamefont {Buenker}\ and\ \citenamefont
  {Peyerimhoff}(1974)}]{buenker1974individualized}%
  \BibitemOpen
  \bibfield  {author} {\bibinfo {author} {\bibfnamefont {R.~J.}\ \bibnamefont
  {Buenker}}\ and\ \bibinfo {author} {\bibfnamefont {S.~D.}\ \bibnamefont
  {Peyerimhoff}},\ }\href
  {https://link.springer.com/article/10.1007/BF02394557} {\bibfield  {journal}
  {\bibinfo  {journal} {Theor. Chim. Acta}\ }\textbf {\bibinfo {volume} {35}},\
  \bibinfo {pages} {33} (\bibinfo {year} {1974})}\BibitemShut {NoStop}%
\bibitem [{\citenamefont {Buenker}\ and\ \citenamefont
  {Peyerimhoff}(1975)}]{buenker1975energy}%
  \BibitemOpen
  \bibfield  {author} {\bibinfo {author} {\bibfnamefont {R.~J.}\ \bibnamefont
  {Buenker}}\ and\ \bibinfo {author} {\bibfnamefont {S.~D.}\ \bibnamefont
  {Peyerimhoff}},\ }\href
  {https://link.springer.com/article/10.1007/bf00555301} {\bibfield  {journal}
  {\bibinfo  {journal} {Theor. Chim. Acta}\ }\textbf {\bibinfo {volume} {39}},\
  \bibinfo {pages} {217} (\bibinfo {year} {1975})}\BibitemShut {NoStop}%
\bibitem [{\citenamefont {Booth}\ \emph {et~al.}(2009)\citenamefont {Booth},
  \citenamefont {Thom},\ and\ \citenamefont {Alavi}}]{booth2009fermion}%
  \BibitemOpen
  \bibfield  {author} {\bibinfo {author} {\bibfnamefont {G.~H.}\ \bibnamefont
  {Booth}}, \bibinfo {author} {\bibfnamefont {A.~J.}\ \bibnamefont {Thom}},\
  and\ \bibinfo {author} {\bibfnamefont {A.}~\bibnamefont {Alavi}},\ }\href
  {https://doi.org/10.1063/1.3193710} {\bibfield  {journal} {\bibinfo
  {journal} {J. Chem. Phys}\ }\textbf {\bibinfo {volume} {131}},\ \bibinfo
  {pages} {054106} (\bibinfo {year} {2009})}\BibitemShut {NoStop}%
\bibitem [{\citenamefont {Cleland}\ \emph {et~al.}(2010)\citenamefont
  {Cleland}, \citenamefont {Booth},\ and\ \citenamefont
  {Alavi}}]{cleland2010communications}%
  \BibitemOpen
  \bibfield  {author} {\bibinfo {author} {\bibfnamefont {D.}~\bibnamefont
  {Cleland}}, \bibinfo {author} {\bibfnamefont {G.~H.}\ \bibnamefont {Booth}},\
  and\ \bibinfo {author} {\bibfnamefont {A.}~\bibnamefont {Alavi}},\ }\href
  {https://doi.org/10.1063/1.3302277} {\bibfield  {journal} {\bibinfo
  {journal} {J. Chem. Phys}\ }\textbf {\bibinfo {volume} {132}},\ \bibinfo
  {pages} {041103} (\bibinfo {year} {2010})}\BibitemShut {NoStop}%
\bibitem [{\citenamefont {Guther}\ \emph {et~al.}(2020)\citenamefont {Guther},
  \citenamefont {Anderson}, \citenamefont {Blunt}, \citenamefont {Bogdanov},
  \citenamefont {Cleland}, \citenamefont {Dattani}, \citenamefont {Dobrautz},
  \citenamefont {Ghanem}, \citenamefont {Jeszenszki}, \citenamefont
  {Liebermann} \emph {et~al.}}]{guther2020neci}%
  \BibitemOpen
  \bibfield  {author} {\bibinfo {author} {\bibfnamefont {K.}~\bibnamefont
  {Guther}}, \bibinfo {author} {\bibfnamefont {R.~J.}\ \bibnamefont
  {Anderson}}, \bibinfo {author} {\bibfnamefont {N.~S.}\ \bibnamefont {Blunt}},
  \bibinfo {author} {\bibfnamefont {N.~A.}\ \bibnamefont {Bogdanov}}, \bibinfo
  {author} {\bibfnamefont {D.}~\bibnamefont {Cleland}}, \bibinfo {author}
  {\bibfnamefont {N.}~\bibnamefont {Dattani}}, \bibinfo {author} {\bibfnamefont
  {W.}~\bibnamefont {Dobrautz}}, \bibinfo {author} {\bibfnamefont
  {K.}~\bibnamefont {Ghanem}}, \bibinfo {author} {\bibfnamefont
  {P.}~\bibnamefont {Jeszenszki}}, \bibinfo {author} {\bibfnamefont
  {N.}~\bibnamefont {Liebermann}}, \emph {et~al.},\ }\href
  {https://doi.org/10.1063/5.0005754} {\bibfield  {journal} {\bibinfo
  {journal} {J. Chem. Phys}\ }\textbf {\bibinfo {volume} {153}},\ \bibinfo
  {pages} {034107} (\bibinfo {year} {2020})}\BibitemShut {NoStop}%
\bibitem [{\citenamefont {Holmes}\ \emph {et~al.}(2016)\citenamefont {Holmes},
  \citenamefont {Tubman},\ and\ \citenamefont {Umrigar}}]{holmes2016heat}%
  \BibitemOpen
  \bibfield  {author} {\bibinfo {author} {\bibfnamefont {A.~A.}\ \bibnamefont
  {Holmes}}, \bibinfo {author} {\bibfnamefont {N.~M.}\ \bibnamefont {Tubman}},\
  and\ \bibinfo {author} {\bibfnamefont {C.}~\bibnamefont {Umrigar}},\ }\href
  {https://doi.org/10.1021/acs.jctc.6b00407} {\bibfield  {journal} {\bibinfo
  {journal} {J. Chem. Theory Comput}\ }\textbf {\bibinfo {volume} {12}},\
  \bibinfo {pages} {3674} (\bibinfo {year} {2016})}\BibitemShut {NoStop}%
\bibitem [{\citenamefont {Sharma}\ \emph {et~al.}(2017)\citenamefont {Sharma},
  \citenamefont {Holmes}, \citenamefont {Jeanmairet}, \citenamefont {Alavi},\
  and\ \citenamefont {Umrigar}}]{sharma2017semistochastic}%
  \BibitemOpen
  \bibfield  {author} {\bibinfo {author} {\bibfnamefont {S.}~\bibnamefont
  {Sharma}}, \bibinfo {author} {\bibfnamefont {A.~A.}\ \bibnamefont {Holmes}},
  \bibinfo {author} {\bibfnamefont {G.}~\bibnamefont {Jeanmairet}}, \bibinfo
  {author} {\bibfnamefont {A.}~\bibnamefont {Alavi}},\ and\ \bibinfo {author}
  {\bibfnamefont {C.~J.}\ \bibnamefont {Umrigar}},\ }\href
  {https://doi.org/10.1021/acs.jctc.6b01028} {\bibfield  {journal} {\bibinfo
  {journal} {J. Chem. Theory Comput}\ }\textbf {\bibinfo {volume} {13}},\
  \bibinfo {pages} {1595} (\bibinfo {year} {2017})}\BibitemShut {NoStop}%
\bibitem [{\citenamefont {Li}\ \emph {et~al.}(2018)\citenamefont {Li},
  \citenamefont {Otten}, \citenamefont {Holmes}, \citenamefont {Sharma},\ and\
  \citenamefont {Umrigar}}]{li2018fast}%
  \BibitemOpen
  \bibfield  {author} {\bibinfo {author} {\bibfnamefont {J.}~\bibnamefont
  {Li}}, \bibinfo {author} {\bibfnamefont {M.}~\bibnamefont {Otten}}, \bibinfo
  {author} {\bibfnamefont {A.~A.}\ \bibnamefont {Holmes}}, \bibinfo {author}
  {\bibfnamefont {S.}~\bibnamefont {Sharma}},\ and\ \bibinfo {author}
  {\bibfnamefont {C.~J.}\ \bibnamefont {Umrigar}},\ }\href
  {https://doi.org/10.1063/1.5055390} {\bibfield  {journal} {\bibinfo
  {journal} {J. Chem. Phys}\ }\textbf {\bibinfo {volume} {149}},\ \bibinfo
  {pages} {214110} (\bibinfo {year} {2018})}\BibitemShut {NoStop}%
\bibitem [{\citenamefont {Liu}\ and\ \citenamefont
  {Hoffmann}(2016{\natexlab{a}})}]{liu2016ici}%
  \BibitemOpen
  \bibfield  {author} {\bibinfo {author} {\bibfnamefont {W.}~\bibnamefont
  {Liu}}\ and\ \bibinfo {author} {\bibfnamefont {M.~R.}\ \bibnamefont
  {Hoffmann}},\ }\href {https://doi.org/10.1021/acs.jctc.5b01099} {\bibfield
  {journal} {\bibinfo  {journal} {J. Chem. Theory Comput}\ }\textbf {\bibinfo
  {volume} {12}},\ \bibinfo {pages} {1169} (\bibinfo {year}
  {2016}{\natexlab{a}})}\BibitemShut {NoStop}%
\bibitem [{\citenamefont {Liu}\ and\ \citenamefont
  {Hoffmann}(2016{\natexlab{b}})}]{liu2016sds}%
  \BibitemOpen
  \bibfield  {author} {\bibinfo {author} {\bibfnamefont {W.}~\bibnamefont
  {Liu}}\ and\ \bibinfo {author} {\bibfnamefont {M.~R.}\ \bibnamefont
  {Hoffmann}},\ }\href
  {https://link.springer.com/article/10.1007/s00214-014-1481-x} {\bibfield
  {journal} {\bibinfo  {journal} {Theor. Chem. Acc}\ }\textbf {\bibinfo
  {volume} {133}},\ \bibinfo {pages} {141} (\bibinfo {year}
  {2016}{\natexlab{b}})}\BibitemShut {NoStop}%
\bibitem [{\citenamefont {Xu}\ \emph {et~al.}(2018)\citenamefont {Xu},
  \citenamefont {Uejima},\ and\ \citenamefont {Ten-No}}]{xu2018full}%
  \BibitemOpen
  \bibfield  {author} {\bibinfo {author} {\bibfnamefont {E.}~\bibnamefont
  {Xu}}, \bibinfo {author} {\bibfnamefont {M.}~\bibnamefont {Uejima}},\ and\
  \bibinfo {author} {\bibfnamefont {S.~L.}\ \bibnamefont {Ten-No}},\ }\href
  {https://journals.aps.org/prl/abstract/10.1103/PhysRevLett.121.113001}
  {\bibfield  {journal} {\bibinfo  {journal} {Phys. Rev. Lett}\ }\textbf
  {\bibinfo {volume} {121}},\ \bibinfo {pages} {113001} (\bibinfo {year}
  {2018})}\BibitemShut {NoStop}%
\bibitem [{\citenamefont {Eriksen}\ and\ \citenamefont
  {Gauss}(2018)}]{eriksen2018many}%
  \BibitemOpen
  \bibfield  {author} {\bibinfo {author} {\bibfnamefont {J.~J.}\ \bibnamefont
  {Eriksen}}\ and\ \bibinfo {author} {\bibfnamefont {J.}~\bibnamefont
  {Gauss}},\ }\href {https://doi.org/10.1021/acs.jctc.8b00680} {\bibfield
  {journal} {\bibinfo  {journal} {J. Chem. Theory Comput}\ }\textbf {\bibinfo
  {volume} {14}},\ \bibinfo {pages} {5180} (\bibinfo {year}
  {2018})}\BibitemShut {NoStop}%
\bibitem [{\citenamefont {Eriksen}\ and\ \citenamefont
  {Gauss}(2019{\natexlab{a}})}]{eriksen2019many}%
  \BibitemOpen
  \bibfield  {author} {\bibinfo {author} {\bibfnamefont {J.~J.}\ \bibnamefont
  {Eriksen}}\ and\ \bibinfo {author} {\bibfnamefont {J.}~\bibnamefont
  {Gauss}},\ }\href {https://doi.org/10.1021/acs.jctc.9b00456} {\bibfield
  {journal} {\bibinfo  {journal} {J. Chem. Theory Comput}\ }\textbf {\bibinfo
  {volume} {15}},\ \bibinfo {pages} {4873} (\bibinfo {year}
  {2019}{\natexlab{a}})}\BibitemShut {NoStop}%
\bibitem [{\citenamefont {Eriksen}\ and\ \citenamefont
  {Gauss}(2019{\natexlab{b}})}]{eriksen2019generalized}%
  \BibitemOpen
  \bibfield  {author} {\bibinfo {author} {\bibfnamefont {J.~J.}\ \bibnamefont
  {Eriksen}}\ and\ \bibinfo {author} {\bibfnamefont {J.}~\bibnamefont
  {Gauss}},\ }\href {https://doi.org/10.1021/acs.jpclett.9b02968} {\bibfield
  {journal} {\bibinfo  {journal} {J. Phys. Chem. Lett}\ }\textbf {\bibinfo
  {volume} {10}},\ \bibinfo {pages} {7910} (\bibinfo {year}
  {2019}{\natexlab{b}})}\BibitemShut {NoStop}%
\bibitem [{\citenamefont {Rowe}(1968)}]{rowe1968equations}%
  \BibitemOpen
  \bibfield  {author} {\bibinfo {author} {\bibfnamefont {D.}~\bibnamefont
  {Rowe}},\ }\href
  {https://journals.aps.org/rmp/abstract/10.1103/RevModPhys.40.153} {\bibfield
  {journal} {\bibinfo  {journal} {Rev. Mod. Phys}\ }\textbf {\bibinfo {volume}
  {40}},\ \bibinfo {pages} {153} (\bibinfo {year} {1968})}\BibitemShut
  {NoStop}%
\bibitem [{\citenamefont {Bohm}\ and\ \citenamefont
  {Pines}(1951)}]{bohm1951collective}%
  \BibitemOpen
  \bibfield  {author} {\bibinfo {author} {\bibfnamefont {D.}~\bibnamefont
  {Bohm}}\ and\ \bibinfo {author} {\bibfnamefont {D.}~\bibnamefont {Pines}},\
  }\href {https://journals.aps.org/pr/abstract/10.1103/PhysRev.82.625}
  {\bibfield  {journal} {\bibinfo  {journal} {Phys. Rev}\ }\textbf {\bibinfo
  {volume} {82}},\ \bibinfo {pages} {625} (\bibinfo {year} {1951})}\BibitemShut
  {NoStop}%
\bibitem [{\citenamefont {Sawada}(1957)}]{sawada1957correlation}%
  \BibitemOpen
  \bibfield  {author} {\bibinfo {author} {\bibfnamefont {K.}~\bibnamefont
  {Sawada}},\ }\href
  {https://journals.aps.org/pr/abstract/10.1103/PhysRev.106.372} {\bibfield
  {journal} {\bibinfo  {journal} {Phys. Rev}\ }\textbf {\bibinfo {volume}
  {106}},\ \bibinfo {pages} {372} (\bibinfo {year} {1957})}\BibitemShut
  {NoStop}%
\bibitem [{\citenamefont {Pines}\ and\ \citenamefont
  {Bohm}(1952)}]{pines1952collective}%
  \BibitemOpen
  \bibfield  {author} {\bibinfo {author} {\bibfnamefont {D.}~\bibnamefont
  {Pines}}\ and\ \bibinfo {author} {\bibfnamefont {D.}~\bibnamefont {Bohm}},\
  }\href {https://journals.aps.org/pr/abstract/10.1103/PhysRev.85.338}
  {\bibfield  {journal} {\bibinfo  {journal} {Phys. Rev}\ }\textbf {\bibinfo
  {volume} {85}},\ \bibinfo {pages} {338} (\bibinfo {year} {1952})}\BibitemShut
  {NoStop}%
\bibitem [{\citenamefont {Bohm}\ and\ \citenamefont
  {Pines}(1953)}]{bohm1953collective}%
  \BibitemOpen
  \bibfield  {author} {\bibinfo {author} {\bibfnamefont {D.}~\bibnamefont
  {Bohm}}\ and\ \bibinfo {author} {\bibfnamefont {D.}~\bibnamefont {Pines}},\
  }\href {https://journals.aps.org/pr/abstract/10.1103/PhysRev.92.609}
  {\bibfield  {journal} {\bibinfo  {journal} {Phys. Rev}\ }\textbf {\bibinfo
  {volume} {92}},\ \bibinfo {pages} {609} (\bibinfo {year} {1953})}\BibitemShut
  {NoStop}%
\bibitem [{\citenamefont {McLachlan}\ and\ \citenamefont
  {Ball}(1964)}]{mclachlan1964time}%
  \BibitemOpen
  \bibfield  {author} {\bibinfo {author} {\bibfnamefont {A.}~\bibnamefont
  {McLachlan}}\ and\ \bibinfo {author} {\bibfnamefont {M.}~\bibnamefont
  {Ball}},\ }\href
  {https://journals.aps.org/rmp/abstract/10.1103/RevModPhys.36.844} {\bibfield
  {journal} {\bibinfo  {journal} {Rev. Mod. Phys}\ }\textbf {\bibinfo {volume}
  {36}},\ \bibinfo {pages} {844} (\bibinfo {year} {1964})}\BibitemShut
  {NoStop}%
\bibitem [{\citenamefont {Bonche}\ \emph {et~al.}(1976)\citenamefont {Bonche},
  \citenamefont {Koonin},\ and\ \citenamefont {Negele}}]{bonche1976one}%
  \BibitemOpen
  \bibfield  {author} {\bibinfo {author} {\bibfnamefont {P.}~\bibnamefont
  {Bonche}}, \bibinfo {author} {\bibfnamefont {S.}~\bibnamefont {Koonin}},\
  and\ \bibinfo {author} {\bibfnamefont {J.}~\bibnamefont {Negele}},\ }\href
  {https://journals.aps.org/prc/abstract/10.1103/PhysRevC.13.1226} {\bibfield
  {journal} {\bibinfo  {journal} {Phys. Rev. C}\ }\textbf {\bibinfo {volume}
  {13}},\ \bibinfo {pages} {1226} (\bibinfo {year} {1976})}\BibitemShut
  {NoStop}%
\bibitem [{\citenamefont {Dreuw}\ and\ \citenamefont
  {Head-Gordon}(2005)}]{dreuw2005single}%
  \BibitemOpen
  \bibfield  {author} {\bibinfo {author} {\bibfnamefont {A.}~\bibnamefont
  {Dreuw}}\ and\ \bibinfo {author} {\bibfnamefont {M.}~\bibnamefont
  {Head-Gordon}},\ }\href {https://pubs.acs.org/doi/10.1021/cr0505627}
  {\bibfield  {journal} {\bibinfo  {journal} {Chem. Rev}\ }\textbf {\bibinfo
  {volume} {105}},\ \bibinfo {pages} {4009} (\bibinfo {year}
  {2005})}\BibitemShut {NoStop}%
\bibitem [{\citenamefont {Stanton}\ and\ \citenamefont
  {Bartlett}(1993)}]{stanton1993equation}%
  \BibitemOpen
  \bibfield  {author} {\bibinfo {author} {\bibfnamefont {J.~F.}\ \bibnamefont
  {Stanton}}\ and\ \bibinfo {author} {\bibfnamefont {R.~J.}\ \bibnamefont
  {Bartlett}},\ }\href {https://doi.org/10.1063/1.464746} {\bibfield  {journal}
  {\bibinfo  {journal} {J. Chem. Phys}\ }\textbf {\bibinfo {volume} {98}},\
  \bibinfo {pages} {7029} (\bibinfo {year} {1993})}\BibitemShut {NoStop}%
\bibitem [{\citenamefont {Prasad}\ \emph {et~al.}(1985)\citenamefont {Prasad},
  \citenamefont {Pal},\ and\ \citenamefont {Mukherjee}}]{prasad1985some}%
  \BibitemOpen
  \bibfield  {author} {\bibinfo {author} {\bibfnamefont {M.~D.}\ \bibnamefont
  {Prasad}}, \bibinfo {author} {\bibfnamefont {S.}~\bibnamefont {Pal}},\ and\
  \bibinfo {author} {\bibfnamefont {D.}~\bibnamefont {Mukherjee}},\ }\href
  {https://journals.aps.org/pra/abstract/10.1103/PhysRevA.31.1287} {\bibfield
  {journal} {\bibinfo  {journal} {Phys. Rev. A}\ }\textbf {\bibinfo {volume}
  {31}},\ \bibinfo {pages} {1287} (\bibinfo {year} {1985})}\BibitemShut
  {NoStop}%
\bibitem [{\citenamefont {Datta}\ \emph {et~al.}(1993)\citenamefont {Datta},
  \citenamefont {Mukhopadhyay},\ and\ \citenamefont
  {Mukherjee}}]{datta1993consistent}%
  \BibitemOpen
  \bibfield  {author} {\bibinfo {author} {\bibfnamefont {B.}~\bibnamefont
  {Datta}}, \bibinfo {author} {\bibfnamefont {D.}~\bibnamefont
  {Mukhopadhyay}},\ and\ \bibinfo {author} {\bibfnamefont {D.}~\bibnamefont
  {Mukherjee}},\ }\href
  {https://journals.aps.org/pra/abstract/10.1103/PhysRevA.47.3632} {\bibfield
  {journal} {\bibinfo  {journal} {Phys. Rev. A}\ }\textbf {\bibinfo {volume}
  {47}},\ \bibinfo {pages} {3632} (\bibinfo {year} {1993})}\BibitemShut
  {NoStop}%
\bibitem [{\citenamefont {Stanton}\ and\ \citenamefont
  {Gauss}(1994)}]{stanton1994analytic}%
  \BibitemOpen
  \bibfield  {author} {\bibinfo {author} {\bibfnamefont {J.~F.}\ \bibnamefont
  {Stanton}}\ and\ \bibinfo {author} {\bibfnamefont {J.}~\bibnamefont
  {Gauss}},\ }\href {https://doi.org/10.1063/1.468022} {\bibfield  {journal}
  {\bibinfo  {journal} {J. Chem. Phys}\ }\textbf {\bibinfo {volume} {101}},\
  \bibinfo {pages} {8938} (\bibinfo {year} {1994})}\BibitemShut {NoStop}%
\bibitem [{\citenamefont {Krylov}(2008)}]{krylov2008equation}%
  \BibitemOpen
  \bibfield  {author} {\bibinfo {author} {\bibfnamefont {A.~I.}\ \bibnamefont
  {Krylov}},\ }\href
  {https://doi.org/10.1146/annurev.physchem.59.032607.093602} {\bibfield
  {journal} {\bibinfo  {journal} {Annu. Rev. Phys. Chem}\ }\textbf {\bibinfo
  {volume} {59}},\ \bibinfo {pages} {433} (\bibinfo {year} {2008})}\BibitemShut
  {NoStop}%
\bibitem [{\citenamefont {Nooijen}\ and\ \citenamefont
  {Bartlett}(1995)}]{nooijen1995equation}%
  \BibitemOpen
  \bibfield  {author} {\bibinfo {author} {\bibfnamefont {M.}~\bibnamefont
  {Nooijen}}\ and\ \bibinfo {author} {\bibfnamefont {R.~J.}\ \bibnamefont
  {Bartlett}},\ }\href {https://doi.org/10.1063/1.468592} {\bibfield  {journal}
  {\bibinfo  {journal} {J. Chem. Phys}\ }\textbf {\bibinfo {volume} {102}},\
  \bibinfo {pages} {3629} (\bibinfo {year} {1995})}\BibitemShut {NoStop}%
\bibitem [{\citenamefont {Wang}\ and\ \citenamefont
  {Berkelbach}(2020)}]{wang2020excitons}%
  \BibitemOpen
  \bibfield  {author} {\bibinfo {author} {\bibfnamefont {X.}~\bibnamefont
  {Wang}}\ and\ \bibinfo {author} {\bibfnamefont {T.~C.}\ \bibnamefont
  {Berkelbach}},\ }\href {https://doi.org/10.1021/acs.jctc.0c00101} {\bibfield
  {journal} {\bibinfo  {journal} {J. Chem. Theory Comput}\ }\textbf {\bibinfo
  {volume} {16}},\ \bibinfo {pages} {3095} (\bibinfo {year}
  {2020})}\BibitemShut {NoStop}%
\bibitem [{\citenamefont {Gao}\ \emph {et~al.}(2020)\citenamefont {Gao},
  \citenamefont {Sun}, \citenamefont {Jason}, \citenamefont {Motta},
  \citenamefont {McClain}, \citenamefont {White}, \citenamefont {Minnich},\
  and\ \citenamefont {Chan}}]{gao2020electronic}%
  \BibitemOpen
  \bibfield  {author} {\bibinfo {author} {\bibfnamefont {Y.}~\bibnamefont
  {Gao}}, \bibinfo {author} {\bibfnamefont {Q.}~\bibnamefont {Sun}}, \bibinfo
  {author} {\bibfnamefont {M.~Y.}\ \bibnamefont {Jason}}, \bibinfo {author}
  {\bibfnamefont {M.}~\bibnamefont {Motta}}, \bibinfo {author} {\bibfnamefont
  {J.}~\bibnamefont {McClain}}, \bibinfo {author} {\bibfnamefont {A.~F.}\
  \bibnamefont {White}}, \bibinfo {author} {\bibfnamefont {A.~J.}\ \bibnamefont
  {Minnich}},\ and\ \bibinfo {author} {\bibfnamefont {G.~K.-L.}\ \bibnamefont
  {Chan}},\ }\href
  {https://journals.aps.org/prb/abstract/10.1103/PhysRevB.101.165138}
  {\bibfield  {journal} {\bibinfo  {journal} {Phys. Rev. B}\ }\textbf {\bibinfo
  {volume} {101}},\ \bibinfo {pages} {165138} (\bibinfo {year}
  {2020})}\BibitemShut {NoStop}%
\bibitem [{\citenamefont {Lange}\ and\ \citenamefont
  {Berkelbach}(2021)}]{lange2021improving}%
  \BibitemOpen
  \bibfield  {author} {\bibinfo {author} {\bibfnamefont {M.~F.}\ \bibnamefont
  {Lange}}\ and\ \bibinfo {author} {\bibfnamefont {T.~C.}\ \bibnamefont
  {Berkelbach}},\ }\href {https://doi.org/10.1063/5.0061242} {\bibfield
  {journal} {\bibinfo  {journal} {J. Chem. Phys}\ }\textbf {\bibinfo {volume}
  {155}},\ \bibinfo {pages} {081101} (\bibinfo {year} {2021})}\BibitemShut
  {NoStop}%
\bibitem [{\citenamefont {Thomas}\ \emph {et~al.}(2021)\citenamefont {Thomas},
  \citenamefont {Hampe}, \citenamefont {Stopkowicz},\ and\ \citenamefont
  {Gauss}}]{thomas2021complex}%
  \BibitemOpen
  \bibfield  {author} {\bibinfo {author} {\bibfnamefont {S.}~\bibnamefont
  {Thomas}}, \bibinfo {author} {\bibfnamefont {F.}~\bibnamefont {Hampe}},
  \bibinfo {author} {\bibfnamefont {S.}~\bibnamefont {Stopkowicz}},\ and\
  \bibinfo {author} {\bibfnamefont {J.}~\bibnamefont {Gauss}},\ }\href
  {https://doi.org/10.1080/00268976.2021.1968056} {\bibfield  {journal}
  {\bibinfo  {journal} {Mol. Phys}\ }\textbf {\bibinfo {volume} {119}},\
  \bibinfo {pages} {e1968056} (\bibinfo {year} {2021})}\BibitemShut {NoStop}%
\bibitem [{\citenamefont {K{\"o}hn}\ and\ \citenamefont
  {Tajti}(2007)}]{kohn2007can}%
  \BibitemOpen
  \bibfield  {author} {\bibinfo {author} {\bibfnamefont {A.}~\bibnamefont
  {K{\"o}hn}}\ and\ \bibinfo {author} {\bibfnamefont {A.}~\bibnamefont
  {Tajti}},\ }\href {https://doi.org/10.1063/1.2755681} {\bibfield  {journal}
  {\bibinfo  {journal} {J. Chem. Phys}\ }\textbf {\bibinfo {volume} {127}},\
  \bibinfo {pages} {044105} (\bibinfo {year} {2007})}\BibitemShut {NoStop}%
\bibitem [{\citenamefont {Yarkony}(2012)}]{yarkony2012nonadiabatic}%
  \BibitemOpen
  \bibfield  {author} {\bibinfo {author} {\bibfnamefont {D.~R.}\ \bibnamefont
  {Yarkony}},\ }\href {https://pubs.acs.org/doi/10.1021/cr2001299} {\bibfield
  {journal} {\bibinfo  {journal} {Chem. Rev}\ }\textbf {\bibinfo {volume}
  {112}},\ \bibinfo {pages} {481} (\bibinfo {year} {2012})}\BibitemShut
  {NoStop}%
\bibitem [{\citenamefont {Bernardi}\ \emph {et~al.}(1997)\citenamefont
  {Bernardi}, \citenamefont {Olivucci},\ and\ \citenamefont
  {Robb}}]{bernardi1997role}%
  \BibitemOpen
  \bibfield  {author} {\bibinfo {author} {\bibfnamefont {F.}~\bibnamefont
  {Bernardi}}, \bibinfo {author} {\bibfnamefont {M.}~\bibnamefont {Olivucci}},\
  and\ \bibinfo {author} {\bibfnamefont {M.~A.}\ \bibnamefont {Robb}},\ }\href
  {https://doi.org/10.1016/S1010-6030(96)04573-X} {\bibfield  {journal}
  {\bibinfo  {journal} {J. Photochem. Photobiol. A}\ }\textbf {\bibinfo
  {volume} {105}},\ \bibinfo {pages} {365} (\bibinfo {year}
  {1997})}\BibitemShut {NoStop}%
\bibitem [{\citenamefont {Schmidt}\ and\ \citenamefont
  {Gordon}(1998)}]{schmidt1998construction}%
  \BibitemOpen
  \bibfield  {author} {\bibinfo {author} {\bibfnamefont {M.~W.}\ \bibnamefont
  {Schmidt}}\ and\ \bibinfo {author} {\bibfnamefont {M.~S.}\ \bibnamefont
  {Gordon}},\ }\href
  {https://www.annualreviews.org/doi/abs/10.1146/annurev.physchem.49.1.233}
  {\bibfield  {journal} {\bibinfo  {journal} {Annu. Rev. Phys. Chem}\ }\textbf
  {\bibinfo {volume} {49}},\ \bibinfo {pages} {233} (\bibinfo {year}
  {1998})}\BibitemShut {NoStop}%
\bibitem [{\citenamefont {K{\"o}hn}\ \emph {et~al.}(2013)\citenamefont
  {K{\"o}hn}, \citenamefont {Hanauer}, \citenamefont {Mueck}, \citenamefont
  {Jagau},\ and\ \citenamefont {Gauss}}]{kohn2013state}%
  \BibitemOpen
  \bibfield  {author} {\bibinfo {author} {\bibfnamefont {A.}~\bibnamefont
  {K{\"o}hn}}, \bibinfo {author} {\bibfnamefont {M.}~\bibnamefont {Hanauer}},
  \bibinfo {author} {\bibfnamefont {L.~A.}\ \bibnamefont {Mueck}}, \bibinfo
  {author} {\bibfnamefont {T.-C.}\ \bibnamefont {Jagau}},\ and\ \bibinfo
  {author} {\bibfnamefont {J.}~\bibnamefont {Gauss}},\ }\href
  {https://wires.onlinelibrary.wiley.com/doi/abs/10.1002/wcms.1120} {\bibfield
  {journal} {\bibinfo  {journal} {WIREs Comput. Mol. Sci}\ }\textbf {\bibinfo
  {volume} {3}},\ \bibinfo {pages} {176} (\bibinfo {year} {2013})}\BibitemShut
  {NoStop}%
\bibitem [{\citenamefont {Mahapatra}\ \emph {et~al.}(1998)\citenamefont
  {Mahapatra}, \citenamefont {Datta}, \citenamefont {Bandyopadhyay},\ and\
  \citenamefont {Mukherjee}}]{mahapatra1998state}%
  \BibitemOpen
  \bibfield  {author} {\bibinfo {author} {\bibfnamefont {U.~S.}\ \bibnamefont
  {Mahapatra}}, \bibinfo {author} {\bibfnamefont {B.}~\bibnamefont {Datta}},
  \bibinfo {author} {\bibfnamefont {B.}~\bibnamefont {Bandyopadhyay}},\ and\
  \bibinfo {author} {\bibfnamefont {D.}~\bibnamefont {Mukherjee}},\ }\href
  {https://www.sciencedirect.com/science/article/abs/pii/S0065327608605079}
  {\bibfield  {journal} {\bibinfo  {journal} {Adv. Quantum Chem}\ }\textbf
  {\bibinfo {volume} {30}},\ \bibinfo {pages} {163} (\bibinfo {year}
  {1998})}\BibitemShut {NoStop}%
\bibitem [{\citenamefont {Roos}\ \emph {et~al.}(1996)\citenamefont {Roos},
  \citenamefont {Andersson}, \citenamefont {F{\"u}lscher}, \citenamefont
  {Malmqvist}, \citenamefont {Serrano-Andr{\'e}s}, \citenamefont {Pierloot},\
  and\ \citenamefont {Merch{\'a}n}}]{roos1996multiconfigurational}%
  \BibitemOpen
  \bibfield  {author} {\bibinfo {author} {\bibfnamefont {B.~O.}\ \bibnamefont
  {Roos}}, \bibinfo {author} {\bibfnamefont {K.}~\bibnamefont {Andersson}},
  \bibinfo {author} {\bibfnamefont {M.~P.}\ \bibnamefont {F{\"u}lscher}},
  \bibinfo {author} {\bibfnamefont {P.-{\AA}.}\ \bibnamefont {Malmqvist}},
  \bibinfo {author} {\bibfnamefont {L.}~\bibnamefont {Serrano-Andr{\'e}s}},
  \bibinfo {author} {\bibfnamefont {K.}~\bibnamefont {Pierloot}},\ and\
  \bibinfo {author} {\bibfnamefont {M.}~\bibnamefont {Merch{\'a}n}},\ }in\
  \href {https://onlinelibrary.wiley.com/doi/abs/10.1002/9780470141526.ch5}
  {\emph {\bibinfo {booktitle} {Advances in chemical physics: new methods in
  computational quantum mechanics}}},\ \bibinfo {editor} {edited by\ \bibinfo
  {editor} {\bibfnamefont {I.}~\bibnamefont {Prigogine}}\ and\ \bibinfo
  {editor} {\bibfnamefont {S.~A.}\ \bibnamefont {Rice}}}\ (\bibinfo
  {publisher} {John Wiley \& Sons},\ \bibinfo {year} {1996})\ Chap.~\bibinfo
  {chapter} {5}, pp.\ \bibinfo {pages} {219--331}\BibitemShut {NoStop}%
\bibitem [{\citenamefont {Zhou}\ \emph
  {et~al.}(2006{\natexlab{a}})\citenamefont {Zhou}, \citenamefont {Saad},
  \citenamefont {Tiago},\ and\ \citenamefont {Chelikowsky}}]{zhou2006parallel}%
  \BibitemOpen
  \bibfield  {author} {\bibinfo {author} {\bibfnamefont {Y.}~\bibnamefont
  {Zhou}}, \bibinfo {author} {\bibfnamefont {Y.}~\bibnamefont {Saad}}, \bibinfo
  {author} {\bibfnamefont {M.~L.}\ \bibnamefont {Tiago}},\ and\ \bibinfo
  {author} {\bibfnamefont {J.~R.}\ \bibnamefont {Chelikowsky}},\ }\href
  {https://doi.org/10.1103/PhysRevE.74.066704} {\bibfield  {journal} {\bibinfo
  {journal} {Phys. Rev. E}\ }\textbf {\bibinfo {volume} {74}},\ \bibinfo
  {pages} {066704} (\bibinfo {year} {2006}{\natexlab{a}})}\BibitemShut
  {NoStop}%
\bibitem [{\citenamefont {Zhou}\ \emph
  {et~al.}(2006{\natexlab{b}})\citenamefont {Zhou}, \citenamefont {Saad},
  \citenamefont {Tiago},\ and\ \citenamefont
  {Chelikowsky}}]{zhou2006selfconsistent}%
  \BibitemOpen
  \bibfield  {author} {\bibinfo {author} {\bibfnamefont {Y.}~\bibnamefont
  {Zhou}}, \bibinfo {author} {\bibfnamefont {Y.}~\bibnamefont {Saad}}, \bibinfo
  {author} {\bibfnamefont {M.~L.}\ \bibnamefont {Tiago}},\ and\ \bibinfo
  {author} {\bibfnamefont {J.~R.}\ \bibnamefont {Chelikowsky}},\ }\href
  {https://doi.org/https://doi.org/10.1016/j.jcp.2006.03.017} {\bibfield
  {journal} {\bibinfo  {journal} {Journal of Computational Physics}\ }\textbf
  {\bibinfo {volume} {219}},\ \bibinfo {pages} {172} (\bibinfo {year}
  {2006}{\natexlab{b}})}\BibitemShut {NoStop}%
\bibitem [{\citenamefont {Saad}(2016)}]{saad2016analysis}%
  \BibitemOpen
  \bibfield  {author} {\bibinfo {author} {\bibfnamefont {Y.}~\bibnamefont
  {Saad}},\ }\href {https://doi.org/10.1137/141002037} {\bibfield  {journal}
  {\bibinfo  {journal} {SIAM J. Matrix Anal}\ }\textbf {\bibinfo {volume}
  {37}},\ \bibinfo {pages} {103} (\bibinfo {year} {2016})},\ \Eprint
  {https://arxiv.org/abs/https://doi.org/10.1137/141002037}
  {https://doi.org/10.1137/141002037} \BibitemShut {NoStop}%
\bibitem [{\citenamefont {Takeshita}\ \emph {et~al.}(2020)\citenamefont
  {Takeshita}, \citenamefont {Rubin}, \citenamefont {Jiang}, \citenamefont
  {Lee}, \citenamefont {Babbush},\ and\ \citenamefont
  {McClean}}]{takeshita2020increasing}%
  \BibitemOpen
  \bibfield  {author} {\bibinfo {author} {\bibfnamefont {T.}~\bibnamefont
  {Takeshita}}, \bibinfo {author} {\bibfnamefont {N.~C.}\ \bibnamefont
  {Rubin}}, \bibinfo {author} {\bibfnamefont {Z.}~\bibnamefont {Jiang}},
  \bibinfo {author} {\bibfnamefont {E.}~\bibnamefont {Lee}}, \bibinfo {author}
  {\bibfnamefont {R.}~\bibnamefont {Babbush}},\ and\ \bibinfo {author}
  {\bibfnamefont {J.~R.}\ \bibnamefont {McClean}},\ }\href
  {https://journals.aps.org/prx/abstract/10.1103/PhysRevX.10.011004} {\bibfield
   {journal} {\bibinfo  {journal} {Phys. Rev. X}\ }\textbf {\bibinfo {volume}
  {10}},\ \bibinfo {pages} {011004} (\bibinfo {year} {2020})}\BibitemShut
  {NoStop}%
\bibitem [{\citenamefont {Cohn}\ \emph {et~al.}(2021)\citenamefont {Cohn},
  \citenamefont {Motta},\ and\ \citenamefont {Parrish}}]{cohn2021quantum}%
  \BibitemOpen
  \bibfield  {author} {\bibinfo {author} {\bibfnamefont {J.}~\bibnamefont
  {Cohn}}, \bibinfo {author} {\bibfnamefont {M.}~\bibnamefont {Motta}},\ and\
  \bibinfo {author} {\bibfnamefont {R.~M.}\ \bibnamefont {Parrish}},\ }\href
  {https://journals.aps.org/prxquantum/abstract/10.1103/PRXQuantum.2.040352}
  {\bibfield  {journal} {\bibinfo  {journal} {PRX Quantum}\ }\textbf {\bibinfo
  {volume} {2}},\ \bibinfo {pages} {040352} (\bibinfo {year}
  {2021})}\BibitemShut {NoStop}%
\bibitem [{\citenamefont {Seki}\ and\ \citenamefont
  {Yunoki}(2021)}]{seki2021powermethod}%
  \BibitemOpen
  \bibfield  {author} {\bibinfo {author} {\bibfnamefont {K.}~\bibnamefont
  {Seki}}\ and\ \bibinfo {author} {\bibfnamefont {S.}~\bibnamefont {Yunoki}},\
  }\href {https://doi.org/10.1103/PRXQuantum.2.010333} {\bibfield  {journal}
  {\bibinfo  {journal} {PRX Quantum}\ }\textbf {\bibinfo {volume} {2}},\
  \bibinfo {pages} {010333} (\bibinfo {year} {2021})}\BibitemShut {NoStop}%
\bibitem [{\citenamefont {Yoshioka}\ \emph {et~al.}(2022)\citenamefont
  {Yoshioka}, \citenamefont {Hakoshima}, \citenamefont {Matsuzaki},
  \citenamefont {Tokunaga}, \citenamefont {Suzuki},\ and\ \citenamefont
  {Endo}}]{yoshioka2022generalized}%
  \BibitemOpen
  \bibfield  {author} {\bibinfo {author} {\bibfnamefont {N.}~\bibnamefont
  {Yoshioka}}, \bibinfo {author} {\bibfnamefont {H.}~\bibnamefont {Hakoshima}},
  \bibinfo {author} {\bibfnamefont {Y.}~\bibnamefont {Matsuzaki}}, \bibinfo
  {author} {\bibfnamefont {Y.}~\bibnamefont {Tokunaga}}, \bibinfo {author}
  {\bibfnamefont {Y.}~\bibnamefont {Suzuki}},\ and\ \bibinfo {author}
  {\bibfnamefont {S.}~\bibnamefont {Endo}},\ }\href
  {https://journals.aps.org/prl/abstract/10.1103/PhysRevLett.129.020502}
  {\bibfield  {journal} {\bibinfo  {journal} {Phys. Rev. Lett}\ }\textbf
  {\bibinfo {volume} {129}},\ \bibinfo {pages} {020502} (\bibinfo {year}
  {2022})}\BibitemShut {NoStop}%
\bibitem [{\citenamefont {Cortes}\ and\ \citenamefont
  {Gray}(2022)}]{cortes2022quantum}%
  \BibitemOpen
  \bibfield  {author} {\bibinfo {author} {\bibfnamefont {C.~L.}\ \bibnamefont
  {Cortes}}\ and\ \bibinfo {author} {\bibfnamefont {S.~K.}\ \bibnamefont
  {Gray}},\ }\href
  {https://journals.aps.org/pra/abstract/10.1103/PhysRevA.105.022417}
  {\bibfield  {journal} {\bibinfo  {journal} {Phys. Rev. A}\ }\textbf {\bibinfo
  {volume} {105}},\ \bibinfo {pages} {022417} (\bibinfo {year}
  {2022})}\BibitemShut {NoStop}%
\bibitem [{\citenamefont {Klymko}\ \emph {et~al.}(2022)\citenamefont {Klymko},
  \citenamefont {Mejuto-Zaera}, \citenamefont {Cotton}, \citenamefont
  {Wudarski}, \citenamefont {Urbanek}, \citenamefont {Hait}, \citenamefont
  {Head-Gordon}, \citenamefont {Whaley}, \citenamefont {Moussa}, \citenamefont
  {Wiebe} \emph {et~al.}}]{klymko2022real}%
  \BibitemOpen
  \bibfield  {author} {\bibinfo {author} {\bibfnamefont {K.}~\bibnamefont
  {Klymko}}, \bibinfo {author} {\bibfnamefont {C.}~\bibnamefont
  {Mejuto-Zaera}}, \bibinfo {author} {\bibfnamefont {S.~J.}\ \bibnamefont
  {Cotton}}, \bibinfo {author} {\bibfnamefont {F.}~\bibnamefont {Wudarski}},
  \bibinfo {author} {\bibfnamefont {M.}~\bibnamefont {Urbanek}}, \bibinfo
  {author} {\bibfnamefont {D.}~\bibnamefont {Hait}}, \bibinfo {author}
  {\bibfnamefont {M.}~\bibnamefont {Head-Gordon}}, \bibinfo {author}
  {\bibfnamefont {K.~B.}\ \bibnamefont {Whaley}}, \bibinfo {author}
  {\bibfnamefont {J.}~\bibnamefont {Moussa}}, \bibinfo {author} {\bibfnamefont
  {N.}~\bibnamefont {Wiebe}}, \emph {et~al.},\ }\href
  {https://journals.aps.org/prxquantum/abstract/10.1103/PRXQuantum.3.020323}
  {\bibfield  {journal} {\bibinfo  {journal} {PRX Quantum}\ }\textbf {\bibinfo
  {volume} {3}},\ \bibinfo {pages} {020323} (\bibinfo {year}
  {2022})}\BibitemShut {NoStop}%
\bibitem [{\citenamefont {Baek}\ \emph {et~al.}(2023)\citenamefont {Baek},
  \citenamefont {Hait}, \citenamefont {Shee}, \citenamefont {Leimkuhler},
  \citenamefont {Huggins}, \citenamefont {Stetina}, \citenamefont
  {Head-Gordon},\ and\ \citenamefont {Whaley}}]{baek2022say}%
  \BibitemOpen
  \bibfield  {author} {\bibinfo {author} {\bibfnamefont {U.}~\bibnamefont
  {Baek}}, \bibinfo {author} {\bibfnamefont {D.}~\bibnamefont {Hait}}, \bibinfo
  {author} {\bibfnamefont {J.}~\bibnamefont {Shee}}, \bibinfo {author}
  {\bibfnamefont {O.}~\bibnamefont {Leimkuhler}}, \bibinfo {author}
  {\bibfnamefont {W.~J.}\ \bibnamefont {Huggins}}, \bibinfo {author}
  {\bibfnamefont {T.~F.}\ \bibnamefont {Stetina}}, \bibinfo {author}
  {\bibfnamefont {M.}~\bibnamefont {Head-Gordon}},\ and\ \bibinfo {author}
  {\bibfnamefont {K.~B.}\ \bibnamefont {Whaley}},\ }\href
  {https://journals.aps.org/prxquantum/abstract/10.1103/PRXQuantum.4.030307}
  {\bibfield  {journal} {\bibinfo  {journal} {PRX Quantum}\ }\textbf {\bibinfo
  {volume} {4}},\ \bibinfo {pages} {030307} (\bibinfo {year}
  {2023})}\BibitemShut {NoStop}%
\bibitem [{\citenamefont {Tkachenko}\ \emph {et~al.}(2022)\citenamefont
  {Tkachenko}, \citenamefont {Zhang}, \citenamefont {Cincio}, \citenamefont
  {Boldyrev}, \citenamefont {Tretiak},\ and\ \citenamefont
  {Dub}}]{tkachenko2022davidson}%
  \BibitemOpen
  \bibfield  {author} {\bibinfo {author} {\bibfnamefont {N.~V.}\ \bibnamefont
  {Tkachenko}}, \bibinfo {author} {\bibfnamefont {Y.}~\bibnamefont {Zhang}},
  \bibinfo {author} {\bibfnamefont {L.}~\bibnamefont {Cincio}}, \bibinfo
  {author} {\bibfnamefont {A.~I.}\ \bibnamefont {Boldyrev}}, \bibinfo {author}
  {\bibfnamefont {S.}~\bibnamefont {Tretiak}},\ and\ \bibinfo {author}
  {\bibfnamefont {P.~A.}\ \bibnamefont {Dub}},\ }\href
  {https://doi.org/10.48550/arXiv.2204.10741} {\bibfield  {journal} {\bibinfo
  {journal} {arXiv preprint, arXiv:2204.10741}\ } (\bibinfo {year}
  {2022})}\BibitemShut {NoStop}%
\bibitem [{\citenamefont {Kirby}\ \emph {et~al.}(2023)\citenamefont {Kirby},
  \citenamefont {Motta},\ and\ \citenamefont {Mezzacapo}}]{kirby2023exact}%
  \BibitemOpen
  \bibfield  {author} {\bibinfo {author} {\bibfnamefont {W.}~\bibnamefont
  {Kirby}}, \bibinfo {author} {\bibfnamefont {M.}~\bibnamefont {Motta}},\ and\
  \bibinfo {author} {\bibfnamefont {A.}~\bibnamefont {Mezzacapo}},\ }\href
  {https://quantum-journal.org/papers/q-2023-05-23-1018/} {\bibfield  {journal}
  {\bibinfo  {journal} {Quantum}\ }\textbf {\bibinfo {volume} {7}},\ \bibinfo
  {pages} {1018} (\bibinfo {year} {2023})}\BibitemShut {NoStop}%
\bibitem [{\citenamefont {Bittel}\ and\ \citenamefont
  {Kliesch}(2021)}]{bittel2021training}%
  \BibitemOpen
  \bibfield  {author} {\bibinfo {author} {\bibfnamefont {L.}~\bibnamefont
  {Bittel}}\ and\ \bibinfo {author} {\bibfnamefont {M.}~\bibnamefont
  {Kliesch}},\ }\href
  {https://journals.aps.org/prl/abstract/10.1103/PhysRevLett.127.120502}
  {\bibfield  {journal} {\bibinfo  {journal} {Phys. Rev. Lett}\ }\textbf
  {\bibinfo {volume} {127}},\ \bibinfo {pages} {120502} (\bibinfo {year}
  {2021})}\BibitemShut {NoStop}%
\bibitem [{\citenamefont {Bharti}\ and\ \citenamefont
  {Haug}(2021)}]{bharti2021iterative}%
  \BibitemOpen
  \bibfield  {author} {\bibinfo {author} {\bibfnamefont {K.}~\bibnamefont
  {Bharti}}\ and\ \bibinfo {author} {\bibfnamefont {T.}~\bibnamefont {Haug}},\
  }\href {https://journals.aps.org/pra/abstract/10.1103/PhysRevA.104.L050401}
  {\bibfield  {journal} {\bibinfo  {journal} {Phys. Rev. A}\ }\textbf {\bibinfo
  {volume} {104}},\ \bibinfo {pages} {L050401} (\bibinfo {year}
  {2021})}\BibitemShut {NoStop}%
\bibitem [{\citenamefont {Lim}\ \emph {et~al.}(2021)\citenamefont {Lim},
  \citenamefont {Haug}, \citenamefont {Kwek},\ and\ \citenamefont
  {Bharti}}]{lim2021fast}%
  \BibitemOpen
  \bibfield  {author} {\bibinfo {author} {\bibfnamefont {K.~H.}\ \bibnamefont
  {Lim}}, \bibinfo {author} {\bibfnamefont {T.}~\bibnamefont {Haug}}, \bibinfo
  {author} {\bibfnamefont {L.~C.}\ \bibnamefont {Kwek}},\ and\ \bibinfo
  {author} {\bibfnamefont {K.}~\bibnamefont {Bharti}},\ }\href
  {https://iopscience.iop.org/article/10.1088/2058-9565/ac2e52/meta} {\bibfield
   {journal} {\bibinfo  {journal} {Quant. Sci. Tech}\ }\textbf {\bibinfo
  {volume} {7}},\ \bibinfo {pages} {015001} (\bibinfo {year}
  {2021})}\BibitemShut {NoStop}%
\bibitem [{\citenamefont {Asthana}\ \emph {et~al.}(2023)\citenamefont
  {Asthana}, \citenamefont {Kumar}, \citenamefont {Abraham}, \citenamefont
  {Grimsley}, \citenamefont {Zhang}, \citenamefont {Cincio}, \citenamefont
  {Tretiak}, \citenamefont {Dub}, \citenamefont {Economou}, \citenamefont
  {Barnes} \emph {et~al.}}]{asthana2023quantum}%
  \BibitemOpen
  \bibfield  {author} {\bibinfo {author} {\bibfnamefont {A.}~\bibnamefont
  {Asthana}}, \bibinfo {author} {\bibfnamefont {A.}~\bibnamefont {Kumar}},
  \bibinfo {author} {\bibfnamefont {V.}~\bibnamefont {Abraham}}, \bibinfo
  {author} {\bibfnamefont {H.}~\bibnamefont {Grimsley}}, \bibinfo {author}
  {\bibfnamefont {Y.}~\bibnamefont {Zhang}}, \bibinfo {author} {\bibfnamefont
  {L.}~\bibnamefont {Cincio}}, \bibinfo {author} {\bibfnamefont
  {S.}~\bibnamefont {Tretiak}}, \bibinfo {author} {\bibfnamefont {P.~A.}\
  \bibnamefont {Dub}}, \bibinfo {author} {\bibfnamefont {S.~E.}\ \bibnamefont
  {Economou}}, \bibinfo {author} {\bibfnamefont {E.}~\bibnamefont {Barnes}},
  \emph {et~al.},\ }\href
  {https://pubs.rsc.org/en/content/articlelanding/2023/sc/d2sc05371c}
  {\bibfield  {journal} {\bibinfo  {journal} {Chem. Sci}\ }\textbf {\bibinfo
  {volume} {14}},\ \bibinfo {pages} {2405} (\bibinfo {year}
  {2023})}\BibitemShut {NoStop}%
\bibitem [{\citenamefont {Low}\ and\ \citenamefont
  {Chuang}(2019)}]{low2019hamiltonian}%
  \BibitemOpen
  \bibfield  {author} {\bibinfo {author} {\bibfnamefont {G.~H.}\ \bibnamefont
  {Low}}\ and\ \bibinfo {author} {\bibfnamefont {I.~L.}\ \bibnamefont
  {Chuang}},\ }\href {https://quantum-journal.org/papers/q-2019-07-12-163/}
  {\bibfield  {journal} {\bibinfo  {journal} {Quantum}\ }\textbf {\bibinfo
  {volume} {3}},\ \bibinfo {pages} {163} (\bibinfo {year} {2019})}\BibitemShut
  {NoStop}%
\bibitem [{\citenamefont {Zhang}\ \emph {et~al.}(2023)\citenamefont {Zhang},
  \citenamefont {Wang}, \citenamefont {Xu},\ and\ \citenamefont
  {Li}}]{ZhangGaussianPowerKrylov}%
  \BibitemOpen
  \bibfield  {author} {\bibinfo {author} {\bibfnamefont {Z.}~\bibnamefont
  {Zhang}}, \bibinfo {author} {\bibfnamefont {A.}~\bibnamefont {Wang}},
  \bibinfo {author} {\bibfnamefont {X.}~\bibnamefont {Xu}},\ and\ \bibinfo
  {author} {\bibfnamefont {Y.}~\bibnamefont {Li}},\ }\href
  {https://arxiv.org/abs/2301.13353} {\bibfield  {journal} {\bibinfo  {journal}
  {arXiv:2301.13353}\ } (\bibinfo {year} {2023})}\BibitemShut {NoStop}%
\bibitem [{\citenamefont {Epperly}\ \emph {et~al.}(2022)\citenamefont
  {Epperly}, \citenamefont {Lin},\ and\ \citenamefont
  {Nakatsukasa}}]{epperly2022theory}%
  \BibitemOpen
  \bibfield  {author} {\bibinfo {author} {\bibfnamefont {E.~N.}\ \bibnamefont
  {Epperly}}, \bibinfo {author} {\bibfnamefont {L.}~\bibnamefont {Lin}},\ and\
  \bibinfo {author} {\bibfnamefont {Y.}~\bibnamefont {Nakatsukasa}},\ }\href
  {https://doi.org/10.1137/21M145954X} {\bibfield  {journal} {\bibinfo
  {journal} {SIAM J. Matrix Anal}\ }\textbf {\bibinfo {volume} {43}},\ \bibinfo
  {pages} {1263} (\bibinfo {year} {2022})}\BibitemShut {NoStop}%
\bibitem [{\citenamefont {Shen}\ \emph
  {et~al.}(2023{\natexlab{a}})\citenamefont {Shen}, \citenamefont {Klymko},
  \citenamefont {Sud}, \citenamefont {Williams-Young}, \citenamefont
  {de~Jong},\ and\ \citenamefont {Tubman}}]{shen2023real}%
  \BibitemOpen
  \bibfield  {author} {\bibinfo {author} {\bibfnamefont {Y.}~\bibnamefont
  {Shen}}, \bibinfo {author} {\bibfnamefont {K.}~\bibnamefont {Klymko}},
  \bibinfo {author} {\bibfnamefont {J.}~\bibnamefont {Sud}}, \bibinfo {author}
  {\bibfnamefont {D.~B.}\ \bibnamefont {Williams-Young}}, \bibinfo {author}
  {\bibfnamefont {W.~A.}\ \bibnamefont {de~Jong}},\ and\ \bibinfo {author}
  {\bibfnamefont {N.~M.}\ \bibnamefont {Tubman}},\ }\href
  {https://quantum-journal.org/papers/q-2023-07-25-1066/} {\bibfield  {journal}
  {\bibinfo  {journal} {Quantum}\ }\textbf {\bibinfo {volume} {7}},\ \bibinfo
  {pages} {1066} (\bibinfo {year} {2023}{\natexlab{a}})}\BibitemShut {NoStop}%
\bibitem [{\citenamefont {Stair}\ \emph {et~al.}(2023)\citenamefont {Stair},
  \citenamefont {Cortes}, \citenamefont {Parrish}, \citenamefont {Cohn},\ and\
  \citenamefont {Motta}}]{stair2023stochastic}%
  \BibitemOpen
  \bibfield  {author} {\bibinfo {author} {\bibfnamefont {N.~H.}\ \bibnamefont
  {Stair}}, \bibinfo {author} {\bibfnamefont {C.~L.}\ \bibnamefont {Cortes}},
  \bibinfo {author} {\bibfnamefont {R.~M.}\ \bibnamefont {Parrish}}, \bibinfo
  {author} {\bibfnamefont {J.}~\bibnamefont {Cohn}},\ and\ \bibinfo {author}
  {\bibfnamefont {M.}~\bibnamefont {Motta}},\ }\href
  {https://journals.aps.org/pra/abstract/10.1103/PhysRevA.107.032414}
  {\bibfield  {journal} {\bibinfo  {journal} {Phys. Rev. A}\ }\textbf {\bibinfo
  {volume} {107}},\ \bibinfo {pages} {032414} (\bibinfo {year}
  {2023})}\BibitemShut {NoStop}%
\bibitem [{\citenamefont {Low}\ and\ \citenamefont
  {Chuang}(2017)}]{low2017optimal}%
  \BibitemOpen
  \bibfield  {author} {\bibinfo {author} {\bibfnamefont {G.~H.}\ \bibnamefont
  {Low}}\ and\ \bibinfo {author} {\bibfnamefont {I.~L.}\ \bibnamefont
  {Chuang}},\ }\href
  {https://journals.aps.org/prl/abstract/10.1103/PhysRevLett.118.010501}
  {\bibfield  {journal} {\bibinfo  {journal} {Phys. Rev. Lett}\ }\textbf
  {\bibinfo {volume} {118}},\ \bibinfo {pages} {010501} (\bibinfo {year}
  {2017})}\BibitemShut {NoStop}%
\bibitem [{\citenamefont {Dong}\ \emph {et~al.}(2021)\citenamefont {Dong},
  \citenamefont {Meng}, \citenamefont {Whaley},\ and\ \citenamefont
  {Lin}}]{dong2021efficient}%
  \BibitemOpen
  \bibfield  {author} {\bibinfo {author} {\bibfnamefont {Y.}~\bibnamefont
  {Dong}}, \bibinfo {author} {\bibfnamefont {X.}~\bibnamefont {Meng}}, \bibinfo
  {author} {\bibfnamefont {K.~B.}\ \bibnamefont {Whaley}},\ and\ \bibinfo
  {author} {\bibfnamefont {L.}~\bibnamefont {Lin}},\ }\href
  {https://journals.aps.org/pra/abstract/10.1103/PhysRevA.103.042419}
  {\bibfield  {journal} {\bibinfo  {journal} {Phys. Rev. A}\ }\textbf {\bibinfo
  {volume} {103}},\ \bibinfo {pages} {042419} (\bibinfo {year}
  {2021})}\BibitemShut {NoStop}%
\bibitem [{\citenamefont {Wang}\ \emph {et~al.}(2022)\citenamefont {Wang},
  \citenamefont {Dong},\ and\ \citenamefont {Lin}}]{wang2022energy}%
  \BibitemOpen
  \bibfield  {author} {\bibinfo {author} {\bibfnamefont {J.}~\bibnamefont
  {Wang}}, \bibinfo {author} {\bibfnamefont {Y.}~\bibnamefont {Dong}},\ and\
  \bibinfo {author} {\bibfnamefont {L.}~\bibnamefont {Lin}},\ }\href
  {https://quantum-journal.org/papers/q-2022-11-03-850/} {\bibfield  {journal}
  {\bibinfo  {journal} {Quantum}\ }\textbf {\bibinfo {volume} {6}},\ \bibinfo
  {pages} {850} (\bibinfo {year} {2022})}\BibitemShut {NoStop}%
\bibitem [{\citenamefont {Lin}(2022)}]{lin2022lecture}%
  \BibitemOpen
  \bibfield  {author} {\bibinfo {author} {\bibfnamefont {L.}~\bibnamefont
  {Lin}},\ }\href {https://arxiv.org/abs/2201.08309} {\bibfield  {journal}
  {\bibinfo  {journal} {arXiv:2201.08309}\ } (\bibinfo {year}
  {2022})}\BibitemShut {NoStop}%
\bibitem [{\citenamefont {McArdle}\ \emph {et~al.}(2019)\citenamefont
  {McArdle}, \citenamefont {Jones}, \citenamefont {Endo}, \citenamefont {Li},
  \citenamefont {Benjamin},\ and\ \citenamefont
  {Yuan}}]{mcardle2019variational}%
  \BibitemOpen
  \bibfield  {author} {\bibinfo {author} {\bibfnamefont {S.}~\bibnamefont
  {McArdle}}, \bibinfo {author} {\bibfnamefont {T.}~\bibnamefont {Jones}},
  \bibinfo {author} {\bibfnamefont {S.}~\bibnamefont {Endo}}, \bibinfo {author}
  {\bibfnamefont {Y.}~\bibnamefont {Li}}, \bibinfo {author} {\bibfnamefont
  {S.~C.}\ \bibnamefont {Benjamin}},\ and\ \bibinfo {author} {\bibfnamefont
  {X.}~\bibnamefont {Yuan}},\ }\href
  {https://www.nature.com/articles/s41534-019-0187-2} {\bibfield  {journal}
  {\bibinfo  {journal} {npj Quantum Inf}\ }\textbf {\bibinfo {volume} {5}},\
  \bibinfo {pages} {1} (\bibinfo {year} {2019})}\BibitemShut {NoStop}%
\bibitem [{\citenamefont {Gomes}\ \emph {et~al.}(2020)\citenamefont {Gomes},
  \citenamefont {Zhang}, \citenamefont {Berthusen}, \citenamefont {Wang},
  \citenamefont {Ho}, \citenamefont {Orth},\ and\ \citenamefont
  {Yao}}]{gomes2020}%
  \BibitemOpen
  \bibfield  {author} {\bibinfo {author} {\bibfnamefont {N.}~\bibnamefont
  {Gomes}}, \bibinfo {author} {\bibfnamefont {F.}~\bibnamefont {Zhang}},
  \bibinfo {author} {\bibfnamefont {N.~F.}\ \bibnamefont {Berthusen}}, \bibinfo
  {author} {\bibfnamefont {C.-Z.}\ \bibnamefont {Wang}}, \bibinfo {author}
  {\bibfnamefont {K.-M.}\ \bibnamefont {Ho}}, \bibinfo {author} {\bibfnamefont
  {P.~P.}\ \bibnamefont {Orth}},\ and\ \bibinfo {author} {\bibfnamefont
  {Y.}~\bibnamefont {Yao}},\ }\href {10.1021/acs.jctc.0c00666} {\bibfield
  {journal} {\bibinfo  {journal} {J. Chem. Theory Comput}\ }\textbf {\bibinfo
  {volume} {16}},\ \bibinfo {pages} {6256–6266} (\bibinfo {year}
  {2020})}\BibitemShut {NoStop}%
\bibitem [{\citenamefont {Peruzzo}\ \emph {et~al.}(2014)\citenamefont
  {Peruzzo}, \citenamefont {McClean}, \citenamefont {Shadbolt}, \citenamefont
  {Yung}, \citenamefont {Zhou}, \citenamefont {Love}, \citenamefont
  {Aspuru-Guzik},\ and\ \citenamefont {O'Brien}}]{peruzzo2014variational}%
  \BibitemOpen
  \bibfield  {author} {\bibinfo {author} {\bibfnamefont {A.}~\bibnamefont
  {Peruzzo}}, \bibinfo {author} {\bibfnamefont {J.}~\bibnamefont {McClean}},
  \bibinfo {author} {\bibfnamefont {P.}~\bibnamefont {Shadbolt}}, \bibinfo
  {author} {\bibfnamefont {M.-H.}\ \bibnamefont {Yung}}, \bibinfo {author}
  {\bibfnamefont {X.-Q.}\ \bibnamefont {Zhou}}, \bibinfo {author}
  {\bibfnamefont {P.~J.}\ \bibnamefont {Love}}, \bibinfo {author}
  {\bibfnamefont {A.}~\bibnamefont {Aspuru-Guzik}},\ and\ \bibinfo {author}
  {\bibfnamefont {J.~L.}\ \bibnamefont {O'Brien}},\ }\href
  {https://www.nature.com/articles/ncomms5213} {\bibfield  {journal} {\bibinfo
  {journal} {Nat. Commun}\ }\textbf {\bibinfo {volume} {5}},\ \bibinfo {pages}
  {1} (\bibinfo {year} {2014})}\BibitemShut {NoStop}%
\bibitem [{\citenamefont {Rice}\ \emph {et~al.}(2021)\citenamefont {Rice},
  \citenamefont {Gujarati}, \citenamefont {Motta}, \citenamefont {Takeshita},
  \citenamefont {Lee}, \citenamefont {Latone},\ and\ \citenamefont
  {Garcia}}]{Rice2021}%
  \BibitemOpen
  \bibfield  {author} {\bibinfo {author} {\bibfnamefont {J.}~\bibnamefont
  {Rice}}, \bibinfo {author} {\bibfnamefont {T.}~\bibnamefont {Gujarati}},
  \bibinfo {author} {\bibfnamefont {M.}~\bibnamefont {Motta}}, \bibinfo
  {author} {\bibfnamefont {T.}~\bibnamefont {Takeshita}}, \bibinfo {author}
  {\bibfnamefont {E.}~\bibnamefont {Lee}}, \bibinfo {author} {\bibfnamefont
  {J.}~\bibnamefont {Latone}},\ and\ \bibinfo {author} {\bibfnamefont
  {J.}~\bibnamefont {Garcia}},\ }\href {10.1063/5.0044068} {\bibfield
  {journal} {\bibinfo  {journal} {J. Chem. Phys.}\ }\textbf {\bibinfo {volume}
  {154}},\ \bibinfo {pages} {134115} (\bibinfo {year} {2021})}\BibitemShut
  {NoStop}%
\bibitem [{\citenamefont {Barkoutsos}\ \emph {et~al.}(2018)\citenamefont
  {Barkoutsos}, \citenamefont {Gonthier}, \citenamefont {Sokolov},
  \citenamefont {Moll}, \citenamefont {Salis}, \citenamefont {Fuhrer},
  \citenamefont {Ganzhorn}, \citenamefont {Egger}, \citenamefont {Troyer},
  \citenamefont {Mezzacapo}, \citenamefont {Filipp},\ and\ \citenamefont
  {Tavernelli}}]{Barkoutsos2018}%
  \BibitemOpen
  \bibfield  {author} {\bibinfo {author} {\bibfnamefont {P.~K.}\ \bibnamefont
  {Barkoutsos}}, \bibinfo {author} {\bibfnamefont {J.~F.}\ \bibnamefont
  {Gonthier}}, \bibinfo {author} {\bibfnamefont {I.}~\bibnamefont {Sokolov}},
  \bibinfo {author} {\bibfnamefont {N.}~\bibnamefont {Moll}}, \bibinfo {author}
  {\bibfnamefont {G.}~\bibnamefont {Salis}}, \bibinfo {author} {\bibfnamefont
  {A.}~\bibnamefont {Fuhrer}}, \bibinfo {author} {\bibfnamefont
  {M.}~\bibnamefont {Ganzhorn}}, \bibinfo {author} {\bibfnamefont {D.~J.}\
  \bibnamefont {Egger}}, \bibinfo {author} {\bibfnamefont {M.}~\bibnamefont
  {Troyer}}, \bibinfo {author} {\bibfnamefont {A.}~\bibnamefont {Mezzacapo}},
  \bibinfo {author} {\bibfnamefont {S.}~\bibnamefont {Filipp}},\ and\ \bibinfo
  {author} {\bibfnamefont {I.}~\bibnamefont {Tavernelli}},\ }\href
  {https://link.aps.org/doi/10.1103/PhysRevA.98.022322} {\bibfield  {journal}
  {\bibinfo  {journal} {Phys. Rev. A}\ }\textbf {\bibinfo {volume} {98}},\
  \bibinfo {pages} {022322} (\bibinfo {year} {2018})}\BibitemShut {NoStop}%
\bibitem [{\citenamefont {Grimsley}\ \emph {et~al.}(2019)\citenamefont
  {Grimsley}, \citenamefont {Economou}, \citenamefont {Barnes},\ and\
  \citenamefont {Mayhall}}]{Grimsley2019}%
  \BibitemOpen
  \bibfield  {author} {\bibinfo {author} {\bibfnamefont {H.}~\bibnamefont
  {Grimsley}}, \bibinfo {author} {\bibfnamefont {S.}~\bibnamefont {Economou}},
  \bibinfo {author} {\bibfnamefont {E.}~\bibnamefont {Barnes}},\ and\ \bibinfo
  {author} {\bibfnamefont {N.}~\bibnamefont {Mayhall}},\ }\href
  {https://www.nature.com/articles/s41467-019-10988-2} {\bibfield  {journal}
  {\bibinfo  {journal} {Nat. Commun}\ }\textbf {\bibinfo {volume} {10}},\
  \bibinfo {pages} {3007} (\bibinfo {year} {2019})}\BibitemShut {NoStop}%
\bibitem [{\citenamefont {Sokolov}\ \emph {et~al.}(2020)\citenamefont
  {Sokolov}, \citenamefont {Barkoutsos}, \citenamefont {Ollitrault},
  \citenamefont {Greenberg}, \citenamefont {Rice}, \citenamefont {Pistoia},\
  and\ \citenamefont {Tavernelli}}]{Sokolov2020ooUCCSD}%
  \BibitemOpen
  \bibfield  {author} {\bibinfo {author} {\bibfnamefont {I.}~\bibnamefont
  {Sokolov}}, \bibinfo {author} {\bibfnamefont {P.}~\bibnamefont {Barkoutsos}},
  \bibinfo {author} {\bibfnamefont {P.}~\bibnamefont {Ollitrault}}, \bibinfo
  {author} {\bibfnamefont {D.}~\bibnamefont {Greenberg}}, \bibinfo {author}
  {\bibfnamefont {J.}~\bibnamefont {Rice}}, \bibinfo {author} {\bibfnamefont
  {M.}~\bibnamefont {Pistoia}},\ and\ \bibinfo {author} {\bibfnamefont
  {I.}~\bibnamefont {Tavernelli}},\ }\href {10.1063/1.5141835} {\bibfield
  {journal} {\bibinfo  {journal} {J. Chem. Phys}\ }\textbf {\bibinfo {volume}
  {152}},\ \bibinfo {pages} {124107} (\bibinfo {year} {2020})}\BibitemShut
  {NoStop}%
\bibitem [{\citenamefont {Gao}\ \emph {et~al.}(2021{\natexlab{a}})\citenamefont
  {Gao}, \citenamefont {Nakamura}, \citenamefont {Gujarati}, \citenamefont
  {Jones}, \citenamefont {Rice}, \citenamefont {Wood}, \citenamefont {Pistoia},
  \citenamefont {Garcia},\ and\ \citenamefont {Yamamoto}}]{Gao2021}%
  \BibitemOpen
  \bibfield  {author} {\bibinfo {author} {\bibfnamefont {Q.}~\bibnamefont
  {Gao}}, \bibinfo {author} {\bibfnamefont {H.}~\bibnamefont {Nakamura}},
  \bibinfo {author} {\bibfnamefont {T.~P.}\ \bibnamefont {Gujarati}}, \bibinfo
  {author} {\bibfnamefont {G.~O.}\ \bibnamefont {Jones}}, \bibinfo {author}
  {\bibfnamefont {J.~E.}\ \bibnamefont {Rice}}, \bibinfo {author}
  {\bibfnamefont {S.~P.}\ \bibnamefont {Wood}}, \bibinfo {author}
  {\bibfnamefont {M.}~\bibnamefont {Pistoia}}, \bibinfo {author} {\bibfnamefont
  {J.~M.}\ \bibnamefont {Garcia}},\ and\ \bibinfo {author} {\bibfnamefont
  {N.}~\bibnamefont {Yamamoto}},\ }\href
  {https://doi.org/10.1021/acs.jpca.0c09530} {\bibfield  {journal} {\bibinfo
  {journal} {J. Phys. Chem. A}\ }\textbf {\bibinfo {volume} {125}},\ \bibinfo
  {pages} {1827} (\bibinfo {year} {2021}{\natexlab{a}})}\BibitemShut {NoStop}%
\bibitem [{\citenamefont {Motta}\ \emph
  {et~al.}(2020{\natexlab{b}})\citenamefont {Motta}, \citenamefont {Gujarati},
  \citenamefont {Rice}, \citenamefont {Kumar}, \citenamefont {Masteran},
  \citenamefont {Latone}, \citenamefont {Lee}, \citenamefont {Valeev},\ and\
  \citenamefont {Takeshita}}]{Motta2020transcorrelated}%
  \BibitemOpen
  \bibfield  {author} {\bibinfo {author} {\bibfnamefont {M.}~\bibnamefont
  {Motta}}, \bibinfo {author} {\bibfnamefont {T.~P.}\ \bibnamefont {Gujarati}},
  \bibinfo {author} {\bibfnamefont {J.~E.}\ \bibnamefont {Rice}}, \bibinfo
  {author} {\bibfnamefont {A.}~\bibnamefont {Kumar}}, \bibinfo {author}
  {\bibfnamefont {C.}~\bibnamefont {Masteran}}, \bibinfo {author}
  {\bibfnamefont {J.~A.}\ \bibnamefont {Latone}}, \bibinfo {author}
  {\bibfnamefont {E.}~\bibnamefont {Lee}}, \bibinfo {author} {\bibfnamefont
  {E.~F.}\ \bibnamefont {Valeev}},\ and\ \bibinfo {author} {\bibfnamefont
  {T.~Y.}\ \bibnamefont {Takeshita}},\ }\href
  {https://doi.org/10.1039/D0CP04106H} {\bibfield  {journal} {\bibinfo
  {journal} {Phys. Chem. Chem. Phys}\ }\textbf {\bibinfo {volume} {22}},\
  \bibinfo {pages} {24270} (\bibinfo {year} {2020}{\natexlab{b}})}\BibitemShut
  {NoStop}%
\bibitem [{\citenamefont {D'Cunha}\ \emph {et~al.}(2023)\citenamefont
  {D'Cunha}, \citenamefont {Crawford}, \citenamefont {Motta},\ and\
  \citenamefont {Rice}}]{d2023challenges}%
  \BibitemOpen
  \bibfield  {author} {\bibinfo {author} {\bibfnamefont {R.}~\bibnamefont
  {D'Cunha}}, \bibinfo {author} {\bibfnamefont {T.~D.}\ \bibnamefont
  {Crawford}}, \bibinfo {author} {\bibfnamefont {M.}~\bibnamefont {Motta}},\
  and\ \bibinfo {author} {\bibfnamefont {J.~E.}\ \bibnamefont {Rice}},\ }\href
  {https://doi.org/10.1021/acs.jpca.2c08430} {\bibfield  {journal} {\bibinfo
  {journal} {J. Phys. Chem. A}\ }\textbf {\bibinfo {volume} {127}},\ \bibinfo
  {pages} {3437} (\bibinfo {year} {2023})}\BibitemShut {NoStop}%
\bibitem [{\citenamefont {Parrish}\ \emph {et~al.}(2021)\citenamefont
  {Parrish}, \citenamefont {Anselmetti},\ and\ \citenamefont
  {Gogolin}}]{parrish2021analytical}%
  \BibitemOpen
  \bibfield  {author} {\bibinfo {author} {\bibfnamefont {R.~M.}\ \bibnamefont
  {Parrish}}, \bibinfo {author} {\bibfnamefont {G.-L.~R.}\ \bibnamefont
  {Anselmetti}},\ and\ \bibinfo {author} {\bibfnamefont {C.}~\bibnamefont
  {Gogolin}},\ }\href {https://arxiv.org/abs/2110.05040} {\bibfield  {journal}
  {\bibinfo  {journal} {arXiv:2110.05040}\ } (\bibinfo {year}
  {2021})}\BibitemShut {NoStop}%
\bibitem [{\citenamefont {Sundstrom}\ and\ \citenamefont
  {Head-Gordon}(2014)}]{sundstrom2014non}%
  \BibitemOpen
  \bibfield  {author} {\bibinfo {author} {\bibfnamefont {E.~J.}\ \bibnamefont
  {Sundstrom}}\ and\ \bibinfo {author} {\bibfnamefont {M.}~\bibnamefont
  {Head-Gordon}},\ }\href {https://doi.org/10.1063/1.4868120} {\bibfield
  {journal} {\bibinfo  {journal} {J. Chem. Phys}\ }\textbf {\bibinfo {volume}
  {140}},\ \bibinfo {pages} {114103} (\bibinfo {year} {2014})}\BibitemShut
  {NoStop}%
\bibitem [{\citenamefont {Yost}\ and\ \citenamefont
  {Head-Gordon}(2016)}]{yost2016size}%
  \BibitemOpen
  \bibfield  {author} {\bibinfo {author} {\bibfnamefont {S.~R.}\ \bibnamefont
  {Yost}}\ and\ \bibinfo {author} {\bibfnamefont {M.}~\bibnamefont
  {Head-Gordon}},\ }\href {https://doi.org/10.1063/1.4959794} {\bibfield
  {journal} {\bibinfo  {journal} {J. Chem. Phys}\ }\textbf {\bibinfo {volume}
  {145}},\ \bibinfo {pages} {054105} (\bibinfo {year} {2016})}\BibitemShut
  {NoStop}%
\bibitem [{\citenamefont {Kanno}\ \emph {et~al.}(2023)\citenamefont {Kanno},
  \citenamefont {Kohda}, \citenamefont {Imai}, \citenamefont {Koh},
  \citenamefont {Mitarai}, \citenamefont {Mizukami},\ and\ \citenamefont
  {Nakagawa}}]{kanno2023quantum}%
  \BibitemOpen
  \bibfield  {author} {\bibinfo {author} {\bibfnamefont {K.}~\bibnamefont
  {Kanno}}, \bibinfo {author} {\bibfnamefont {M.}~\bibnamefont {Kohda}},
  \bibinfo {author} {\bibfnamefont {R.}~\bibnamefont {Imai}}, \bibinfo {author}
  {\bibfnamefont {S.}~\bibnamefont {Koh}}, \bibinfo {author} {\bibfnamefont
  {K.}~\bibnamefont {Mitarai}}, \bibinfo {author} {\bibfnamefont
  {W.}~\bibnamefont {Mizukami}},\ and\ \bibinfo {author} {\bibfnamefont
  {Y.~O.}\ \bibnamefont {Nakagawa}},\ }\href {https://arxiv.org/abs/2302.11320}
  {\bibfield  {journal} {\bibinfo  {journal} {arXiv:2302.11320}\ } (\bibinfo
  {year} {2023})}\BibitemShut {NoStop}%
\bibitem [{\citenamefont {Francis}\ \emph {et~al.}(2022)\citenamefont
  {Francis}, \citenamefont {Agrawal}, \citenamefont {Howard}, \citenamefont
  {K{\"o}kc{\"u}},\ and\ \citenamefont {Kemper}}]{francis2022subspace}%
  \BibitemOpen
  \bibfield  {author} {\bibinfo {author} {\bibfnamefont {A.}~\bibnamefont
  {Francis}}, \bibinfo {author} {\bibfnamefont {A.~A.}\ \bibnamefont
  {Agrawal}}, \bibinfo {author} {\bibfnamefont {J.~H.}\ \bibnamefont {Howard}},
  \bibinfo {author} {\bibfnamefont {E.}~\bibnamefont {K{\"o}kc{\"u}}},\ and\
  \bibinfo {author} {\bibfnamefont {A.}~\bibnamefont {Kemper}},\ }\href
  {https://arxiv.org/pdf/2209.10571.pdf} {\bibfield  {journal} {\bibinfo
  {journal} {arXiv:2209.10571}\ } (\bibinfo {year} {2022})}\BibitemShut
  {NoStop}%
\bibitem [{\citenamefont {Mejuto-Zaera}\ and\ \citenamefont
  {Kemper}(2023)}]{mejuto2023quantum}%
  \BibitemOpen
  \bibfield  {author} {\bibinfo {author} {\bibfnamefont {C.}~\bibnamefont
  {Mejuto-Zaera}}\ and\ \bibinfo {author} {\bibfnamefont {A.~F.}\ \bibnamefont
  {Kemper}},\ }\href {https://doi.org/10.1088/2516-1075/ad018f} {\bibfield
  {journal} {\bibinfo  {journal} {Electronic Structure}\ }\textbf {\bibinfo
  {volume} {5}},\ \bibinfo {pages} {045007} (\bibinfo {year}
  {2023})}\BibitemShut {NoStop}%
\bibitem [{\citenamefont {Wick}(1950)}]{wick1950evaluation}%
  \BibitemOpen
  \bibfield  {author} {\bibinfo {author} {\bibfnamefont {G.-C.}\ \bibnamefont
  {Wick}},\ }\href
  {https://journals.aps.org/pr/abstract/10.1103/PhysRev.80.268} {\bibfield
  {journal} {\bibinfo  {journal} {Phys. Rev}\ }\textbf {\bibinfo {volume}
  {80}},\ \bibinfo {pages} {268} (\bibinfo {year} {1950})}\BibitemShut
  {NoStop}%
\bibitem [{\citenamefont {Urbanek}\ \emph {et~al.}(2020)\citenamefont
  {Urbanek}, \citenamefont {Camps}, \citenamefont {Van~Beeumen},\ and\
  \citenamefont {de~Jong}}]{urbanek2020chemistry}%
  \BibitemOpen
  \bibfield  {author} {\bibinfo {author} {\bibfnamefont {M.}~\bibnamefont
  {Urbanek}}, \bibinfo {author} {\bibfnamefont {D.}~\bibnamefont {Camps}},
  \bibinfo {author} {\bibfnamefont {R.}~\bibnamefont {Van~Beeumen}},\ and\
  \bibinfo {author} {\bibfnamefont {W.~A.}\ \bibnamefont {de~Jong}},\ }\href
  {https://doi.org/10.1021/acs.jctc.0c00447} {\bibfield  {journal} {\bibinfo
  {journal} {J. Chem. Theory Comput}\ }\textbf {\bibinfo {volume} {16}},\
  \bibinfo {pages} {5425} (\bibinfo {year} {2020})}\BibitemShut {NoStop}%
\bibitem [{\citenamefont {Tammaro}\ \emph {et~al.}(2023)\citenamefont
  {Tammaro}, \citenamefont {Galli}, \citenamefont {Rice},\ and\ \citenamefont
  {Motta}}]{tammaro2023n}%
  \BibitemOpen
  \bibfield  {author} {\bibinfo {author} {\bibfnamefont {A.}~\bibnamefont
  {Tammaro}}, \bibinfo {author} {\bibfnamefont {D.~E.}\ \bibnamefont {Galli}},
  \bibinfo {author} {\bibfnamefont {J.~E.}\ \bibnamefont {Rice}},\ and\
  \bibinfo {author} {\bibfnamefont {M.}~\bibnamefont {Motta}},\ }\href
  {https://doi.org/10.1021/acs.jpca.2c07653} {\bibfield  {journal} {\bibinfo
  {journal} {J. Phys. Chem. A}\ }\textbf {\bibinfo {volume} {127}},\ \bibinfo
  {pages} {817} (\bibinfo {year} {2023})}\BibitemShut {NoStop}%
\bibitem [{\citenamefont {Angeli}\ \emph
  {et~al.}(2001{\natexlab{a}})\citenamefont {Angeli}, \citenamefont
  {Cimiraglia}, \citenamefont {Evangelisti}, \citenamefont {Leininger},\ and\
  \citenamefont {Malrieu}}]{angeli2001introduction}%
  \BibitemOpen
  \bibfield  {author} {\bibinfo {author} {\bibfnamefont {C.}~\bibnamefont
  {Angeli}}, \bibinfo {author} {\bibfnamefont {R.}~\bibnamefont {Cimiraglia}},
  \bibinfo {author} {\bibfnamefont {S.}~\bibnamefont {Evangelisti}}, \bibinfo
  {author} {\bibfnamefont {T.}~\bibnamefont {Leininger}},\ and\ \bibinfo
  {author} {\bibfnamefont {J.-P.}\ \bibnamefont {Malrieu}},\ }\href
  {https://aip.scitation.org/doi/10.1063/1.1361246} {\bibfield  {journal}
  {\bibinfo  {journal} {J. Chem. Phys}\ }\textbf {\bibinfo {volume} {114}},\
  \bibinfo {pages} {10252} (\bibinfo {year} {2001}{\natexlab{a}})}\BibitemShut
  {NoStop}%
\bibitem [{\citenamefont {Angeli}\ \emph
  {et~al.}(2001{\natexlab{b}})\citenamefont {Angeli}, \citenamefont
  {Cimiraglia},\ and\ \citenamefont {Malrieu}}]{angeli2001n}%
  \BibitemOpen
  \bibfield  {author} {\bibinfo {author} {\bibfnamefont {C.}~\bibnamefont
  {Angeli}}, \bibinfo {author} {\bibfnamefont {R.}~\bibnamefont {Cimiraglia}},\
  and\ \bibinfo {author} {\bibfnamefont {J.-P.}\ \bibnamefont {Malrieu}},\
  }\href
  {https://www.sciencedirect.com/science/article/abs/pii/S0009261401013033}
  {\bibfield  {journal} {\bibinfo  {journal} {Chem. Phys. Lett}\ }\textbf
  {\bibinfo {volume} {350}},\ \bibinfo {pages} {297} (\bibinfo {year}
  {2001}{\natexlab{b}})}\BibitemShut {NoStop}%
\bibitem [{\citenamefont {Sokolov}\ and\ \citenamefont
  {Chan}(2016)}]{sokolov2016time}%
  \BibitemOpen
  \bibfield  {author} {\bibinfo {author} {\bibfnamefont {A.~Y.}\ \bibnamefont
  {Sokolov}}\ and\ \bibinfo {author} {\bibfnamefont {G.~K.-L.}\ \bibnamefont
  {Chan}},\ }\href {https://aip.scitation.org/doi/10.1063/1.4941606} {\bibfield
   {journal} {\bibinfo  {journal} {J. Chem. Phys}\ }\textbf {\bibinfo {volume}
  {144}},\ \bibinfo {pages} {064102} (\bibinfo {year} {2016})}\BibitemShut
  {NoStop}%
\bibitem [{\citenamefont {Sokolov}\ \emph {et~al.}(2017)\citenamefont
  {Sokolov}, \citenamefont {Guo}, \citenamefont {Ronca},\ and\ \citenamefont
  {Chan}}]{sokolov2017time}%
  \BibitemOpen
  \bibfield  {author} {\bibinfo {author} {\bibfnamefont {A.~Y.}\ \bibnamefont
  {Sokolov}}, \bibinfo {author} {\bibfnamefont {S.}~\bibnamefont {Guo}},
  \bibinfo {author} {\bibfnamefont {E.}~\bibnamefont {Ronca}},\ and\ \bibinfo
  {author} {\bibfnamefont {G.~K.-L.}\ \bibnamefont {Chan}},\ }\href
  {https://aip.scitation.org/doi/10.1063/1.4986975} {\bibfield  {journal}
  {\bibinfo  {journal} {J. Chem. Phys}\ }\textbf {\bibinfo {volume} {146}},\
  \bibinfo {pages} {244102} (\bibinfo {year} {2017})}\BibitemShut {NoStop}%
\bibitem [{\citenamefont {Dyall}(1995)}]{dyall1995choice}%
  \BibitemOpen
  \bibfield  {author} {\bibinfo {author} {\bibfnamefont {K.~G.}\ \bibnamefont
  {Dyall}},\ }\href {https://aip.scitation.org/doi/10.1063/1.469539} {\bibfield
   {journal} {\bibinfo  {journal} {J. Chem. Phys}\ }\textbf {\bibinfo {volume}
  {102}},\ \bibinfo {pages} {4909} (\bibinfo {year} {1995})}\BibitemShut
  {NoStop}%
\bibitem [{\citenamefont {Krompiec}\ and\ \citenamefont
  {Ramo}(2022)}]{krompiec2022strongly}%
  \BibitemOpen
  \bibfield  {author} {\bibinfo {author} {\bibfnamefont {M.}~\bibnamefont
  {Krompiec}}\ and\ \bibinfo {author} {\bibfnamefont {D.~M.}\ \bibnamefont
  {Ramo}},\ }\href {https://arxiv.org/abs/2210.05702} {\bibfield  {journal}
  {\bibinfo  {journal} {arXiv:2210.05702}\ } (\bibinfo {year}
  {2022})}\BibitemShut {NoStop}%
\bibitem [{\citenamefont {Gordon}(1965)}]{gordon1965molecular}%
  \BibitemOpen
  \bibfield  {author} {\bibinfo {author} {\bibfnamefont {R.}~\bibnamefont
  {Gordon}},\ }\href {https://aip.scitation.org/doi/10.1063/1.1696920}
  {\bibfield  {journal} {\bibinfo  {journal} {J. Chem. Phys}\ }\textbf
  {\bibinfo {volume} {43}},\ \bibinfo {pages} {1307} (\bibinfo {year}
  {1965})}\BibitemShut {NoStop}%
\bibitem [{\citenamefont {Boulet}\ and\ \citenamefont
  {Robert}(1982)}]{boulet1982short}%
  \BibitemOpen
  \bibfield  {author} {\bibinfo {author} {\bibfnamefont {C.}~\bibnamefont
  {Boulet}}\ and\ \bibinfo {author} {\bibfnamefont {D.}~\bibnamefont
  {Robert}},\ }\href {https://aip.scitation.org/doi/10.1063/1.444430}
  {\bibfield  {journal} {\bibinfo  {journal} {J. Chem. Phys}\ }\textbf
  {\bibinfo {volume} {77}},\ \bibinfo {pages} {4288} (\bibinfo {year}
  {1982})}\BibitemShut {NoStop}%
\bibitem [{\citenamefont {Clerk}\ \emph {et~al.}(2010)\citenamefont {Clerk},
  \citenamefont {Devoret}, \citenamefont {Girvin}, \citenamefont {Marquardt},\
  and\ \citenamefont {Schoelkopf}}]{clerk2010introduction}%
  \BibitemOpen
  \bibfield  {author} {\bibinfo {author} {\bibfnamefont {A.~A.}\ \bibnamefont
  {Clerk}}, \bibinfo {author} {\bibfnamefont {M.~H.}\ \bibnamefont {Devoret}},
  \bibinfo {author} {\bibfnamefont {S.~M.}\ \bibnamefont {Girvin}}, \bibinfo
  {author} {\bibfnamefont {F.}~\bibnamefont {Marquardt}},\ and\ \bibinfo
  {author} {\bibfnamefont {R.~J.}\ \bibnamefont {Schoelkopf}},\ }\href
  {https://journals.aps.org/rmp/pdf/10.1103/RevModPhys.82.1155} {\bibfield
  {journal} {\bibinfo  {journal} {Rev. Mod. Phys}\ }\textbf {\bibinfo {volume}
  {82}},\ \bibinfo {pages} {1155} (\bibinfo {year} {2010})}\BibitemShut
  {NoStop}%
\bibitem [{\citenamefont {Vitale}\ \emph {et~al.}(2015)\citenamefont {Vitale},
  \citenamefont {Dziedzic}, \citenamefont {Dubois}, \citenamefont {Fangohr},\
  and\ \citenamefont {Skylaris}}]{vitale2015anharmonic}%
  \BibitemOpen
  \bibfield  {author} {\bibinfo {author} {\bibfnamefont {V.}~\bibnamefont
  {Vitale}}, \bibinfo {author} {\bibfnamefont {J.}~\bibnamefont {Dziedzic}},
  \bibinfo {author} {\bibfnamefont {S.~M.-M.}\ \bibnamefont {Dubois}}, \bibinfo
  {author} {\bibfnamefont {H.}~\bibnamefont {Fangohr}},\ and\ \bibinfo {author}
  {\bibfnamefont {C.-K.}\ \bibnamefont {Skylaris}},\ }\href
  {https://pubs.acs.org/doi/10.1021/acs.jctc.5b00391} {\bibfield  {journal}
  {\bibinfo  {journal} {J. Chem. Theory Comput}\ }\textbf {\bibinfo {volume}
  {11}},\ \bibinfo {pages} {3321} (\bibinfo {year} {2015})}\BibitemShut
  {NoStop}%
\bibitem [{\citenamefont {Nascimento}\ and\ \citenamefont
  {DePrince~III}(2016)}]{nascimento2016linear}%
  \BibitemOpen
  \bibfield  {author} {\bibinfo {author} {\bibfnamefont {D.~R.}\ \bibnamefont
  {Nascimento}}\ and\ \bibinfo {author} {\bibfnamefont {A.~E.}\ \bibnamefont
  {DePrince~III}},\ }\href {https://pubs.acs.org/doi/10.1021/acs.jctc.6b00796}
  {\bibfield  {journal} {\bibinfo  {journal} {J. Chem. Theory Comput}\ }\textbf
  {\bibinfo {volume} {12}},\ \bibinfo {pages} {5834} (\bibinfo {year}
  {2016})}\BibitemShut {NoStop}%
\bibitem [{\citenamefont {Goings}\ \emph {et~al.}(2018)\citenamefont {Goings},
  \citenamefont {Lestrange},\ and\ \citenamefont {Li}}]{goings2018real}%
  \BibitemOpen
  \bibfield  {author} {\bibinfo {author} {\bibfnamefont {J.~J.}\ \bibnamefont
  {Goings}}, \bibinfo {author} {\bibfnamefont {P.~J.}\ \bibnamefont
  {Lestrange}},\ and\ \bibinfo {author} {\bibfnamefont {X.}~\bibnamefont
  {Li}},\ }\href {https://pubs.acs.org/doi/10.1021/acs.chemrev.0c00223}
  {\bibfield  {journal} {\bibinfo  {journal} {WIREs Comput. Mol. Sci}\ }\textbf
  {\bibinfo {volume} {8}},\ \bibinfo {pages} {e1341} (\bibinfo {year}
  {2018})}\BibitemShut {NoStop}%
\bibitem [{\citenamefont {Li}\ \emph {et~al.}(2020)\citenamefont {Li},
  \citenamefont {Govind}, \citenamefont {Isborn}, \citenamefont
  {DePrince~III},\ and\ \citenamefont {Lopata}}]{li2020real}%
  \BibitemOpen
  \bibfield  {author} {\bibinfo {author} {\bibfnamefont {X.}~\bibnamefont
  {Li}}, \bibinfo {author} {\bibfnamefont {N.}~\bibnamefont {Govind}}, \bibinfo
  {author} {\bibfnamefont {C.}~\bibnamefont {Isborn}}, \bibinfo {author}
  {\bibfnamefont {A.~E.}\ \bibnamefont {DePrince~III}},\ and\ \bibinfo {author}
  {\bibfnamefont {K.}~\bibnamefont {Lopata}},\ }\href
  {https://pubs.acs.org/doi/full/10.1021/acs.chemrev.0c00223} {\bibfield
  {journal} {\bibinfo  {journal} {Chem. Rev}\ }\textbf {\bibinfo {volume}
  {120}},\ \bibinfo {pages} {9951} (\bibinfo {year} {2020})}\BibitemShut
  {NoStop}%
\bibitem [{\citenamefont {Damascelli}(2004)}]{damascelli2004probing}%
  \BibitemOpen
  \bibfield  {author} {\bibinfo {author} {\bibfnamefont {A.}~\bibnamefont
  {Damascelli}},\ }\href
  {https://iopscience.iop.org/article/10.1238/Physica.Topical.109a00061/meta}
  {\bibfield  {journal} {\bibinfo  {journal} {Phys. Scripta}\ }\textbf
  {\bibinfo {volume} {2004}},\ \bibinfo {pages} {61} (\bibinfo {year}
  {2004})}\BibitemShut {NoStop}%
\bibitem [{\citenamefont {Nazarov}\ and\ \citenamefont
  {Blanter}(2009)}]{nazarov2009quantum}%
  \BibitemOpen
  \bibfield  {author} {\bibinfo {author} {\bibfnamefont {Y.~V.}\ \bibnamefont
  {Nazarov}}\ and\ \bibinfo {author} {\bibfnamefont {Y.~M.}\ \bibnamefont
  {Blanter}},\ }\href@noop {} {\emph {\bibinfo {title} {Quantum transport:
  introduction to nanoscience}}}\ (\bibinfo  {publisher} {Cambridge University
  Press},\ \bibinfo {year} {2009})\BibitemShut {NoStop}%
\bibitem [{\citenamefont {Patterson}\ \emph {et~al.}(2010)\citenamefont
  {Patterson}, \citenamefont {Abela}, \citenamefont {Braun}, \citenamefont
  {Flechsig}, \citenamefont {Ganter}, \citenamefont {Kim}, \citenamefont
  {Kirk}, \citenamefont {Oppelt}, \citenamefont {Pedrozzi}, \citenamefont
  {Reiche} \emph {et~al.}}]{patterson2010coherent}%
  \BibitemOpen
  \bibfield  {author} {\bibinfo {author} {\bibfnamefont {B.}~\bibnamefont
  {Patterson}}, \bibinfo {author} {\bibfnamefont {R.}~\bibnamefont {Abela}},
  \bibinfo {author} {\bibfnamefont {H.}~\bibnamefont {Braun}}, \bibinfo
  {author} {\bibfnamefont {U.}~\bibnamefont {Flechsig}}, \bibinfo {author}
  {\bibfnamefont {R.}~\bibnamefont {Ganter}}, \bibinfo {author} {\bibfnamefont
  {Y.}~\bibnamefont {Kim}}, \bibinfo {author} {\bibfnamefont {E.}~\bibnamefont
  {Kirk}}, \bibinfo {author} {\bibfnamefont {A.}~\bibnamefont {Oppelt}},
  \bibinfo {author} {\bibfnamefont {M.}~\bibnamefont {Pedrozzi}}, \bibinfo
  {author} {\bibfnamefont {S.}~\bibnamefont {Reiche}}, \emph {et~al.},\ }\href
  {https://iopscience.iop.org/article/10.1088/1367-2630/12/3/035012} {\bibfield
   {journal} {\bibinfo  {journal} {New J. Phys}\ }\textbf {\bibinfo {volume}
  {12}},\ \bibinfo {pages} {035012} (\bibinfo {year} {2010})}\BibitemShut
  {NoStop}%
\bibitem [{\citenamefont {Weathersby}\ \emph {et~al.}(2015)\citenamefont
  {Weathersby}, \citenamefont {Brown}, \citenamefont {Centurion}, \citenamefont
  {Chase}, \citenamefont {Coffee}, \citenamefont {Corbett}, \citenamefont
  {Eichner}, \citenamefont {Frisch}, \citenamefont {Fry}, \citenamefont
  {G{\"u}hr} \emph {et~al.}}]{weathersby2015mega}%
  \BibitemOpen
  \bibfield  {author} {\bibinfo {author} {\bibfnamefont {S.}~\bibnamefont
  {Weathersby}}, \bibinfo {author} {\bibfnamefont {G.}~\bibnamefont {Brown}},
  \bibinfo {author} {\bibfnamefont {M.}~\bibnamefont {Centurion}}, \bibinfo
  {author} {\bibfnamefont {T.}~\bibnamefont {Chase}}, \bibinfo {author}
  {\bibfnamefont {R.}~\bibnamefont {Coffee}}, \bibinfo {author} {\bibfnamefont
  {J.}~\bibnamefont {Corbett}}, \bibinfo {author} {\bibfnamefont
  {J.}~\bibnamefont {Eichner}}, \bibinfo {author} {\bibfnamefont
  {J.}~\bibnamefont {Frisch}}, \bibinfo {author} {\bibfnamefont
  {A.}~\bibnamefont {Fry}}, \bibinfo {author} {\bibfnamefont {M.}~\bibnamefont
  {G{\"u}hr}}, \emph {et~al.},\ }\href {https://doi.org/10.1063/1.4926994}
  {\bibfield  {journal} {\bibinfo  {journal} {Rev. Sci. Instrum}\ }\textbf
  {\bibinfo {volume} {86}},\ \bibinfo {pages} {073702} (\bibinfo {year}
  {2015})}\BibitemShut {NoStop}%
\bibitem [{\citenamefont {Fischer}\ \emph {et~al.}(2016)\citenamefont
  {Fischer}, \citenamefont {Wilson}, \citenamefont {Robles},\ and\
  \citenamefont {Warren}}]{fischer2016invited}%
  \BibitemOpen
  \bibfield  {author} {\bibinfo {author} {\bibfnamefont {M.~C.}\ \bibnamefont
  {Fischer}}, \bibinfo {author} {\bibfnamefont {J.~W.}\ \bibnamefont {Wilson}},
  \bibinfo {author} {\bibfnamefont {F.~E.}\ \bibnamefont {Robles}},\ and\
  \bibinfo {author} {\bibfnamefont {W.~S.}\ \bibnamefont {Warren}},\ }\href
  {https://doi.org/10.1063/1.4943211} {\bibfield  {journal} {\bibinfo
  {journal} {Rev. Sci. Instrum}\ }\textbf {\bibinfo {volume} {87}},\ \bibinfo
  {pages} {031101} (\bibinfo {year} {2016})}\BibitemShut {NoStop}%
\bibitem [{\citenamefont {Buzzi}\ \emph {et~al.}(2018)\citenamefont {Buzzi},
  \citenamefont {F{\"o}rst}, \citenamefont {Mankowsky},\ and\ \citenamefont
  {Cavalleri}}]{buzzi2018probing}%
  \BibitemOpen
  \bibfield  {author} {\bibinfo {author} {\bibfnamefont {M.}~\bibnamefont
  {Buzzi}}, \bibinfo {author} {\bibfnamefont {M.}~\bibnamefont {F{\"o}rst}},
  \bibinfo {author} {\bibfnamefont {R.}~\bibnamefont {Mankowsky}},\ and\
  \bibinfo {author} {\bibfnamefont {A.}~\bibnamefont {Cavalleri}},\ }\href
  {https://www.nature.com/articles/s41578-018-0024-9} {\bibfield  {journal}
  {\bibinfo  {journal} {Nature Rev. Mater}\ }\textbf {\bibinfo {volume} {3}},\
  \bibinfo {pages} {299} (\bibinfo {year} {2018})}\BibitemShut {NoStop}%
\bibitem [{\citenamefont {Rizzo}\ \emph {et~al.}(2022)\citenamefont {Rizzo},
  \citenamefont {Libbi}, \citenamefont {Tacchino}, \citenamefont {Ollitrault},
  \citenamefont {Marzari},\ and\ \citenamefont {Tavernelli}}]{rizzo2022one}%
  \BibitemOpen
  \bibfield  {author} {\bibinfo {author} {\bibfnamefont {J.}~\bibnamefont
  {Rizzo}}, \bibinfo {author} {\bibfnamefont {F.}~\bibnamefont {Libbi}},
  \bibinfo {author} {\bibfnamefont {F.}~\bibnamefont {Tacchino}}, \bibinfo
  {author} {\bibfnamefont {P.~J.}\ \bibnamefont {Ollitrault}}, \bibinfo
  {author} {\bibfnamefont {N.}~\bibnamefont {Marzari}},\ and\ \bibinfo {author}
  {\bibfnamefont {I.}~\bibnamefont {Tavernelli}},\ }\href
  {https://journals.aps.org/prresearch/abstract/10.1103/PhysRevResearch.4.043011}
  {\bibfield  {journal} {\bibinfo  {journal} {Phys. Rev. Research}\ }\textbf
  {\bibinfo {volume} {4}},\ \bibinfo {pages} {043011} (\bibinfo {year}
  {2022})}\BibitemShut {NoStop}%
\bibitem [{\citenamefont {Jamet}\ \emph {et~al.}(2022)\citenamefont {Jamet},
  \citenamefont {Agarwal},\ and\ \citenamefont {Rungger}}]{jamet2022quantum}%
  \BibitemOpen
  \bibfield  {author} {\bibinfo {author} {\bibfnamefont {F.}~\bibnamefont
  {Jamet}}, \bibinfo {author} {\bibfnamefont {A.}~\bibnamefont {Agarwal}},\
  and\ \bibinfo {author} {\bibfnamefont {I.}~\bibnamefont {Rungger}},\ }\href
  {https://arxiv.org/abs/2205.00094} {\bibfield  {journal} {\bibinfo  {journal}
  {arXiv:2205.00094}\ } (\bibinfo {year} {2022})}\BibitemShut {NoStop}%
\bibitem [{\citenamefont {Motta}\ \emph
  {et~al.}(2023{\natexlab{b}})\citenamefont {Motta}, \citenamefont {Jones},
  \citenamefont {Rice}, \citenamefont {Gujarati}, \citenamefont {Sakuma},
  \citenamefont {Liepuoniute}, \citenamefont {Garcia},\ and\ \citenamefont
  {Ohnishi}}]{motta2023quantum}%
  \BibitemOpen
  \bibfield  {author} {\bibinfo {author} {\bibfnamefont {M.}~\bibnamefont
  {Motta}}, \bibinfo {author} {\bibfnamefont {G.~O.}\ \bibnamefont {Jones}},
  \bibinfo {author} {\bibfnamefont {J.~E.}\ \bibnamefont {Rice}}, \bibinfo
  {author} {\bibfnamefont {T.~P.}\ \bibnamefont {Gujarati}}, \bibinfo {author}
  {\bibfnamefont {R.}~\bibnamefont {Sakuma}}, \bibinfo {author} {\bibfnamefont
  {I.}~\bibnamefont {Liepuoniute}}, \bibinfo {author} {\bibfnamefont {J.~M.}\
  \bibnamefont {Garcia}},\ and\ \bibinfo {author} {\bibfnamefont
  {Y.}~\bibnamefont {Ohnishi}},\ }\href
  {https://pubs.rsc.org/en/content/articlehtml/2023/sc/d2sc06019a} {\bibfield
  {journal} {\bibinfo  {journal} {Chem. Sci}\ }\textbf {\bibinfo {volume}
  {14}},\ \bibinfo {pages} {2915} (\bibinfo {year}
  {2023}{\natexlab{b}})}\BibitemShut {NoStop}%
\bibitem [{\citenamefont {McCullough~Jr}\ and\ \citenamefont
  {Wyatt}(1969)}]{mccullough1969quantum}%
  \BibitemOpen
  \bibfield  {author} {\bibinfo {author} {\bibfnamefont {E.~A.}\ \bibnamefont
  {McCullough~Jr}}\ and\ \bibinfo {author} {\bibfnamefont {R.~E.}\ \bibnamefont
  {Wyatt}},\ }\href {https://doi.org/10.1063/1.1672133} {\bibfield  {journal}
  {\bibinfo  {journal} {J. Chem. Phys}\ }\textbf {\bibinfo {volume} {51}},\
  \bibinfo {pages} {1253} (\bibinfo {year} {1969})}\BibitemShut {NoStop}%
\bibitem [{\citenamefont {Kosloff}(1988)}]{kosloff1988time}%
  \BibitemOpen
  \bibfield  {author} {\bibinfo {author} {\bibfnamefont {R.}~\bibnamefont
  {Kosloff}},\ }\href {https://pubs.acs.org/doi/10.1021/j100319a003} {\bibfield
   {journal} {\bibinfo  {journal} {J. Phys. Chem}\ }\textbf {\bibinfo {volume}
  {92}},\ \bibinfo {pages} {2087} (\bibinfo {year} {1988})}\BibitemShut
  {NoStop}%
\bibitem [{\citenamefont {Huber}\ and\ \citenamefont
  {Klamroth}(2011)}]{huber2011explicitly}%
  \BibitemOpen
  \bibfield  {author} {\bibinfo {author} {\bibfnamefont {C.}~\bibnamefont
  {Huber}}\ and\ \bibinfo {author} {\bibfnamefont {T.}~\bibnamefont
  {Klamroth}},\ }\href {https://doi.org/10.1063/1.3530807} {\bibfield
  {journal} {\bibinfo  {journal} {J. Chem. Phys}\ }\textbf {\bibinfo {volume}
  {134}} (\bibinfo {year} {2011})}\BibitemShut {NoStop}%
\bibitem [{\citenamefont {Kristiansen}\ \emph {et~al.}(2020)\citenamefont
  {Kristiansen}, \citenamefont {Sch{\o}yen}, \citenamefont {Kvaal},\ and\
  \citenamefont {Pedersen}}]{kristiansen2020numerical}%
  \BibitemOpen
  \bibfield  {author} {\bibinfo {author} {\bibfnamefont {H.~E.}\ \bibnamefont
  {Kristiansen}}, \bibinfo {author} {\bibfnamefont {{\O}.~S.}\ \bibnamefont
  {Sch{\o}yen}}, \bibinfo {author} {\bibfnamefont {S.}~\bibnamefont {Kvaal}},\
  and\ \bibinfo {author} {\bibfnamefont {T.~B.}\ \bibnamefont {Pedersen}},\
  }\href {https://doi.org/10.1063/1.5142276} {\bibfield  {journal} {\bibinfo
  {journal} {J. Chem. Phys}\ }\textbf {\bibinfo {volume} {152}} (\bibinfo
  {year} {2020})}\BibitemShut {NoStop}%
\bibitem [{\citenamefont {Berry}\ \emph {et~al.}(2007)\citenamefont {Berry},
  \citenamefont {Ahokas}, \citenamefont {Cleve},\ and\ \citenamefont
  {Sanders}}]{berry2007efficient}%
  \BibitemOpen
  \bibfield  {author} {\bibinfo {author} {\bibfnamefont {D.~W.}\ \bibnamefont
  {Berry}}, \bibinfo {author} {\bibfnamefont {G.}~\bibnamefont {Ahokas}},
  \bibinfo {author} {\bibfnamefont {R.}~\bibnamefont {Cleve}},\ and\ \bibinfo
  {author} {\bibfnamefont {B.~C.}\ \bibnamefont {Sanders}},\ }\href
  {https://arxiv.org/abs/quant-ph/0508139} {\bibfield  {journal} {\bibinfo
  {journal} {Comm. Math. Phys}\ }\textbf {\bibinfo {volume} {270}},\ \bibinfo
  {pages} {359} (\bibinfo {year} {2007})}\BibitemShut {NoStop}%
\bibitem [{\citenamefont {Childs}\ and\ \citenamefont
  {Kothari}(2009)}]{childs2009limitations}%
  \BibitemOpen
  \bibfield  {author} {\bibinfo {author} {\bibfnamefont {A.~M.}\ \bibnamefont
  {Childs}}\ and\ \bibinfo {author} {\bibfnamefont {R.}~\bibnamefont
  {Kothari}},\ }\href {https://dl.acm.org/doi/10.5555/2011373.2011380}
  {\bibfield  {journal} {\bibinfo  {journal} {Quant. Info. Comput}\ }\textbf
  {\bibinfo {volume} {10}} (\bibinfo {year} {2009})}\BibitemShut {NoStop}%
\bibitem [{\citenamefont {Gu}\ \emph {et~al.}(2021)\citenamefont {Gu},
  \citenamefont {Somma},\ and\ \citenamefont
  {{\c{S}}ahino{\u{g}}lu}}]{gu2021fast}%
  \BibitemOpen
  \bibfield  {author} {\bibinfo {author} {\bibfnamefont {S.}~\bibnamefont
  {Gu}}, \bibinfo {author} {\bibfnamefont {R.~D.}\ \bibnamefont {Somma}},\ and\
  \bibinfo {author} {\bibfnamefont {B.}~\bibnamefont {{\c{S}}ahino{\u{g}}lu}},\
  }\href {https://quantum-journal.org/papers/q-2021-11-15-577/} {\bibfield
  {journal} {\bibinfo  {journal} {Quantum}\ }\textbf {\bibinfo {volume} {5}},\
  \bibinfo {pages} {577} (\bibinfo {year} {2021})}\BibitemShut {NoStop}%
\bibitem [{\citenamefont {Atia}\ and\ \citenamefont
  {Aharonov}(2017)}]{atia2017fast}%
  \BibitemOpen
  \bibfield  {author} {\bibinfo {author} {\bibfnamefont {Y.}~\bibnamefont
  {Atia}}\ and\ \bibinfo {author} {\bibfnamefont {D.}~\bibnamefont
  {Aharonov}},\ }\href {https://www.nature.com/articles/s41467-017-01637-7}
  {\bibfield  {journal} {\bibinfo  {journal} {Nat. Commun}\ }\textbf {\bibinfo
  {volume} {8}},\ \bibinfo {pages} {1572} (\bibinfo {year} {2017})}\BibitemShut
  {NoStop}%
\bibitem [{\citenamefont {Cirstoiu}\ \emph {et~al.}(2020)\citenamefont
  {Cirstoiu}, \citenamefont {Holmes}, \citenamefont {Iosue}, \citenamefont
  {Cincio}, \citenamefont {Coles},\ and\ \citenamefont
  {Sornborger}}]{cirstoiu2020variational}%
  \BibitemOpen
  \bibfield  {author} {\bibinfo {author} {\bibfnamefont {C.}~\bibnamefont
  {Cirstoiu}}, \bibinfo {author} {\bibfnamefont {Z.}~\bibnamefont {Holmes}},
  \bibinfo {author} {\bibfnamefont {J.}~\bibnamefont {Iosue}}, \bibinfo
  {author} {\bibfnamefont {L.}~\bibnamefont {Cincio}}, \bibinfo {author}
  {\bibfnamefont {P.~J.}\ \bibnamefont {Coles}},\ and\ \bibinfo {author}
  {\bibfnamefont {A.}~\bibnamefont {Sornborger}},\ }\href
  {https://www.nature.com/articles/s41534-020-00302-0} {\bibfield  {journal}
  {\bibinfo  {journal} {npj Quantum Inf}\ }\textbf {\bibinfo {volume} {6}},\
  \bibinfo {pages} {82} (\bibinfo {year} {2020})}\BibitemShut {NoStop}%
\bibitem [{\citenamefont {Commeau}\ \emph {et~al.}(2020)\citenamefont
  {Commeau}, \citenamefont {Cerezo}, \citenamefont {Holmes}, \citenamefont
  {Cincio}, \citenamefont {Coles},\ and\ \citenamefont
  {Sornborger}}]{commeau2020variational}%
  \BibitemOpen
  \bibfield  {author} {\bibinfo {author} {\bibfnamefont {B.}~\bibnamefont
  {Commeau}}, \bibinfo {author} {\bibfnamefont {M.}~\bibnamefont {Cerezo}},
  \bibinfo {author} {\bibfnamefont {Z.}~\bibnamefont {Holmes}}, \bibinfo
  {author} {\bibfnamefont {L.}~\bibnamefont {Cincio}}, \bibinfo {author}
  {\bibfnamefont {P.~J.}\ \bibnamefont {Coles}},\ and\ \bibinfo {author}
  {\bibfnamefont {A.}~\bibnamefont {Sornborger}},\ }\href
  {https://arxiv.org/abs/2009.02559} {\bibfield  {journal} {\bibinfo  {journal}
  {arXiv:2009.02559}\ } (\bibinfo {year} {2020})}\BibitemShut {NoStop}%
\bibitem [{\citenamefont {Cortes}\ \emph {et~al.}(2022)\citenamefont {Cortes},
  \citenamefont {DePrince~III},\ and\ \citenamefont {Gray}}]{cortes2022fast}%
  \BibitemOpen
  \bibfield  {author} {\bibinfo {author} {\bibfnamefont {C.~L.}\ \bibnamefont
  {Cortes}}, \bibinfo {author} {\bibfnamefont {A.~E.}\ \bibnamefont
  {DePrince~III}},\ and\ \bibinfo {author} {\bibfnamefont {S.~K.}\ \bibnamefont
  {Gray}},\ }\href
  {https://journals.aps.org/pra/abstract/10.1103/PhysRevA.106.042409}
  {\bibfield  {journal} {\bibinfo  {journal} {Phys. Rev. A}\ }\textbf {\bibinfo
  {volume} {106}},\ \bibinfo {pages} {042409} (\bibinfo {year}
  {2022})}\BibitemShut {NoStop}%
\bibitem [{\citenamefont {Heya}\ \emph {et~al.}(2019)\citenamefont {Heya},
  \citenamefont {Nakanishi}, \citenamefont {Mitarai},\ and\ \citenamefont
  {Fujii}}]{heya2019subspace}%
  \BibitemOpen
  \bibfield  {author} {\bibinfo {author} {\bibfnamefont {K.}~\bibnamefont
  {Heya}}, \bibinfo {author} {\bibfnamefont {K.~M.}\ \bibnamefont {Nakanishi}},
  \bibinfo {author} {\bibfnamefont {K.}~\bibnamefont {Mitarai}},\ and\ \bibinfo
  {author} {\bibfnamefont {K.}~\bibnamefont {Fujii}},\ }\href
  {https://arxiv.org/abs/1904.08566} {\bibfield  {journal} {\bibinfo  {journal}
  {arXiv:1904.08566}\ } (\bibinfo {year} {2019})}\BibitemShut {NoStop}%
\bibitem [{\citenamefont {Gibbs}\ \emph {et~al.}(2022)\citenamefont {Gibbs},
  \citenamefont {Gili}, \citenamefont {Holmes}, \citenamefont {Commeau},
  \citenamefont {Arrasmith}, \citenamefont {Cincio}, \citenamefont {Coles},\
  and\ \citenamefont {Sornborger}}]{gibbs2021long}%
  \BibitemOpen
  \bibfield  {author} {\bibinfo {author} {\bibfnamefont {J.}~\bibnamefont
  {Gibbs}}, \bibinfo {author} {\bibfnamefont {K.}~\bibnamefont {Gili}},
  \bibinfo {author} {\bibfnamefont {Z.}~\bibnamefont {Holmes}}, \bibinfo
  {author} {\bibfnamefont {B.}~\bibnamefont {Commeau}}, \bibinfo {author}
  {\bibfnamefont {A.}~\bibnamefont {Arrasmith}}, \bibinfo {author}
  {\bibfnamefont {L.}~\bibnamefont {Cincio}}, \bibinfo {author} {\bibfnamefont
  {P.~J.}\ \bibnamefont {Coles}},\ and\ \bibinfo {author} {\bibfnamefont
  {A.}~\bibnamefont {Sornborger}},\ }\href
  {https://www.nature.com/articles/s41534-022-00625-0} {\bibfield  {journal}
  {\bibinfo  {journal} {npj Quantum Inf}\ }\textbf {\bibinfo {volume} {8}},\
  \bibinfo {pages} {135} (\bibinfo {year} {2022})}\BibitemShut {NoStop}%
\bibitem [{\citenamefont {Gao}\ \emph {et~al.}(2021{\natexlab{b}})\citenamefont
  {Gao}, \citenamefont {Jones}, \citenamefont {Motta}, \citenamefont
  {Sugawara}, \citenamefont {Watanabe}, \citenamefont {Kobayashi},
  \citenamefont {Watanabe}, \citenamefont {Ohnishi}, \citenamefont {Nakamura},\
  and\ \citenamefont {Yamamoto}}]{gao2021applications}%
  \BibitemOpen
  \bibfield  {author} {\bibinfo {author} {\bibfnamefont {Q.}~\bibnamefont
  {Gao}}, \bibinfo {author} {\bibfnamefont {G.~O.}\ \bibnamefont {Jones}},
  \bibinfo {author} {\bibfnamefont {M.}~\bibnamefont {Motta}}, \bibinfo
  {author} {\bibfnamefont {M.}~\bibnamefont {Sugawara}}, \bibinfo {author}
  {\bibfnamefont {H.~C.}\ \bibnamefont {Watanabe}}, \bibinfo {author}
  {\bibfnamefont {T.}~\bibnamefont {Kobayashi}}, \bibinfo {author}
  {\bibfnamefont {E.}~\bibnamefont {Watanabe}}, \bibinfo {author}
  {\bibfnamefont {Y.}~\bibnamefont {Ohnishi}}, \bibinfo {author} {\bibfnamefont
  {H.}~\bibnamefont {Nakamura}},\ and\ \bibinfo {author} {\bibfnamefont
  {N.}~\bibnamefont {Yamamoto}},\ }\href
  {https://www.nature.com/articles/s41524-021-00540-6} {\bibfield  {journal}
  {\bibinfo  {journal} {npj Comput. Mater}\ }\textbf {\bibinfo {volume} {7}},\
  \bibinfo {pages} {70} (\bibinfo {year} {2021}{\natexlab{b}})}\BibitemShut
  {NoStop}%
\bibitem [{\citenamefont {Bravyi}\ \emph {et~al.}(2017)\citenamefont {Bravyi},
  \citenamefont {Gambetta}, \citenamefont {Mezzacapo},\ and\ \citenamefont
  {Temme}}]{bravyi2017tapering}%
  \BibitemOpen
  \bibfield  {author} {\bibinfo {author} {\bibfnamefont {S.}~\bibnamefont
  {Bravyi}}, \bibinfo {author} {\bibfnamefont {J.~M.}\ \bibnamefont
  {Gambetta}}, \bibinfo {author} {\bibfnamefont {A.}~\bibnamefont
  {Mezzacapo}},\ and\ \bibinfo {author} {\bibfnamefont {K.}~\bibnamefont
  {Temme}},\ }\href {https://arxiv.org/abs/1701.08213} {\bibfield  {journal}
  {\bibinfo  {journal} {arXiv:1701.08213}\ } (\bibinfo {year}
  {2017})}\BibitemShut {NoStop}%
\bibitem [{\citenamefont {Setia}\ \emph {et~al.}(2020)\citenamefont {Setia},
  \citenamefont {Chen}, \citenamefont {Rice}, \citenamefont {Mezzacapo},
  \citenamefont {Pistoia},\ and\ \citenamefont
  {Whitfield}}]{setia2020reducing}%
  \BibitemOpen
  \bibfield  {author} {\bibinfo {author} {\bibfnamefont {K.}~\bibnamefont
  {Setia}}, \bibinfo {author} {\bibfnamefont {R.}~\bibnamefont {Chen}},
  \bibinfo {author} {\bibfnamefont {J.~E.}\ \bibnamefont {Rice}}, \bibinfo
  {author} {\bibfnamefont {A.}~\bibnamefont {Mezzacapo}}, \bibinfo {author}
  {\bibfnamefont {M.}~\bibnamefont {Pistoia}},\ and\ \bibinfo {author}
  {\bibfnamefont {J.~D.}\ \bibnamefont {Whitfield}},\ }\href
  {https://pubs.acs.org/doi/10.1021/acs.jctc.0c00113} {\bibfield  {journal}
  {\bibinfo  {journal} {J. Chem. Theory Comput}\ }\textbf {\bibinfo {volume}
  {16}},\ \bibinfo {pages} {6091} (\bibinfo {year} {2020})}\BibitemShut
  {NoStop}%
\bibitem [{\citenamefont {Huang}\ \emph {et~al.}(2023)\citenamefont {Huang},
  \citenamefont {Sheng}, \citenamefont {Govoni},\ and\ \citenamefont
  {Galli}}]{huang2023quantum}%
  \BibitemOpen
  \bibfield  {author} {\bibinfo {author} {\bibfnamefont {B.}~\bibnamefont
  {Huang}}, \bibinfo {author} {\bibfnamefont {N.}~\bibnamefont {Sheng}},
  \bibinfo {author} {\bibfnamefont {M.}~\bibnamefont {Govoni}},\ and\ \bibinfo
  {author} {\bibfnamefont {G.}~\bibnamefont {Galli}},\ }\href
  {https://doi.org/10.1021/acs.jctc.2c01119} {\bibfield  {journal} {\bibinfo
  {journal} {J. Chem. Theory Comput}\ }\textbf {\bibinfo {volume} {19}},\
  \bibinfo {pages} {1487} (\bibinfo {year} {2023})}\BibitemShut {NoStop}%
\bibitem [{\citenamefont {Castellanos}\ \emph {et~al.}(2023)\citenamefont
  {Castellanos}, \citenamefont {Motta},\ and\ \citenamefont
  {Rice}}]{castellanos2023quantum}%
  \BibitemOpen
  \bibfield  {author} {\bibinfo {author} {\bibfnamefont {M.~A.}\ \bibnamefont
  {Castellanos}}, \bibinfo {author} {\bibfnamefont {M.}~\bibnamefont {Motta}},\
  and\ \bibinfo {author} {\bibfnamefont {J.~E.}\ \bibnamefont {Rice}},\ }\href
  {https://doi.org/10.1080/00268976.2023.2282736} {\bibfield  {journal}
  {\bibinfo  {journal} {Mol. Phys}\ ,\ \bibinfo {pages} {e2282736}} (\bibinfo
  {year} {2023})}\BibitemShut {NoStop}%
\bibitem [{\citenamefont {Eddins}\ \emph {et~al.}(2022)\citenamefont {Eddins},
  \citenamefont {Motta}, \citenamefont {Gujarati}, \citenamefont {Bravyi},
  \citenamefont {Mezzacapo}, \citenamefont {Hadfield},\ and\ \citenamefont
  {Sheldon}}]{eddins2021doubling}%
  \BibitemOpen
  \bibfield  {author} {\bibinfo {author} {\bibfnamefont {A.}~\bibnamefont
  {Eddins}}, \bibinfo {author} {\bibfnamefont {M.}~\bibnamefont {Motta}},
  \bibinfo {author} {\bibfnamefont {T.~P.}\ \bibnamefont {Gujarati}}, \bibinfo
  {author} {\bibfnamefont {S.}~\bibnamefont {Bravyi}}, \bibinfo {author}
  {\bibfnamefont {A.}~\bibnamefont {Mezzacapo}}, \bibinfo {author}
  {\bibfnamefont {C.}~\bibnamefont {Hadfield}},\ and\ \bibinfo {author}
  {\bibfnamefont {S.}~\bibnamefont {Sheldon}},\ }\href
  {https://journals.aps.org/prxquantum/abstract/10.1103/PRXQuantum.3.010309}
  {\bibfield  {journal} {\bibinfo  {journal} {PRX Quantum}\ }\textbf {\bibinfo
  {volume} {3}},\ \bibinfo {pages} {010309} (\bibinfo {year}
  {2022})}\BibitemShut {NoStop}%
\bibitem [{\citenamefont {Dhawan}\ \emph {et~al.}(2023)\citenamefont {Dhawan},
  \citenamefont {Zgid},\ and\ \citenamefont {Motta}}]{dhawan2023quantum}%
  \BibitemOpen
  \bibfield  {author} {\bibinfo {author} {\bibfnamefont {D.}~\bibnamefont
  {Dhawan}}, \bibinfo {author} {\bibfnamefont {D.}~\bibnamefont {Zgid}},\ and\
  \bibinfo {author} {\bibfnamefont {M.}~\bibnamefont {Motta}},\ }\href
  {https://arxiv.org/abs/2309.09914} {\bibfield  {journal} {\bibinfo  {journal}
  {arXiv:2309.09914}\ } (\bibinfo {year} {2023})}\BibitemShut {NoStop}%
\bibitem [{\citenamefont {Huang}\ \emph {et~al.}(2022)\citenamefont {Huang},
  \citenamefont {Govoni},\ and\ \citenamefont {Galli}}]{Huang2022simulating}%
  \BibitemOpen
  \bibfield  {author} {\bibinfo {author} {\bibfnamefont {B.}~\bibnamefont
  {Huang}}, \bibinfo {author} {\bibfnamefont {M.}~\bibnamefont {Govoni}},\ and\
  \bibinfo {author} {\bibfnamefont {G.}~\bibnamefont {Galli}},\ }\href
  {https://doi.org/10.1103/PRXQuantum.3.010339} {\bibfield  {journal} {\bibinfo
   {journal} {PRX Quantum}\ }\textbf {\bibinfo {volume} {3}},\ \bibinfo {pages}
  {010339} (\bibinfo {year} {2022})}\BibitemShut {NoStop}%
\bibitem [{\citenamefont {Khan}\ \emph {et~al.}(2023)\citenamefont {Khan},
  \citenamefont {Tudorovskaya}, \citenamefont {Kirsopp}, \citenamefont
  {Muñoz~Ramo}, \citenamefont {Warrier}, \citenamefont {Papanastasiou},\ and\
  \citenamefont {Singh}}]{Khan2022sim}%
  \BibitemOpen
  \bibfield  {author} {\bibinfo {author} {\bibfnamefont {I.~T.}\ \bibnamefont
  {Khan}}, \bibinfo {author} {\bibfnamefont {M.}~\bibnamefont {Tudorovskaya}},
  \bibinfo {author} {\bibfnamefont {J.~J.~M.}\ \bibnamefont {Kirsopp}},
  \bibinfo {author} {\bibfnamefont {D.}~\bibnamefont {Muñoz~Ramo}}, \bibinfo
  {author} {\bibfnamefont {P.}~\bibnamefont {Warrier}}, \bibinfo {author}
  {\bibfnamefont {D.~K.}\ \bibnamefont {Papanastasiou}},\ and\ \bibinfo
  {author} {\bibfnamefont {R.}~\bibnamefont {Singh}},\ }\href
  {https://doi.org/10.1063/5.0144680} {\bibfield  {journal} {\bibinfo
  {journal} {J. Chem. Phys}\ }\textbf {\bibinfo {volume} {158}},\ \bibinfo
  {pages} {214114} (\bibinfo {year} {2023})}\BibitemShut {NoStop}%
\bibitem [{\citenamefont {Lee}\ \emph {et~al.}(2023)\citenamefont {Lee},
  \citenamefont {Lee},\ and\ \citenamefont {Huh}}]{lee2023sampling}%
  \BibitemOpen
  \bibfield  {author} {\bibinfo {author} {\bibfnamefont {G.}~\bibnamefont
  {Lee}}, \bibinfo {author} {\bibfnamefont {D.}~\bibnamefont {Lee}},\ and\
  \bibinfo {author} {\bibfnamefont {J.}~\bibnamefont {Huh}},\ }\href
  {https://arxiv.org/abs/2307.16279} {\bibfield  {journal} {\bibinfo  {journal}
  {arXiv:2307.16279}\ } (\bibinfo {year} {2023})}\BibitemShut {NoStop}%
\bibitem [{\citenamefont {Bauer}\ and\ \citenamefont
  {Fike}(1960)}]{bauer1960norms}%
  \BibitemOpen
  \bibfield  {author} {\bibinfo {author} {\bibfnamefont {F.~L.}\ \bibnamefont
  {Bauer}}\ and\ \bibinfo {author} {\bibfnamefont {C.~T.}\ \bibnamefont
  {Fike}},\ }\href {https://link.springer.com/article/10.1007/BF01386217}
  {\bibfield  {journal} {\bibinfo  {journal} {Num. Math}\ }\textbf {\bibinfo
  {volume} {2}},\ \bibinfo {pages} {137} (\bibinfo {year} {1960})}\BibitemShut
  {NoStop}%
\bibitem [{\citenamefont {Mathias}\ and\ \citenamefont
  {Li}(2004)}]{mathias2004definite}%
  \BibitemOpen
  \bibfield  {author} {\bibinfo {author} {\bibfnamefont {R.}~\bibnamefont
  {Mathias}}\ and\ \bibinfo {author} {\bibfnamefont {C.-K.}\ \bibnamefont
  {Li}},\ }\href {https://api.semanticscholar.org/CorpusID:4160336} {\bibinfo
  {title} {The definite generalized eigenvalue problem : A new perturbation
  theory}} (\bibinfo {year} {2004})\BibitemShut {NoStop}%
\bibitem [{\citenamefont {Ding}\ and\ \citenamefont
  {Lin}(2023{\natexlab{a}})}]{ding2023even}%
  \BibitemOpen
  \bibfield  {author} {\bibinfo {author} {\bibfnamefont {Z.}~\bibnamefont
  {Ding}}\ and\ \bibinfo {author} {\bibfnamefont {L.}~\bibnamefont {Lin}},\
  }\href
  {https://journals.aps.org/prxquantum/abstract/10.1103/PRXQuantum.4.020331}
  {\bibfield  {journal} {\bibinfo  {journal} {PRX Quantum}\ }\textbf {\bibinfo
  {volume} {4}},\ \bibinfo {pages} {020331} (\bibinfo {year}
  {2023}{\natexlab{a}})}\BibitemShut {NoStop}%
\bibitem [{\citenamefont {Ding}\ and\ \citenamefont
  {Lin}(2023{\natexlab{b}})}]{ding2023simultaneous}%
  \BibitemOpen
  \bibfield  {author} {\bibinfo {author} {\bibfnamefont {Z.}~\bibnamefont
  {Ding}}\ and\ \bibinfo {author} {\bibfnamefont {L.}~\bibnamefont {Lin}},\
  }\href {https://quantum-journal.org/papers/q-2023-10-11-1136/} {\bibfield
  {journal} {\bibinfo  {journal} {Quantum}\ }\textbf {\bibinfo {volume} {3}},\
  \bibinfo {pages} {1136} (\bibinfo {year} {2023}{\natexlab{b}})}\BibitemShut
  {NoStop}%
\bibitem [{\citenamefont {Shen}\ \emph
  {et~al.}(2023{\natexlab{b}})\citenamefont {Shen}, \citenamefont {Camps},
  \citenamefont {Darbha}, \citenamefont {Szasz}, \citenamefont {Klymko},
  \citenamefont {Tubman}, \citenamefont {Van~Beeumen} \emph
  {et~al.}}]{shen2023estimating}%
  \BibitemOpen
  \bibfield  {author} {\bibinfo {author} {\bibfnamefont {Y.}~\bibnamefont
  {Shen}}, \bibinfo {author} {\bibfnamefont {D.}~\bibnamefont {Camps}},
  \bibinfo {author} {\bibfnamefont {S.}~\bibnamefont {Darbha}}, \bibinfo
  {author} {\bibfnamefont {A.}~\bibnamefont {Szasz}}, \bibinfo {author}
  {\bibfnamefont {K.}~\bibnamefont {Klymko}}, \bibinfo {author} {\bibfnamefont
  {N.~M.}\ \bibnamefont {Tubman}}, \bibinfo {author} {\bibfnamefont
  {R.}~\bibnamefont {Van~Beeumen}}, \emph {et~al.},\ }\href
  {https://arxiv.org/abs/2306.01858} {\bibfield  {journal} {\bibinfo  {journal}
  {arXiv:2306.01858}\ } (\bibinfo {year} {2023}{\natexlab{b}})}\BibitemShut
  {NoStop}%
\bibitem [{\citenamefont {Mezi{\'c}}\ and\ \citenamefont
  {Banaszuk}(2004)}]{mezic2004comparison}%
  \BibitemOpen
  \bibfield  {author} {\bibinfo {author} {\bibfnamefont {I.}~\bibnamefont
  {Mezi{\'c}}}\ and\ \bibinfo {author} {\bibfnamefont {A.}~\bibnamefont
  {Banaszuk}},\ }\href
  {https://www.sciencedirect.com/science/article/abs/pii/S0167278904002507}
  {\bibfield  {journal} {\bibinfo  {journal} {Phys. D: Nonlinear Phenom}\
  }\textbf {\bibinfo {volume} {197}},\ \bibinfo {pages} {101} (\bibinfo {year}
  {2004})}\BibitemShut {NoStop}%
\bibitem [{\citenamefont {Mezi{\'c}}(2005)}]{mezic2005spectral}%
  \BibitemOpen
  \bibfield  {author} {\bibinfo {author} {\bibfnamefont {I.}~\bibnamefont
  {Mezi{\'c}}},\ }\href
  {https://link.springer.com/article/10.1007/s11071-005-2824-x} {\bibfield
  {journal} {\bibinfo  {journal} {Nonlinear Dyn}\ }\textbf {\bibinfo {volume}
  {41}},\ \bibinfo {pages} {309} (\bibinfo {year} {2005})}\BibitemShut
  {NoStop}%
\bibitem [{\citenamefont {McClean}\ \emph {et~al.}(2020)\citenamefont
  {McClean}, \citenamefont {Jiang}, \citenamefont {Rubin}, \citenamefont
  {Babbush},\ and\ \citenamefont {Neven}}]{mcclean2020decoding}%
  \BibitemOpen
  \bibfield  {author} {\bibinfo {author} {\bibfnamefont {J.~R.}\ \bibnamefont
  {McClean}}, \bibinfo {author} {\bibfnamefont {Z.}~\bibnamefont {Jiang}},
  \bibinfo {author} {\bibfnamefont {N.~C.}\ \bibnamefont {Rubin}}, \bibinfo
  {author} {\bibfnamefont {R.}~\bibnamefont {Babbush}},\ and\ \bibinfo {author}
  {\bibfnamefont {H.}~\bibnamefont {Neven}},\ }\href
  {https://www.nature.com/articles/s41467-020-14341-w} {\bibfield  {journal}
  {\bibinfo  {journal} {Nat. Commun}\ }\textbf {\bibinfo {volume} {11}},\
  \bibinfo {pages} {636} (\bibinfo {year} {2020})}\BibitemShut {NoStop}%
\bibitem [{\citenamefont {Xiong}\ \emph {et~al.}(2021)\citenamefont {Xiong},
  \citenamefont {Ng},\ and\ \citenamefont {Hanzo}}]{xiong2021quantum}%
  \BibitemOpen
  \bibfield  {author} {\bibinfo {author} {\bibfnamefont {Y.}~\bibnamefont
  {Xiong}}, \bibinfo {author} {\bibfnamefont {S.~X.}\ \bibnamefont {Ng}},\ and\
  \bibinfo {author} {\bibfnamefont {L.}~\bibnamefont {Hanzo}},\ }\href
  {https://ieeexplore.ieee.org/document/9638483/} {\bibfield  {journal}
  {\bibinfo  {journal} {IEEE Trans. Comm}\ }\textbf {\bibinfo {volume} {70}},\
  \bibinfo {pages} {1927} (\bibinfo {year} {2021})}\BibitemShut {NoStop}%
\bibitem [{\citenamefont {Ohkura}\ \emph {et~al.}(2023)\citenamefont {Ohkura},
  \citenamefont {Endo}, \citenamefont {Satah}, \citenamefont {Van~Meter},\ and\
  \citenamefont {Yoshioka}}]{ohkura2023leveraging}%
  \BibitemOpen
  \bibfield  {author} {\bibinfo {author} {\bibfnamefont {Y.}~\bibnamefont
  {Ohkura}}, \bibinfo {author} {\bibfnamefont {S.}~\bibnamefont {Endo}},
  \bibinfo {author} {\bibfnamefont {T.}~\bibnamefont {Satah}}, \bibinfo
  {author} {\bibfnamefont {R.}~\bibnamefont {Van~Meter}},\ and\ \bibinfo
  {author} {\bibfnamefont {N.}~\bibnamefont {Yoshioka}},\ }\href
  {https://arxiv.org/abs/2303.07660} {\bibfield  {journal} {\bibinfo  {journal}
  {arXiv:2303.07660}\ } (\bibinfo {year} {2023})}\BibitemShut {NoStop}%
\bibitem [{\citenamefont {Koczor}(2021)}]{koczor2021exponential}%
  \BibitemOpen
  \bibfield  {author} {\bibinfo {author} {\bibfnamefont {B.}~\bibnamefont
  {Koczor}},\ }\href
  {https://journals.aps.org/prx/abstract/10.1103/PhysRevX.11.031057} {\bibfield
   {journal} {\bibinfo  {journal} {Phys. Rev. X}\ }\textbf {\bibinfo {volume}
  {11}},\ \bibinfo {pages} {031057} (\bibinfo {year} {2021})}\BibitemShut
  {NoStop}%
\bibitem [{\citenamefont {Yang}\ \emph {et~al.}(2023)\citenamefont {Yang},
  \citenamefont {Yoshioka}, \citenamefont {Harada}, \citenamefont {Hakkaku},
  \citenamefont {Tokunaga}, \citenamefont {Hakoshima}, \citenamefont
  {Yamamoto},\ and\ \citenamefont {Endo}}]{yang2023dual}%
  \BibitemOpen
  \bibfield  {author} {\bibinfo {author} {\bibfnamefont {B.}~\bibnamefont
  {Yang}}, \bibinfo {author} {\bibfnamefont {N.}~\bibnamefont {Yoshioka}},
  \bibinfo {author} {\bibfnamefont {H.}~\bibnamefont {Harada}}, \bibinfo
  {author} {\bibfnamefont {S.}~\bibnamefont {Hakkaku}}, \bibinfo {author}
  {\bibfnamefont {Y.}~\bibnamefont {Tokunaga}}, \bibinfo {author}
  {\bibfnamefont {H.}~\bibnamefont {Hakoshima}}, \bibinfo {author}
  {\bibfnamefont {K.}~\bibnamefont {Yamamoto}},\ and\ \bibinfo {author}
  {\bibfnamefont {S.}~\bibnamefont {Endo}},\ }\href
  {https://arxiv.org/pdf/2309.14171.pdf} {\bibfield  {journal} {\bibinfo
  {journal} {arXiv:2309.14171}\ } (\bibinfo {year} {2023})}\BibitemShut
  {NoStop}%
\bibitem [{\citenamefont {Aharonov}\ \emph {et~al.}(2009)\citenamefont
  {Aharonov}, \citenamefont {Jones},\ and\ \citenamefont
  {Landau}}]{aharonov2009polynomial}%
  \BibitemOpen
  \bibfield  {author} {\bibinfo {author} {\bibfnamefont {D.}~\bibnamefont
  {Aharonov}}, \bibinfo {author} {\bibfnamefont {V.}~\bibnamefont {Jones}},\
  and\ \bibinfo {author} {\bibfnamefont {Z.}~\bibnamefont {Landau}},\ }\href
  {https://doi.org/https://doi.org/10.1007/s00453-008-9168-0} {\bibfield
  {journal} {\bibinfo  {journal} {Algorithmica}\ }\textbf {\bibinfo {volume}
  {55}},\ \bibinfo {pages} {395} (\bibinfo {year} {2009})}\BibitemShut
  {NoStop}%
\bibitem [{\citenamefont {Mazziotti}(1998)}]{mazziotti1998approximate}%
  \BibitemOpen
  \bibfield  {author} {\bibinfo {author} {\bibfnamefont {D.~A.}\ \bibnamefont
  {Mazziotti}},\ }\href
  {https://www.sciencedirect.com/science/article/abs/pii/S0009261498004709}
  {\bibfield  {journal} {\bibinfo  {journal} {Chem. Phys. Lett}\ }\textbf
  {\bibinfo {volume} {289}},\ \bibinfo {pages} {419} (\bibinfo {year}
  {1998})}\BibitemShut {NoStop}%
\bibitem [{\citenamefont {Choi}\ and\ \citenamefont
  {Izmaylov}(2023)}]{choi2023measurement}%
  \BibitemOpen
  \bibfield  {author} {\bibinfo {author} {\bibfnamefont {S.}~\bibnamefont
  {Choi}}\ and\ \bibinfo {author} {\bibfnamefont {A.~F.}\ \bibnamefont
  {Izmaylov}},\ }\href {https://doi.org/10.1021/acs.jctc.3c00218} {\bibfield
  {journal} {\bibinfo  {journal} {J. Chem. Theory Comput}\ } (\bibinfo {year}
  {2023})}\BibitemShut {NoStop}%
\bibitem [{\citenamefont {Wootters}\ and\ \citenamefont
  {Fields}(1989)}]{wootters1989optimal}%
  \BibitemOpen
  \bibfield  {author} {\bibinfo {author} {\bibfnamefont {W.~K.}\ \bibnamefont
  {Wootters}}\ and\ \bibinfo {author} {\bibfnamefont {B.~D.}\ \bibnamefont
  {Fields}},\ }\href
  {https://www.sciencedirect.com/science/article/abs/pii/0003491689903229}
  {\bibfield  {journal} {\bibinfo  {journal} {Ann. Phys}\ }\textbf {\bibinfo
  {volume} {191}},\ \bibinfo {pages} {363} (\bibinfo {year}
  {1989})}\BibitemShut {NoStop}%
\bibitem [{\citenamefont {Yen}\ \emph {et~al.}(2020)\citenamefont {Yen},
  \citenamefont {Verteletskyi},\ and\ \citenamefont
  {Izmaylov}}]{yen2020measuring}%
  \BibitemOpen
  \bibfield  {author} {\bibinfo {author} {\bibfnamefont {T.-C.}\ \bibnamefont
  {Yen}}, \bibinfo {author} {\bibfnamefont {V.}~\bibnamefont {Verteletskyi}},\
  and\ \bibinfo {author} {\bibfnamefont {A.~F.}\ \bibnamefont {Izmaylov}},\
  }\href {https://pubs.acs.org/doi/10.1021/acs.jctc.0c00008} {\bibfield
  {journal} {\bibinfo  {journal} {J. Chem. Theory Comput}\ }\textbf {\bibinfo
  {volume} {16}},\ \bibinfo {pages} {2400} (\bibinfo {year}
  {2020})}\BibitemShut {NoStop}%
\bibitem [{\citenamefont {Huang}\ \emph {et~al.}(2020)\citenamefont {Huang},
  \citenamefont {Kueng},\ and\ \citenamefont {Preskill}}]{huang2020predicting}%
  \BibitemOpen
  \bibfield  {author} {\bibinfo {author} {\bibfnamefont {H.-Y.}\ \bibnamefont
  {Huang}}, \bibinfo {author} {\bibfnamefont {R.}~\bibnamefont {Kueng}},\ and\
  \bibinfo {author} {\bibfnamefont {J.}~\bibnamefont {Preskill}},\ }\href
  {https://www.nature.com/articles/s41567-020-0932-7} {\bibfield  {journal}
  {\bibinfo  {journal} {Nat. Phys}\ }\textbf {\bibinfo {volume} {16}},\
  \bibinfo {pages} {1050} (\bibinfo {year} {2020})}\BibitemShut {NoStop}%
\bibitem [{\citenamefont {Zhao}\ \emph {et~al.}(2021)\citenamefont {Zhao},
  \citenamefont {Rubin},\ and\ \citenamefont {Miyake}}]{zhao2021fermionic}%
  \BibitemOpen
  \bibfield  {author} {\bibinfo {author} {\bibfnamefont {A.}~\bibnamefont
  {Zhao}}, \bibinfo {author} {\bibfnamefont {N.~C.}\ \bibnamefont {Rubin}},\
  and\ \bibinfo {author} {\bibfnamefont {A.}~\bibnamefont {Miyake}},\ }\href
  {https://journals.aps.org/prl/abstract/10.1103/PhysRevLett.127.110504}
  {\bibfield  {journal} {\bibinfo  {journal} {Phys. Rev. Lett}\ }\textbf
  {\bibinfo {volume} {127}},\ \bibinfo {pages} {110504} (\bibinfo {year}
  {2021})}\BibitemShut {NoStop}%
\bibitem [{\citenamefont {Huang}\ \emph {et~al.}(2021)\citenamefont {Huang},
  \citenamefont {Kueng},\ and\ \citenamefont {Preskill}}]{huang2021efficient}%
  \BibitemOpen
  \bibfield  {author} {\bibinfo {author} {\bibfnamefont {H.-Y.}\ \bibnamefont
  {Huang}}, \bibinfo {author} {\bibfnamefont {R.}~\bibnamefont {Kueng}},\ and\
  \bibinfo {author} {\bibfnamefont {J.}~\bibnamefont {Preskill}},\ }\href
  {https://journals.aps.org/prl/abstract/10.1103/PhysRevLett.127.030503}
  {\bibfield  {journal} {\bibinfo  {journal} {Phys. Rev. Lett}\ }\textbf
  {\bibinfo {volume} {127}},\ \bibinfo {pages} {030503} (\bibinfo {year}
  {2021})}\BibitemShut {NoStop}%
\bibitem [{\citenamefont {Crawford}\ \emph {et~al.}(2021)\citenamefont
  {Crawford}, \citenamefont {van Straaten}, \citenamefont {Wang}, \citenamefont
  {Parks}, \citenamefont {Campbell},\ and\ \citenamefont
  {Brierley}}]{crawford2021efficient}%
  \BibitemOpen
  \bibfield  {author} {\bibinfo {author} {\bibfnamefont {O.}~\bibnamefont
  {Crawford}}, \bibinfo {author} {\bibfnamefont {B.}~\bibnamefont {van
  Straaten}}, \bibinfo {author} {\bibfnamefont {D.}~\bibnamefont {Wang}},
  \bibinfo {author} {\bibfnamefont {T.}~\bibnamefont {Parks}}, \bibinfo
  {author} {\bibfnamefont {E.}~\bibnamefont {Campbell}},\ and\ \bibinfo
  {author} {\bibfnamefont {S.}~\bibnamefont {Brierley}},\ }\href
  {https://quantum-journal.org/papers/q-2021-01-20-385/} {\bibfield  {journal}
  {\bibinfo  {journal} {Quantum}\ }\textbf {\bibinfo {volume} {5}},\ \bibinfo
  {pages} {385} (\bibinfo {year} {2021})}\BibitemShut {NoStop}%
\bibitem [{\citenamefont {Yen}\ \emph {et~al.}(2023)\citenamefont {Yen},
  \citenamefont {Ganeshram},\ and\ \citenamefont
  {Izmaylov}}]{yen2023deterministic}%
  \BibitemOpen
  \bibfield  {author} {\bibinfo {author} {\bibfnamefont {T.-C.}\ \bibnamefont
  {Yen}}, \bibinfo {author} {\bibfnamefont {A.}~\bibnamefont {Ganeshram}},\
  and\ \bibinfo {author} {\bibfnamefont {A.~F.}\ \bibnamefont {Izmaylov}},\
  }\href {https://www.nature.com/articles/s41534-023-00683-y} {\bibfield
  {journal} {\bibinfo  {journal} {npj Quantum Inf}\ }\textbf {\bibinfo {volume}
  {9}},\ \bibinfo {pages} {14} (\bibinfo {year} {2023})}\BibitemShut {NoStop}%
\bibitem [{\citenamefont {Bonet-Monroig}\ \emph {et~al.}(2020)\citenamefont
  {Bonet-Monroig}, \citenamefont {Babbush},\ and\ \citenamefont
  {O’Brien}}]{bonet2020nearly}%
  \BibitemOpen
  \bibfield  {author} {\bibinfo {author} {\bibfnamefont {X.}~\bibnamefont
  {Bonet-Monroig}}, \bibinfo {author} {\bibfnamefont {R.}~\bibnamefont
  {Babbush}},\ and\ \bibinfo {author} {\bibfnamefont {T.~E.}\ \bibnamefont
  {O’Brien}},\ }\href
  {https://journals.aps.org/prx/abstract/10.1103/PhysRevX.10.031064} {\bibfield
   {journal} {\bibinfo  {journal} {Phys. Rev. X}\ }\textbf {\bibinfo {volume}
  {10}},\ \bibinfo {pages} {031064} (\bibinfo {year} {2020})}\BibitemShut
  {NoStop}%
\bibitem [{\citenamefont {Choi}\ \emph {et~al.}(2023)\citenamefont {Choi},
  \citenamefont {Loaiza},\ and\ \citenamefont {Izmaylov}}]{choi2023fluid}%
  \BibitemOpen
  \bibfield  {author} {\bibinfo {author} {\bibfnamefont {S.}~\bibnamefont
  {Choi}}, \bibinfo {author} {\bibfnamefont {I.}~\bibnamefont {Loaiza}},\ and\
  \bibinfo {author} {\bibfnamefont {A.~F.}\ \bibnamefont {Izmaylov}},\ }\href
  {https://quantum-journal.org/papers/q-2023-01-03-889/} {\bibfield  {journal}
  {\bibinfo  {journal} {Quantum}\ }\textbf {\bibinfo {volume} {7}},\ \bibinfo
  {pages} {889} (\bibinfo {year} {2023})}\BibitemShut {NoStop}%
\end{thebibliography}


\end{document}